\documentclass[usegraphicx,usenatbib,useAMS]{mn2e}
\usepackage{times}
\newcommand{\be}{\begin{equation}}
\newcommand{\ba}{\begin{eqnarray}}
\newcommand{\ee}{\end{equation}}
\newcommand{\ea}{\end{eqnarray}}  

\def\lesssim{\mathrel{\hbox{\rlap{\hbox{\lower4pt\hbox{$\sim$}}}\hbox{$<$}}}}
\def\gtrsim{\mathrel{\hbox{\rlap{\hbox{\lower4pt\hbox{$\sim$}}}\hbox{$>$}}}}

\def\gtsima{$\; \buildrel > \over \sim \;$}
\def\ltsima{$\; \buildrel < \over \sim \;$}
\def\gsim{\lower.5ex\hbox{\gtsima}}
\def\lsim{\lower.5ex\hbox{\ltsima}}
\def\simgt{\lower.5ex\hbox{\gtsima}}
\def\simlt{\lower.5ex\hbox{\ltsima}}
\def\simpr{\lower.5ex\hbox{\prosima}}

\def\ga{\gsim}

\def\simless{\mathbin{\lower 3pt\hbox
   {$\rlap{\raise 5pt\hbox{$\char'074$}}\mathchar''7218$}}}   
\def\simgreat{\mathbin{\lower 3pt\hbox
   {$\rlap{\raise 5pt\hbox{$\char'076$}}\mathchar''7218$}}}   

\begin{document}

\title[Cosmological Radiative Transfer Codes Comparison Project]
{Cosmological Radiative Transfer Codes Comparison Project I: The 
Static Density Field Tests 
}
\author[I. T. Iliev, et al.]{Ilian~T.~Iliev$^1$\thanks{e-mail: 
  iliev@cita.utoronto.ca}, Benedetta Ciardi$^2$, 
  Marcelo A. Alvarez$^3$,  Antonella Maselli$^2$,\newauthor   
Andrea Ferrara$^{4}$, Nickolay~Y.~Gnedin$^{5,6}$, 
Garrelt Mellema$^{7,8}$, Taishi Nakamoto$^9$, 
 \newauthor   Michael~L.~Norman$^{10}$, Alexei~O.~Razoumov$^{11}$,  
  Erik-Jan Rijkhorst$^8$, Jelle Ritzerveld$^8$,\newauthor  
  Paul~R.~Shapiro$^3$, Hajime Susa$^{12}$, Masayuki Umemura$^9$, 
  Daniel~J.~Whalen$^{10,13}$
\\
$^1$ Canadian Institute for Theoretical Astrophysics, University
  of Toronto, 60 St. George Street, Toronto, ON M5S 3H8, Canada\\
$^2$ Max-Planck-Institut f\"{u}r Astrophysik, 85741 Garching, Germany\\
$^3$ Department of Astronomy, University of Texas, Austin, TX 78712-1083, 
U.S.A.\\
$^{4}$ SISSA/International School for Advanced Studies, Via Beirut 4, 
34014 Trieste, Italy\\
$^5$ Fermilab, MS209, P.O. 500, Batavia, IL 60510, U.S.A.\\
$^6$ Department of Astronomy \& Astrophysics, The University of Chicago, 
Chicago, IL 60637, U.S.A.\\
$^7$ ASTRON, P.O. Box 1, NL-7990 AA Dwingeloo, The Netherlands\\
$^8$ Sterrewacht Leiden, P.O. Box 9513, NL-2300 RA Leiden, The Netherlands\\
$^9$ Center for Computational Sciences, University of Tsukuba, Tsukuba, 
Ibaraki 305-8577, Japan\\
$^{10}$ Center for Astrophysics and Space Sciences, University of California, 
San Diego, 9500 Gilman Drive, La Jolla, CA 92093-0424, U.S.A.\\
$^{11}$ Physics Division, Oak Ridge National Laboratory, Oak Ridge, TN
37831-6354, U.S.A.\\
$^{12}$ Department of Physics, College of Science, Rikkyo University, 
3-34-1 Nishi-Ikebukuro, Toshimaku, Tokyo, Japan \\
$^{13}$ T-6 Theoretical Astrophysics, Los Alamos National Laboratory, Los
Alamos, NM 87545, U.S.A.}
\pubyear{2006} \volume{000} \pagerange{1}
\twocolumn
\maketitle
\label{firstpage}

\begin{abstract}
Radiative transfer simulations are now at the forefront of numerical 
astrophysics. 
They are becoming crucial for an increasing number of astrophysical and
cosmological problems; at the same time their computational cost has come to the
reach of currently available computational power. Further progress is retarded
by the considerable number of different algorithms (including various flavours
of ray-tracing and moment schemes) developed, which makes the selection of the
most suitable technique for a given problem a non-trivial task. Assessing the
validity ranges, accuracy and performances of these schemes is the main aim of
this paper, for which we have compared 11 independent RT codes on 5 test
problems: (0) basic physics, (1) isothermal H~II region expansion and
(2) H~II region expansion with evolving temperature, (3) I-front trapping 
and shadowing by a dense clump, (4) multiple
sources in a cosmological density field. The outputs of these tests have been
compared and differences analyzed. The agreement between the various codes is
satisfactory although not perfect. The main source of discrepancy appears to
reside in the multi-frequency treatment approach, resulting in  different
thicknesses of the ionized-neutral transition regions and the temperature structure. 
The present results and
tests represent the most complete benchmark available for the development of
new codes and improvement of existing ones. To this aim all test inputs and
outputs are made publicly available in digital form.  

\end{abstract}


\section{Introduction}
Numerous physical problems require a detailed understanding of radiative
transfer (RT) of photons in different environments, ranging from intergalactic
and interstellar medium to stellar or planetary atmospheres. In particular, a
number of problems of cosmological interest cannot be solved without
incorporating RT calculations, e.g. modeling and understanding of the
Ly$\alpha$ forest, absorption lines in spectra of high-z quasars, radiative
feedback effects, the reionization of the intergalactic medium (IGM) and star
formation, just to mention a few \citep[see e.g.][for a recent
review on some of these topics]{2005SSRv..116..625C}. In many of these
situations, the physical conditions are such that the gas in which photons
propagate is optically thick; also, the geometry of the problem often is quite
complex. As a consequence, approaches relying on optically thin or geometrical
approximations yield unsatisfactory (and sometimes incorrect) results. 

The RT equation in 3D space has seven (three spatial,
two angular, one frequency, one time) dimensions. Although in specific cases
certain kinds of symmetry or approximations can be exploited, leading to a
partial simplification, most problems of astrophysical and cosmological
interest remain very complex. For this reason, although the basic physics
involved is well understood, the detailed solution of the complete radiative
transfer equation is presently beyond the available computational
capabilities. In addition, the technical implementation of the RT equation in
numerical codes is a very young and immature subject in astrophysics. 

RT approaches have been attempted in the past to study specific problems as
the light curves of supernovae, radiation from proto-stellar and active
galactic nuclei accretion disks, line radiation from collapsing molecular
clouds, continuum photon escape from galaxies and the effects of dust
obscuration in galaxies. Typically, either the low dimensionality of these
approaches or the simplified physics allowed a numerical treatment that
resulted in a sufficiently low computational cost given the available
machines. As this last constraint has become less demanding, the number and
type of desirable applications has expanded extremely rapidly, particularly
spreading to galaxy formation and cosmology fields.    

Following this wave of excitement, several groups then attacked the problem
from a variety of perspectives by using completely different, independent,
and dedicated numerical algorithms.  
It was immediately clear that the validation and assessment 
of the various codes was crucial in order not to waste (always) limited
computational and human resources.  
At that point the community faced the problem that, in contrast with
e.g. gas-dynamic studies, very few simple radiative transfer problems admit
exact analytical solutions that could be used as benchmarks. 
Just a few years ago only a few RT codes were available 
and these were still mostly in the testing/optimization phase, and therefore 
lacking the necessary degree of stability required to isolate truly 
scientific results from uncertainties due to internal programming bugs. In the
last few years the subject has rapidly changed. Not only has the reliability 
of existing codes matured, a new crop of codes based on novel techniques has 
been developed, making a comparison of them timely. 
The time is now ripe to do this comparison project in a fairly complete and
meaningful form and this paper presents the detailed outcome of our effort.  
     
The aim of the present comparison is to determine the type of problems the
codes are (un)able to solve, to understand the origin of the differences
inevitably found in the results, to stimulate improvements and further
developments of the existing codes and, finally, to serve as a benchmark
to testing future ones. We therefore invite interested RT researchers to make
use of our results, including tests descriptions, input and output data, all
of which can be found in digital form at the project website {\tt
  http://www.mpa-garching.mpg.de/tsu3/}.  At this stage our interest is not 
focused on the performances of the codes in terms of speed and optimization.


This project is a collaboration of most of the community and includes a wide
range of different methods (e.g. various versions of ray-tracing, as well as
moment schemes), some of which have already been applied to study a variety of
astrophysical and cosmological problems. The comparison is made among 11
independent codes, each of which is described concisely in \S~\ref{codes_sect}
(and more extensively in the corresponding methodology paper, whenever 
available). We would like to emphasize that the interaction among the
various participating groups has been characterized by a very constructive
and scientifically honest attitude, and resulted in many improvements of the
codes. 

Here we present the results from a set of tests on fixed density fields, both
homogeneous and inhomogeneous, which verify the radiative transfer methods
themselves. In a follow-up paper we plan to discuss the direct coupling
to gas-dynamics and compare the results on a set of several radiative
hydrodynamics problems. 

\section{The Codes}
\label{codes_sect}
In this section we briefly describe the eleven radiative transfer codes which
are taking part in this comparison project. The descriptions point to the more
detailed methodology papers whenever available. All the codes, their names,
authors and current features are summarized in Table~\ref{summary}. 

\begin{table*}
\caption{Participating codes and their current features.}
\label{summary}
\begin{tabular}{llllll}
\hline
Code (Authors)& Grid                   & Parallelization& gasdynamics & Helium & Rec. radiation\\ [2mm]\hline\\
$C^2$-Ray (G. Mellema, I. Iliev, P. Shapiro, M. Alvarez)    & fixed/AMR              & shared         & yes&no&no\\
OTVET (N. Gnedin, T. Abel)      & fixed                  & shared             & yes&yes&yes\\
CRASH (A. Maselli, A. Ferrara, B. Ciardi)       & fixed                  & no             & no &yes&yes\\
RSPH      (H. Susa, M. Umemura)    & no grid, particle-based& distributed    & yes&no&no\\
ART (T. Nakamoto, H. Susa, K. Hiroi, M. Umemura)     & fixed                  & distributed    & no&no&yes\\
FTTE     (A. Razoumov)    & fixed/AMR              & no             & yes&yes&yes\\
SimpleX  (J. Ritzerveld, V. Icke, E.-J. Rijkhorst)    & unstructured           & no             & no&no&yes\\
Zeus-MP  (D. Whalen, M. Norman)    & fixed                  & distributed    & yes&no&no\\
FLASH-HC (E.-J. Rijkhorst, T. Plewa, A. Dubey, G. Mellema)    & fixed/AMR              & distributed    & yes&no&no\\
IFT  (M. Alvarez, P. Shapiro)       & fixed/AMR              & no             & no&no&no\\
Coral (I. Iliev, A. Raga, G. Mellema, P. Shapiro)        & AMR                    & no             & yes&yes&no\\
\hline\\
\end{tabular}
\end{table*}

\subsection{$C^2$-Ray: Photon-conserving transport of ionizing radiation 
(G. Mellema, I. Iliev, P. Shapiro, M. Alvarez)}

C$^2$-Ray is a grid-based short characteristics ray-tracing code
which is photon-conserving and causally traces the rays away from
the ionizing sources up to each cell. Explicit photon-conservation is assured 
by taking a finite-volume approach when calculating the photoionization rates, 
and by using time-averaged optical depths. The latter property allows for 
integration time steps much larger than the ionization time scale, which leads 
to a considerable speed-up of the calculation and facilitates the coupling of
our code to gasdynamic evolution. The code is described and tested in detail 
in \citet{methodpaper}.

The frequency dependence of the photoionization rates and the photoionization
heating rates is dealt with by using frequency-integrated rates, stored as 
functions of the optical depth at the ionization threshold. In its current 
version the code includes only hydrogen and does not include the effects of
helium, although they could be added in a relatively straightforward way. 

The transfer calculation is done using short characteristics, where
the optical depth is calculated by interpolating values of grid cells
lying closer to the source.
Because of the causal nature of the ray-tracing, the calculation cannot
easily be parallelized through domain decomposition. However, using
OpenMP the code is efficiently parallelized over the sources.
The code is currently used for large-scale simulations of cosmic
reionization and its observability \citep{topologypaper,21cmreionpaper}
on grid sizes up to $406^3$ and up to more than $10^5$ ionizing sources.

There are 1D, 2D and 3D versions of the code available. It was developed 
to be directly coupled with hydrodynamics calculations. The large time steps 
allowed for the radiative transfer enable the use of the hydrodynamic time 
step for evolving the combined system. The first gasdynamic application of our
code is presented in \citet{2005astro.ph.12554M}.

\subsection{OTVET: Optically-Thin Variable Eddington Tensor Code (N. Gnedin, 
T. Abel)}

The Optically Thin Variable Eddington Tensor (OTVET) approximation 
\citep{2001NewA....6..437G} is based on the moment formulation of the 
radiative transfer equation:
\ba
       {a\over c}{\partial E_\nu\over\partial t} +
        {\partial F^i_\nu\over \partial x^i} & = & - \hat\kappa_\nu E_\nu
        + S_\nu,\nonumber\\
        {a\over c}{\partial F^j_\nu\over\partial t} +
        {\partial \over \partial x^i} E_\nu h^{ij}_\nu & = & -
        \hat\kappa_\nu F^j_\nu,
        \label{otveteq}
\ea
where $E_\nu$ and $F^j_\nu$ are the energy density and the flux of
radiation respectively, $\kappa_\nu$ is the absorption coefficient,
$S_\nu$ is the source function, and $h^{ij}_\nu$ is a unit
trace tensor normally called the Eddington tensor.

It is important to underscore that the source function $S_\nu(\vec{x})$ is
considered to be an arbitrary function of position, so that it may
contain both the delta-function contributions from any number of
individual point sources and the smoothly varying contributions from
the diffuse sources.

Equations (\ref{otveteq}) form an open system of two partial
differential equations, because the Eddington tensor can not be
determined from them. In the OTVET approximation the Eddington tensor
is computed from all sources of radiation as if they were optically
thin, $h_\nu^{ij} = {P_\nu^{ij}/{\rm Tr\,}P_\nu^{ij}}$, where
\[
        P_\nu^{ij} = \int d^3 x_1 \rho_*(\vec{x}_1)
        {(x^i-x_1^i)(x^j-x_1^j)\over (\vec{x}-\vec{x}_1)^4},
\]
where $\rho_*$ is the mass density of the sources e.g. stars.

Thus, the OTVET approximation conserves the number density of photons
(in the absence of absorption) and the flux, but may introduce an
error in the direction of the flux propagation. It can be shown
rigorously that the OTVET approximation is exact for a single
point-like source and for uniformly distributed sources, but is not
exact in other cases.

While it is difficult to prove rigorously, it appears that the largest
error is introduced for the case
of two sources, one much stronger than the other. In that case the H~II
region around the strong source is modeled highly precisely, but the
shape of the H~II region around the weaker source (before the two H~II
regions merge) becomes ellipsoidal with the deviation from the
spherical symmetry never exceeding 17\% (1/6) in any direction.

\subsection{CRASH: Cosmological RAdiative transfer Scheme for
Hydrodynamics (A. Maselli, A. Ferrara, B. Ciardi)}

{ CRASH} is a 3D ray-tracing radiative-transfer code based on the Monte
Carlo (MC) technique for sampling distribution functions. The grid-based 
algorithm follows the propagation of the ionizing radiation through an 
arbitrary H/He static density field and calculates the time-evolving 
temperature and ionization structure of the gas.

The MC approach to RT requires that the radiation field is discretized into
photon packets.  The radiation field is thus reproduced by emitting packets,
according to the configuration under analysis, and by following their
propagation accounting for the opacity of the gas.  For each emitted photon
packet, the emission location, frequency and propagation direction are
determined by randomly sampling the appropriate probability distribution
functions (PDFs), which are assigned as initial conditions. It is possible to
include an arbitrary number of point/extended sources and/or diffuse
background radiation in a single simulation. This approach thus allows 
a straightforward and self-consistent treatment of the diffuse radiation
produced by H/He recombinations in the ionized gas.

The relevant radiation-matter interactions are accounted for during the photon
packet's propagation. At each cell crossed, each packet deposits a fraction of
its photon content according to the cell's opacity which determines the
absorption probability. Once the number of photons absorbed in the cell is
calculated, we find the effect on the temperature and on the ionization 
state of the gas, by solving the discretized non-equilibrium chemistry and 
energy equations. Recombinations, collisional ionizations and cooling are
treated as continuous processes.

The detailed description of the { CRASH} implementation is given in
\citet{2001MNRAS.324..381C} and \citet{2003MNRAS.345..379M}, and an improved 
algorithm for dealing with a background diffuse ionizing radiation is described in
\citet{2005astro.ph.10258M}. The tests described in this paper have
been performed using the most updated version of the code 
\citep{2003MNRAS.345..379M}. 

The code has been primarily developed to study a number of cosmological
problems, such as hydrogen and helium reionization, the physical state of the
Ly$\alpha$ forest, the escape fraction of Lyman continuum photons from
galaxies, the diffuse Ly$\alpha$ emission from recombining gas. However,
its flexibility allows applications that could be relevant to a wide range of
astrophysical problems.  The code architecture is sufficiently simple that
additional physics can be easily added using the algorithms already
implemented.  For example, dust absorption/re-emission can be included with
minimum effort; molecular opacity and line emission, although more
complicated, do not represent a particular challenge given the numerical
scheme adopted. Obviously, were such processes added, the computational time
could become so long that parallelization would be necessary. This would be
required also when {CRASH} will be coupled to a hydrodynamical code to
study the feedback of photo-processes onto the (thermo-)dynamics of the
system.

\subsection{RSPH: SPH coupled with radiative transfer (H. Susa, M. Umemura)}

The Radiation-SPH scheme is designed to investigate the
formation and evolution of the first generation objects at $z \ga 10$ 
\citep{2006astro.ph..1642S},
where the radiative feedback from various sources play important
roles. The code can compute the fraction of chemical species e, H$^+$, H,
H$^-$, H$_2$,  and H$_2^+$ by fully implicit time integration. It also
can deal with multiple sources of ionizing radiation, as well as
the radiation in the Lyman-Werner band.

Hydrodynamics is calculated by the Smoothed Particle Hydrodynamics (SPH)
method. We use the version of SPH by \citet{1993ApJ...406..361U} with 
the modification by \citet{1993A&A...268..391S}, and we also
adopt the particle resizing formalism by \citet{2000MNRAS.319..619T}. 
In the present version, we do not use the  entropy formalism.

The non-equilibrium chemistry and radiative cooling 
for primordial gas are calculated by the code
developed by \citet{2000MNRAS.317..175S}, where H$_2$ cooling and
reaction rates are mostly taken from \citet{1998A&A...335..403G}.

As for the photoionization process, 
we employ the  on-the-spot approximation
\citep{1978ppim.book.....S}. We solve the transfer of ionizing photons
directly from the 
source but do not solve the transfer of diffuse photons. Instead, it is
assumed that the recombination photons are absorbed in the 
neighbourhood of the spatial position where they are emitted. 
The absence of source terms in this approximation greatly simplifies 
the radiation transfer equation. Solving the transfer equation reduces 
to the determination of the optical depth from the source to every SPH 
particle.

The optical depth is integrated utilizing the 
neighbour lists of SPH particles. 
It is similar to the code described
in \citet{2004ApJ...600....1S}, but now we can deal with multiple point
sources. In our new scheme, we do not create so many grid points on the light
ray as the previous code \citep{2004ApJ...600....1S} did.  
Instead, we just create one grid point per SPH particle in its
neighbor. We find the 'upstream' particle for each SPH particle on its
line of sight to the source. Then the optical depth from the source to
the SPH particle is obtained by summing up the optical depth at the 'upstream'
particle and the differential optical depth between the two particles.

The code is already parallelized using the MPI library. The computational
domain is divided by the Orthogonal Recursive Bisection method. 
The parallelization method for radiation transfer part
is similar to the Multiple Wave Front method developed by \citet{2001MNRAS.321..593N} and
\citet{2005astro.ph..3510H}, but it is changed to fit the RSPH code. The details
are described in \citet{2006astro.ph..1642S}.
The code also is able to handle gravity with a Barnes-Hut tree, which is
also parallelized.

\subsection{ART: Authentic Radiative Transfer with
  Discretized Long Beams (T. Nakamoto, H. Susa, K. Hiroi, M. Umemura)} 

ART is a grid based code
designed to solve the transfer of radiation from point sources
as well as diffuse radiation. On the photo-ionization problem,
ART solves time-dependent ionization states and energies.
Hydrodynamics is not incorporated into the current version.

In the first step of the ART scheme, a ray, on which photons propagate,
is cast from the origin by specifying the propagation angles.
When the distance from the ray to a grid point is smaller than a
certain value, a segment is located at the grid point.
A collection of all the segments along the ray is considered
to be a decomposition of the ray into segments.
The radiative transfer calculation is sequentially done from
the origin towards the downstream side on each segment.
From one segment to another, the optical
depth is calculated and added.
Finally, changing the angle and shifting the origin, we can obtain
intensities at all the grid points directed along all the angles.

ART has two versions for the integration of intensities over
the angle.  The first is a simple summation of intensities with
finite solid angles. This scheme is fit to diffuse radiation cases.
The second one, designed to fit point sources,
uses a solid angle of the source at the grid point to evaluate
the dilution factor.
Multiplying the representing intensity at the grid point
and the dilution factor,
we can obtain the integration of the intensity over the angle.
Generally, the radiation field can be divided into two parts:
one is the direct incident radiation from point sources and the
other is the diffuse radiation.
ART can treat both radiation fields appropriately by using
two schemes simultaneously.

The integration over the frequency is done using the one-frequency
method, which is similar to the six-frequency method devised
by \citet{2001MNRAS.321..593N}. Since in our present
problems both the spectrum of the source and the frequency-dependence
of the absorption coefficient are known in advance, once we
calculate the optical depth at the Lyman limit frequency,
the amount of absorption at any frequency can be obtained
without carrying out integration along the ray for the different
frequency.

The parallelization of ART can be done based not only
on the angle-frequency decomposition but also
on the spatial domain decomposition using the Multiple Wave
Front (MWF) method \citep{2001MNRAS.321..593N}.

\subsection{FTTE: Fully threaded transport engine (A. Razoumov)}
FTTE is a new method for transport
of both diffuse and point source radiation on refined grids developed
at Oak Ridge National Laboratory. The diffuse part of the solver has been
described in \citet{2005MNRAS.362.1413R}. Transfer around point sources is
done in a separate module acting on the same fully threaded data
structure (3D fields of density, temperature, etc.) as the diffuse
part. The point source algorithm is an extension of the adaptive
ray-splitting scheme of \citet{abel.02} to a model with variable grid
resolution, with all discretization done on the grid. Sources of
radiation can be hosted by cells of any level of refinement, although
usually in cosmological applications sources reside in the
deepest level of refinement. Around each point source we build a
system of radial rays which split either when we move further away
from the source, or when we enter a refined cell, to match the local
minimum required angular resolution. Once any radial ray is refined it
stays refined (even if we leave the high spatial resolution patch)
until further angular refinement is necessary.

All ray segments are stored as elements of their host cells, and
actual transport just follows these interconnected data
structures. Whenever possible, an attempt is made to have the most
generic ray pattern possible, as each ray pattern needs to be computed
only once. The simplest example is an unigrid calculation (no
refinement), where there is a single ray pattern which can be used for
all sources. Another computationally trivial example is a collection
of halos within refined patches with the same local grid geometry as
seen from each source.

For multiple sources located in the same H~II region, we can merge
their respective ray trees when the distance from the sources to a ray
segment far exceeds the source separation, and the optical depth to
both sources is negligible -- see \citet{razoumov...02} for details.

The transport quantity in the diffuse solver is the specific (per unit
frequency) intensity, whereas in the point source solver it is the
specific photon luminosity -- the number of photons per unit frequency
entering a particular ray segment per unit time. For discretization in
angles both modules use the HEALPix algorithm \citep{gorski...02},
dividing the entire sphere into $12\times4^{n-1}$ equal area pixels,
where $n=1,2,...$ is the local angular resolution. For point source
transfer, $n$ is a function of the local grid resolution,
the physical distance to the source, and the local ray pattern in a
cell. As we go from one ray segment to another, in each cell we
accumulate the mean diffuse intensity and all photo-reaction rates due
to point source radiation. For rate equations, we use the
time-dependent chemistry solver from \citet{1997NewA....2..209A}.

Currently there is only a serial version of the code, although there
is a project underway at the San Diego Supercomputing Center to
parallelize the diffuse part of the algorithm angle-by-angle. Even in
the serial mode the code is very fast, limited in practice by the
memory available to hold ray patterns. We have run various test
problems up to grid sizes $256^3$ with five levels of refinement
($8192^3$ effective spatial resolution), and up to the angular
resolution level $n=13$.

\subsection{SimpleX: Radiative Transfer on Unstructured Grids 
(J. Ritzerveld, V. Icke, E.-J. Rijkhorst)}

SimpleX \citep{2003astro.ph.12301R} is a mesoscopic particle method,
using unstructured Lagrangian
grids to solve the Boltzmann equation for a photon gas. It has many
similarities with Lattice Boltzmann Solvers, which are used in complex
fluid flow simulations, with the exception that it uses an adaptive
grid based on the criterion that the local grid step is chosen to
correlate with the mean free path of the photon. In the optimal case,
this last modification results in an
operation count for our method which does \emph{not} scale with the number
of sources.

More specifically, SimpleX does not use a grid in the usual sense, but
has as a basis a point distribution, which follows the medium density,
or the opacity. Given a regular grid with medium density values, we
use a Monte Carlo method to sample our points according to this
density distribution. We construct an unstructured grid from this
point distribution by using the Delaunay tessellation technique. This
recipe was intentionally chosen this way to ensure that the local
optical mean free paths correlate linearly with the line lengths
between points. This way, the radiation-matter interactions, which 
determine the collision term on the rhs of the Boltzmann equation., 
can be incorporated by
introducing a set of `interaction coefficients' $\{c_i\}$, one for
each interaction. These coefficients are exactly the linear
correlation coefficients between the optical mean free path and the
local line lengths.

The transport of radiation through a medium can subsequently be
defined and implemented as a walk on this resultant graph, with the
interaction taking place at each node. The operation count of this
resultant method is $O\left(N^{1+1/m}\right)$, in which $N$ is the
number of points, or resolution, and $m$ is the dimension. This is
\emph{in}dependent of the number of sources, which makes it ideal to
do large scale reionization calculations, in which a large number of
sources is needed.

The generality of the method's setup defines its versatility. Boltzmann-like
transport equations describe not only the flow of a photon gas, but
also that of a fluid
or of a plasma. It is therefore straightforward to define a SimpleX method
which solves both the radiative transfer equations and the
hydrodynamics equations
self-consistently. At the same time, we are in the process of making a
dynamic coupling of SimpleX with the grid based hydro codes FLASH
\citep{2000ApJS..131..273F} and  GADGET-2 \citep{2005MNRAS.364.1105S}.
The code easily runs on a single desktop machine, but will be
parallelized in order to accommodate this coupling.

For this comparison project, we used a SimpleX method set up to do cosmological
radiative transfer calculations. The result is a method which is designed to be
photon conserving, updating the ionization fraction and the resultant
exact local optical depth
dynamically throughout the simulation. Moreover, diffuse recombination
radiation, shown to
be quite important for the overall result \citep[e.g.][]{2005A&A...439L..23R},
can readily be implemented self-consistently and without loss of
computational speed.

For now, we do not solve the energy equation, and do simulations
H-only. We ignore the effect
of spectral hardening, by which we can analytically derive the
fraction of the flux above
the Lyman limit, given that the source radiates as a black body. We account
for H-absorption by using a blackbody averaged absorption coefficient.
The final results are interpolated from the unstructured grid cells
onto the $128^3$ datacube to accommodate this comparison.

\subsection{ZEUS-MP with radiative transfer (D. Whalen, M. Norman)}

The ZEUS-MP hydrocode in this comparison project solves explicit 
finite-difference approximations to Euler's equations of fluid dynamics 
together with a 9-species 
reactive network that utilizes photoionization rate coefficients computed by a 
ray-casting radiative transfer module \citep{2006ApJS..162..281W}. Ionization
fronts thus arise as an emergent feature of reactive flow and radiative
transfer in our simulations and are not tracked by computing equilibria 
positions along lines of sight. 
The hydrodynamical variables ($\rho$, e, and the $\rho$v$_{i}$)  
are updated term by term in operator-split and directionally-split substeps,
with a given substep incorporating the partial update from the previous
substep. The gradient (force) terms in the Euler equations are computed in 
source routines and the divergence  
terms are calculated in advection routines \citep{1992ApJS...80..753S}.

The primordial species added to ZEUS-MP (H, H$^{+}$, He, He$^{+}$, He$^{2+}$, H$^{-}$, 
H$^{+}_{2}$, H$_{2}$, and e$^{-}$) are evolved by nine additional continuity equations 
and the nonequilibrium rate equations of \citet{1997NewA....2..209A}. The
divergence term for each   
species is evaluated in the advection routines, while the other terms form a reaction 
network that is solved separately from the source and advective updates. Although the 
calculations performed for this comparison study take the gas to be hydrogen
only, in general we sequentially  
advance each n$_{i}$ in the network, building the i$^{th}$ species' update from the 
i - 1 (and earlier) updated species while applying rate coefficients evaluated at the 
current problem time. Charge and baryon conservation are enforced at the end of each 
hydrodynamic cycle and microphysical cooling and heating processes are included by an 
isochoric operator-split update to the energy density computed each time the reaction 
network is advanced.  The radiative transfer module computes the photoionization rate 
coefficients required by the reaction network by solving the static equation of transfer 
recast into flux form along radial rays outward from a point source centered in a 
spherical-coordinate geometry.  The number of ionizations in a zone is calculated in a 
photon-conserving manner to be the number of photons entering the zone minus the number 
exiting.

The order of execution of the algorithm is as follows: first, the radiative transfer module 
is called to calculate ionization rates in order to determine the smallest heating/cooling 
time on the grid. The grid minimum of the Courant time is then computed and
the hydrodynamics 
equations are evolved over the smaller of the two timescales.  Next, the shortest chemistry 
timescale of the grid is calculated 
\begin{equation}
\Delta t_{chem} = 0.1 \, \displaystyle\frac{n_{e}}{{\dot{n}}_{e}}
\end{equation}
which is formulated to ensure that the fastest reaction operating at any place or time in 
the problem governs the maximum time by which the reaction network may be accurately 
advanced. The species concentrations and gas energy are then advanced over this timestep, 
the transfer module is called again to compute a new chemistry timestep, and the network
and energy updates are performed again. The n$_{i}$ and energy are subcycled
over successive  
chemistry timesteps until the hydrodynamical timestep has been covered, at which point full 
updates of velocities, energies, and densities by the source and advection routines are 
computed. A new hydrodynamical timestep is then determined and the cycle repeats.  

This algorithm has been extensively tested with a comprehensive suite of quantitative static 
and hydrodynamical benchmarks complementing those appearing in this paper.  The tests are 
described in detail in \citet{2006ApJS..162..281W}

\subsection{FLASH-HC: Hybrid Characteristics (E.-J. Rijkhorst, T. Plewa,
  A. Dubey, G. Mellema)} 
The Hybrid Characteristics (HC) method \citep{rijkhorst2005, rijkhorst2006} is
a three-dimensional radiative transfer algorithm designed specifically for use
with parallel adaptive mesh refinement (AMR) hydrodynamics codes. 
It introduces a novel form of ray tracing that can neither be classified as
long, nor as short characteristics. 
It however does apply the underlying principles, i.e. efficient execution
through interpolation and parallelizability, of both these approaches.  

Primary applications of the HC method are radiation hydrodynamics problems
that take into account the effects of photoionization and heating due to point
sources of radiation. 
The method is implemented into the hydrodynamics package Flash
\citep{2000ApJS..131..273F}. 
The ionization, heating, and cooling processes are modeled using the Doric
package \citep{1994A&A...289..937F}. 
Upon comparison with the long characteristics method, it was found that the HC
method calculates shadows with similarly high accuracy. 
Although the method is developed for problems involving photoionization due to
point sources, the algorithm can easily be adapted to the case of more general
radiation fields. 

The Hybrid Characteristics algorithm can be summarized as follows.
Consider an AMR hierarchy of grids that is distributed over a number of
processors. 
Rays are traced over these different grids and must, to make this a parallel
algorithm, be split up into independent ray sections. 
Naturally these sections are in the first place defined by the boundaries of
each processor's sub-domain, and in the second place by the boundaries of the
grids contained within that sub-domain. 
At the start of a time step, each processor checks if its sub-domain contains
the source. 
The processor that owns the source stores its grid and processor id and makes
it available to all other processors. 
Then each processor ray traces the grids it owns to obtain local column
density contributions. 
Since in general rays traverse more than one processor domain, these local
contributions are made available on all processors through a global
communication operation. 
By interpolating and accumulating all contributions for all rays (using a so
called `grid-mapping') the total column density for each cell is obtained. 
The coefficients used in the interpolation are chosen such that the exact
solution for the column density is retrieved when there are no gradients in
the density distribution.  
Tests with a $1/r^2$ density distribution resulted in errors $<0.5\%$ in the
value for the  total column density as compared to a long characteristics
method. 

Once the column density from the source up to each cell face is known, the
ionization fractions and temperature can be computed. For this we use the
Doric package \citep[see][]{2002A&A...394..901M,1994A&A...289..937F}. 
These routines calculate the photo- and collisional ionization, the
photoheating, and the radiative cooling rate. 
Using an analytical solution to the rate equation for the ionization
fractions, the temperature and ionization fractions are found through an
iterative process. 
Since evaluating the integrals for the photoionization and heating rate is too
time consuming to perform for every value of the optical depth, they are
stored in look-up tables and are interpolated when needed. 

The hydrodynamics and ionization calculations are coupled through operator
splitting. 
To avoid having to take time steps that are the minimum of the hydrodynamics, ionization, and heating/cooling time scales, the fact that the equations for the ionization and heating/cooling can be iterated to convergence is used.
This means that the only restriction on the time step comes from the
hydrodynamics (i.e. the Courant condition). 
Note however that, when one needs to follow R-type ionization fronts, an
additional time step constraint is used to find the correct propagation
velocity for this type of front. 

An assessment of the parallel performance of the HC method was presented by
\citet{rijkhorst2005}. 
It was found that the ray tracing part takes less time to execute than other
parts of the calculation (e.g. hydrodynamics and AMR.). 
Tests involving randomly distributed sources show that the algorithm scales
linearly with the number of sources. 
Weak scaling tests, where the amount of work per processor is kept constant,
as well as strong scaling tests, where the total amount of work is kept
constant, were performed as well. 
By carefully choosing the amount of work per processor it was shown that the
HC algorithm scales well for at least $\sim 100$ processors on an SGI Altix,
and $\sim 1000$ processors on an IBM BlueGene/L system. 

The HC method will be made publicly available in a future Flash release.

\subsection{IFT: Ionization Front Tracking (M. Alvarez and P. Shapiro)}

We have developed a ray-tracing code which explicitly follows the
progress of an ionization front (I-front) around a point source of ionizing
radiation in an arbitrary three-dimensional density field of atomic
hydrogen.  Because we do not solve the non-equilibrium chemical and 
energy rate equations and use a simplified treatment of radiative
transfer, our method is extremely fast.  While our code is currently
capable of handling only one source, we are in the process of
generalizing it to handle an arbitrary number of point sources.
More detailed descriptions of various aspects of this code can be found in
\S~5.1.3 of \citet{methodpaper} and \S3 of \citet{2005astro.ph..7684A}.

The fundamental assumption we make is that the I-front is sharp.
Behind the front, the ionized fraction and temperature are assumed to
take their equilibrium values, while ahead of the front they are
assigned their initial values. The assumption of equilibrium behind
the I-front is justified because the equilibration time on the ionized
side of the I-front is shorter than the recombination time by a factor
of the neutral fraction, which is small behind the I-front. Because the 
progress of the I-front
will be different in different directions, it is necessary to solve
for its time-dependent position along rays that emanates from the source, 
the angular orientation of which are chosen to lie at the centers of
HEALPixels\footnote{http://healpix.jpl.nasa.gov}
\citep{2005ApJ...622..759G}.   Typically, we choose a sufficient
resolution of rays in the sky such that there is approximately one ray
per edge cell.  Along a given ray, we solve the fully-relativistic 
equation for the propagation of the I-front \citep{IASS05}:
\begin{equation}
\frac{dR}{dt}=\frac{cQ(R)}{Q(R)+4\pi R^2cn(R)},
\label{diffeq}
\end{equation}
where $Q(R)$ is the ionizing photon luminosity at the surface of the
front, $R$ is the distance along the ray, and $c$ is the speed of
light.  This equation correctly accounts for the finite travel time of
ionizing photons, i.e. as $Q(R)\longrightarrow \infty$, $dR/dt
\longrightarrow c$.   
The arrival rate of ionizing photons is given by
\begin{equation}
Q(R)=Q_*-4\pi\int_0^R \alpha_B(r)n^2(r)r^2dr,
\end{equation}
where $Q_*$ is the ionizing photon luminosity of the source, $n(r)$ is
the density along the ray, and $\alpha_B(r)$ is the ``case B''
recombination coefficient along the ray.  The value of $\alpha_B(r)$
is determined by the equilibrium temperature, which varies along the
ray.  

A brief outline of the algorithm is as follows.  First, we interpolate
the density from the grid to discrete points along each ray, where
the spacing between points along each ray is approximately the same as
the cell size of the grid.  Next, we compute the equilibrium profile
of ionized fraction and temperature along each ray by moving outward
from the source, using the equilibrium neutral hydrogen
density of the previous zones to attenuate the flux to each successive
zone.  Equation
(\ref{diffeq}) is then solved for each ray, which gives the I-front
position in that direction.  The values of ionized fraction
and temperature along each ray are set to their equilibrium values
inside of the I-front, and set to their initial values on the
outside.  Finally, the ionized fraction and temperature are
interpolated back from the rays to the grid.  

\subsection{CORAL (I. Iliev, A. Raga, G. Mellema, P. Shapiro)}
CORAL is a 2-D, axisymmetric Eulerian fluid dynamics adaptive mesh refinement
(AMR) code \citep[see][and references therein for detailed
description]{1998A&A...331..335M,2004MNRAS.348..753S}. It solves the Euler
equations in their conservative finite-volume form using the second-order 
method of van~Leer flux-splitting, which allows for correct and precise
treatment of shocks. The grid refinement and de-refinement criteria are based
on the gradients of all code variables. When the gradient of any variable is
larger than a pre-defined value the cell is refined, while when the criterion
for refinement is not met the cell is de-refined.

The code follows, by a semi-implicit method, the non-equilibrium chemistry of
multiple species (H, He, C II-VI, N I-VI, O I-VI, Ne I-VI, and S II-VI) and
the corresponding cooling \citep{1997ApJS..109..517R,1998A&A...331..335M}, as
well as Compton cooling. The photoheating rate used is a sum of the
photoionization heating rates for H~I, He~I and He~II. For computational
efficiency all heating and cooling rates are pre-computed and stored in tables.
The microphysical processes -- chemical reactions, radiative processes, 
transfer of radiation, heating and cooling -- are implemented though the 
standard approach of operator-splitting (i.e. solved each time-step, side-by-side with 
the hydrodynamics and coupled to it through the energy equation).
The latest versions of the code also include the effects of an external
gravity force. 

Currently the code uses a black-body, or a power-law ionizing source spectra,
although any other spectrum can be accommodated. Radiative transfer of the
ionizing photons is treated explicitly by taking into account the bound-free
opacity of H and He in the photoionization and photoheating rates. The 
photoionization and
photoheating rates of H~I, He~I and He~II are pre-computed for the given
spectrum and stored in tables vs. the optical depths at the ionizing
thresholds of these species, which are then used to obtain the total optical depths.
The code correctly tracks both fast (by evolving on an ionization timestep,
$\Delta t\sim\dot{n_H}/n_H$) and slow I-fronts. 

The code has been tested extensively and has been applied to many
astrophysical problems, e.g. photoevaporation of clumps in planetary nebulae 
\citep{1998A&A...331..335M}, cosmological minihalo photoevaporation during
reionization \citep{2004MNRAS.348..753S,2005MNRAS...361..405I}, and studies of
the radiative feedback from propagating ionization fronts on dense clumps in
Damped Lyman-$\alpha$ systems \citep{2005astro.ph..9261I}.

\begin{figure}
\begin{center}
  \includegraphics[width=3.5in]{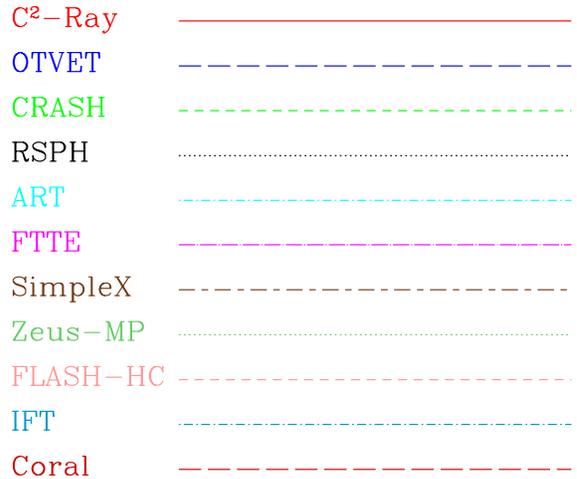}
\vspace{-0.8in}
\caption{Legend for the line plots.
\label{legend_fig}}
\vspace{-0.5cm}
\end{center}
\end{figure}

\section{Tests and Results}
\label{tests_sect}

\begin{figure*}
\begin{center}
  \includegraphics[width=7in]{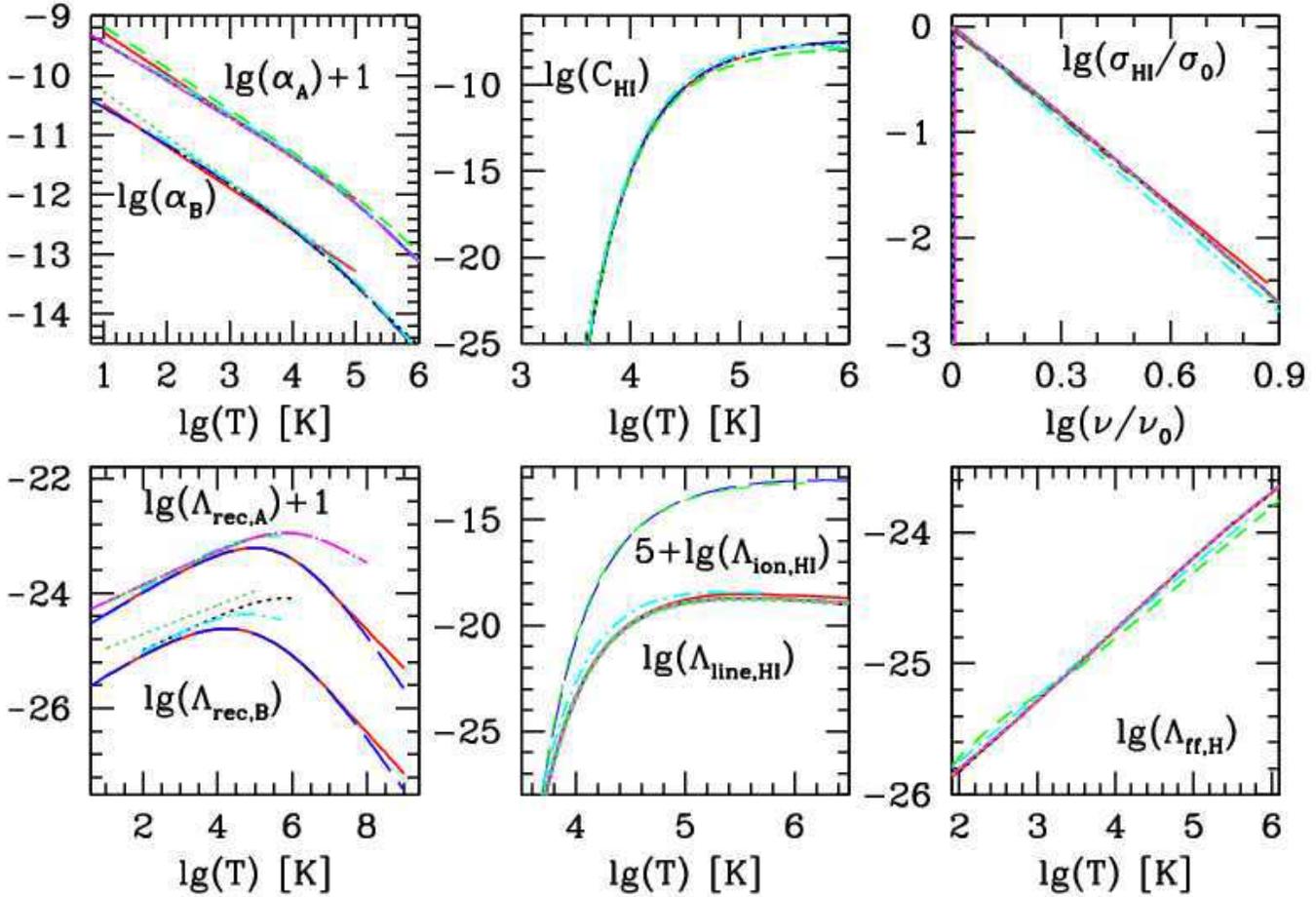}
\caption{Test 0, part 1: Hydrogen rates, cooling and cross-sections used by 
the participating codes. 
\label{rates_fig}}
\end{center}
\end{figure*}

In this section we describe the tests we have performed, along with their 
detailed parameters, geometry and setup, and the results we obtained. 
When constructing these tests, we aimed for the simplest and cleanest, but 
nonetheless cosmologically-interesting problems. We designed them in a way 
which allows us to test and compare all the important aspects of any
radiative-transfer code. These include correct tracking of both slow and 
fast I-fronts, in homogeneous and inhomogeneous density fields, formation 
of shadows, spectrum hardening, and solving for the gas temperature state.
In a companion paper (Paper~II) we will present the tests which include 
the interaction with fluid flows and radiative feedback on the gas. For 
simplicity, in all tests the gas is assumed to be hydrogen-only. The gas 
density distribution is fixed. Finally, in order to be included in the 
comparison, each test had to be done by at least three or more of the 
participating codes.

Figure~\ref{legend_fig} provides a legend allowing the reader to identify 
which line corresponds to which code in the figures throughout the paper.
The images we present are identified in the corresponding figure caption.

All test problems are solved in three dimensions (3D), with grid dimensions 
$128^3$ cells. For the 1D (Zeus-MP) and 2D (Coral) codes the data is
interpolated on the same-size 3D grid and analyzed the same way. Unless
otherwise noted, the sources of ionizing radiation are assumed to have a 
black-body spectrum with effective temperature $T_{\rm eff}=100,000$ K, 
except for 
Test~1 where the assumed spectrum is monochromatic with all photons having 
energy $h\nu=13.6$ eV, the ionization threshold of hydrogen. In all tests 
the temperature is allowed to vary due to atomic heating and cooling processes 
in accordance with the energy equation, again 
with the exception of { Test~1 where we fix} the temperature at $T=10^4$~K.
The few codes that do not yet include an energy equation use a constant 
temperature value.

\subsection{Test 0: The basic physics}

\begin{table*}
\caption[]{Rates adopted by the different radiative transfer codes. The 
columns are, from left to right, name of the code and reference for:
case A recombination rate (RRA) of H~II, He~II and He~III;
case B recombination rate (RRB) of H~II, He~II and He~III;
dielectronic recombination rate (DRR) of He~II;
collisional ionization rate (CIR) of H~I, He~I and He~III;
collisional ionization cooling rate (CICR) of H~I, He~I and He~II;
case A recombination cooling rate (RCRA) of H~II, He~II and He~III;
case B recombination cooling rate (RCRB) of H~II, He~II and He~III;
dielectronic recombination cooling rate (DRCR) of He~II;
collisional excitation cooling rate (CECR) of H~I, He~I and He~II;
Bremsstrahlung cooling rate (BCR);
Compton cooling rate (CCR);
cross-section (CS) of H~I, He~I and He~II.
Units of RRA, RRB, DRR and CIR are [cm$^3$s$^{-1}$],
units of CICR, RCRA, RCRB, DRCR, CECR, BCR, CCR are [erg~cm$^3$s$^{-1}$] 
and units of CS are [cm$^2$]. Note that for those codes in which the 
treatment of He is not included, only the references to the H rates
are given.}
\label{tablerates}
\begin{tabular}{lllllllllllll}
\hline
CODE        & RRA   & RRB      & DRR & CIR   & CICR     & RCRA  & RCRB     & DRCR & CECR  & BCR & CCR & CS \\
\hline
C$^2$-Ray   &       & 10,11,10 &     & 6     &          &       & 10,11,10 &      & 23,-,-      & 10  & 18  & 15      \\
OTVET       & 9,4,9 & 9,4,9    & 2   & 9,9,9 & 9,9,9    & 9,4,9 & 9,4,9    & 2    & 5,-,5 &     & 17  & 22\\
CRASH       & 5,5,5 & 20,20,20 & 5   & 5,5,5 & 5,5,5    & 5,5,5 & 20,20,20 & 5    & 5,5,5 & 5   & 8   & 15\\
RSPH         &       & 20       &     & 18    &          &       & 20       &      & 7     &     &     & 18      \\
ART     & 19    & 20       &     & 19    &          & 19    & 20       &      & 19    & 19  & 19  & 13      \\
FTTE        & 1     & 9,9,9    & 1   & 1,1,1 & 18,18,18 & 5,5,5 & 10,11,10 & 5    & 3,5,5 & 3   & 16  & 15      \\
SimpleX     &       & 20       &     & 6     &          &       &          &      &       &     &     & 15      \\
ZEUS-MP     & 1     & 21       &     & 12    & 18       & 5     &          &      & 5,-,-     & 3   & 16  & 14      \\
FLASH-HC    &       & 10,11,10 &     & 6     &          &       & 10,11,10 &      & 23,-,-      & 10  &     & 15      \\
IFT        &       & 9        &     & 25    & 5        &       & 18       &      & 5,-,-      & 3   &     & 22      \\
Coral       &10,11,10& 10,11,10& 24  & 6     &23,23,23  &10,11,10& 10,11,10& 24   & 23,23,23& 10& 18  & 15      \\
\hline
\end{tabular}
\\
(1) \citet{abel...97}; 
(2) \citet{1973A&A....25..137A}; 
(3) \citet{1981MNRAS.197..553B}; 
(4) \citet{1960MNRAS.121..471B}; 
(5) \citet{1992ApJS...78..341C}; 
(6) \citet{1970PhDT.........2C}; 
(7) \citet{1994MNRAS.269..563F}; 
(8) \citet{1996ApJ...464..523H}; 
(9) \citet{1997MNRAS.292...27H}; 
(10) \citet{1994MNRAS.268..109H}; 
(11) \citet{1998MNRAS.297.1073H}; 
(12) \citet{1987ephh.book.....J}; 
(13) \citet{1974afcp.book.....L}; 
(14) \citet{1974agn..book.....O}; 
(15) \citet{1989agna.book.....O}; 
(16) \citet{1971phco.book.....P}; 
(17) \citet{1993ppc..book.....P}; 
(18) \citet{1987ApJ...318...32S}; 
(19) \citet{1979ApJ...232....1S}; 
(20) \citet{1978ppim.book.....S}; 
(21) \citet{1986MNRAS.221..635T}; 
(22) \citet{1996ApJ...465..487V}; 
(23) \citet{1983MNRAS.202P..15A}; 
(24) \citet{1997ApJS..109..517R};
(25) \citet{1997ADNDT..65....1V};  
\\
Notes: the codes $C^2$-Ray, Flash-HC and Coral all share the same
nonequilibrium chemistry module (DORIC, developed by G. Mellema), so they have
the same hydrogen chemistry and heating rates and cross-section, and the same 
chemistry solver. However, Coral also includes the chemistry of helium and a
number of metals.
\end{table*}

The solution of the radiative transfer equation is intimately related to the
ionization and thermal states of the gas. These depend on the atomic physics
reaction rates, { photoionization cross-sections}, as well as cooling and heating 
rates used. { As there is a variety of rates available in the literature, for
the sake of clarity we have summarized those used in our codes in
Table~\ref{tablerates} and plotted them in Figure~\ref{rates_fig}.  
The table columns indicate, from left to right, the name of the code and the 
reference for: case A recombination rate of H~II, He~II and He~III;
case B recombination rate of H~II, He~II and He~III;
dielectronic recombination rate of He~II;
collisional ionization rate of H~I, He~I and He~III;
collisional ionization cooling rate of H~I, He~I and He~II;
case A recombination cooling rate of H~II, He~II and He~III;
case B recombination cooling rate of H~II, He~II and He~III;
dielectronic recombination cooling rate of He~II;
collisional excitation cooling rate of H~I, He~I and He~II;
Bremsstrahlung cooling rate; Compton cooling rate; cross-section of H~I, He~I and He~II.
Note that for those codes in which the treatment of He is not included, only the 
references to the H rates are given.} 

The first thing to note in Figure~\ref{rates_fig} is that although the rates
come from a wide variety of sources, they largely agree. The main differences 
are in our recombination cooling rates, particularly at very high
temperatures, beyond the typical range of gas temperatures achieved by
photoionization heating ($T\lesssim10^5$~K). It should be noted, however, that
e.g. shock-heated gas can reach much higher temperatures, in which case the
differences in our rates become very large and caution should be excersised in
choosing the appropriate rates. However, even at the typical photoionization
temperatures there are differences between the rates by up to factors of
$\sim2$. The origin of these discrepancies is currently unclear, and it is
also unclear which fit to the experimental data is more precise.   
There are also a few cases in which particular cooling rates (e.g. Zeus Case B
recombination cooling, ART line cooling) are notably different from the
rest.

\begin{figure}
\begin{center}
  \includegraphics[width=3.5in]{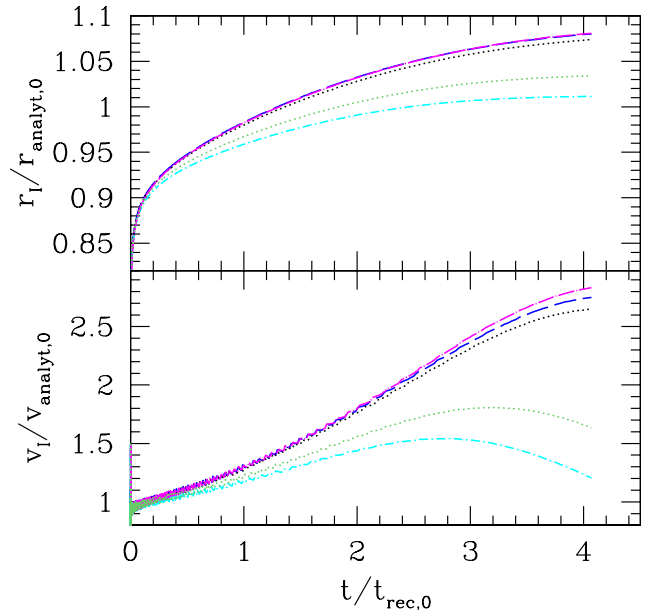}
\caption{Test 0, part 2: I-front expansion in uniform-density field with 
temperature evolution (same as Test 2 below). Plotted are the I-front 
position (top) and velocity (bottom) all derived from the same code (1D, 
spherically-symmetric version of ART code) and using the photoionization 
cross-section from that code, but with the rest of the microphysics 
(chemistry and cooling rates) taken from several of the participating 
codes, as indicated by line-type and color. All results are normalized 
to the analytical ones (which assume fixed temperature, $T=10^4$ K) given
in equations~(\ref{strom0}) below (see text for details).
\label{T0_front_fig}}
\end{center}
\end{figure}

\begin{figure*}
\begin{center}
  \includegraphics[width=3.2in]{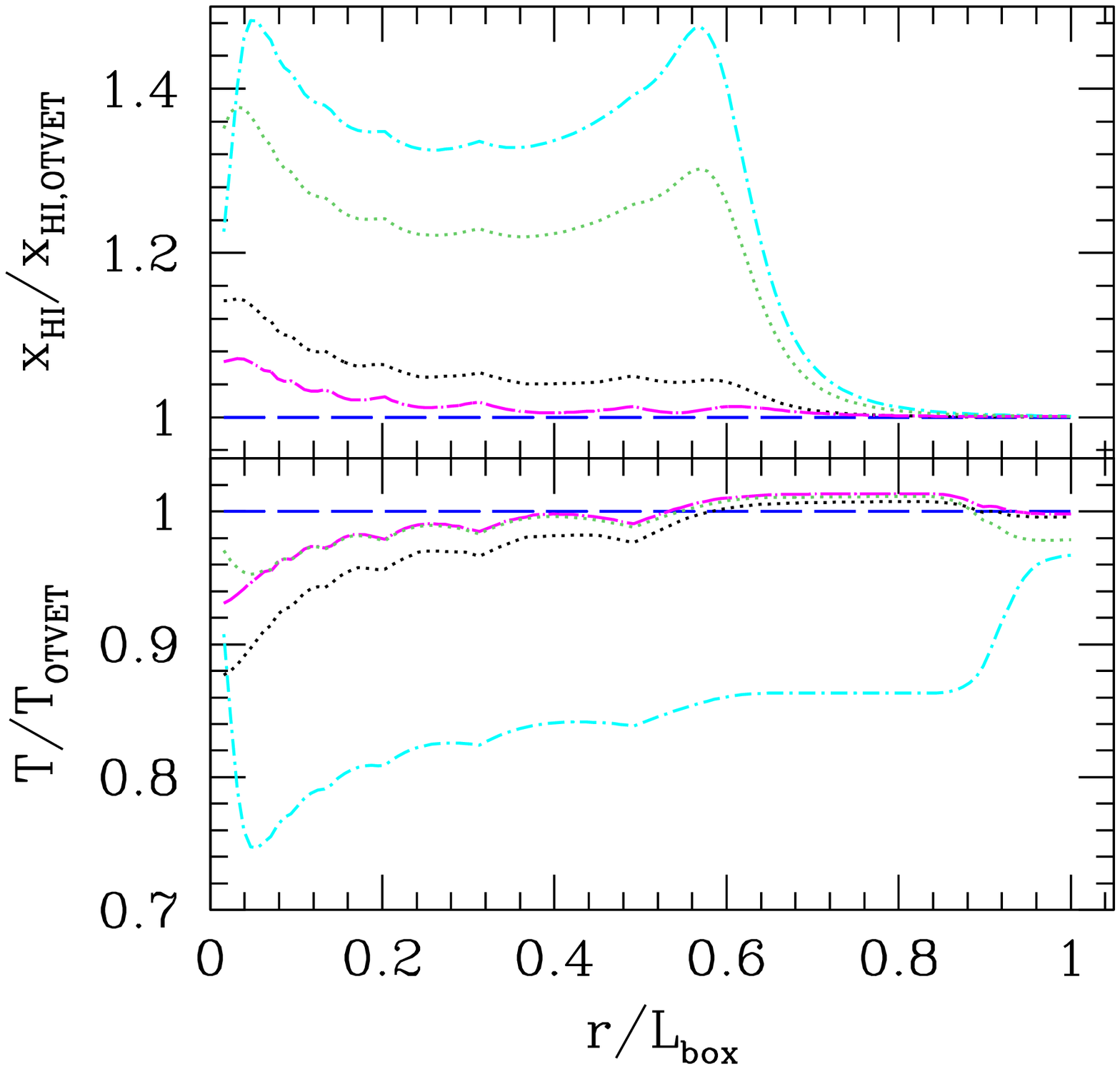}
  \includegraphics[width=3.2in]{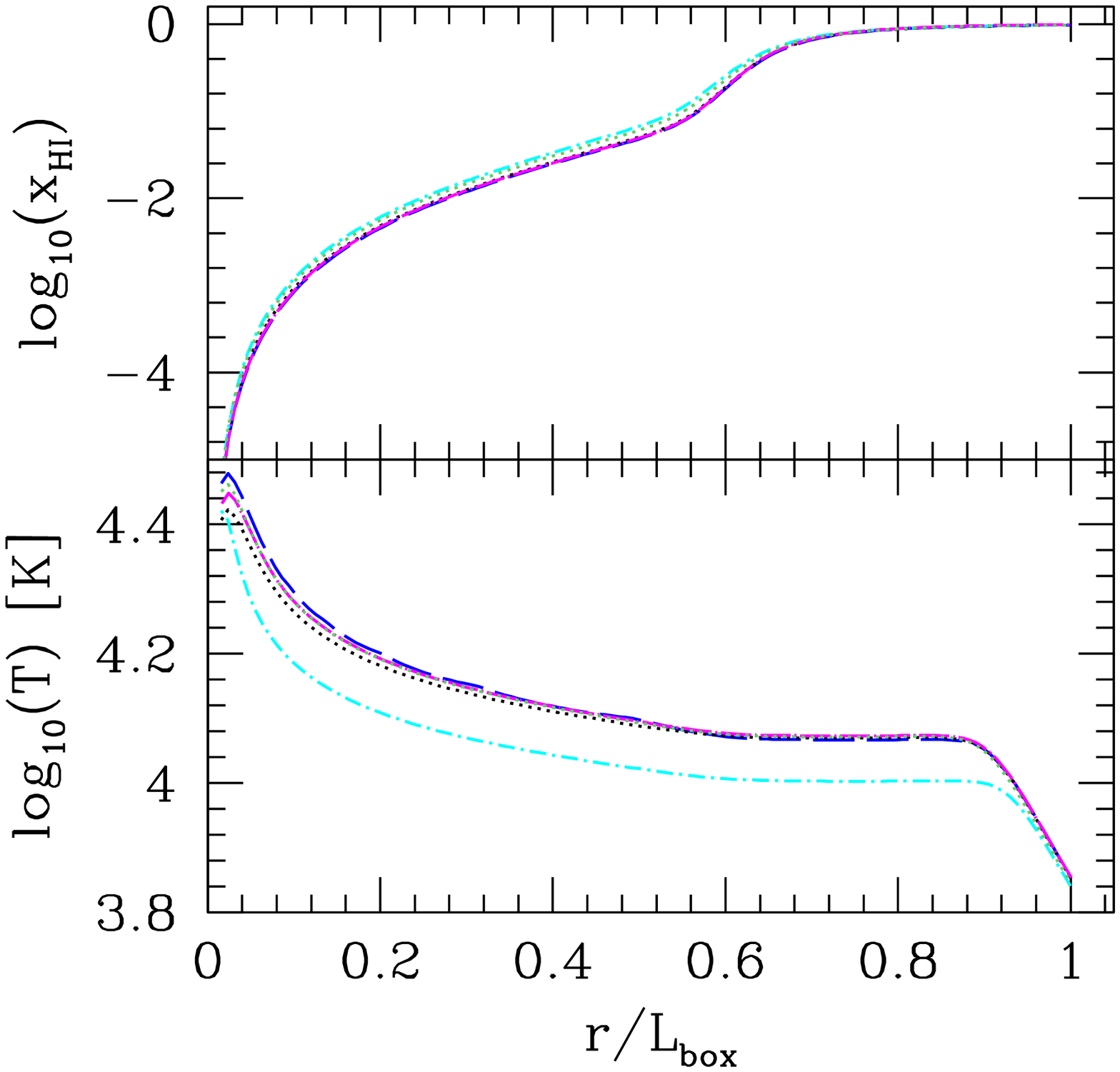}
\caption{Test 0, part 2: Same test as in Fig.~\ref{T0_front_fig}. 
(left) (a) The radial profiles of the neutral fraction (top, normalized to the 
result which uses the OTVET rates), temperature (middle, again normalized to 
the results for OTVET rates) at time $t=100$~Myr$=0.82t_{\rm rec,0}$.
(right) (b) The same as in (a) but in absolute units.
\label{T0_cuts_fig}}
\end{center}
\end{figure*}

\begin{figure}
\begin{center}
  \includegraphics[width=3.2in]{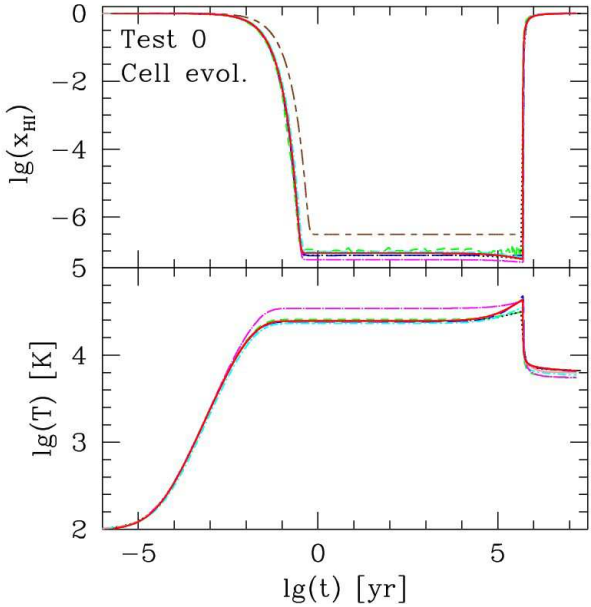}
\caption{Test 0, part 3: Single-zone ionizing up and cooling and recombining. 
\label{T0_evol_fig}}
\end{center}
\end{figure}

As a next step, we did several numerical experiments to assess the offsets
between our results that can arise solely based on our different chemistry and
cooling rates. We did this by implementing the rates from several codes
representative of the full range of rates present above into a single code
(1-D version of ART code) and running the same problem (Test 2 described 
below, which is expansion of an H~II region in uniform density gas, see
\S~\ref{T2_sect} for detailed test definition and solution features). In all
cases the same photoionization cross-section (the one of ART) is used and
only the rates are varied. The code OTVET here stands also for the codes 
$C^2$-Ray, CRASH, Flash-HC and Coral, since all these codes have either 
identical, or closely-matching rates. Results are shown in 
Figures~\ref{T0_front_fig} and \ref{T0_cuts_fig}. 
In Figure~\ref{T0_front_fig} we show the I-front position,
$r_I$, and the I-front velocity, $v_I$ (Both quantities are normalized to the 
analytical solutions of that problem obtained at temperature $T=10^4$ K, while 
the time is in units of the recombination time at the same temperature). Note 
that the results for Zeus use the recombination cooling rate of ART since its 
currently-implemented rate is overly-simplified. We show the results using
the rates of OTVET, FTTE, RSPH, Zeus-MP and ART. 

Initially, when the
I-front is fast and still far away from reaching its Str\"omgren sphere the
results for all codes agree fairly well, as expected since recombinations are
still unimportant. Once recombinations do become important, at $t>t_{\rm
  rec}$, the results start diverging. The results for OTVET, FTTE and RSPH 
remain in close agreement, within a fraction of once per cent
in the I-front radius and within $\sim2\%$ in the I-front velocity. The
results using the Zeus and ART rates however depart noticeably from the
others, by up to 4\% and 6, respectively in radius. The corresponding
velocities are different even more, by up to factor of $\sim 2$ at the end,
when the I-front is close to stationary. In Figure~\ref{T0_cuts_fig} we show
the radial profiles of the ionized fraction $x$, and temperature $T$,
normalized to the results using OTVET rates (left), and in absolute units
(right). The ionized fraction profiles for ART and Zeus again are fairly
different from the rest, by 20-40\%, while the rest of the codes agree between
themselves much better, to less than 10\%. Agreement is slightly worse close
to the ionizing source. In terms of temperature profiles the codes agree to
better than 10\%, with the exception of ART, in which case the resulting
temperature is noticeably lower, by $\sim20\%$.

\begin{figure*}
\begin{center}\hspace{6mm}
  \includegraphics[width=2in]{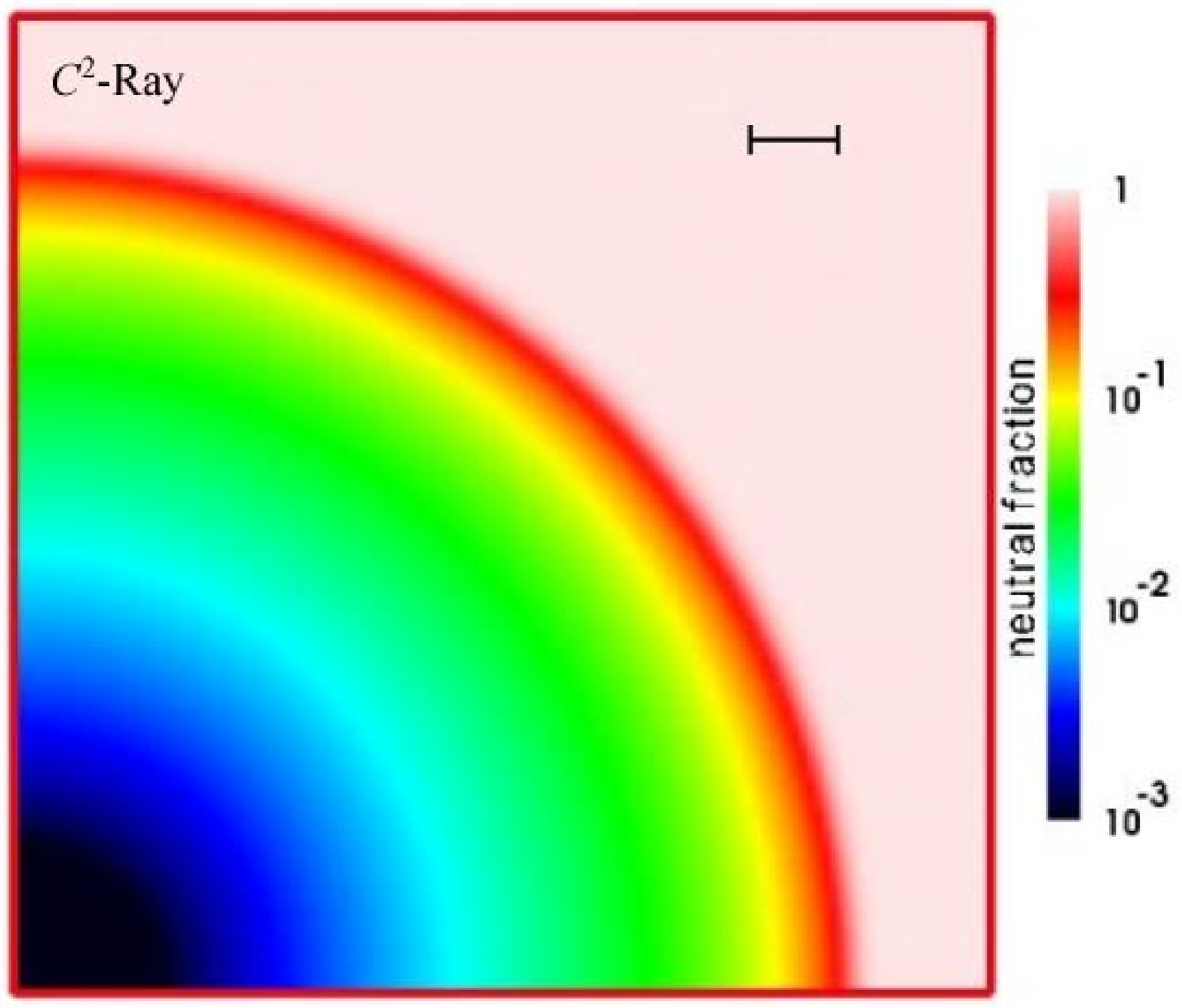}
  \includegraphics[width=2.3in]{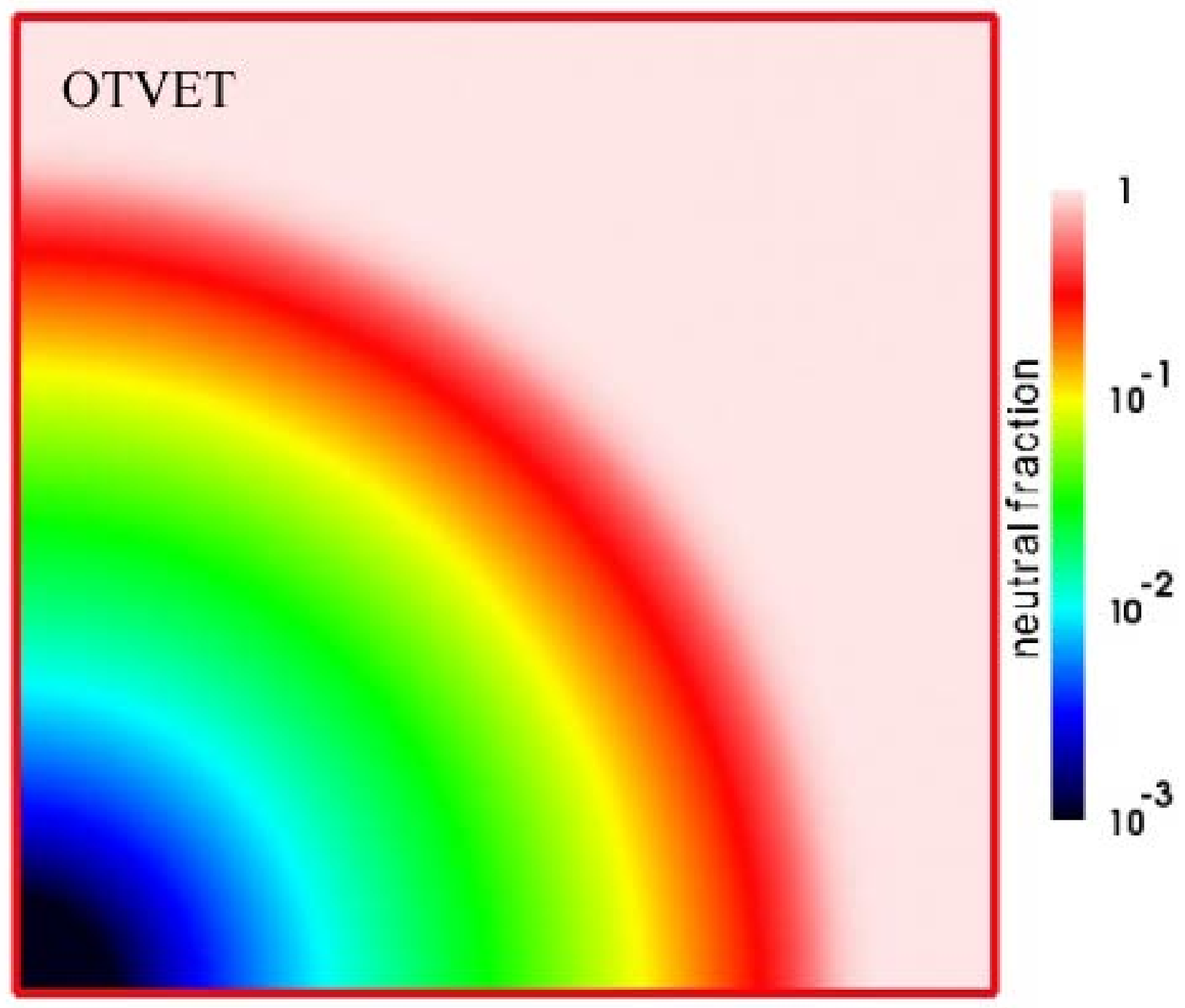}
  \includegraphics[width=2.3in]{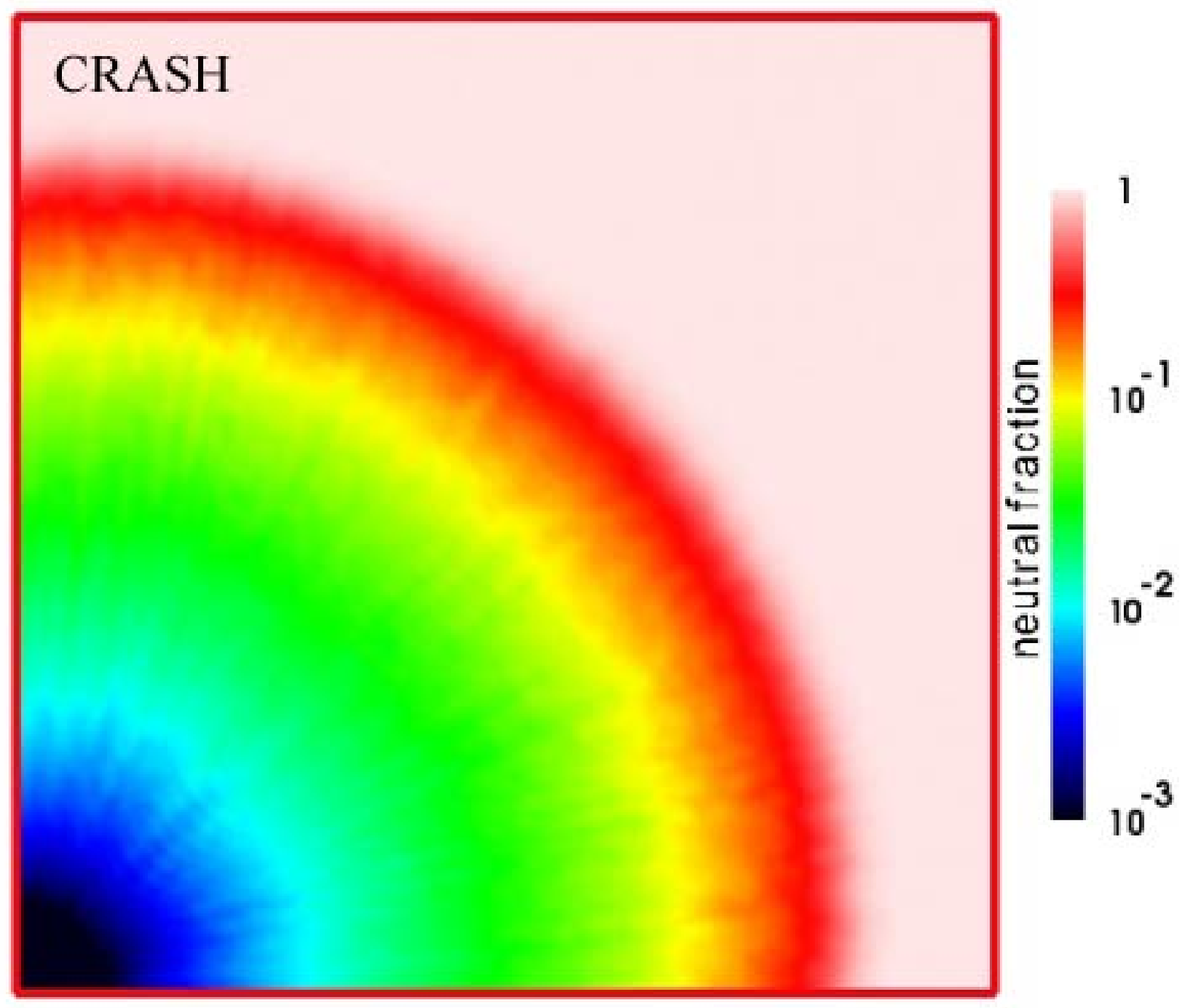}
  \includegraphics[width=2.3in]{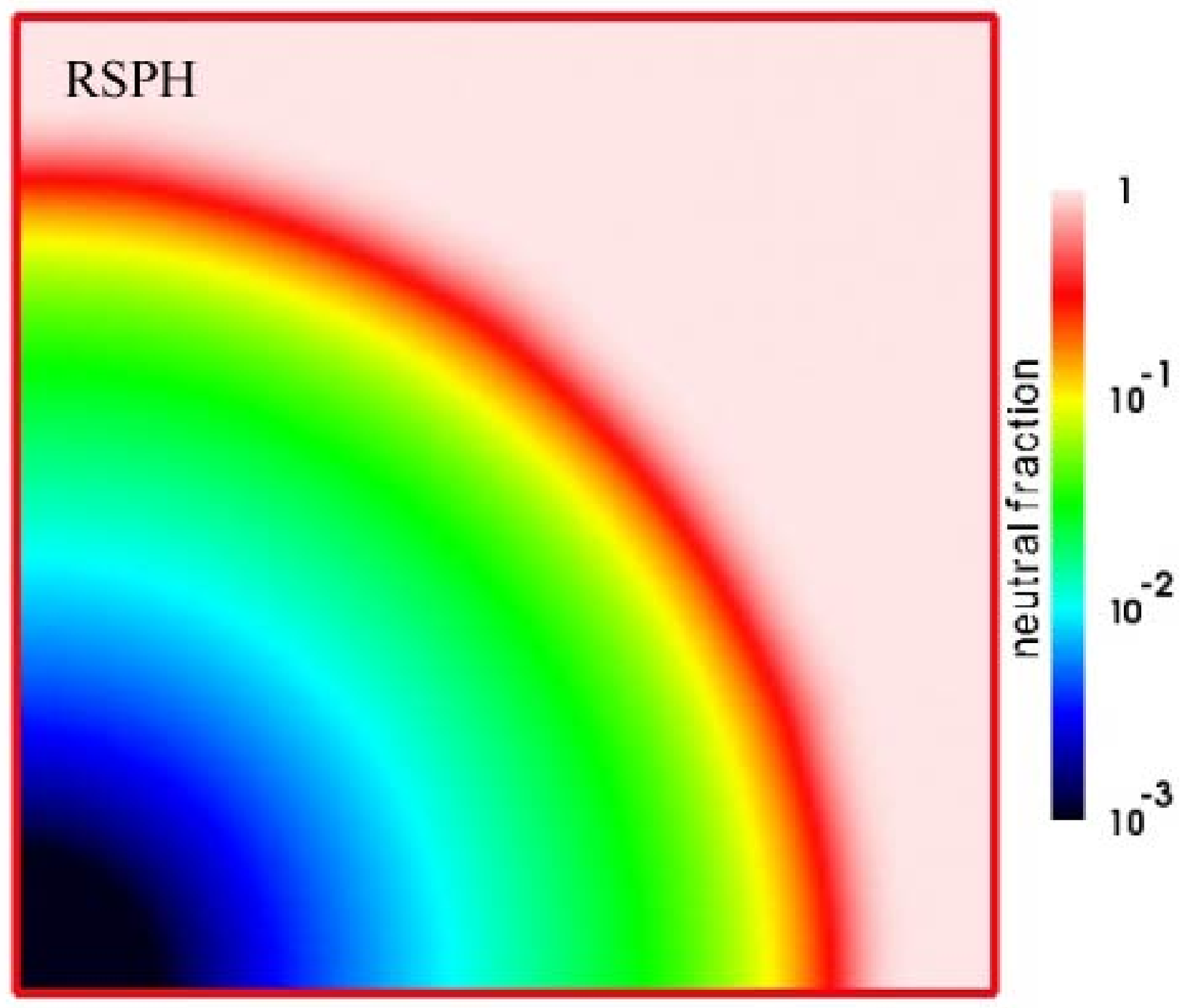}
  \includegraphics[width=2.3in]{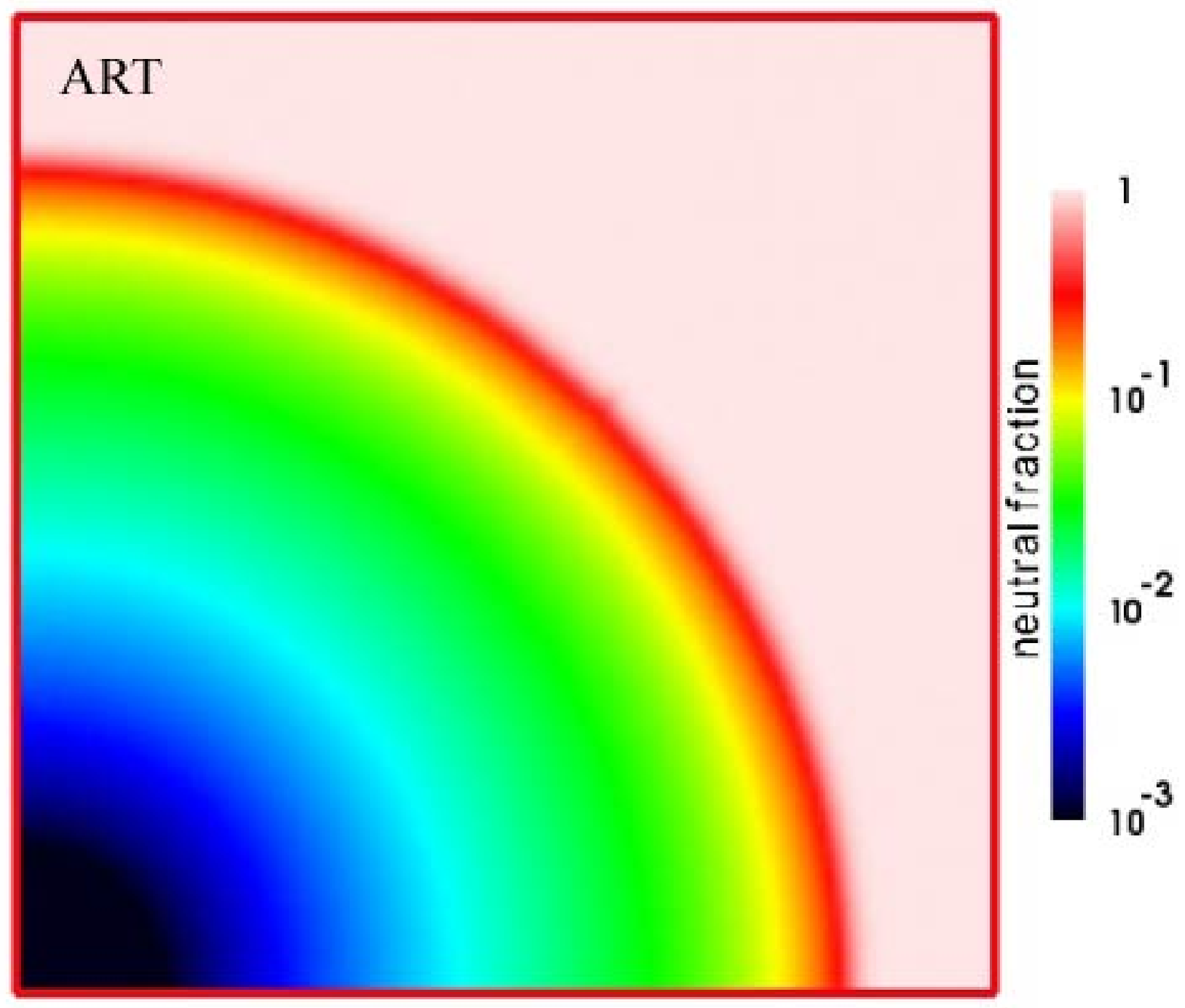}
  \includegraphics[width=2.3in]{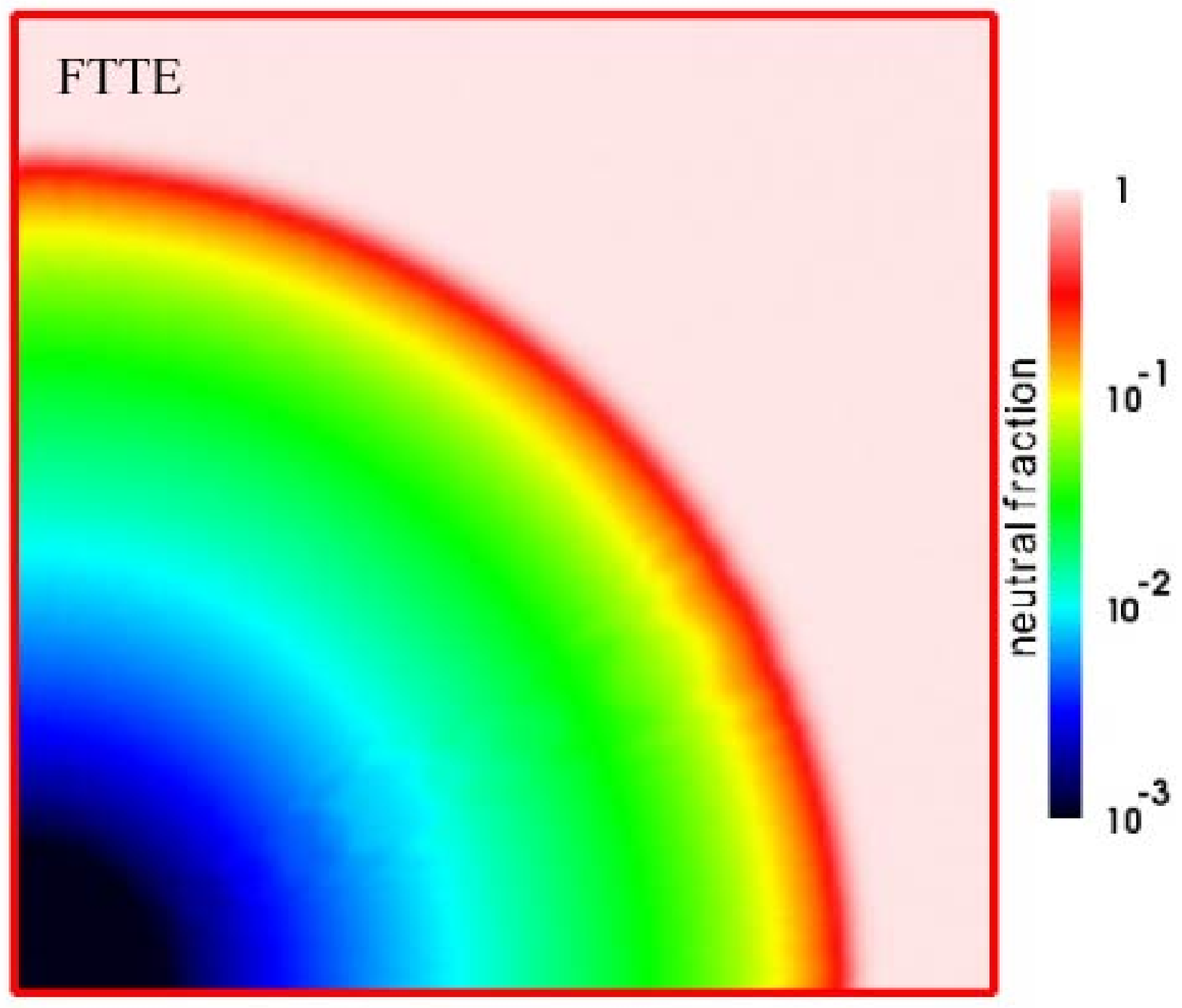}
  \includegraphics[width=2.3in]{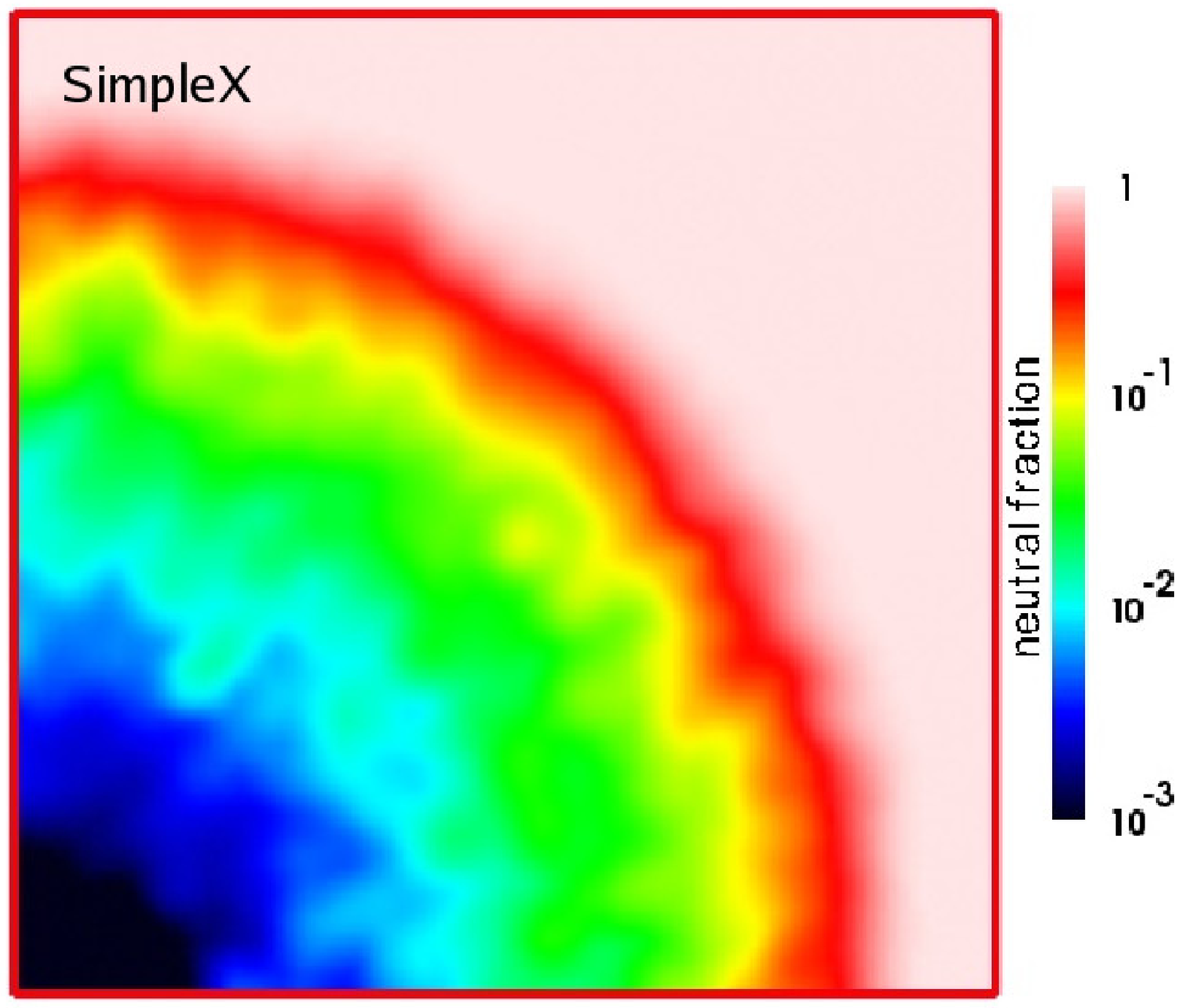}
  \includegraphics[width=2.3in]{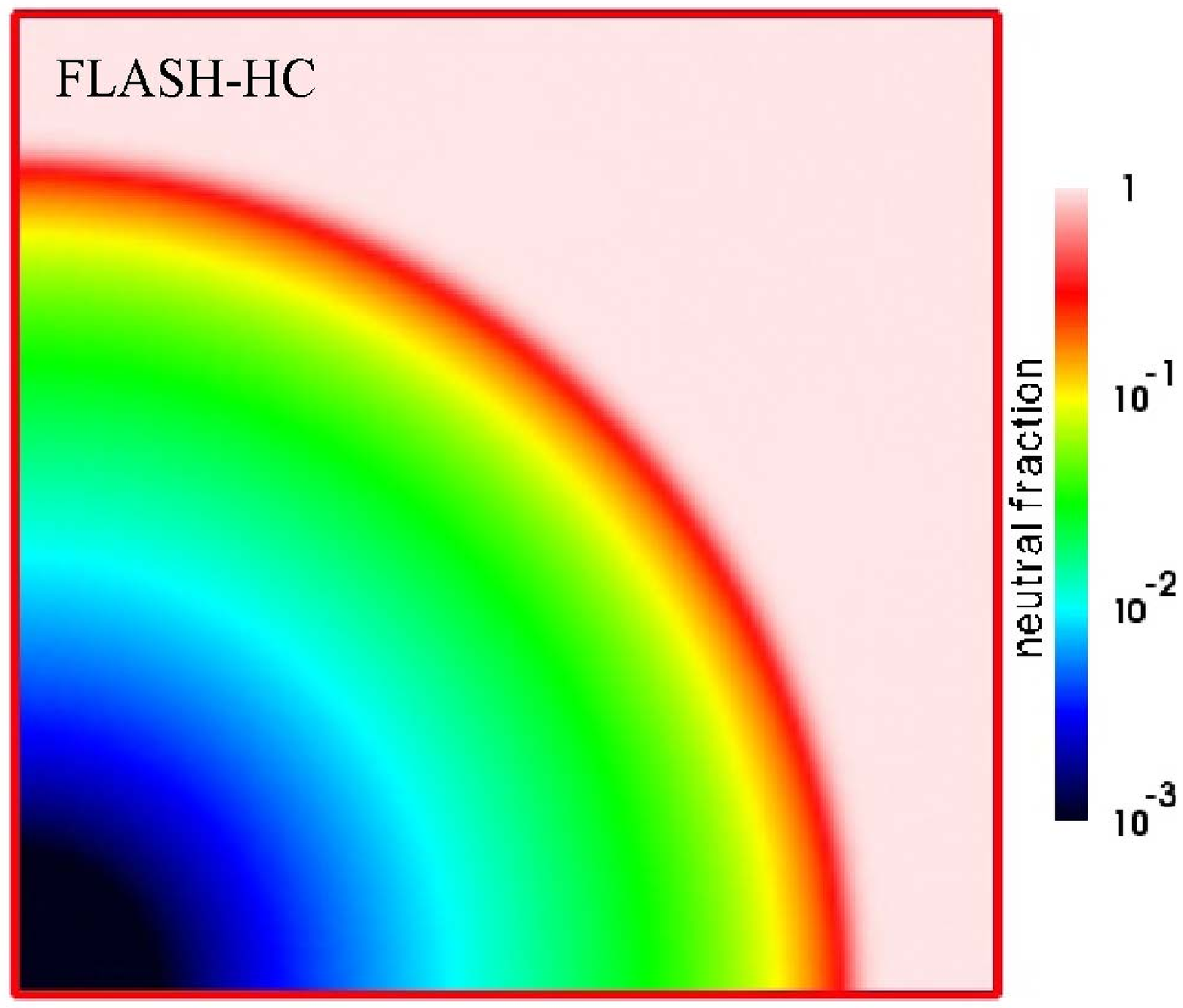}
  \includegraphics[width=2.3in]{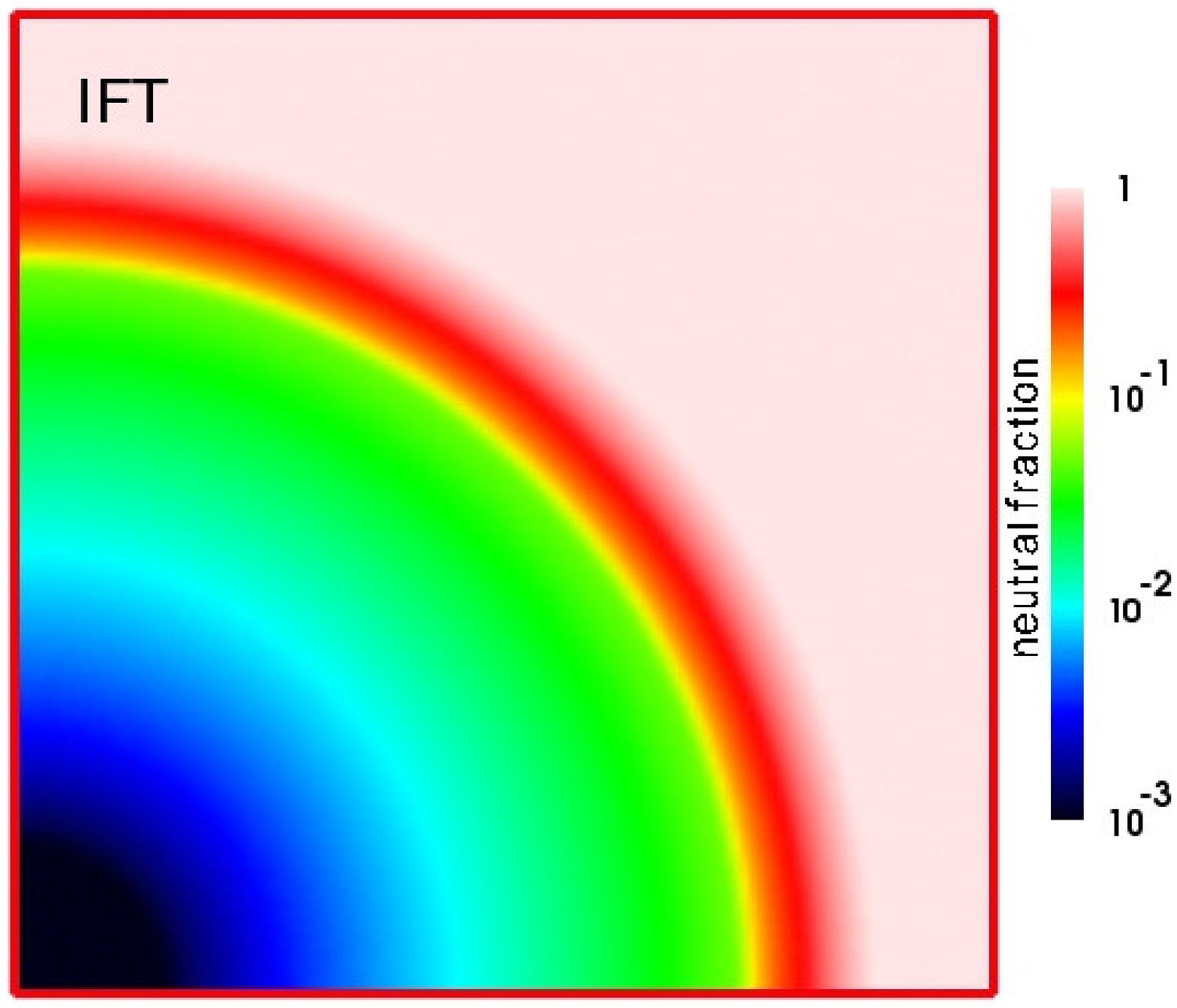}
\caption{Test 1 (H~II region expansion in an uniform gas at fixed 
temperature): Images of the H~I fraction, cut through the 
simulation volume at coordinate $z=0$ at time $t=500$ Myr 
(final Str\"omgren sphere) for (left to right and top to 
bottom) $C^2$-Ray, OTVET, CRASH, RSPH, ART, FTTE, SimpleX, 
  FLASH-HC, and IFT.
\label{T1_images_fig}}
\end{center}
\end{figure*}

The reasons for these discrepancies lie largely in differences in the
recombination rates and the recombination and line cooling rates. 
The line cooling rate of ART is larger than the ones for most of the 
other codes by a factor of 2 - 5 in the temperature range $10^4 - 10^5$~K,
while the recombination rate used by that code is about 10-15
percent larger in the same temperature range, thus the resulting gas
temperature is correspondingly lower. This lower temperature results in a
higher recombination rate and hence the slower propagation of the I-front we
observed. The reason for the discrepancy with Zeus is mostly due to its
higher recombination rate. This again results in a somewhat slower I-front
propagation, but not in significant temperature differences. Accordingly, for
both of these codes the neutral gas fraction is significantly higher at all
radii. The slightly higher recombination cooling of RSPH results in slightly
lower temperature and proportionally higher neutral fraction, although both
are off by only a few percent, and up to 10\% close to the ionizing source.
 
The last important element in this basic physics comparison is to assess the
accuracy and robustness of the methods we use for solving the non-equilibrium
chemistry equations. These equations are stiff and thus generally require
implicit solution methods. Such methods are generally expensive, however, so 
often certain approximations are used to speed-up the calculations.   
In order to test them, we performed the following simple test with a 
single, optically thin zone. We start with a completely neutral zone 
at time $t=0$. We then applied photoionizing flux of 
$F=10^{12}\,\rm photons/s/cm^2$, with $10^5$ K black-body spectrum for 
0.5 Myr, which results in the gas parcel becoming heated and highly 
ionized. Thereafter, the ionizing flux is switched off and the zone 
cools down and recombines for further 5 Myr. The zone contains 
only hydrogen gas with number density of $n=1\, cm^{-3}$, and initial 
temperature of $T_i=100$~K. Our results are shown in 
Figure~\ref{T0_evol_fig}.

All codes agree very well in terms of the evolution of the neutral fraction 
(top panel), with the sole exception of SimpleX, in which case both the speed
with which the gas parcel ionizes up and the achieved level of ionization are 
significantly different from the rest. The reason for this discrepancy is that
currently this code does not solve the energy equation to find the gas
temperature but has to assume a value instead ($T=10^4$ K in this test). 
FTTE finds slightly higher temperatures after its initial rise and
correspondingly lower neutral fractions.  

Some differences are also seen after time $t\sim0.1$~Myr, at which point there
is a slight rise in temperature and corresponding dip in the neutral
fraction. These occur around the time when the recombinations start becoming
important, since $t_{\rm rec}\sim0.1$~Myr, which gives rise to slight
additional heating. About half of the codes predict somewhat lower 
temperature rise then the rest.  
The cooling/recombination phase after source turn off demonstrates good
agreement between the codes, although there is small difference in the final
temperatures reached which is due to small differences in the hydrogen line
cooling rates, resulting in slightly different temperatures at which the
cooling becomes inefficient.

\subsection{Test 1: Pure-hydrogen isothermal H~II region expansion}

\begin{figure} 
\begin{center}
  \includegraphics[width=3.2in]{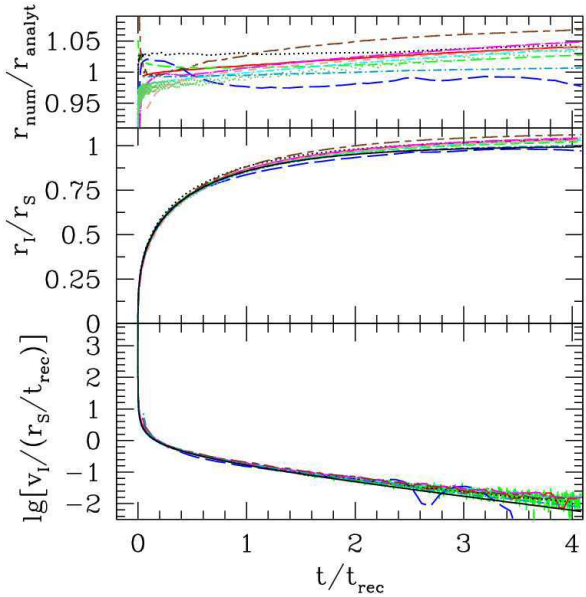}
\caption{Test 1 (H~II region expansion in an uniform gas at fixed 
temperature): The evolution of the position and velocity of the I-front. 
\label{T1_Ifront_evol_fig}}
\end{center}
\end{figure}

\begin{figure*}
\begin{center}
  \includegraphics[width=3.2in]{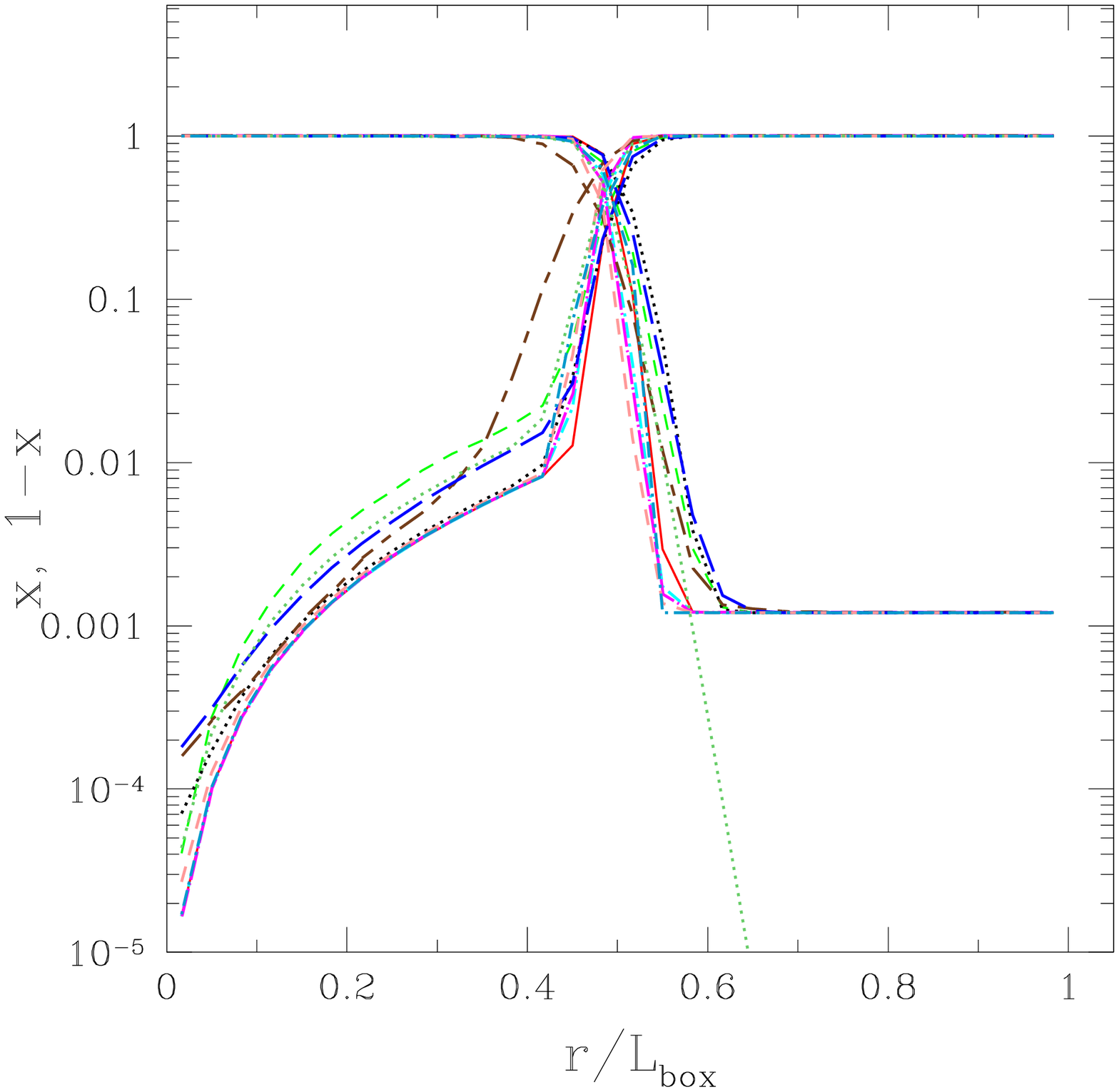}
  \includegraphics[width=3.2in]{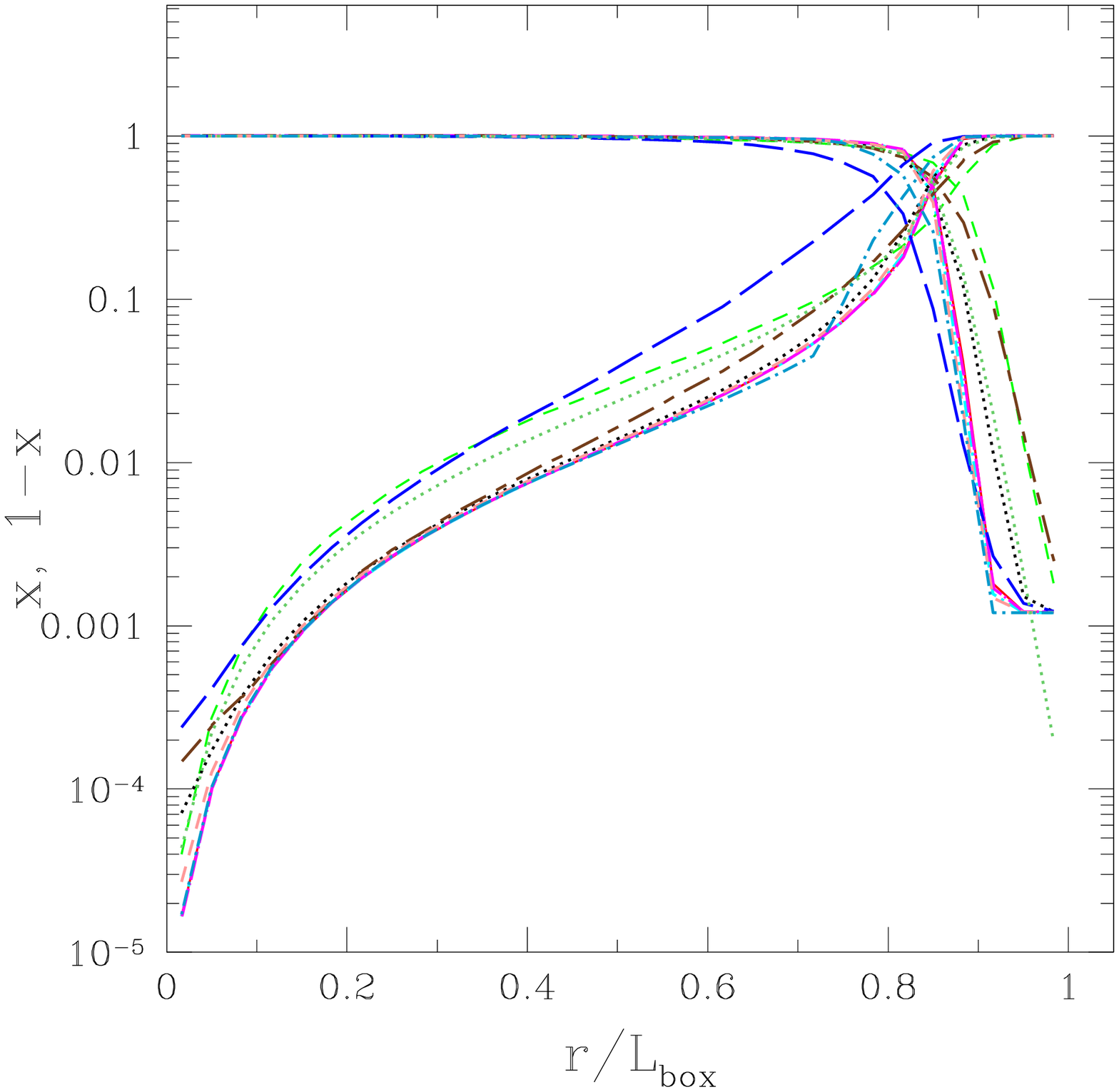}
\caption{Test 1 (H~II region expansion in an uniform gas at fixed 
temperature): Spherically-averaged profiles for ionized fractions $x$ 
  and neutral fractions $x_{\rm HI}=1-x$ at times $t=30$ Myr and 500 Myr
  vs. dimensionless radius (in units of the box size). 
\label{T1_profs_fig}}
\end{center}
\end{figure*}

This test is the classical problem of an H~II region expansion in an uniform
gas around a single ionizing source 
\citep{1939ApJ....89..526S,1978ppim.book.....S}. A steady, monochromatic
($h\nu=13.6$ eV) source emitting $\dot{N}_\gamma$ ionizing photons per unit
time is turning on in an initially-neutral, uniform-density, static
environment with hydrogen number density $n_H$. For this test we assume that
the temperature is fixed at $T=10^4$ K. Under these conditions, and if we 
assume that the front is sharp (i.e. that it is infinitely-thin, with the 
gas inside fully-ionized and the gas outside fully-neutral) there is a 
well-known analytical solution for the evolution of the I-front radius, 
$r_I$, and velocity, $\rm v_I$, given by 
\ba
r_I&=&r_{\rm S}\left[1-\exp(-t/t_{\rm rec})\right]^{1/3}\,,\\
\rm v_I&=&\frac{r_{\rm S}}{3t_{\rm rec}}\frac{\exp{(-t/t_{\rm rec})}}
{\left[1-\exp(-t/t_{\rm rec})\right]^{2/3}}\,,
\label{strom0}
\ea 
where
\begin{equation}
r_{\rm S}=\left[{3\dot{N}_\gamma\over 4\pi \alpha_B(T) n_{\rm
  H}^2}\right]^{1/3}\,,
\end{equation}
is the Str\"omgren radius, i.e. the final, maximum size of the ionized region
at which point recombinations inside it balance the incoming photons and the
H~II region expansion stops. The Str\"omgren radius is obtained from 
\be 
F=\int_0^{r_S}d\ell n_e
n_H\alpha_B(T),
\label{strom}
\ee 
i.e. by balancing the number of recombinations with the number of ionizing
photons arriving along a given line of sight (LOS). Here $n_e$ is the electron 
density, 
\begin{equation}
t_{\rm rec}=\left[\alpha_B(T) n_{\rm H}\right]^{-1}\,,
\end{equation}
is the recombination time, and $\alpha_B(T)$ is the Case B recombination
coefficient of hydrogen in the ionized region at temperature $T$. The H~II
region initially expands quickly and then slows considerably as the evolution 
time approaches the recombination time, $t\sim t_{\rm rec}$, at which point
the recombinations start balancing the ionizations and the H~II region
approaches its Str\"omgren radius. At a few recombination times I-front
stops at radius $r_I=r_S$ and in absence of gas motions remains static
thereafter. The photon mean-free-path is given by 
\be
\lambda_{\rm mfp}=\frac{1}{n_H\sigma_0}=0.041\,\rm pc.
\ee

The particular numerical parameters we used for this test are as follows: 
computational box dimension $L=6.6$~kpc, gas number density 
$n_H=10^{-3}$~cm$^{-3}$, initial ionization fraction (given by collisional 
equilibrium) $x=1.2\times10^{-3}$, and ionization rate 
$\dot{N}_\gamma=5\times10^{48}$ photons\,s$^{-1}$. The source is at the 
corner of the box). For these parameters the recombination time is 
$t_{\rm rec}=3.86\times10^{15}\,\rm s=122.4$ Myr. Assuming a recombination 
rate $\alpha_B(T)=2.59\times10^{-13}\rm\,cm^{3}s^{-1}$ at $T=10^4$ K, then
$r_S=5.4$ kpc.  The simulation time is $t_{\rm sim}=500$ Myr
$\approx4\,t_{\rm rec}$.  The required outputs are the neutral fraction of
hydrogen on the whole grid at times $t=10,30,100,200,$ and 500 Myr, and the
I-front position (defined by the 50\% neutral fraction) and velocity vs. time
along the $x$-axis.

\begin{figure*}
\begin{center}
  \includegraphics[width=2.2in]{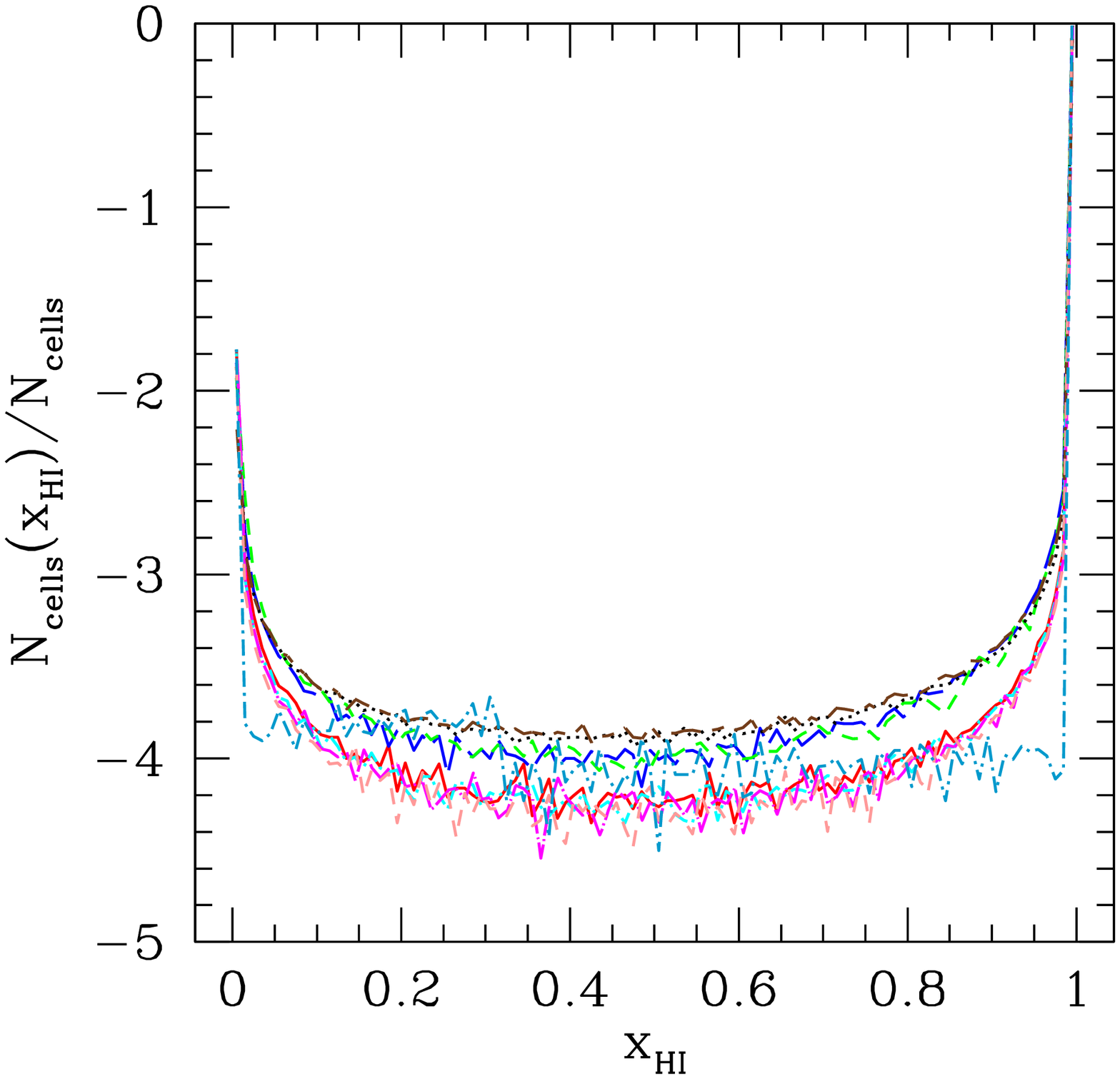}
  \includegraphics[width=2.2in]{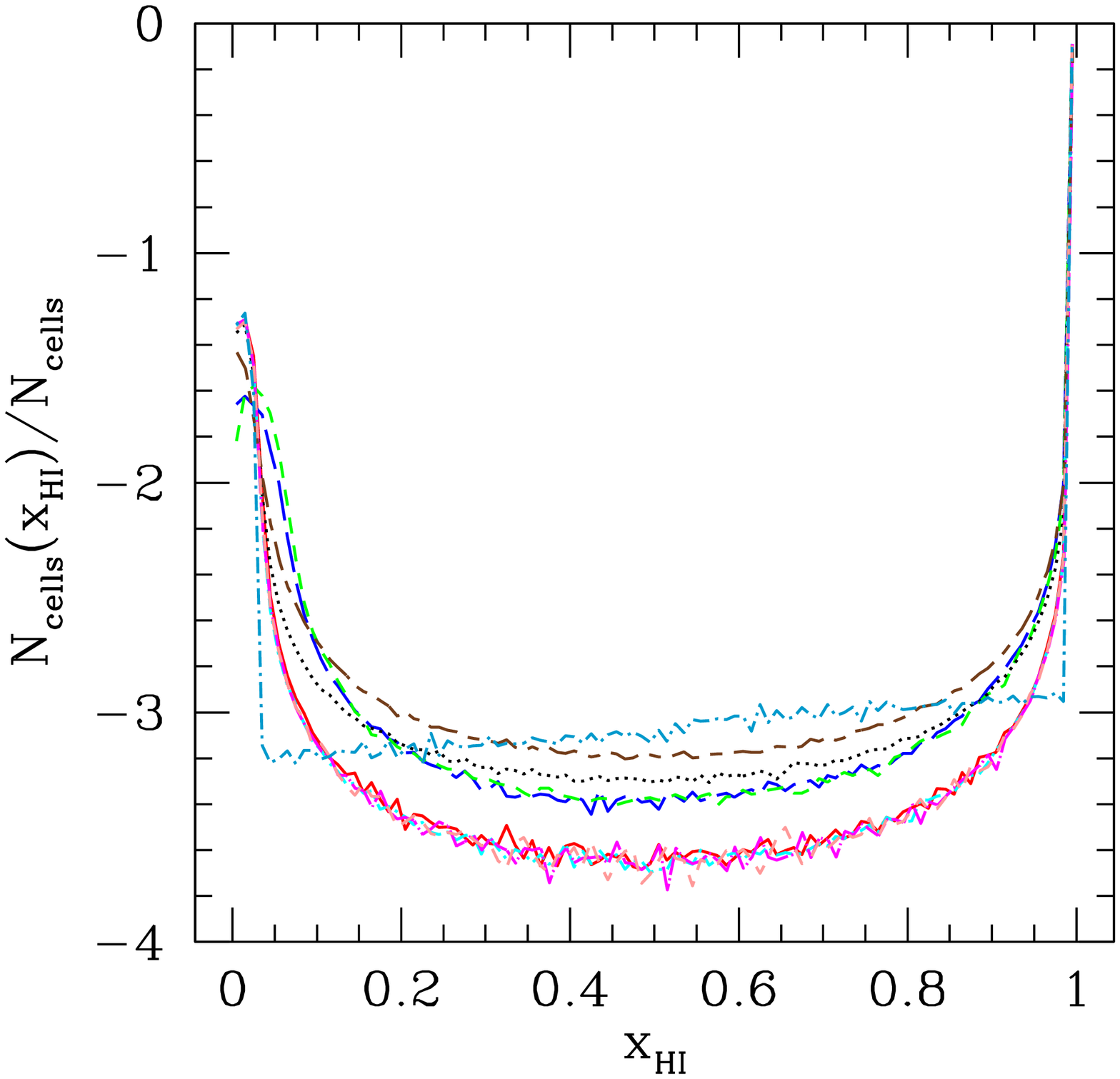}
  \includegraphics[width=2.2in]{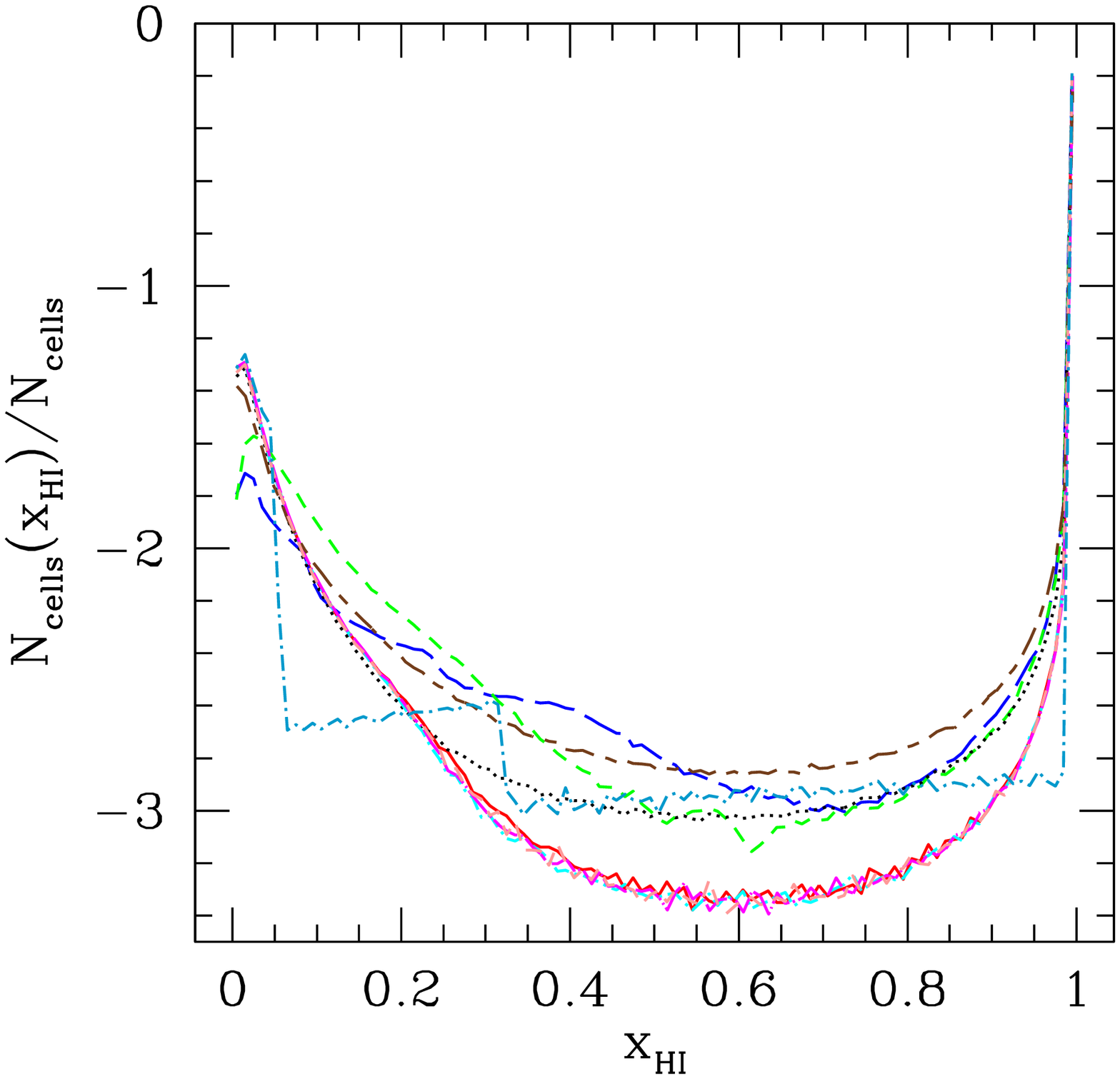}
\caption{Test 1 (H~II region expansion in an uniform gas at fixed 
temperature): Fraction of cells with a given neutral fraction, $x_{\rm HI}=1-x$ at
   times (left) $t=10$ Myr, (middle) 100 Myr and (right) 500 Myr. 
\label{T1_hist_fig}}
\end{center}
\end{figure*}

In Figure~\ref{T1_images_fig} we show images of the neutral fraction in the
$z=0$ plane at time $t=500$~Myr, at which point the equilibrium 
Str\"omgren sphere is reached. The size of the final ionized region is in 
very good agreement between the codes. In most cases the H~II region is nicely
spherical, although some anisotropies exist in the CRASH and SimpleX
results. In the first case these are due to the Monte-Carlo random sampling 
nature of this code, while in the second case it is due to the unstructured
grid used by that code, which had to be interpolated on the regular grid
format used for this comparison. There are also certain differences in the
H~II region ionized structure, e.g. in the thickness of the ionized-neutral
transition at the Str\"omgren sphere boundary. The inherent thickness of this
transition (defined as the radial distance between 0.1 and 0.9 ionized
fraction points) for monochromatic spectrum is 
$\approx 18\lambda_{\rm mfp}=0.74$ kpc, or about 14 simulation cells, equal to
11\% of the simulation box size. This thickness is indicated in the
upper left panel of Figure~\ref{T1_images_fig}. Most codes find widths which
are very close to this expected value. Only the OTVET, CRASH and SimpleX codes 
find thicker transitions due to the inherently greater diffusivity of these methods, 
which spreads out the transition. For the same reason the highly-ionized
proximity region of the source (blue-black colors) is noticeably smaller for
the same two codes.

\begin{figure}
\begin{center}
  \includegraphics[width=3.5in]{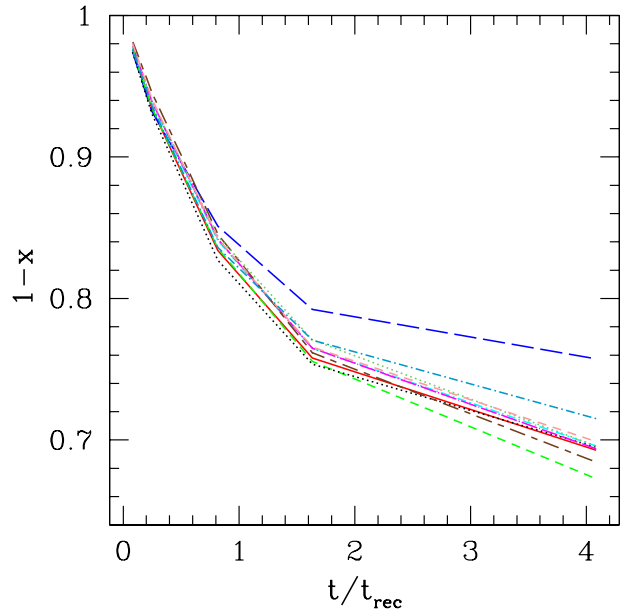}
\caption{Test 1 (H~II region expansion in an uniform gas at fixed 
temperature): Evolution of the total neutral fraction. 
\label{T1_fracs_fig}}
\end{center}
\end{figure}

In Figure~\ref{T1_Ifront_evol_fig} we show the evolution of the I-front
position and velocity. The analytical results in equation~(\ref{strom0}) are
shown as well (black, solid lines). All codes track the I-front correctly, with 
the position never varying by more than 5\% from the analytical solution. 
These small differences are partly due to differences in our recombination 
rates, as discussed above, and partly a consequence of our (somewhat 
arbitrary) definition of the I-front position as the point of 50\%
ionization. Our chosen parameters are such that the I-front internal  
structure is well-resolved, and the I-front intrinsic thickness is larger than
the discrepancies between the different codes. The IFT code in particular 
tracks the 
I-front almost perfectly, as is expected for this code by construction. The
ray-tracing codes agree between themselves a bit better than they do with the 
moment-based method OTVET. This is again related to the different, somewhat 
more diffusive nature of the last code. The I-front velocities also show 
excellent agreement with the analytical result, at least until late times (at
few recombination times), at which point the I-front essentially stops and 
its remaining slow motion forward is not possible to resolve with the 
relatively coarse resolution adopted for our test. The I-front at this point 
is moving so slowly that most of its remaining motion takes place within a 
single grid cell for extended periods of time and thus falls below our
resolution there.  

In Figure~\ref{T1_profs_fig} we plot the spherically-averaged radial profiles
of the ionized and the neutral fraction. In the left panel we show these
profiles at $t=30$~Myr, during the early, fast expansion of the I-front. Most 
of the ray-tracing codes ($C^2$-Ray, ART, FLASH-HC and IFT) agree excellently 
at all radii. The OTVET, CRASH and SimpleX codes appear more diffusive, finding a 
thicker I-front transition and lower ionized fraction inside the H~II region. 
The Zeus code also derives lower ionized fractions inside the H~II region due
to its slightly higher recombinational coefficient. The RSPH code is
intermediate between the two groups of codes, finding essentially the same 
neutral gas profile inside the H~II region as the ray-tracing codes, but a
slightly thicker I-front, i.e. the ionized fraction drops more slowly ahead of
the I-front. 

\begin{figure*}
\begin{center}
  \includegraphics[width=2.3in]{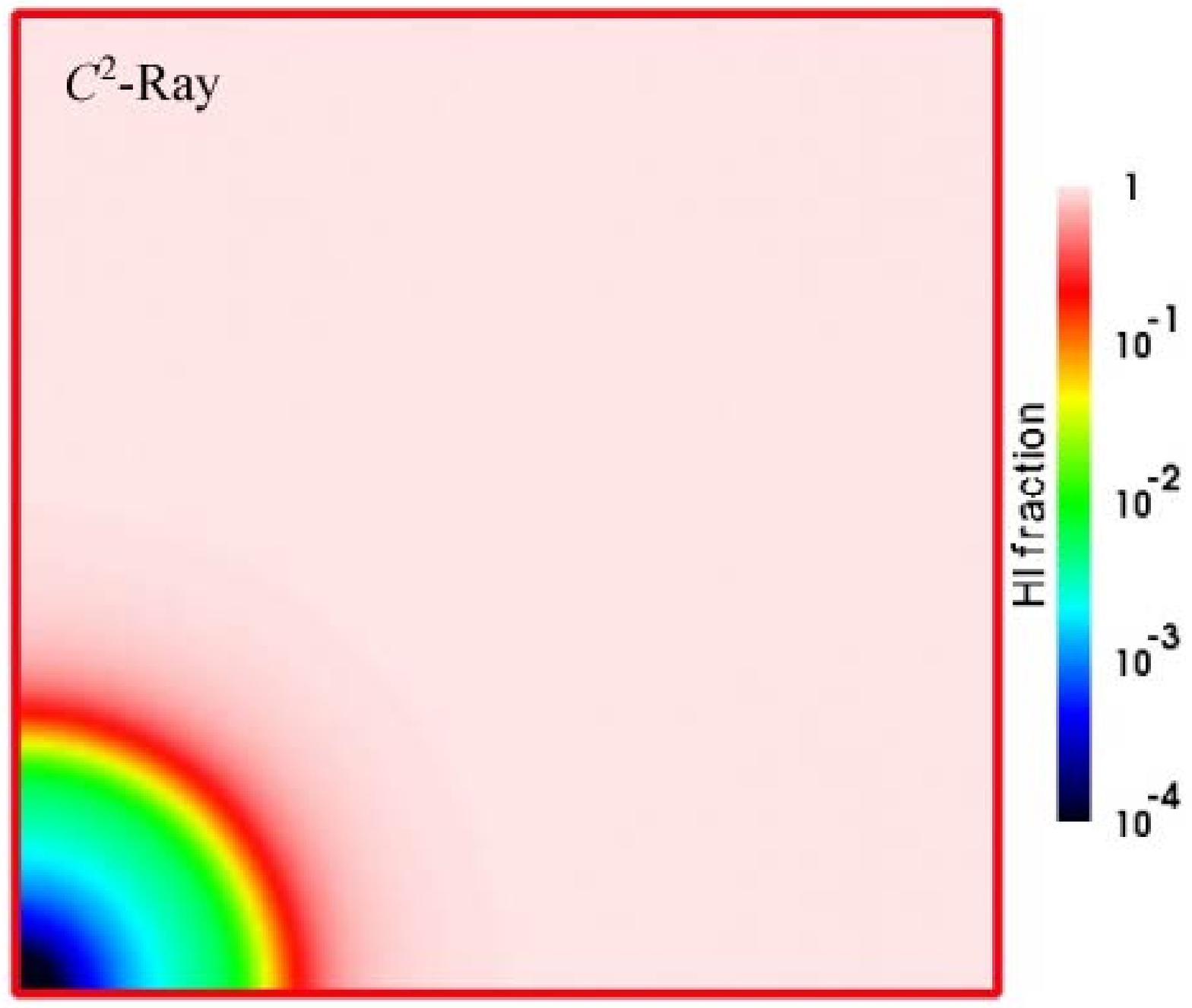}
  \includegraphics[width=2.3in]{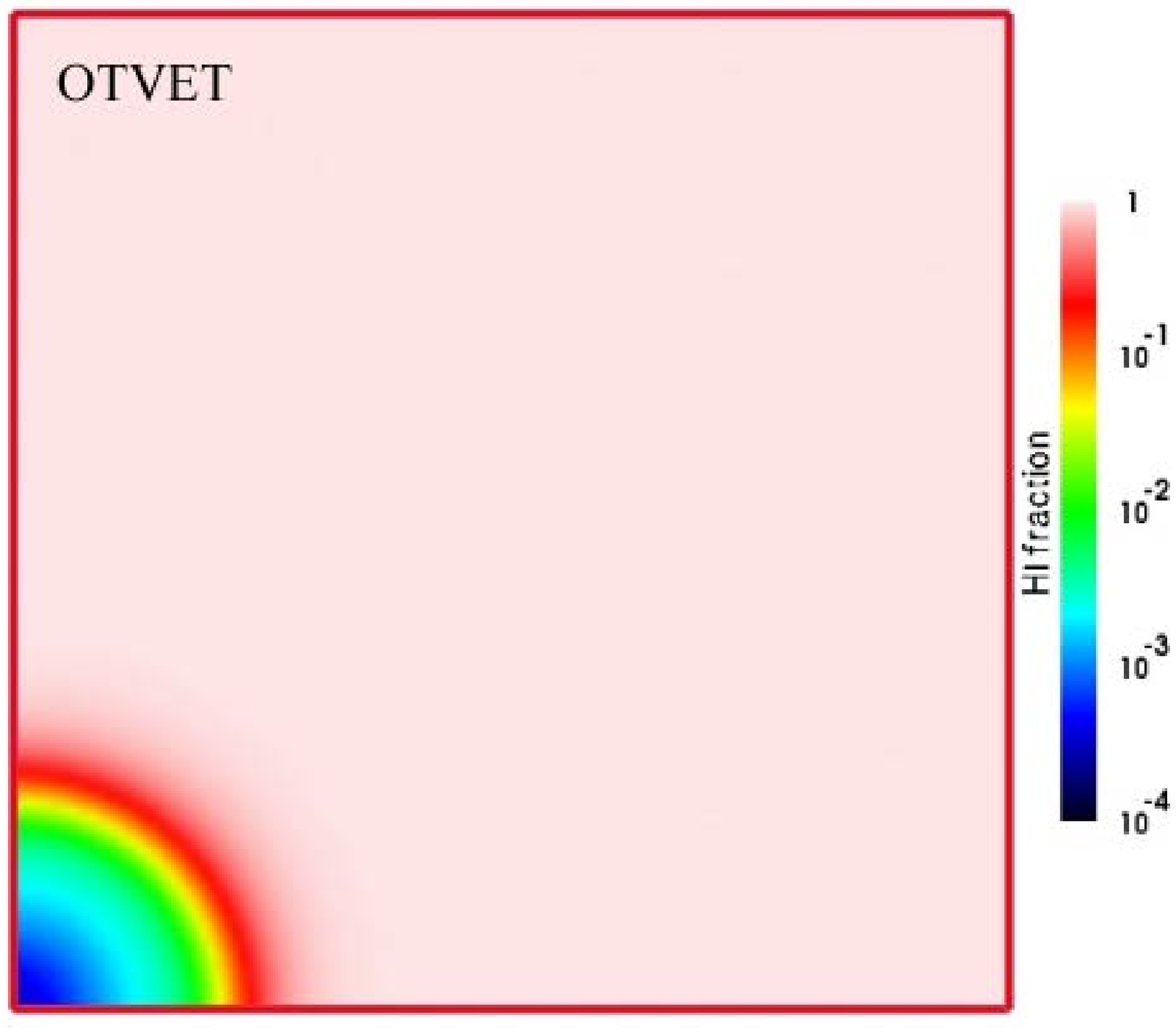}
  \includegraphics[width=2.3in]{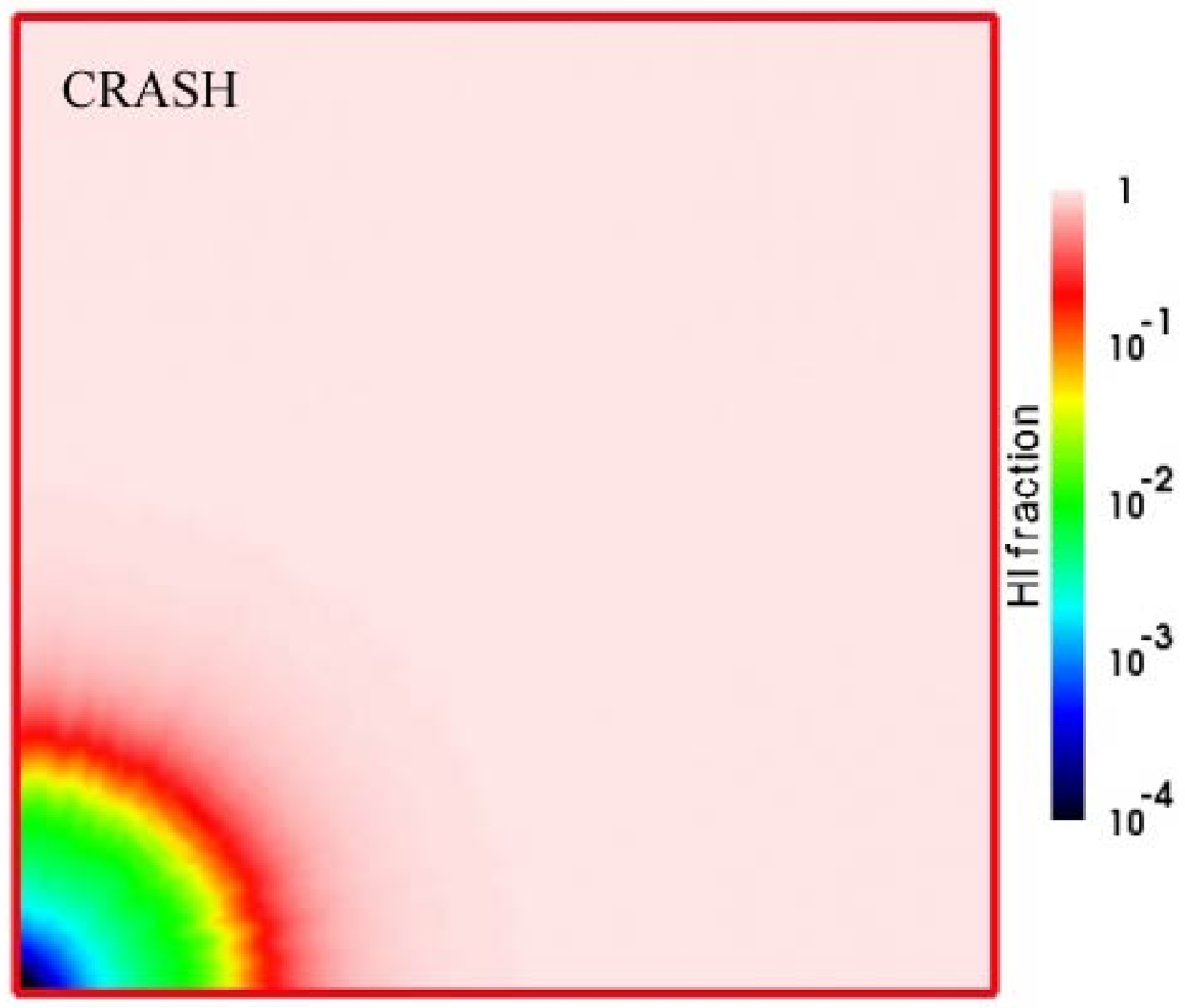}
  \includegraphics[width=2.3in]{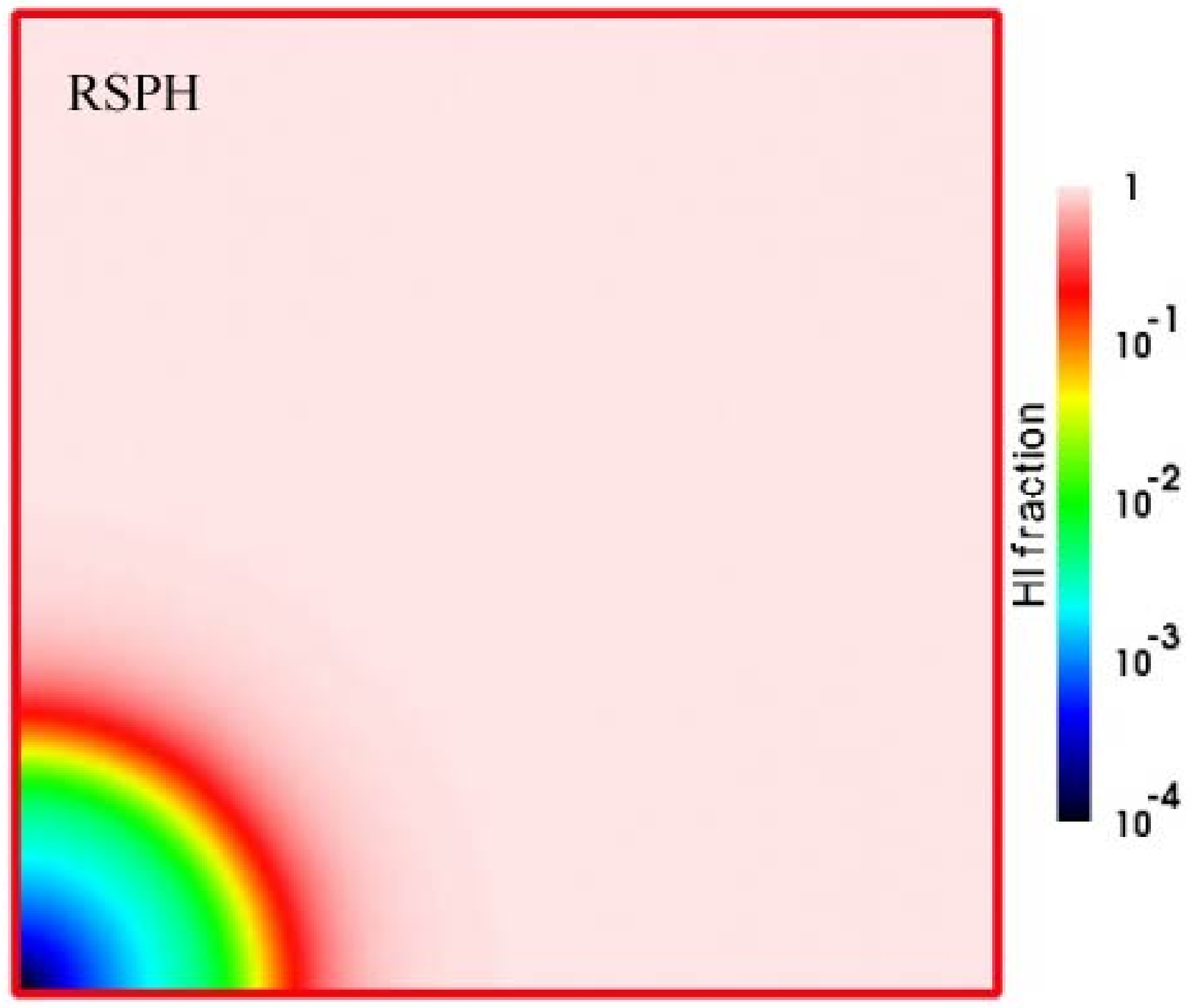}
  \includegraphics[width=2.3in]{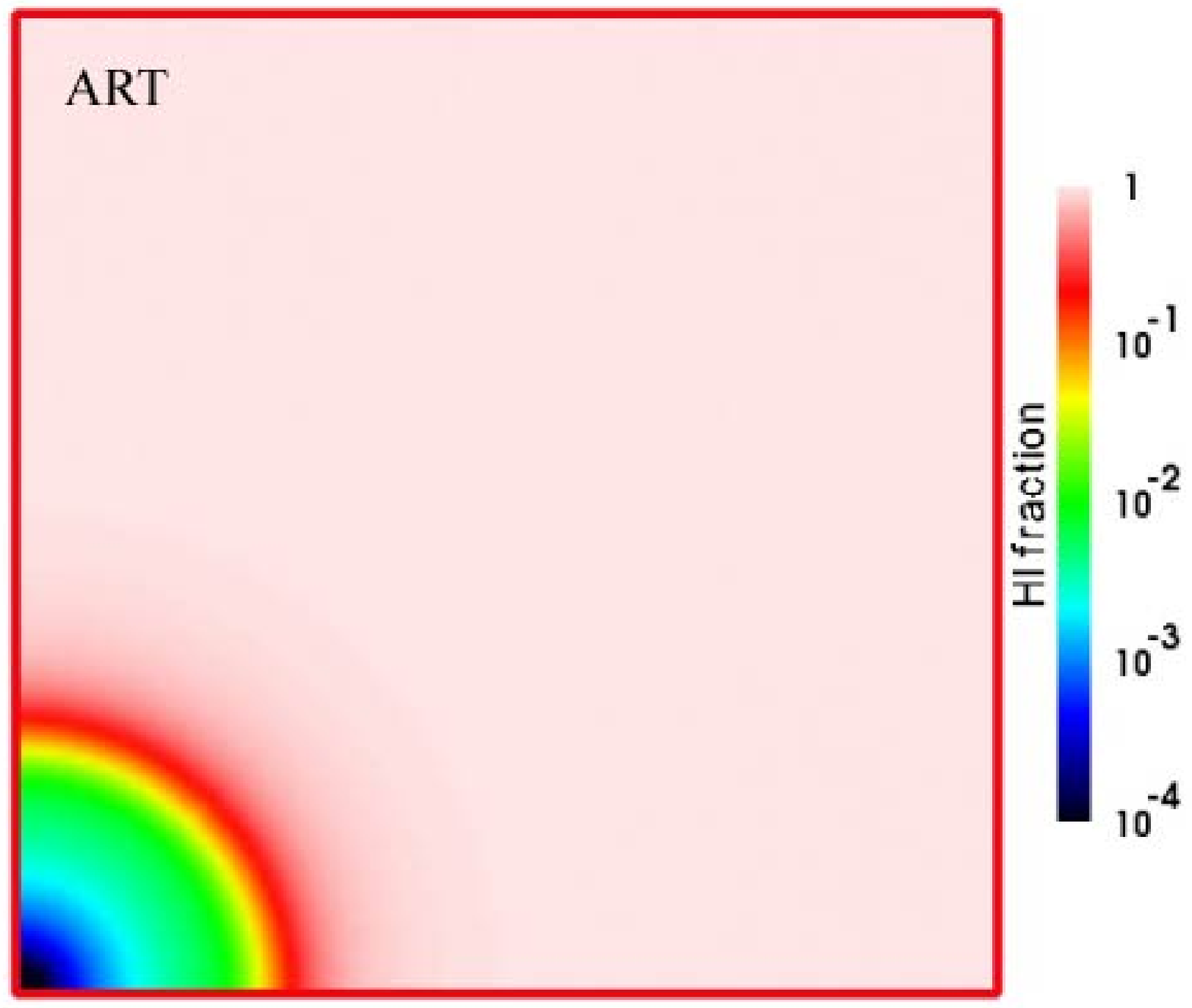}
  \includegraphics[width=2.3in]{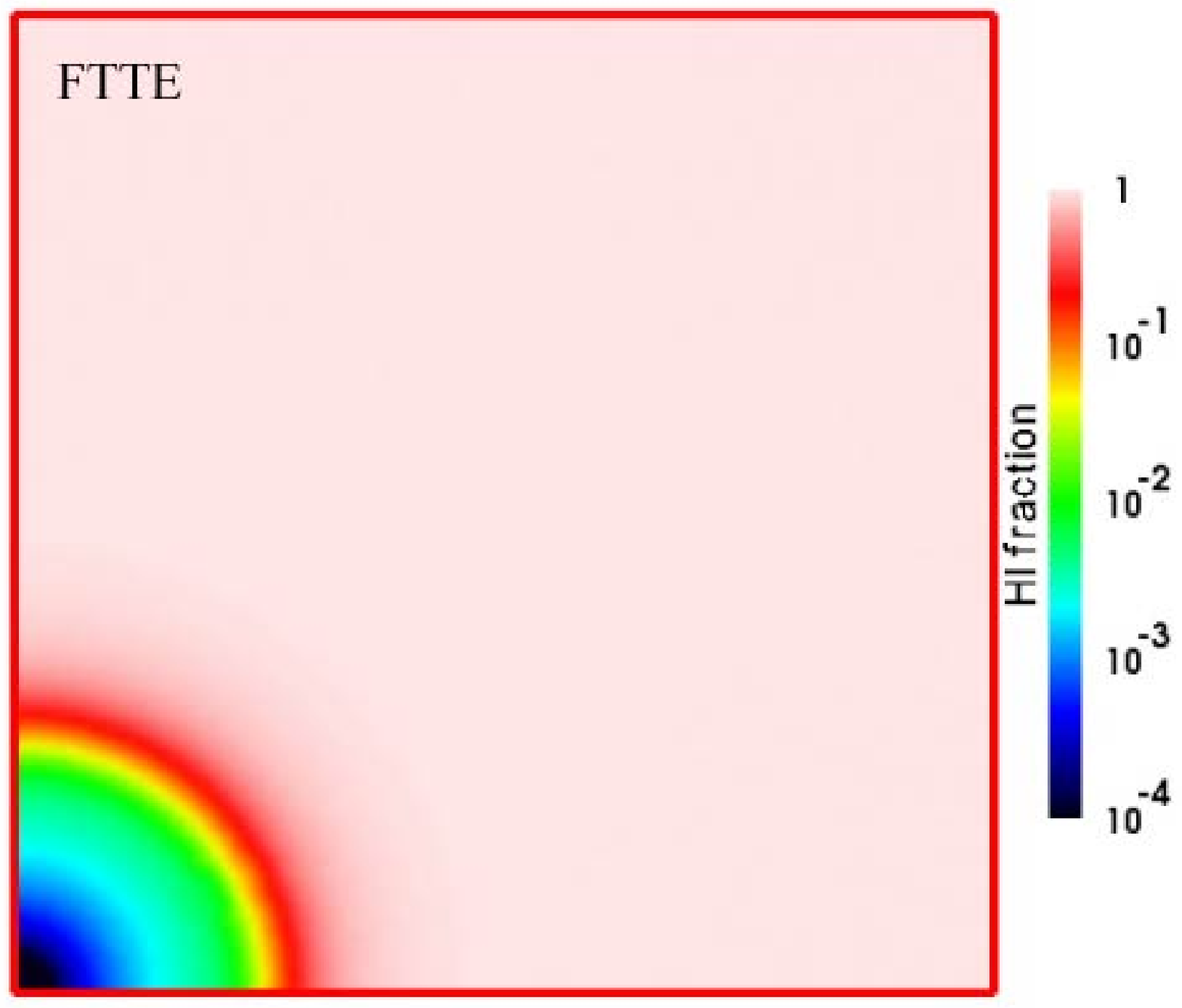}
  \includegraphics[width=2.3in]{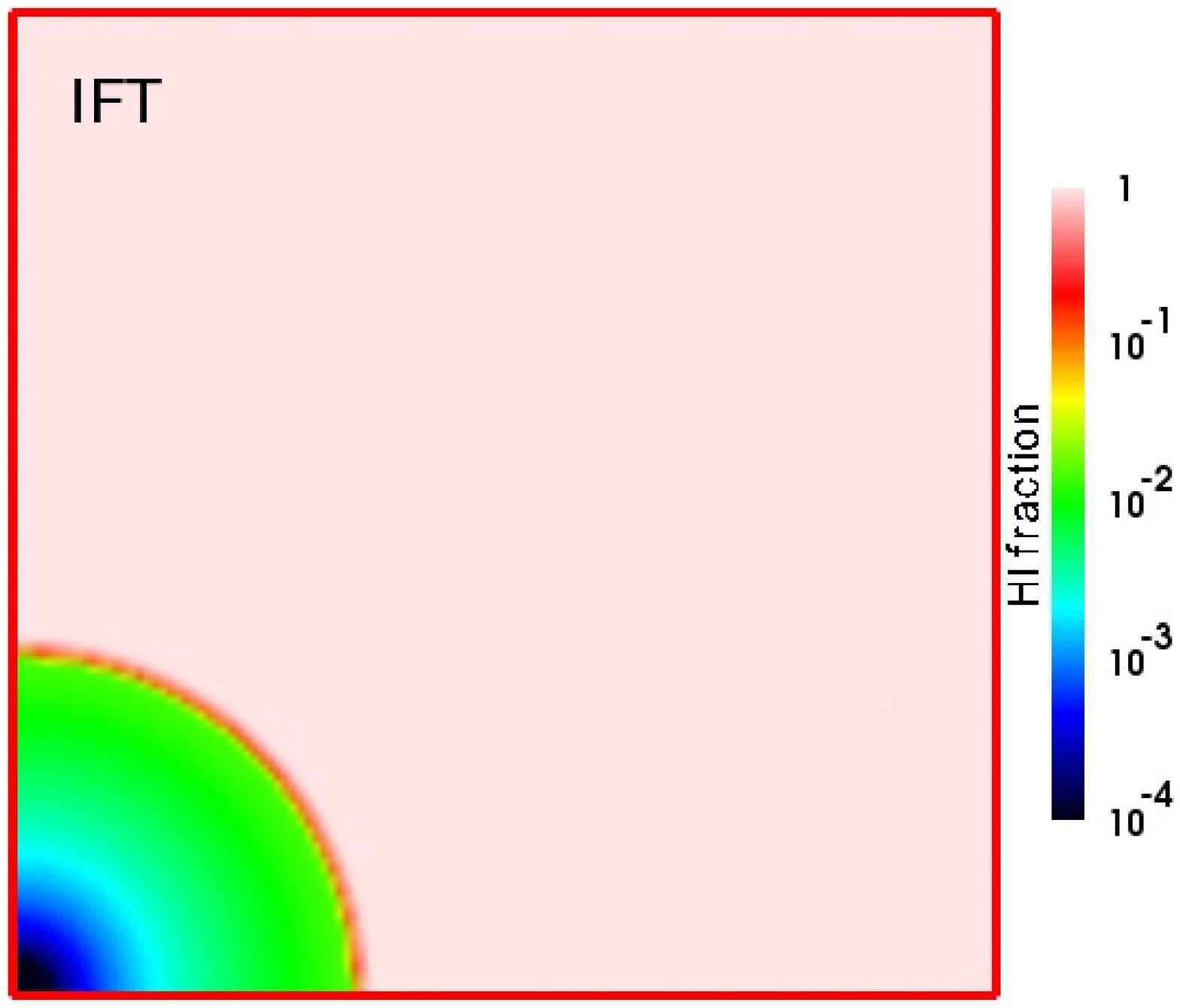}
\caption{Test 2 (H~II region expansion in an uniform gas with varying 
temperature): Images of the H~I fraction, cut through the simulation 
volume at coordinate $z=0$ at time $t=10$ Myr for  (left to
  right and top to bottom) $C^2$-Ray, OTVET, 
  CRASH, RSPH, ART, FTTE, 
and IFT.
\label{T2_images1_HI_fig}}
\end{center}
\end{figure*} 

The same differences persist in the ionized structure of the final Str\"omgren
sphere at $t=500$~Myr (Figure~\ref{T1_profs_fig}, left panel). The majority of
the ray-tracing codes again agree fairly well. The IFT code is based on the
exact analytical solution of this particular problem, and thus to a
significant extent could be considered a substitute for the analytical H~II
region structure. Its differences from the exact solution are only close to
the I-front, where the non-equilibrium effects dominate, while IFT currently
assumes equilibrium chemistry. Away from the I-front, however, the ionized
state of the gas is in equilibrium and there all ray-tracing codes agree
perfectly.  The SimpleX, OTVET and CRASH
codes find thicker sphere boundaries and lower ionized fractions inside, but the
first two codes find slightly smaller Str\"omgren spheres, while the last finds
a slightly larger one. The Zeus code finds lower ionized fractions inside the
ionized region and a somewhat thicker I-front, but an overall H~II region size
that agrees with the other ray-tracing codes. The lower ionization is due to
the current restriction of Zeus to monochromatic radiative transfer, with its
lower postfront temperatures and hence higher recombination rates.

In Figure~\ref{T1_hist_fig} we show histograms of the fraction of cells with a
given neutral fraction during the early, fast expansion phase (at time
$t=10$~Myr; left), when it starts slowing down ($t=100$~Myr, close to one 
recombination time; middle), and when the final Str\"omgren sphere is reached
($t=500$~Myr; right). These histograms reflect the differences in the I-front
transition thickness and internal structure. All codes predict a transitional
region of similar size, which contains a few percent of the total volume. 
In detail, however, once again the results fall into two main groups. One
group includes most of the ray-tracing codes, which agree perfectly 
at all times and predict thin I-fronts close to the
analytical prediction. The other group includes the more diffusive schemes,
namely OTVET, CRASH, RSPH and SimpleX, which all find somewhat thicker I-fronts. 
During the expansion phase of the H~II region these three codes agree well
between themselves, but they disagree somewhat on the structure of the final 
equilibrium Str\"omgren sphere, particularly in the proximity region of the
source. The IFT code histograms differ significantly from the rest, due to
its assumed equilibrium chemistry, which is not quite correct at the I-front. 
 
Finally, the evolution of the globally-averaged neutral fractions is shown in
Figure~\ref{T1_fracs_fig}. The same trends are evident, with the ray-tracing
codes agreeing closely among themselves, while the OTVET finds
about 10\% more neutral material at the final time, due to the different
ionization structure obtained by this method.

\subsection{Test 2: H~II region expansion: the temperature state}
\label{T2_sect}

\begin{figure*}
\begin{center}
  \includegraphics[width=2.3in]{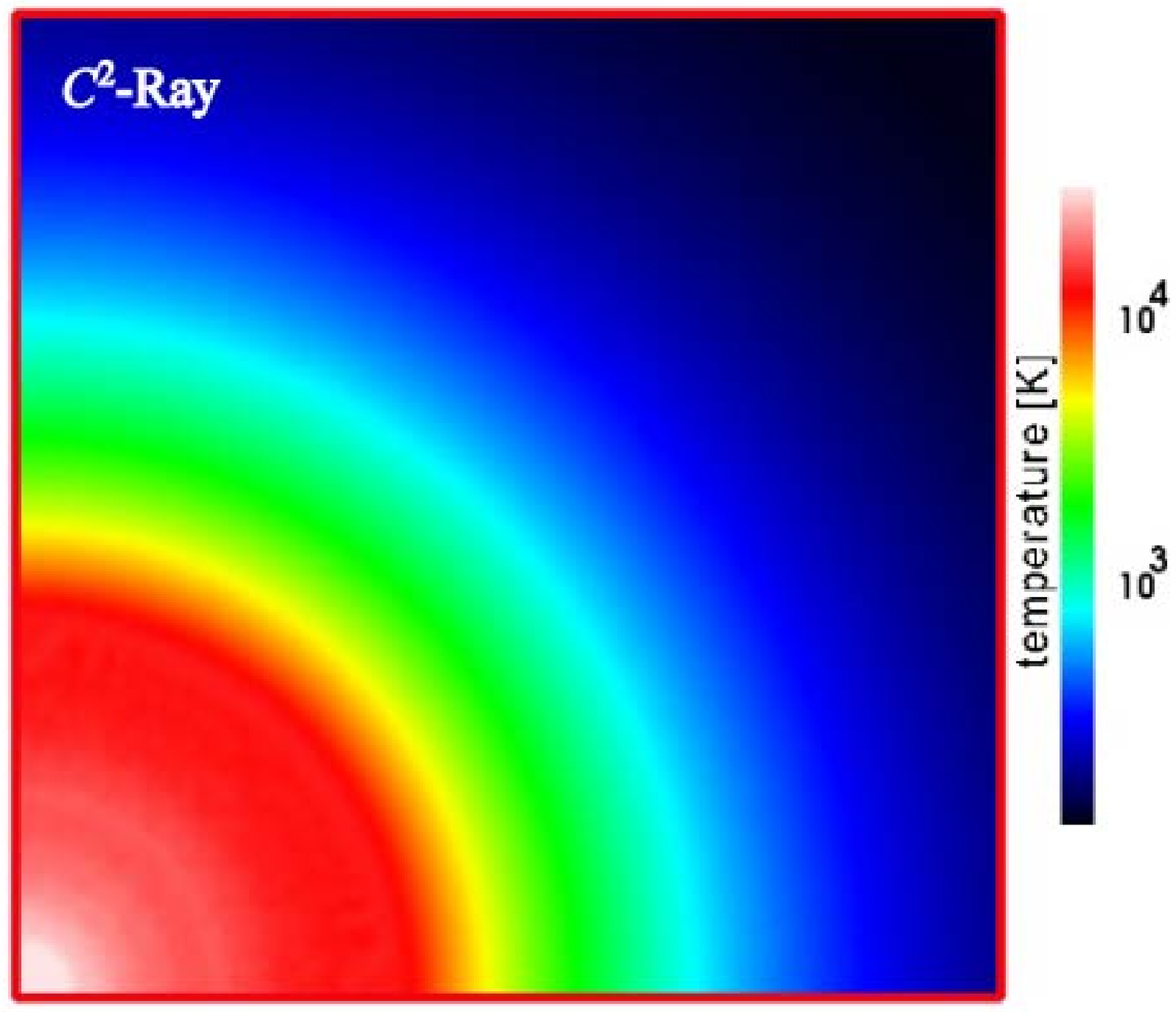}
  \includegraphics[width=2.3in]{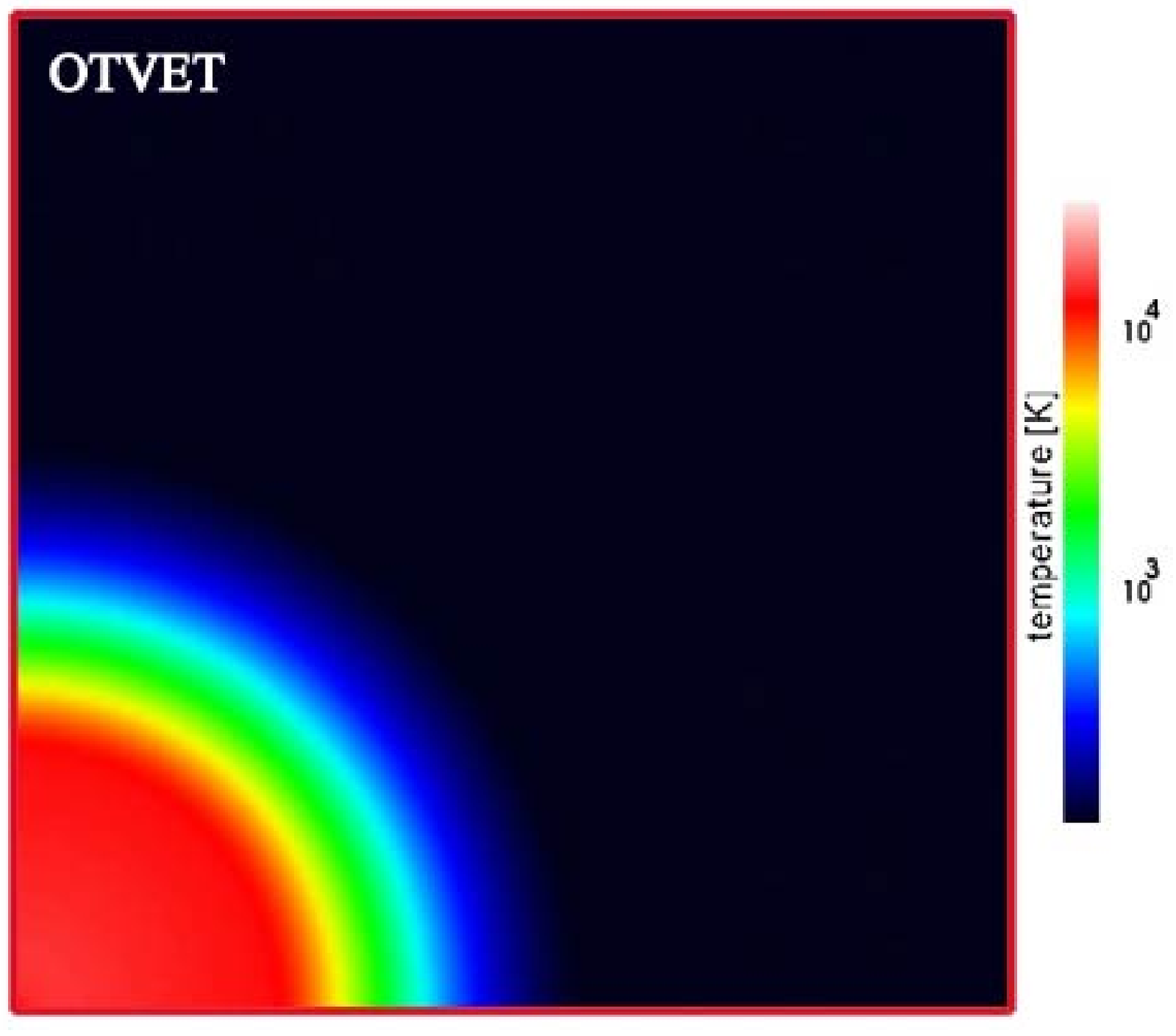}
  \includegraphics[width=2.3in]{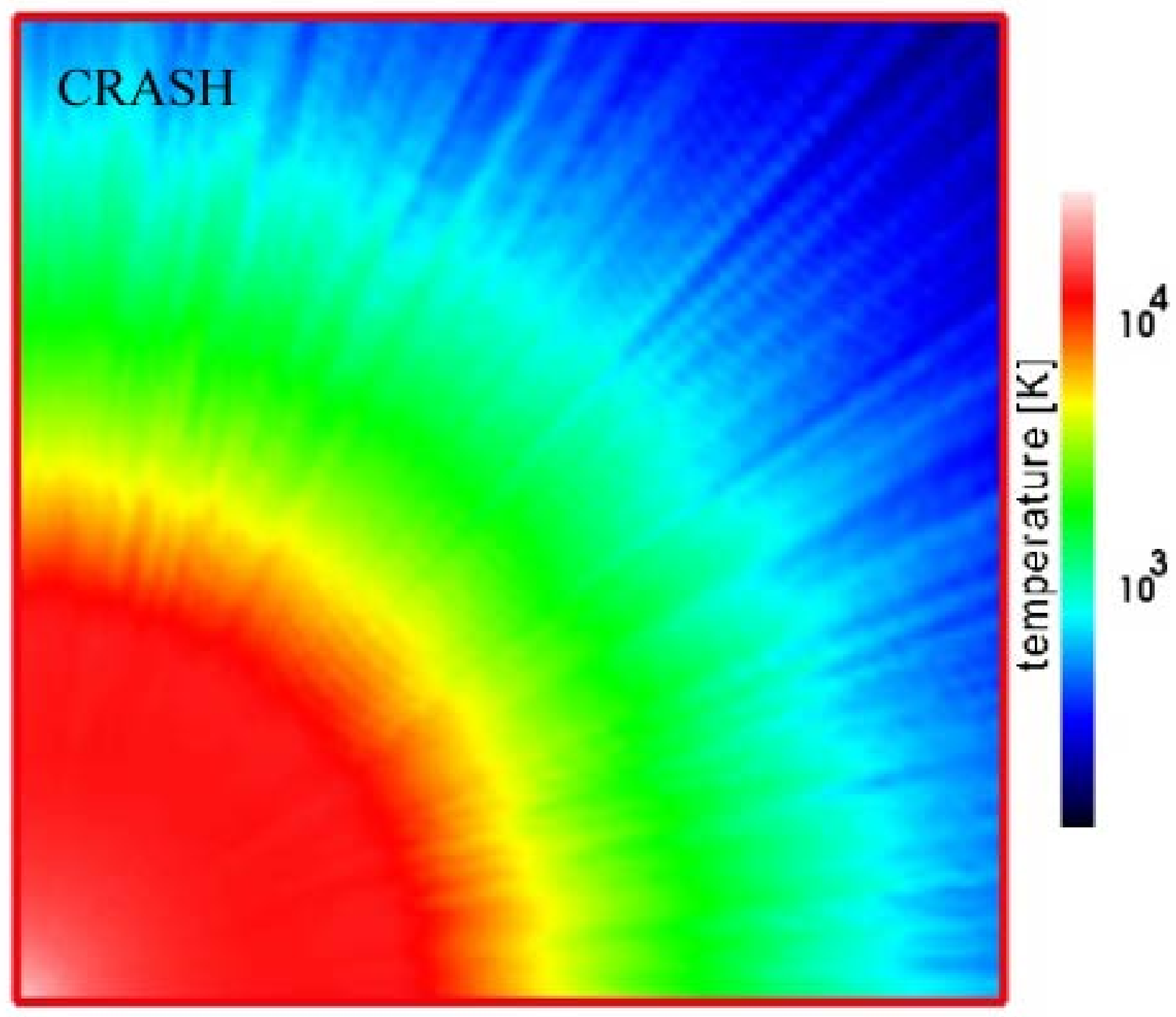}
  \includegraphics[width=2.3in]{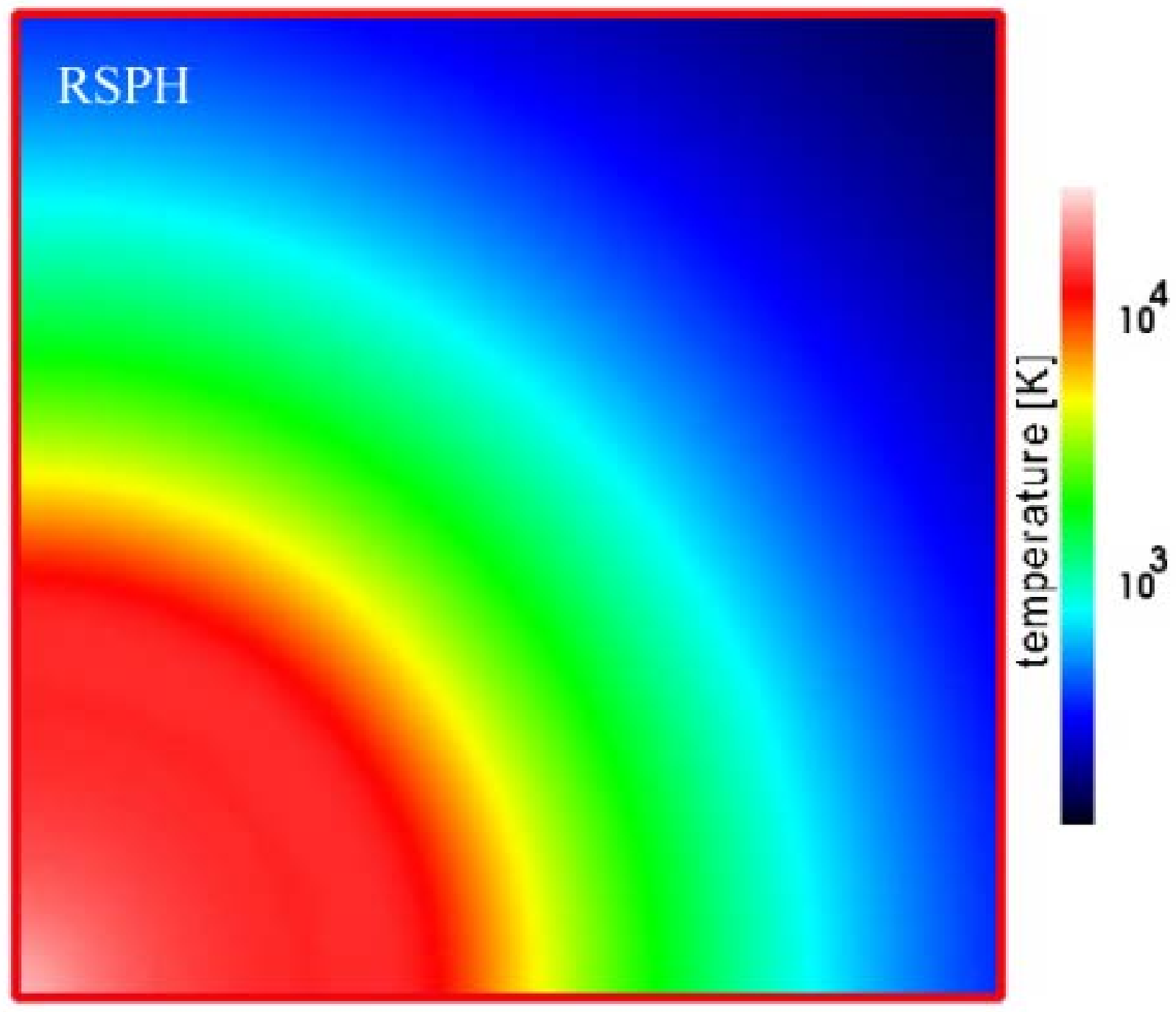}
  \includegraphics[width=2.3in]{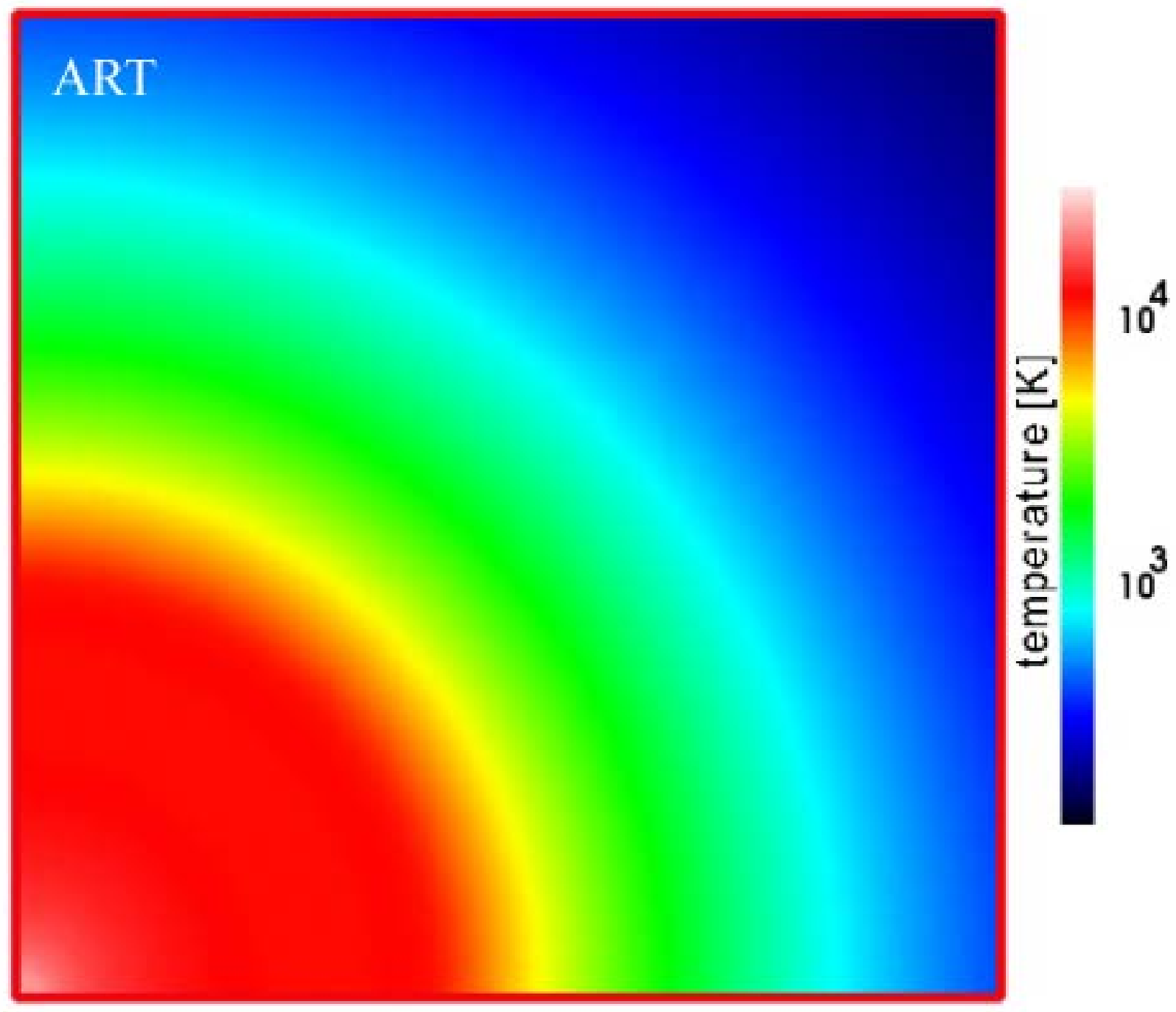}
  \includegraphics[width=2.3in]{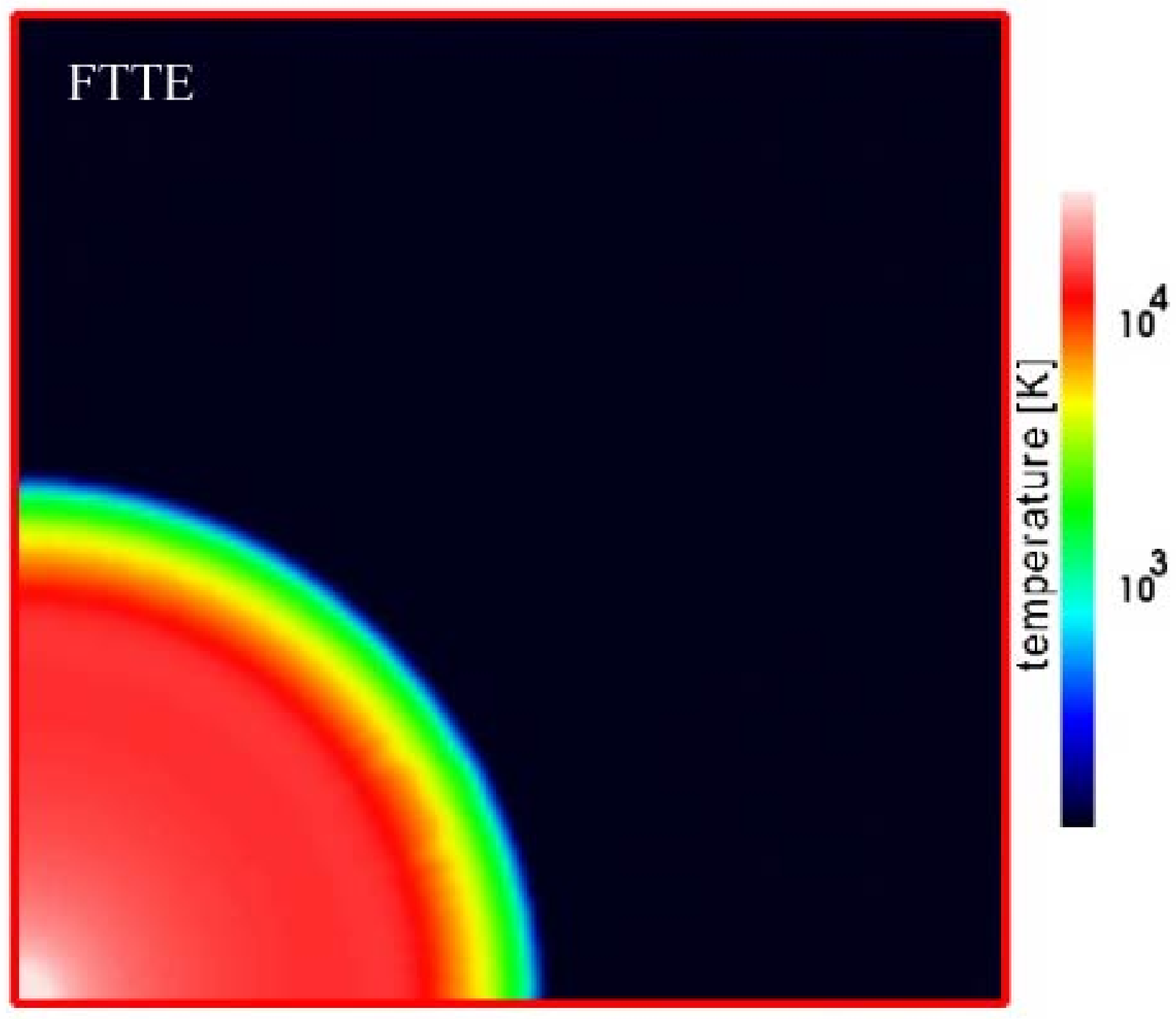}
  \includegraphics[width=2.3in]{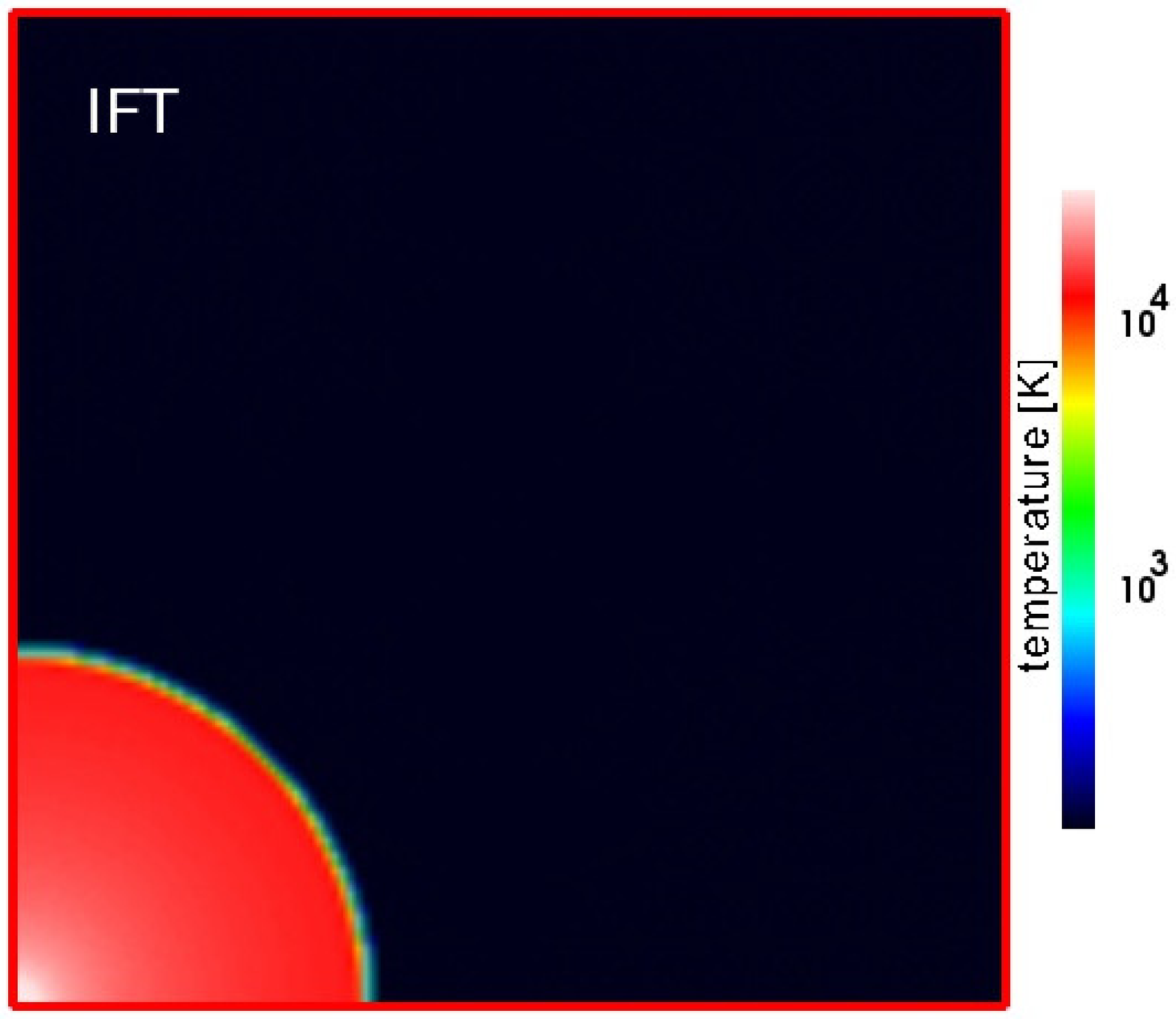}
\caption{Test 2 (H~II region expansion in an uniform gas with varying 
temperature): Images of the temperature, cut through the simulation 
  volume at coordinate $z=0$ at time $t=10$ Myr for (left to
  right and top to bottom) $C^2$-Ray, OTVET, 
  CRASH, RSPH, ART, FTTE, 
and IFT.
\label{T2_images1_T_fig}}
\end{center}
\end{figure*}

\begin{figure*}
\begin{center}
  \includegraphics[width=2.3in]{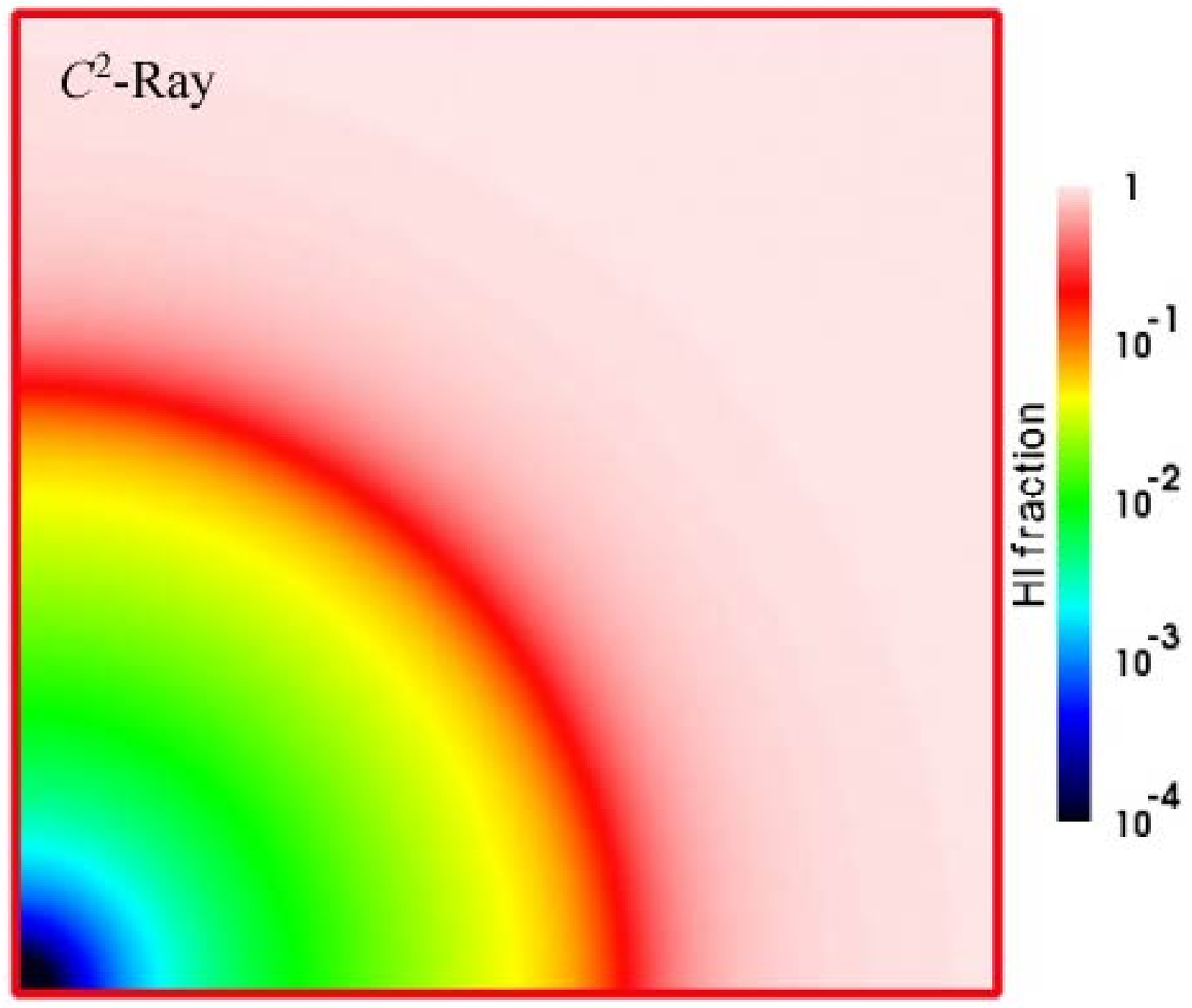}
  \includegraphics[width=2.3in]{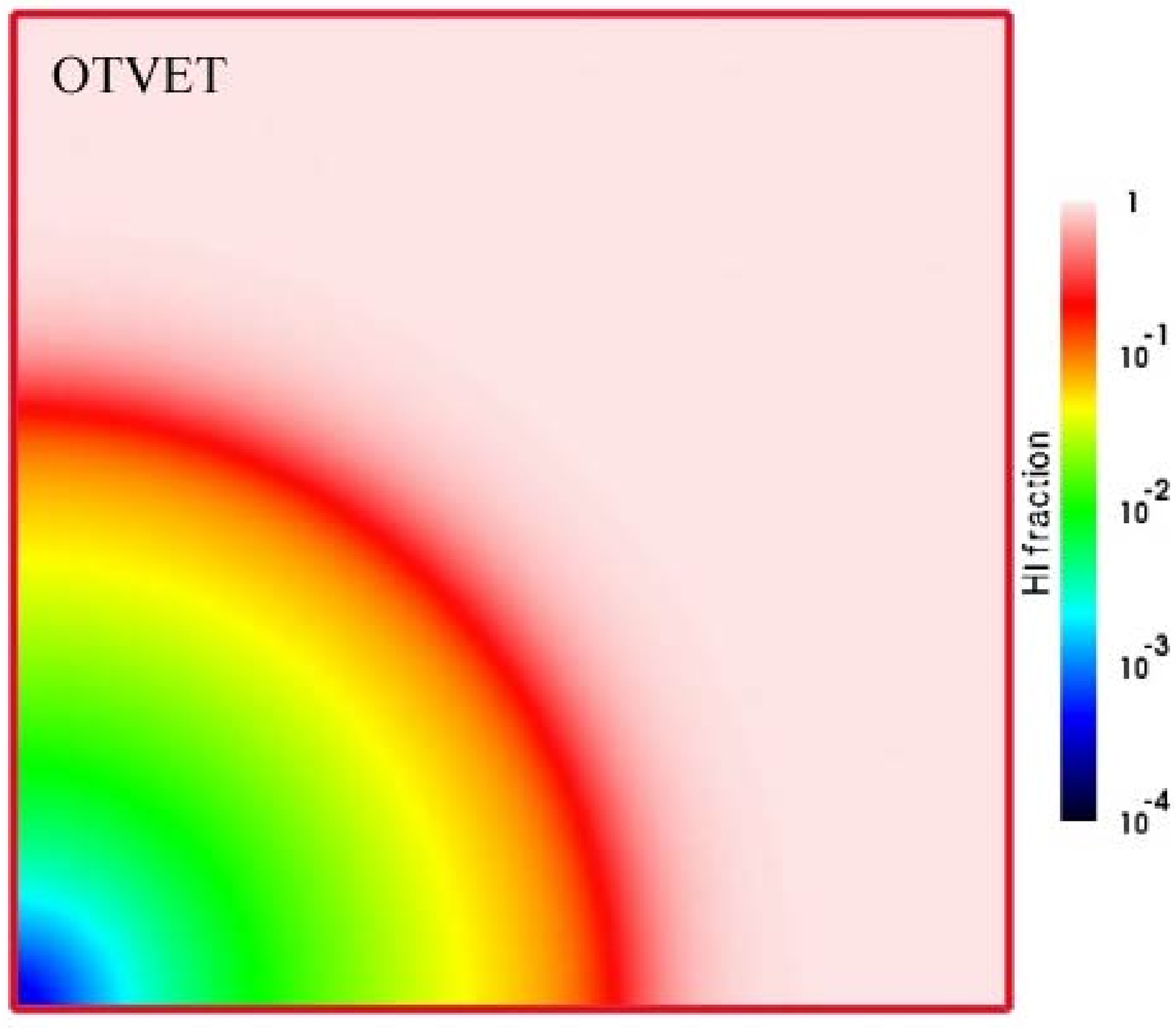}
  \includegraphics[width=2.3in]{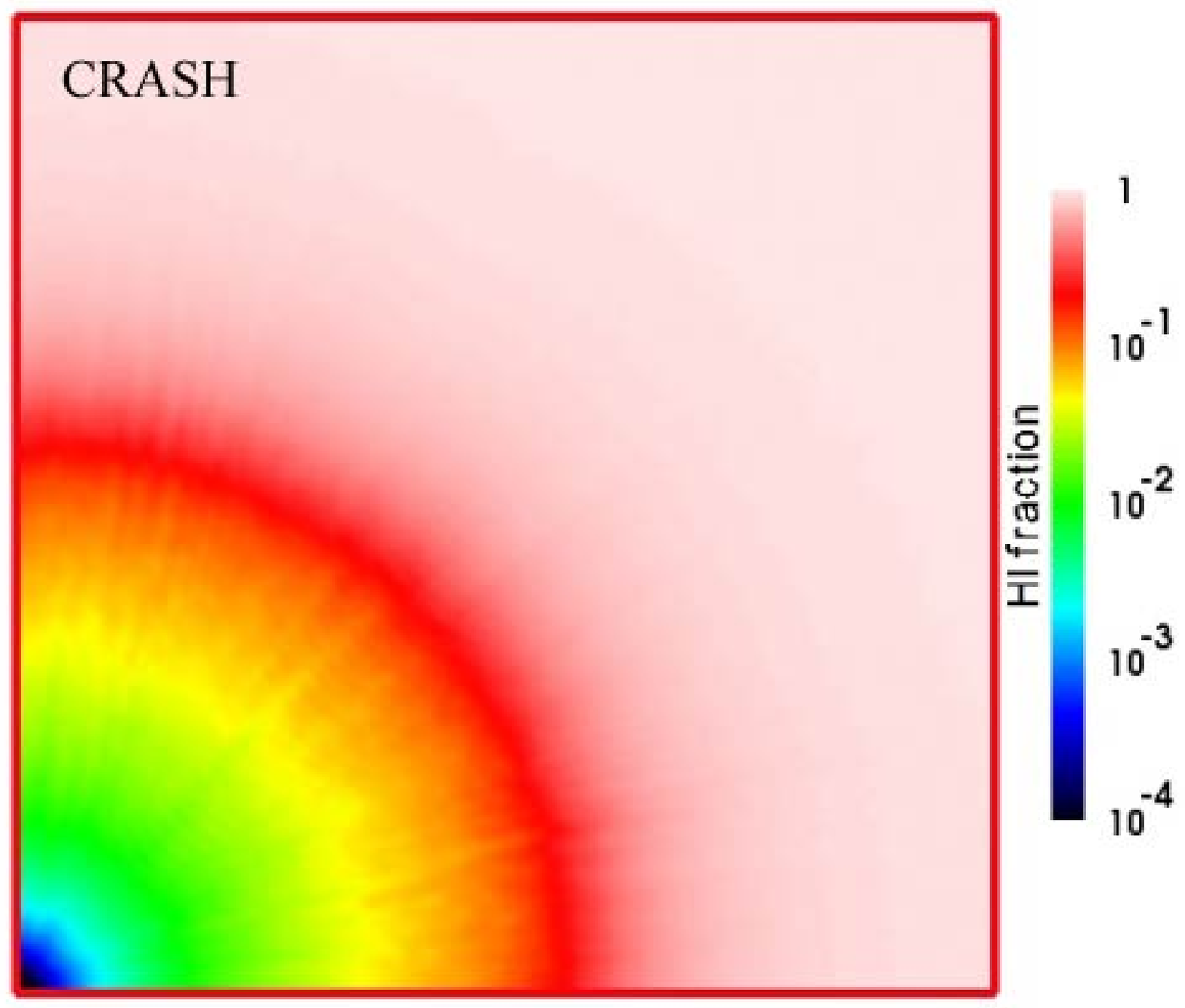}
  \includegraphics[width=2.3in]{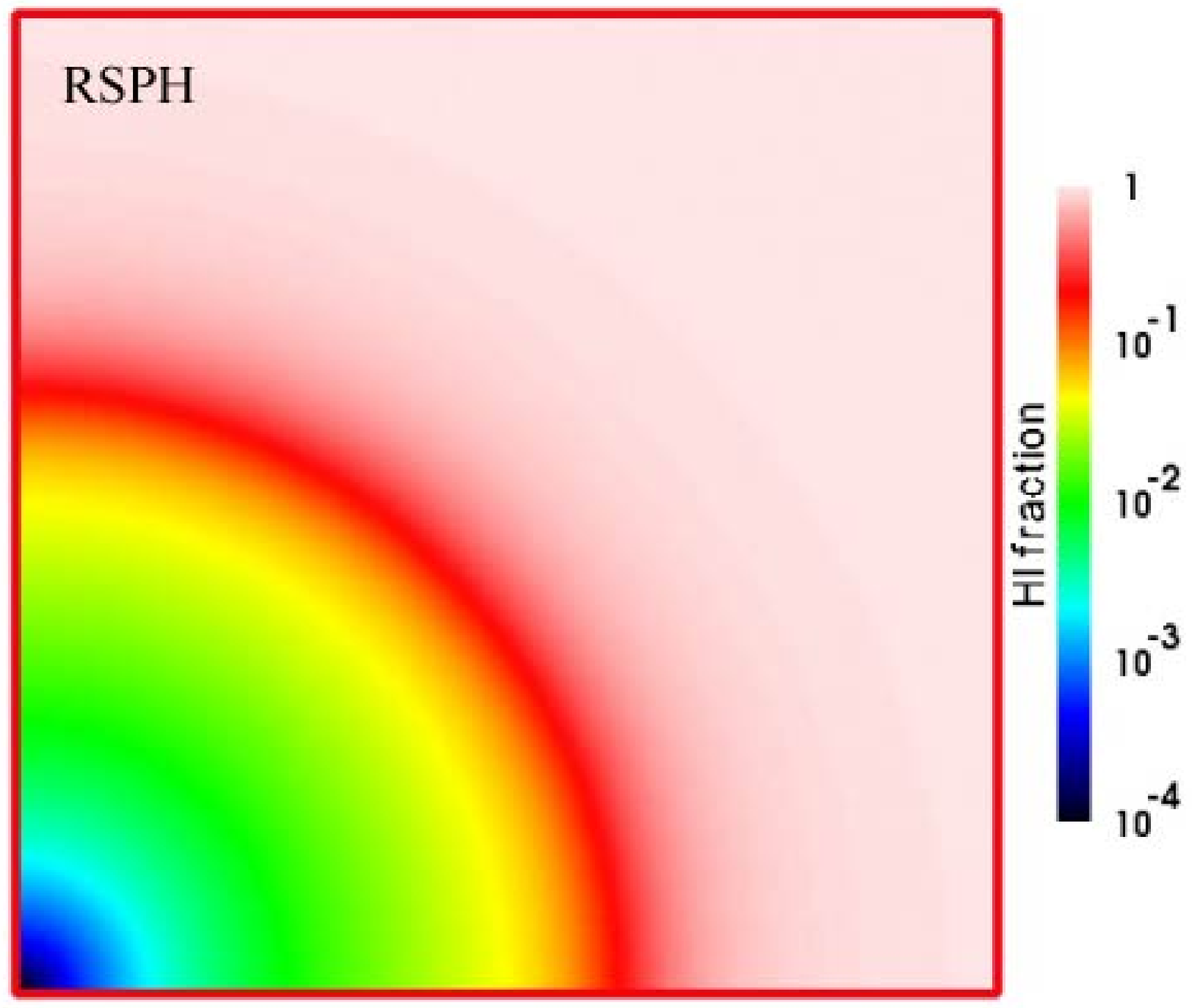}
  \includegraphics[width=2.3in]{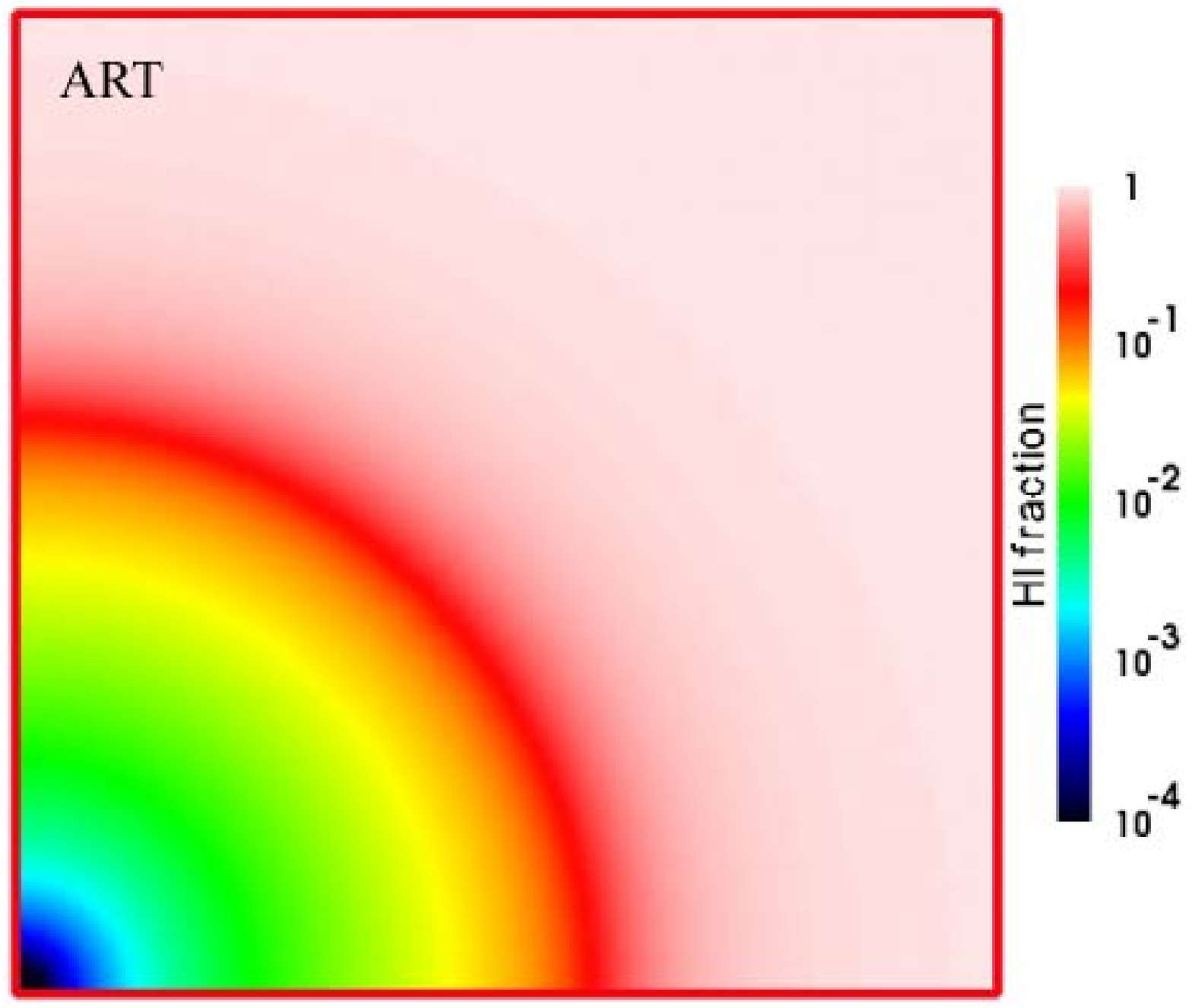}
  \includegraphics[width=2.3in]{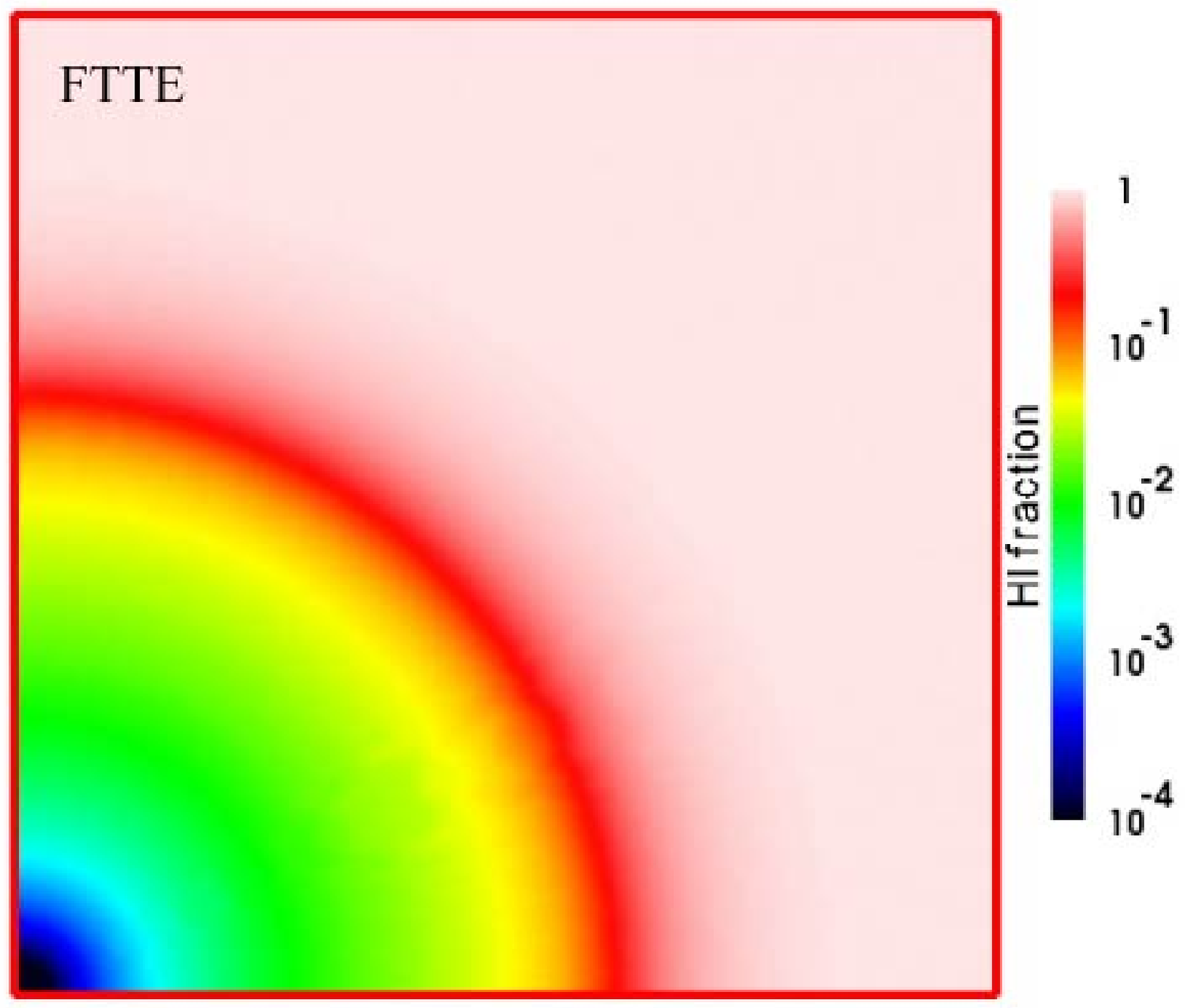}
  \includegraphics[width=2.3in]{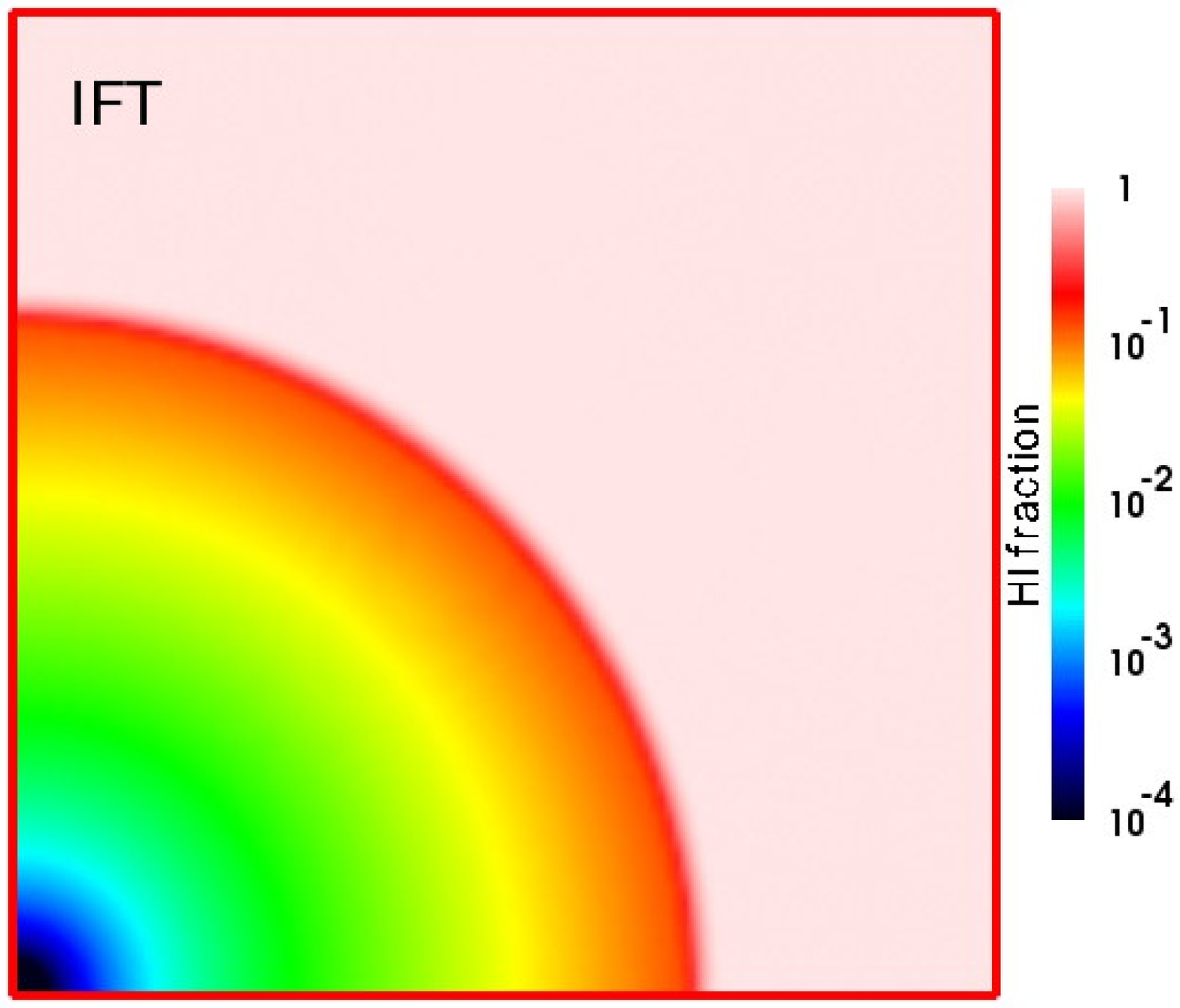}
\caption{Test 2 (H~II region expansion in an uniform gas with varying 
temperature): Images of the H~I fraction, cut through the 
simulation volume at coordinate $z=0$ at time $t=100$ Myr for  
(left to right and top to bottom) $C^2$-Ray, OTVET, 
  CRASH, RSPH, ART, FTTE, 
and IFT.
\label{T2_images3_HI_fig}}
\end{center}
\end{figure*}

\begin{figure*}
\begin{center}
  \includegraphics[width=2.3in]{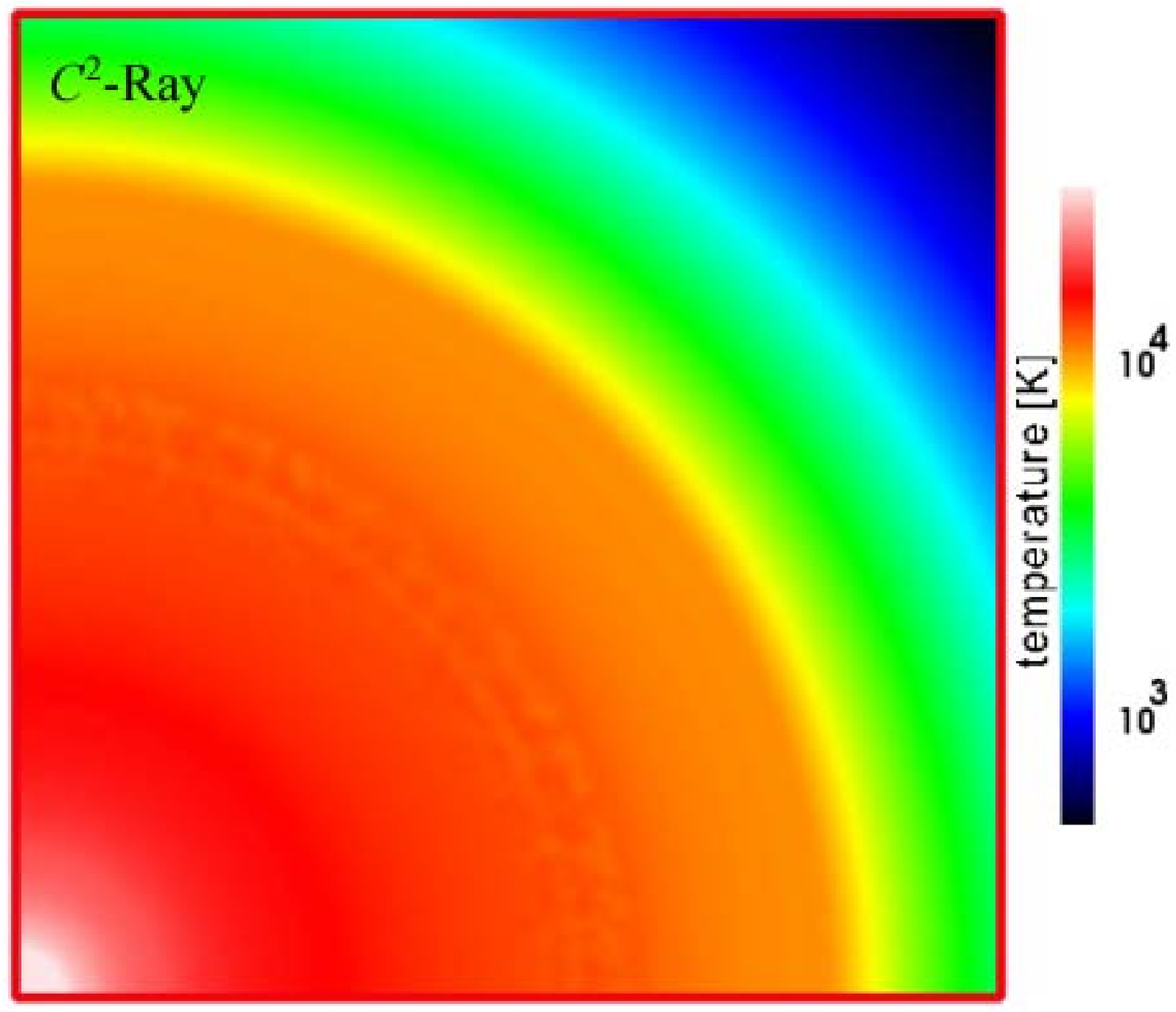}
  \includegraphics[width=2.3in]{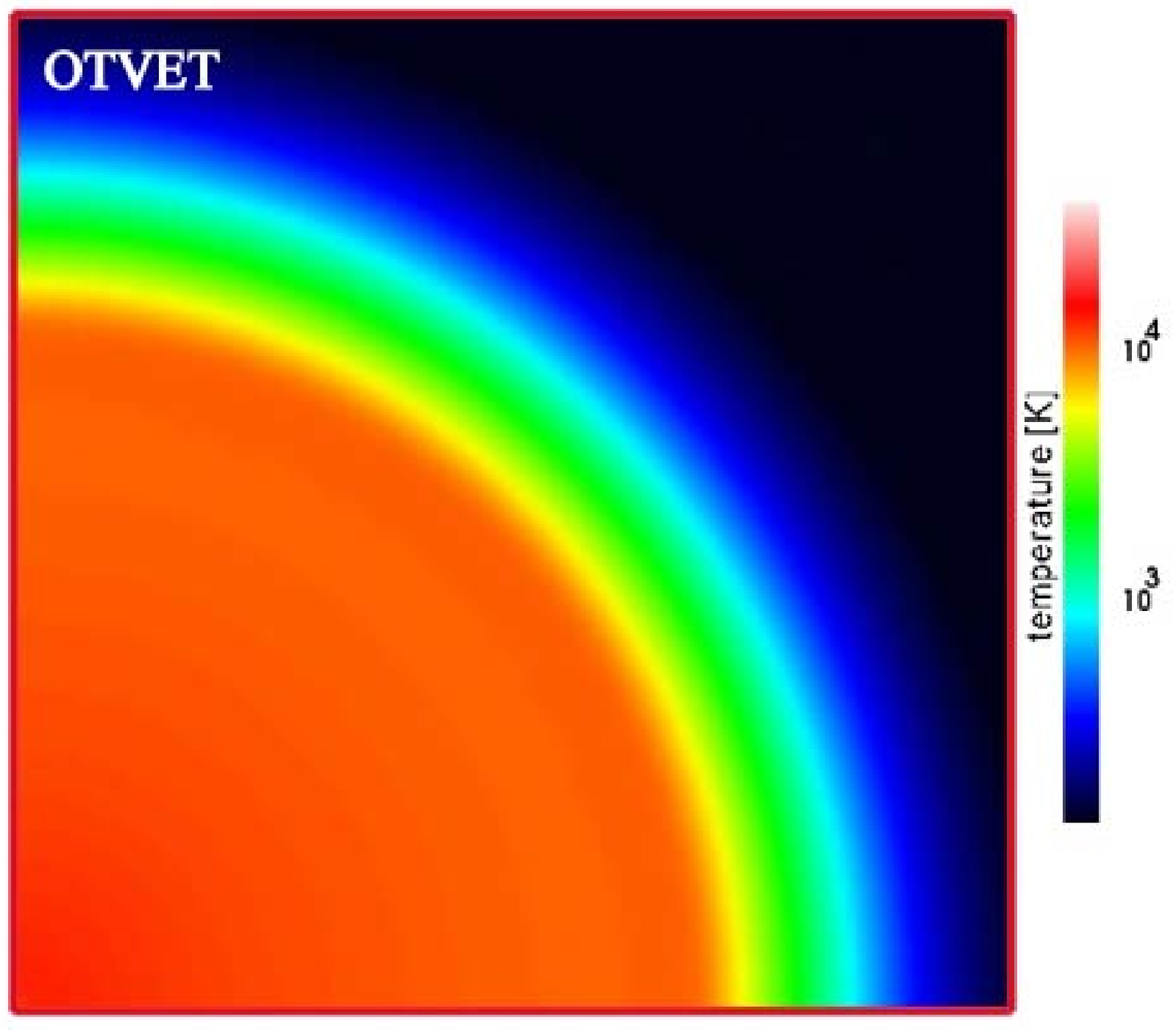}
  \includegraphics[width=2.3in]{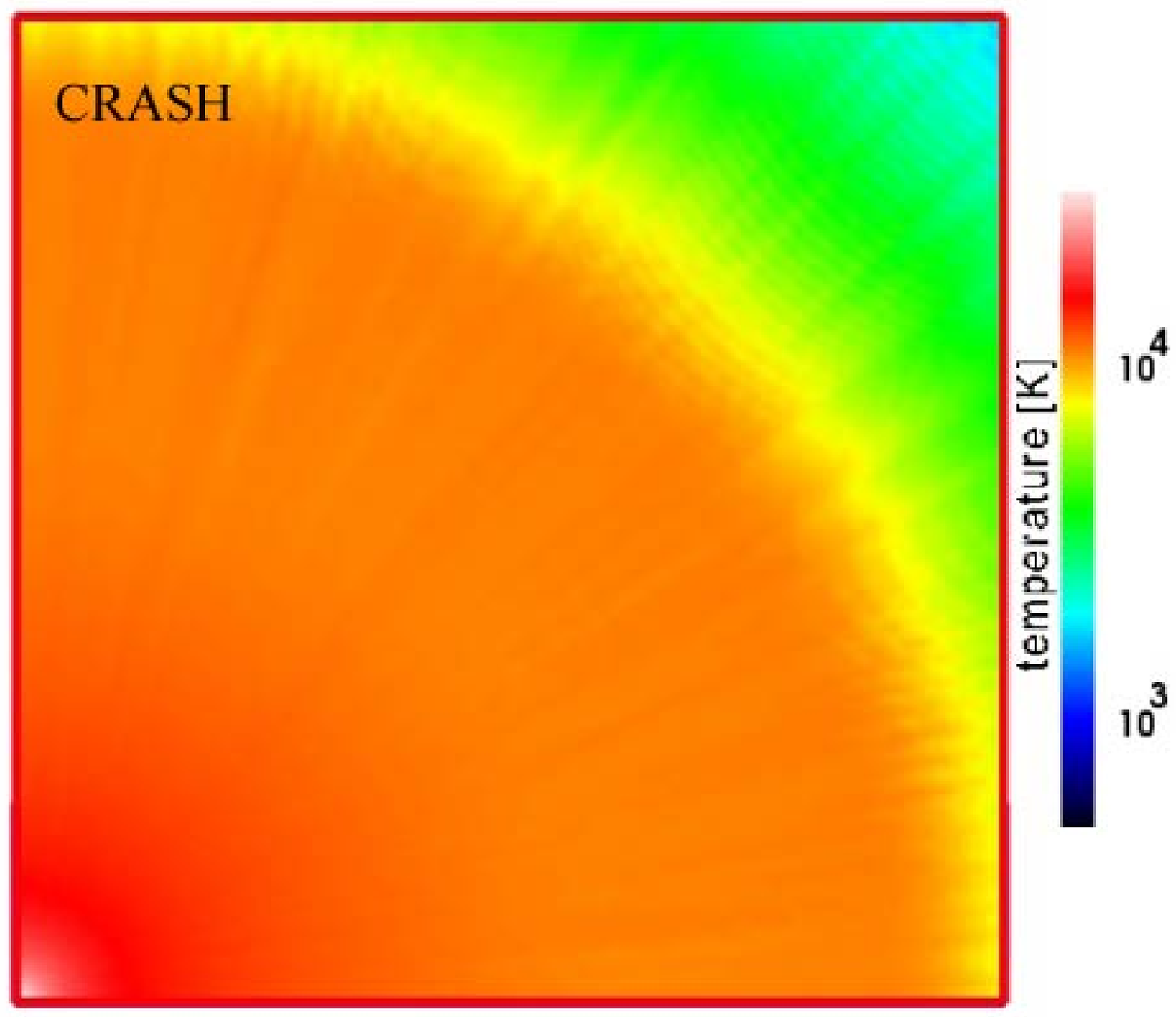}
  \includegraphics[width=2.3in]{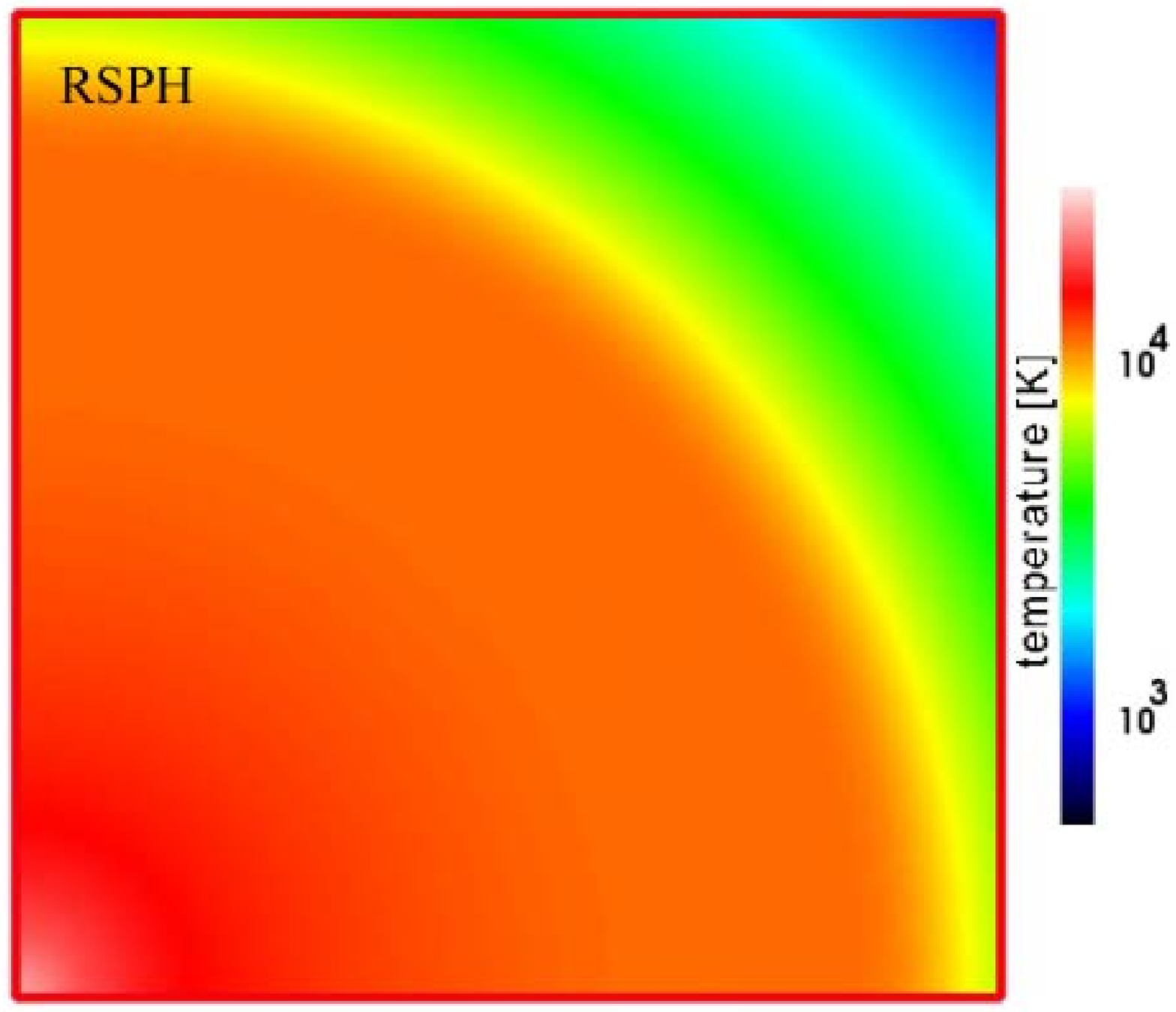}
  \includegraphics[width=2.3in]{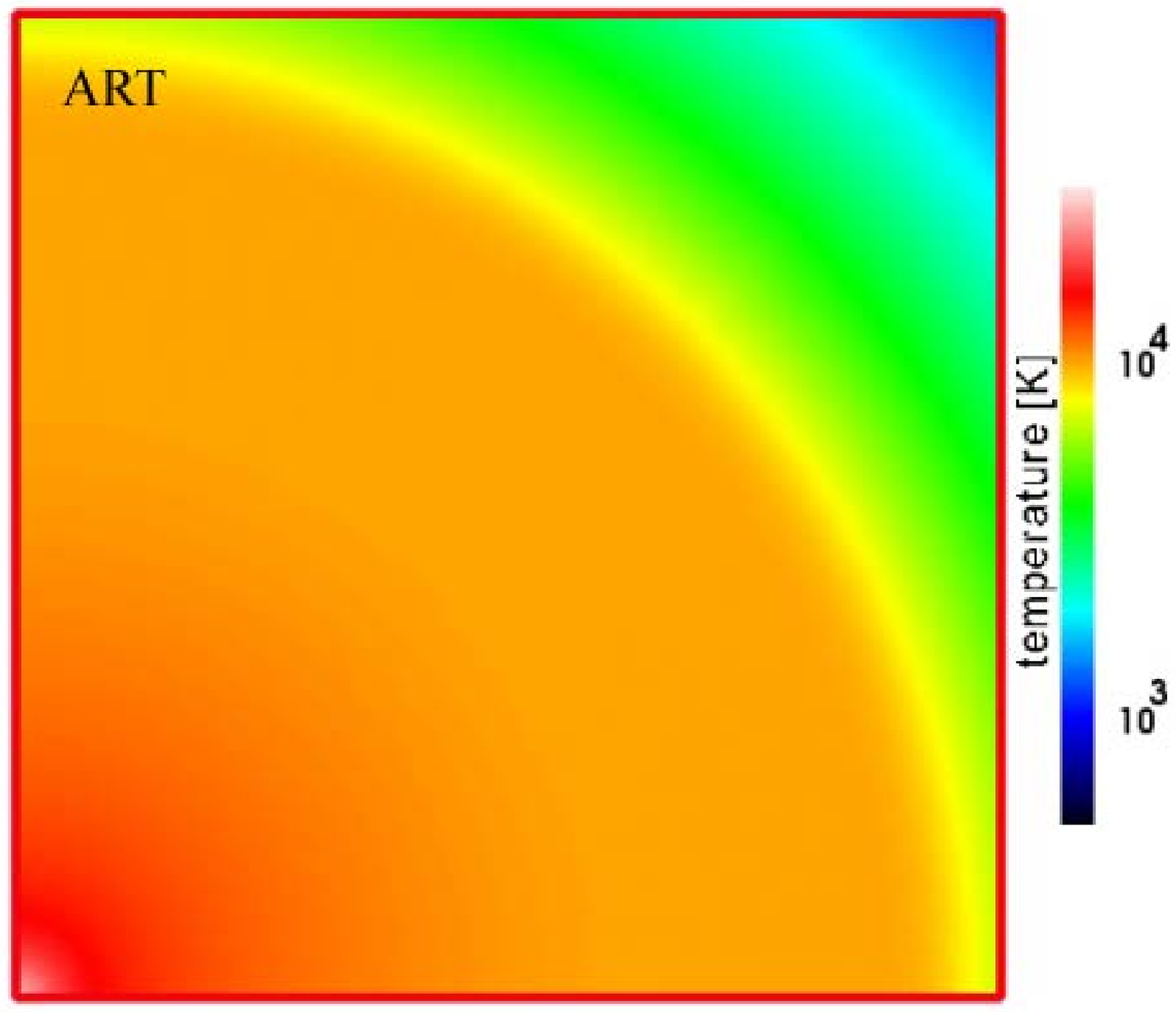}
  \includegraphics[width=2.3in]{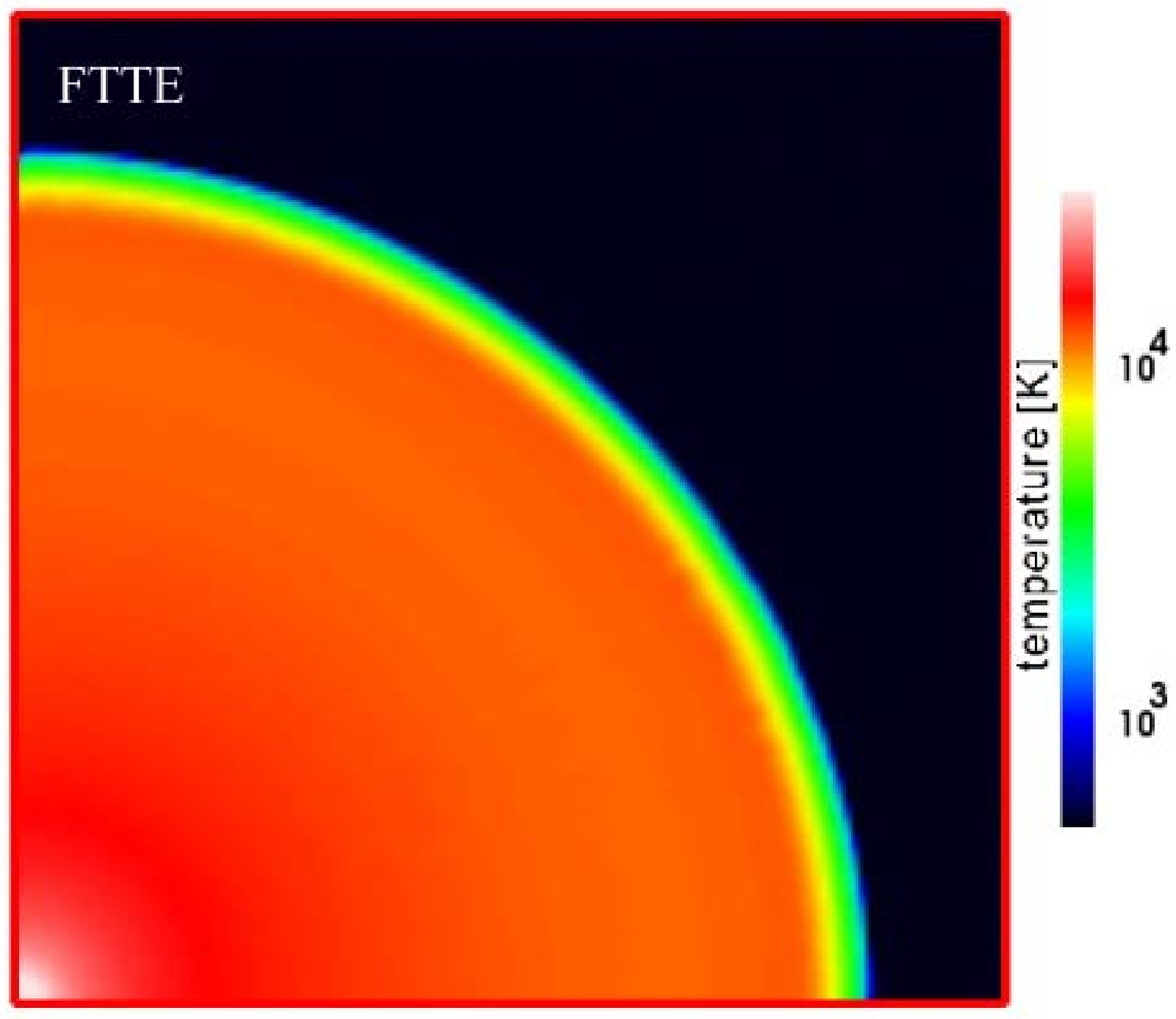}
  \includegraphics[width=2.3in]{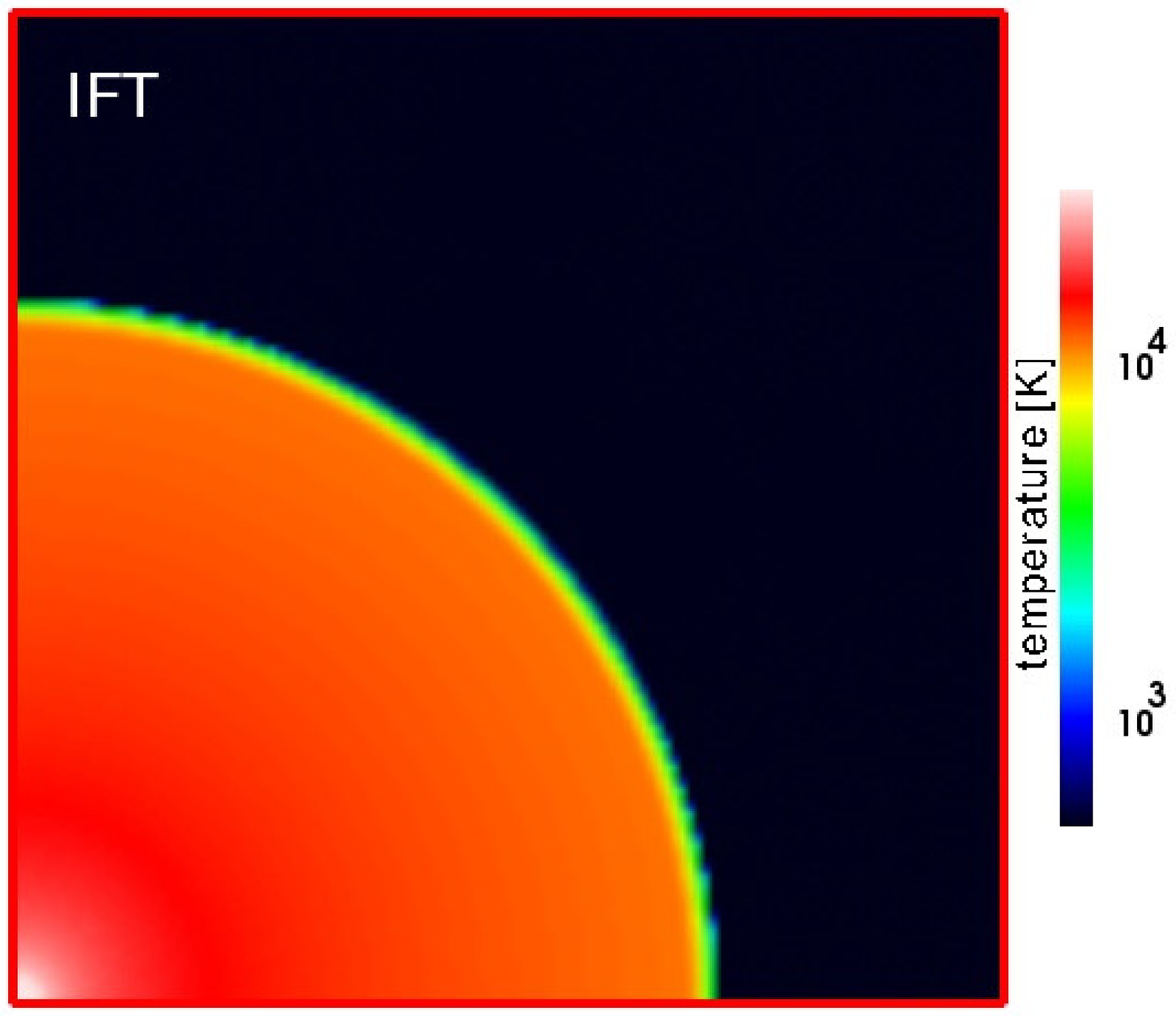}
\caption{Test 2 (H~II region expansion in an uniform gas with varying 
temperature): Images of the temperature, cut through the simulation 
  volume at coordinate $z=0$ at time $t=100$ Myr for (left to
  right and top to bottom) $C^2$-Ray, OTVET, 
  CRASH, RSPH, ART, FTTE, 
and IFT.
\label{T2_images3_T_fig}}
\end{center}
\end{figure*}

\begin{figure*}
\begin{center}
  \includegraphics[width=3.5in]{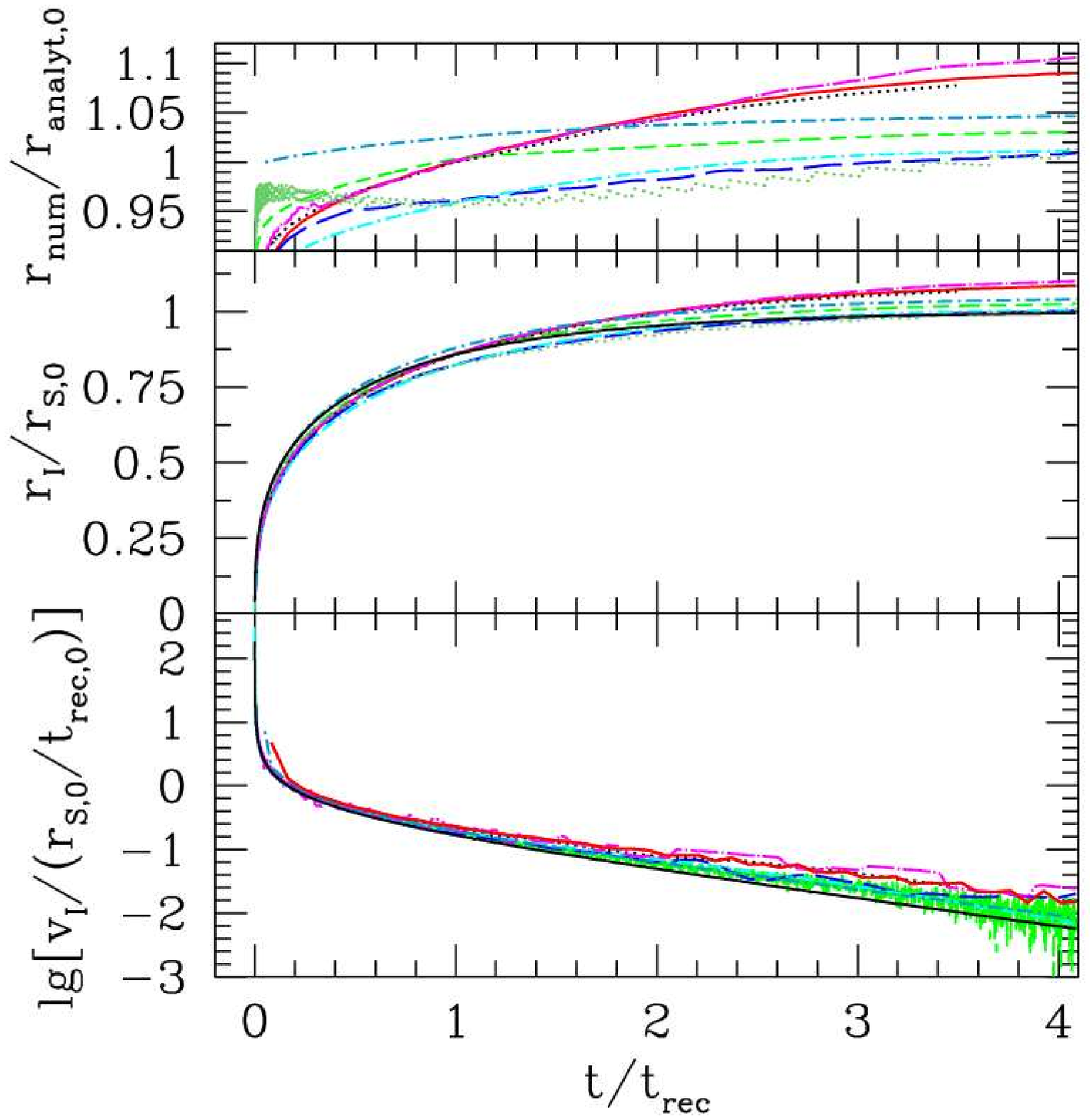}
\caption{Test 2 (H~II region expansion in an uniform gas with varying 
temperature): The evolution of the position and velocity of the I-front. 
\label{T2_Ifront_evol_fig}}
\end{center}
\end{figure*}

Test 2 solves essentially the same problem as Test 1, but the ionizing source
is assumed to have a $10^5$~K black-body spectrum and we allow the gas 
temperature to vary due to heating and cooling processes,  as determined by
the energy equation. The test geometry and gas density are the same as in
Test~1. The gas is initially fully neutral and has a temperature of $T=100$~K. 

In Figure~\ref{T2_images1_HI_fig} we show images of the neutral fraction on
the $z=0$ plane at time $t=10$~Myr, during the initial fast expansion phase of
the H~II region. All of the results agree fairly well on the overall size of
the ionized region and its internal structure. Again there are modest 
differences in the thickness of the I-front and the ionizing source proximity
region, e.g. the CRASH code again produces a somewhat thicker transition and
smaller proximity region. The IFT code finds significantly sharper I-fronts
due to its equilibrium chemistry, but the internal H~II region structure away
from the front (which is close to equilibrium) and overall size of the ionized
region both agree well with the rest of the codes. The temperature structure
of the H~II region, on the other hand, demonstrates significant differences
(Figure~\ref{T2_images1_T_fig}). These stem largely from the different way the
codes handle spectrum hardening, i.e. the long mean free paths of the
high-energy photons due to the much lower photoionization cross-section at
high frequencies. These long mean free paths result in a much thicker I-front
transition and a significant pre-heating ahead of the actual I-front, since
the high energy photons heat the gas, but there are not enough of them to
ionize it. The CRASH code, which follows multiple bins in frequency, finds a
larger pre-heated region than the other codes. There are also significant
anisotropies in the CRASH results, due to the Monte-Carlo sampling method
used (since not many high-energy photon packets are sent, leading
to undersampling in angle). In production runs, a multifrequency treatment 
of the single photon packets has been introduced, reducing the anisotropies 
in the results. The temperature results of $C^2$-Ray, RSPH and
ART codes agree fairly well among themselves, while OTVET and FTTE give much
less spectrum hardening. Finally, IFT assumes the I-front is sharp, and does
not have spectrum hardening by construction.

The same trends persist at later times, when the I-front is approaching the
Str\"omgren sphere (Figures~\ref{T2_images3_HI_fig} and
\ref{T2_images3_T_fig}). Once again the H~II regions predicted by all the codes
are similar in size and internal structure, but with a little different
I-front thickness in terms of neutral fraction and significant differences in
terms of spectral hardening. The FTTE still gives a very sharp I-front, while
OTVET finds somewhat less hardening, but its later-time result is more similar
to the other codes than at early times.

In Figure~\ref{T2_Ifront_evol_fig} we plot the the position and velocity of
the I-front vs. time.  Unlike Test 1, in this case there is no closed-form
analytical solution since the recombination coefficients vary with the
spatially-varying temperature. Nevertheless, as a point of reference we have
again shown the analytical solution in equations~(\ref{strom0}) (assuming
$T=10^4$~K). All the codes find slightly larger H~II regions and slightly
faster I-front propagation compared to this analytical solution. This is to
be expected due to the temperature being higher than $10^4$~K and the inverse
temperature dependence of the recombination coefficient. The $C^2$-Ray, RSPH
and FTTE results agree perfectly among themselves, to $\sim1\%$, as do the
results from OTVET, ART, and Zeus, again among themselves. These two groups of
results differ by $\sim10\%$, however, while CRASH and IFT find an H~II
region size intermediate between the two groups.    

\begin{figure*}
\begin{center}
  \includegraphics[width=2.2in]{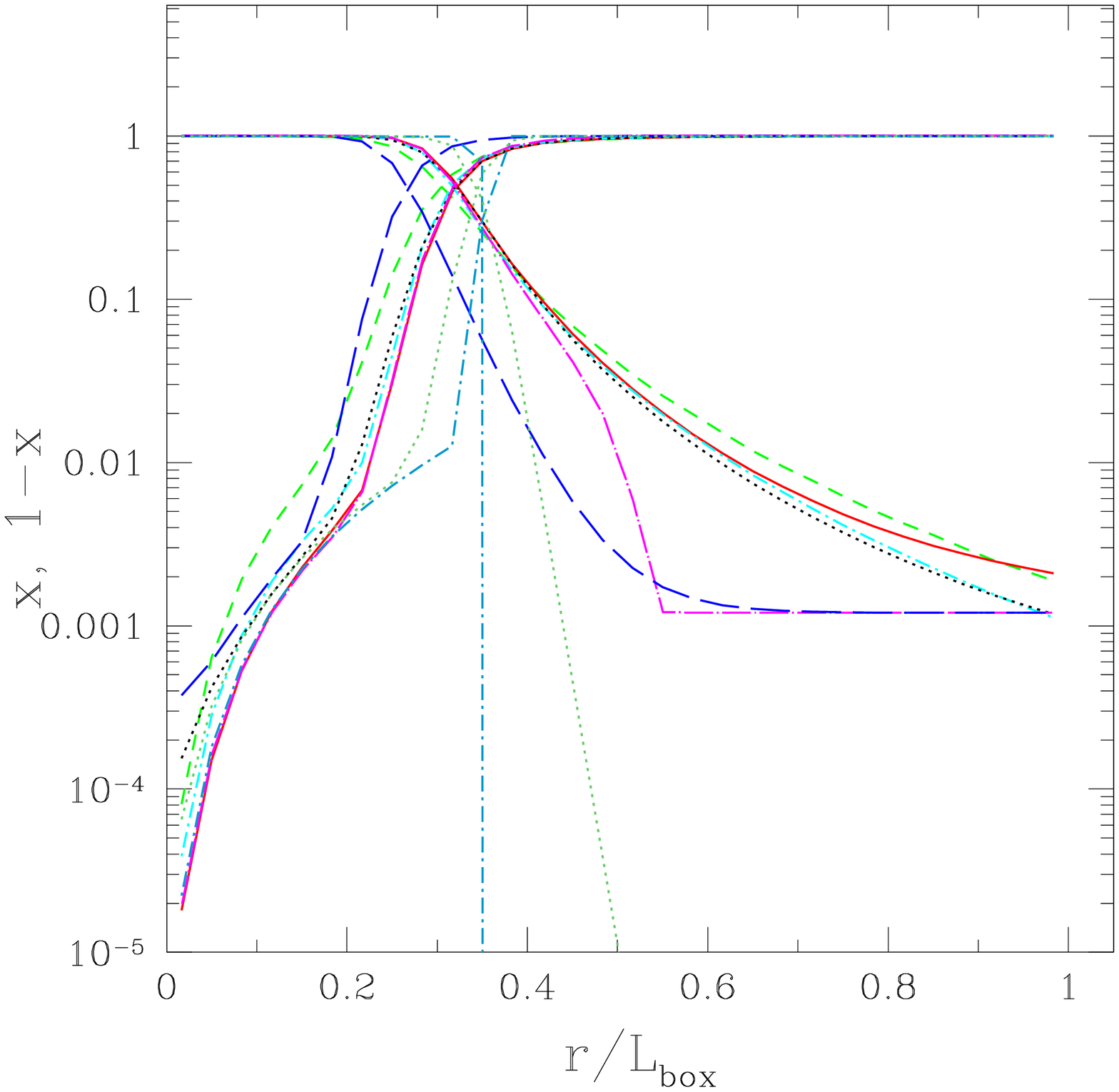}
  \includegraphics[width=2.2in]{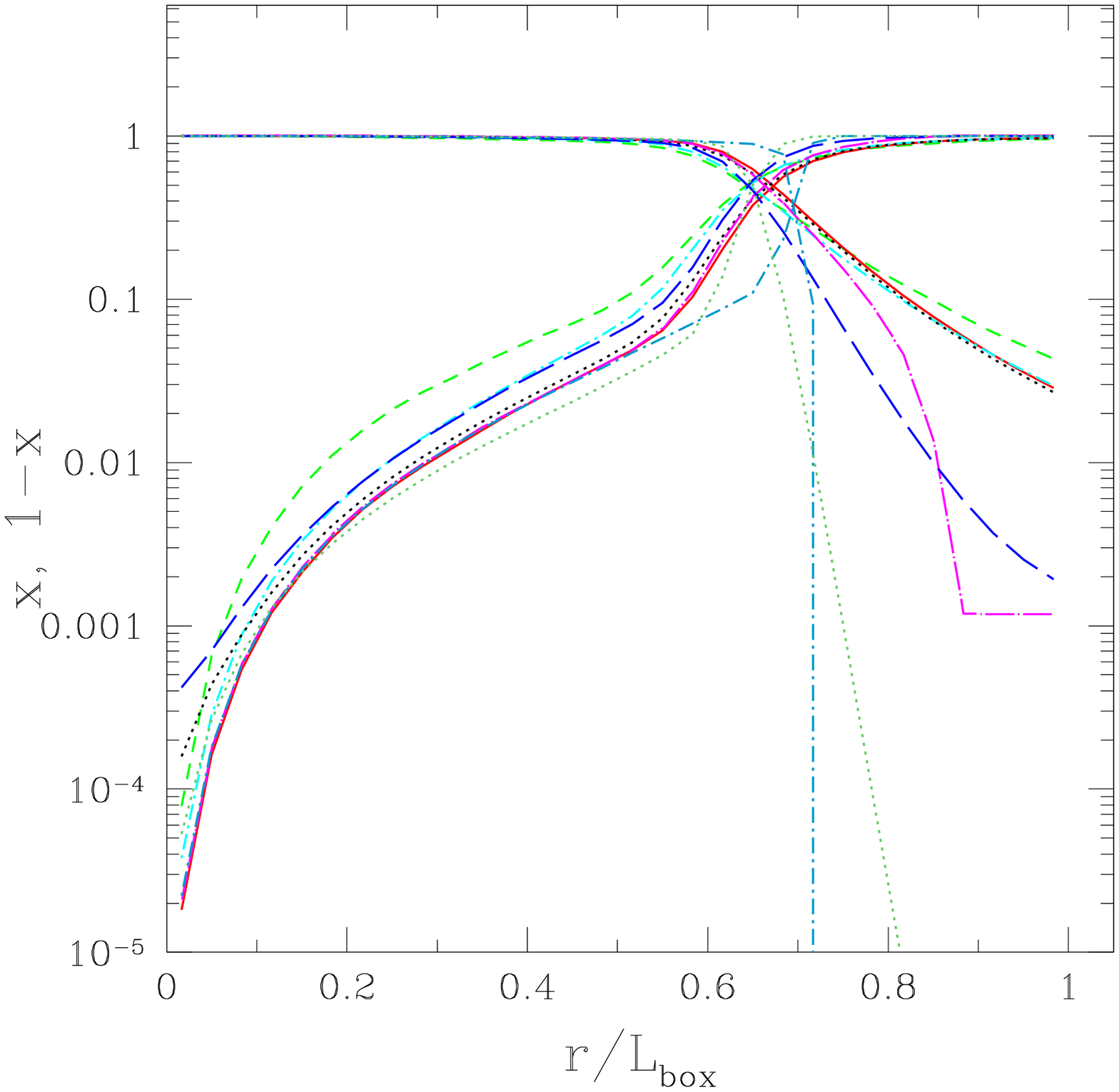}
  \includegraphics[width=2.2in]{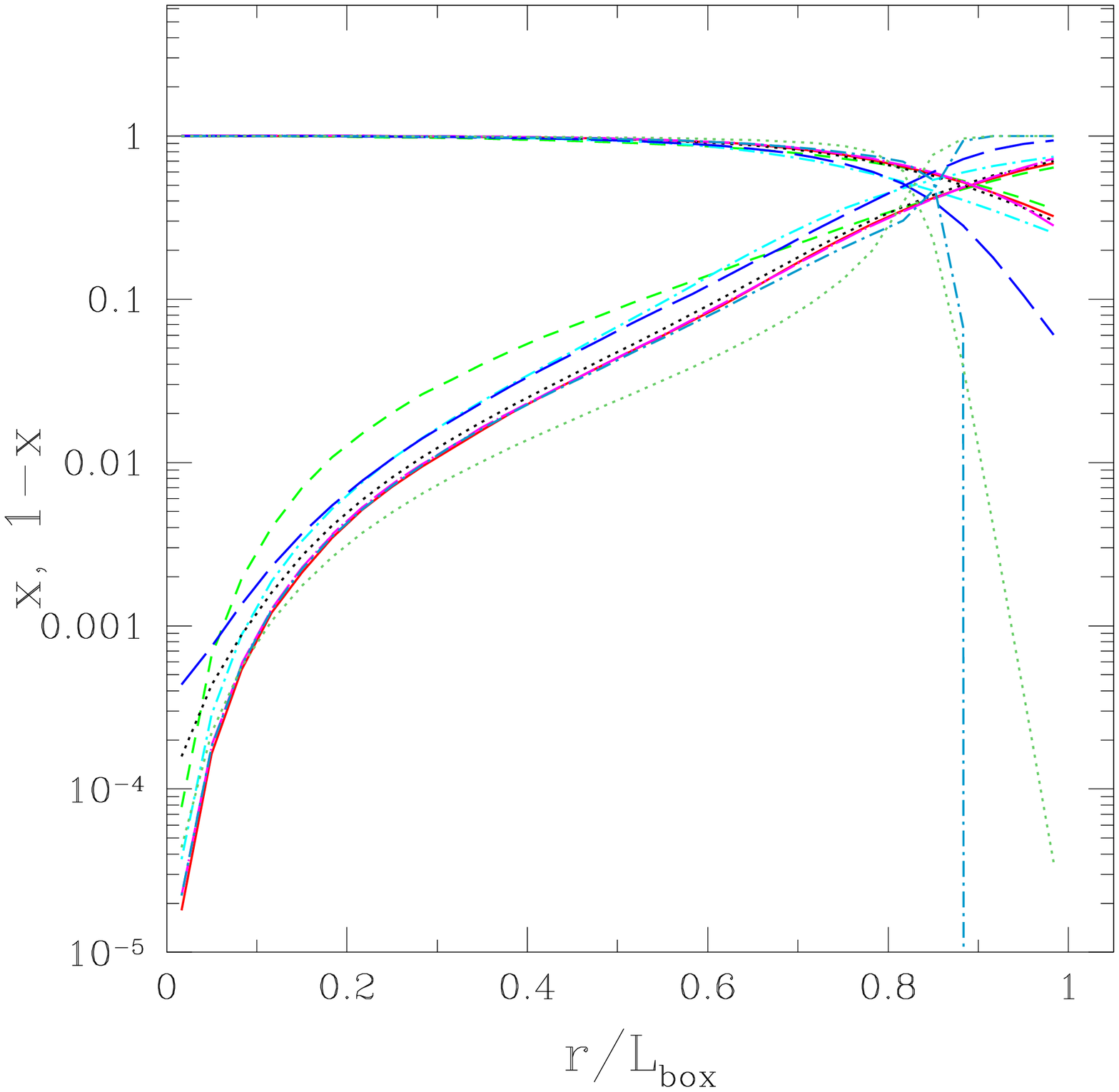}
\caption{Test 2 (H~II region expansion in an uniform gas with varying 
temperature): Spherically-averaged  ionized fraction $x$ 
  and neutral fraction $1-x$ profiles at times $t=10$ Myr, 100 Myr and 
500 Myr. 
\label{T2_profs_fig}}
\end{center}
\end{figure*}

In Figure~\ref{T2_profs_fig} we show the spherically-averaged radial profiles
of the neutral and ionized fractions during the fast expansion phase
($t=10$~Myr, left), the slowing-down phase ($t=100$~Myr, middle) and the final
Str\"omgren sphere ($t=500$~Myr, right). These confirm, in a more quantitative
way, the trends already noted based on the 2D images above. The profiles from
$C^2$-Ray and RSPH codes are in excellent agreement at all radii and all
times. The IFT code closely agrees with them in the source proximity region,
where the gas ionized state is at equilibrium, but diverges around the I-front
(due to its assumed equilibrium chemistry) and ahead of the I-front (due to
its assumption that the front is sharp). Compared to these codes, CRASH
and ART find slightly higher neutral fractions close to the ionizing
source, but agree well with $C^2$-Ray and RSPH ahead of the I-front in
the spectral hardening region. The FTTE code is in excellent agreement with 
$C^2$-Ray and RSPH codes close to the source, but its I-front is much
sharper. The OTVET code also finds a somewhat sharper I-front, but to a much
lesser extent than the FTTE code, while close to the source its neutral
fraction is only slightly higher than the majority of codes, and in close
agreement with the ART code. Finally, the I-front derived by the Zeus code is
very sharp. This is due to the use of only single-energy photons by this
code, which does not allow for spectral hardening.

\begin{figure*}
\begin{center}
  \includegraphics[width=2.2in]{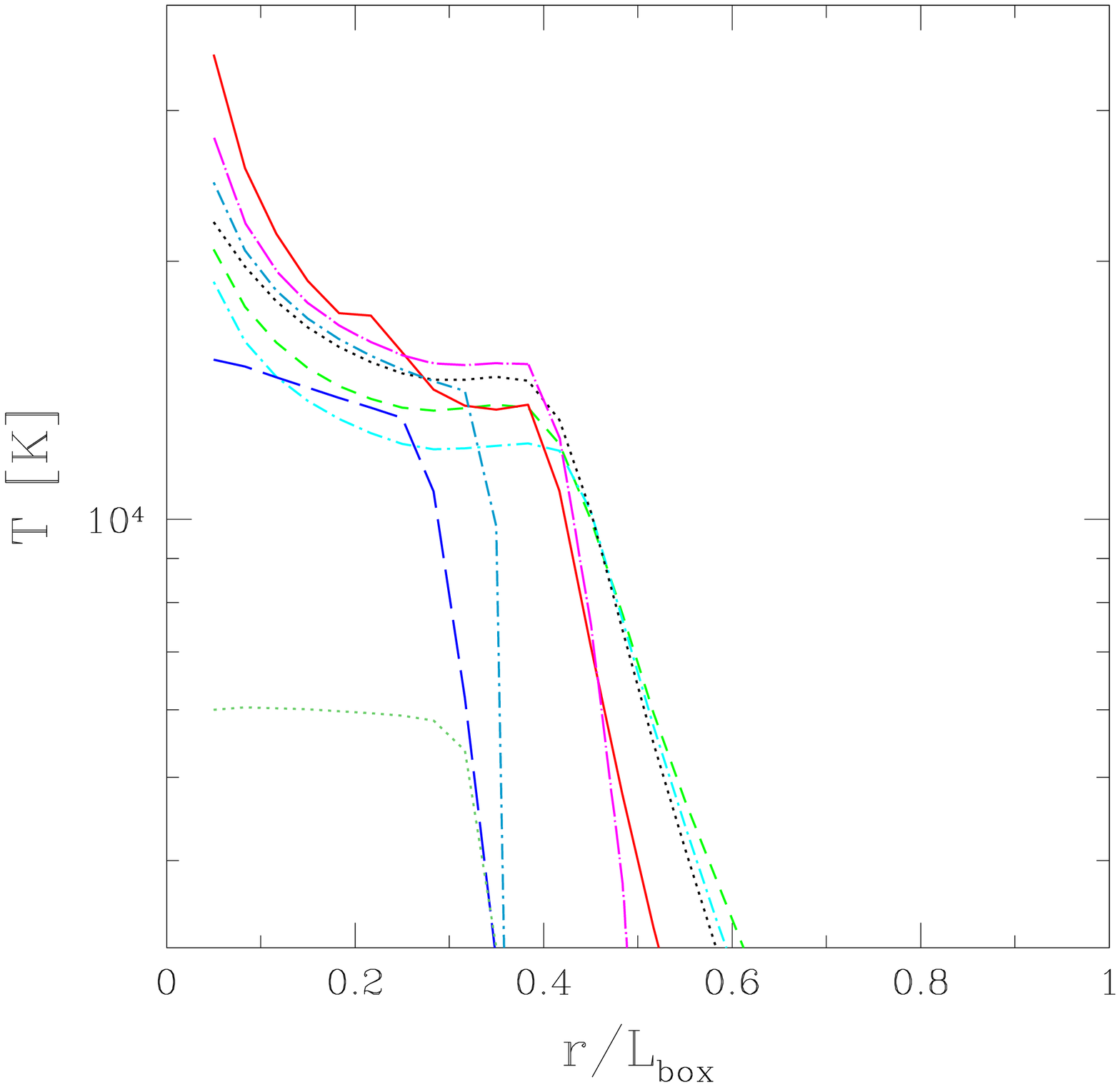}
  \includegraphics[width=2.2in]{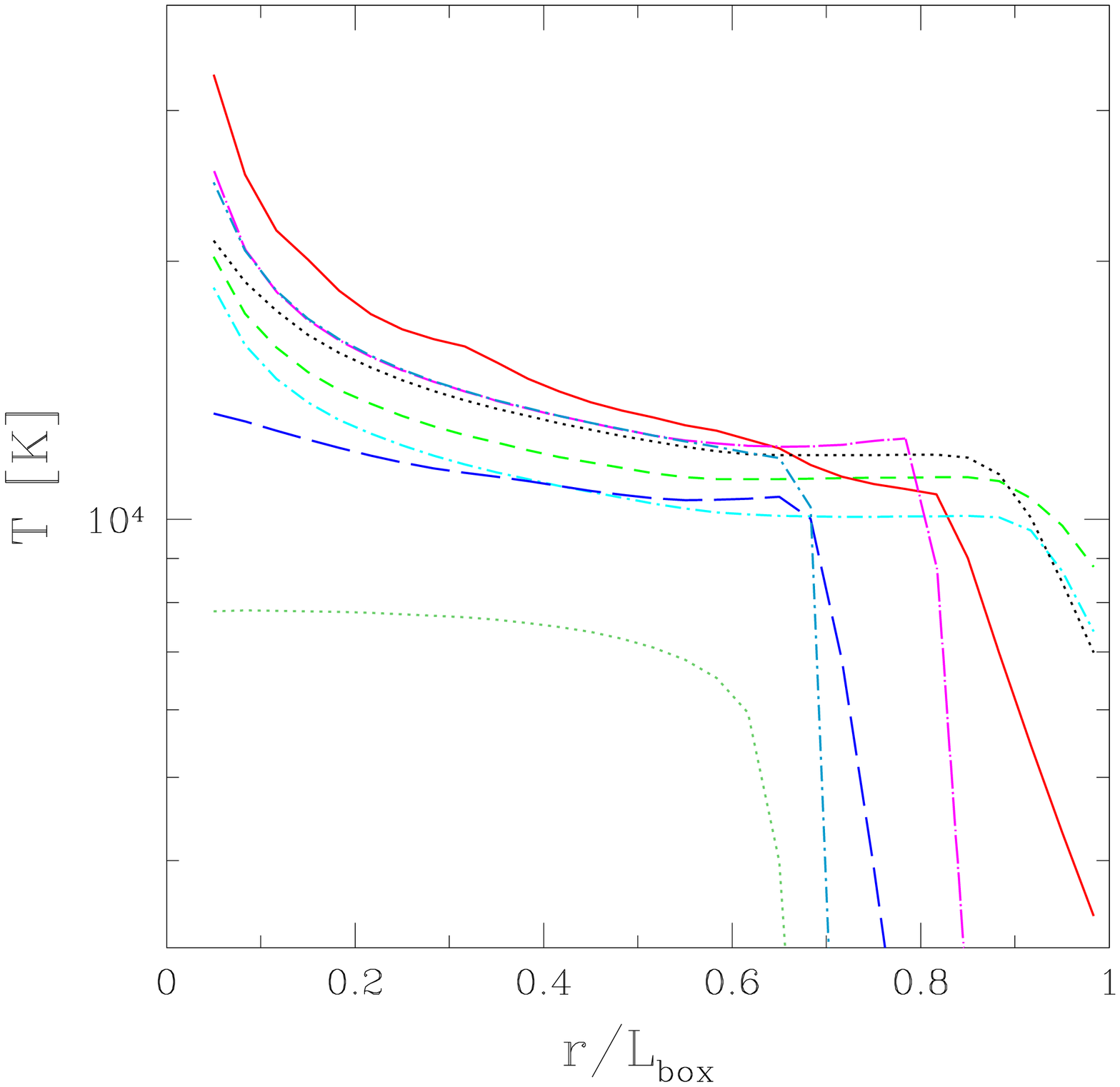}
  \includegraphics[width=2.2in]{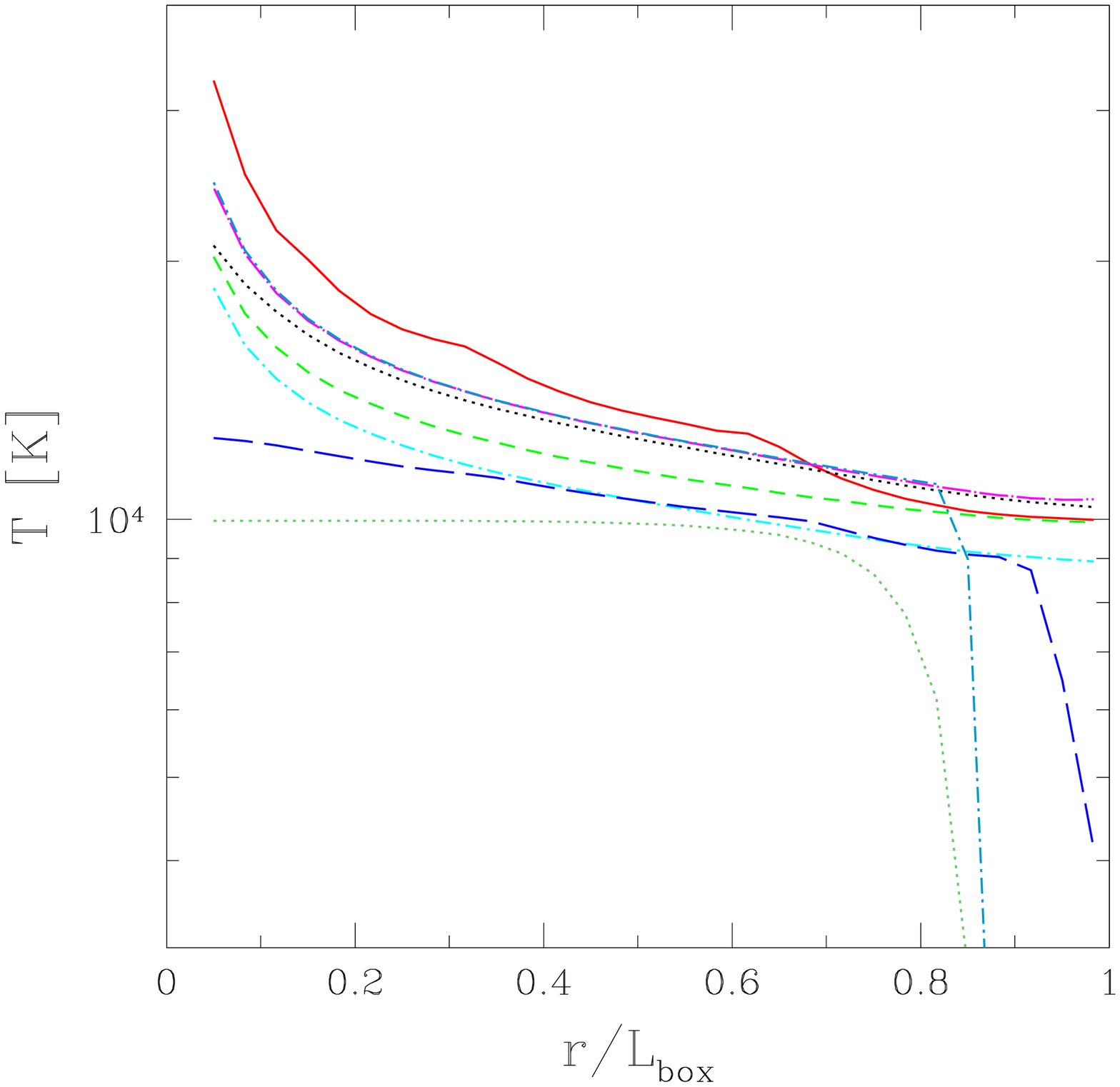}
\caption{Test 2 (H~II region expansion in an uniform gas with varying 
temperature): Spherically-averaged temperature profiles at times 
$t=10$ Myr, 100 Myr and 500 Myr. 
\label{T2_profsT_fig}}
\end{center}
\end{figure*}

The corresponding spherically-averaged radial temperature profiles at the same
three I-front evolutionary phases are shown in Figure~\ref{T2_profsT_fig}. All
results (except the one from Zeus, due to the monochromatic spectrum it used)
agree well inside the ionized region, with the differences arising largely due
to slight differences in the cooling rates adopted. The more diffusive OTVET
code does not show as sharp temperature rise in the source proximity as the
other codes, but elsewhere the temperature structure it finds agrees with the 
majority of the codes. Again at the I-front and ahead of it the differences
between the results are significant, reflecting the different handling of hard
photons by the codes.  

\begin{figure*}
\begin{center}
  \includegraphics[width=2.2in]{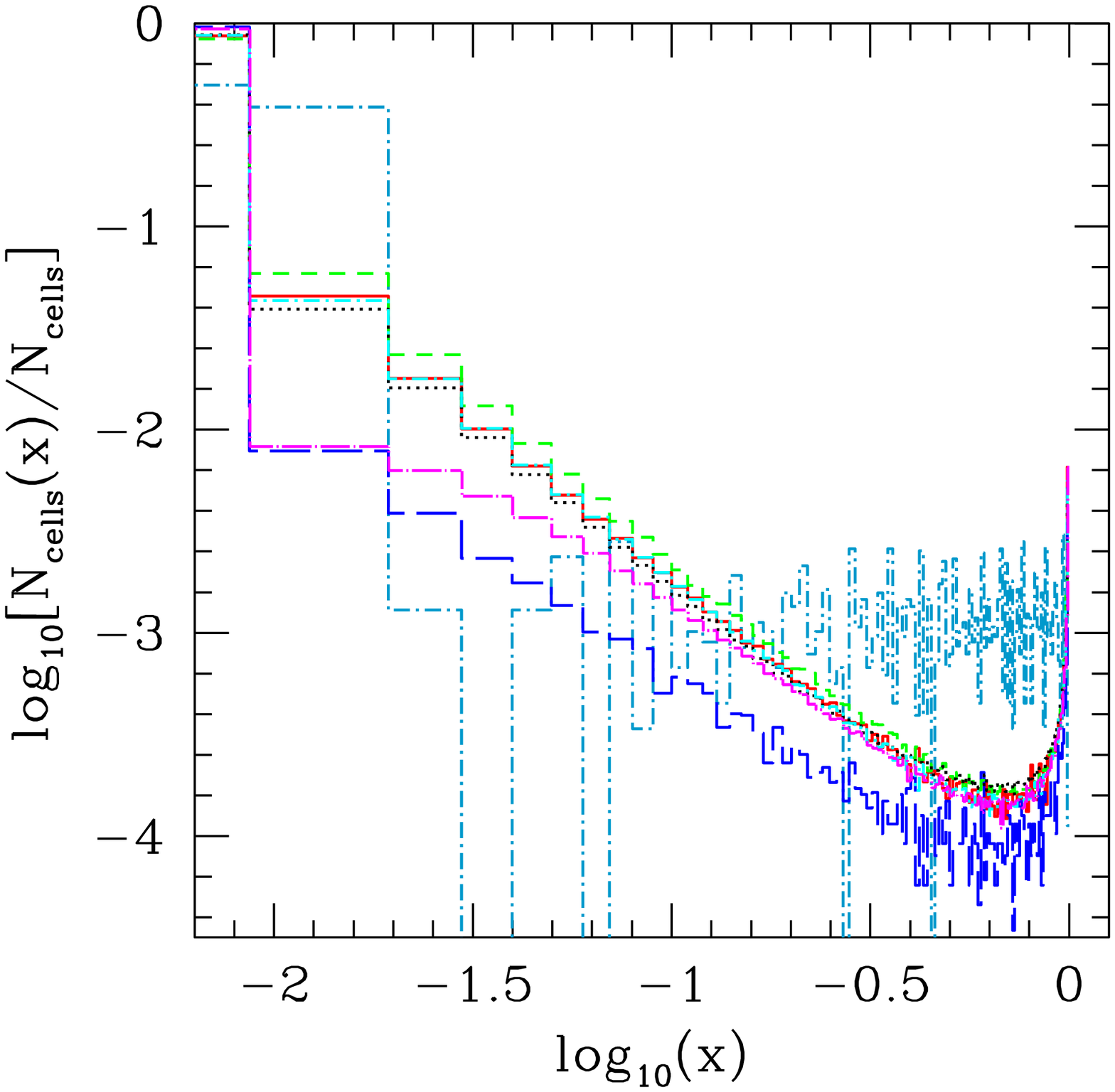}
  \includegraphics[width=2.2in]{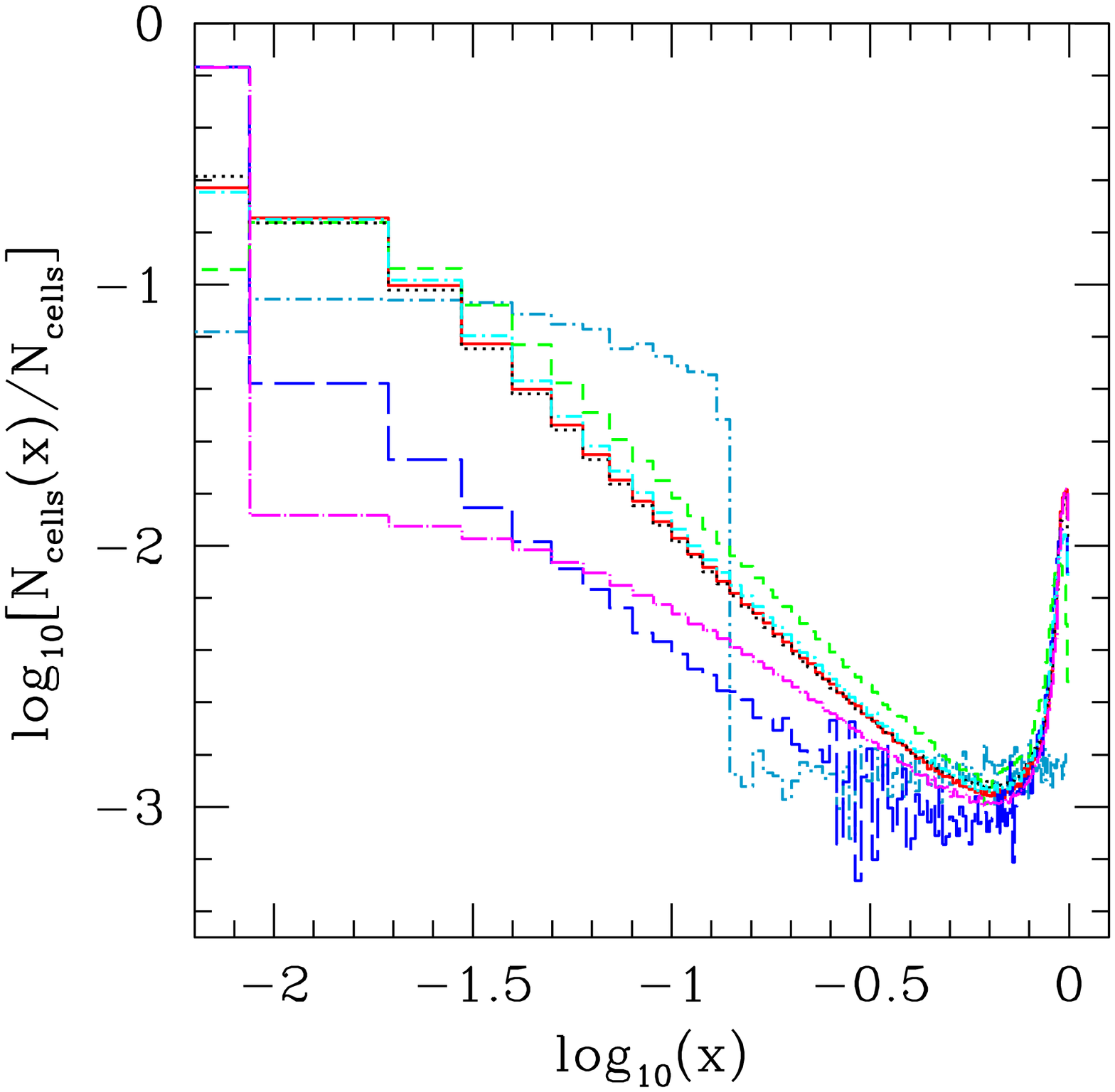}
  \includegraphics[width=2.2in]{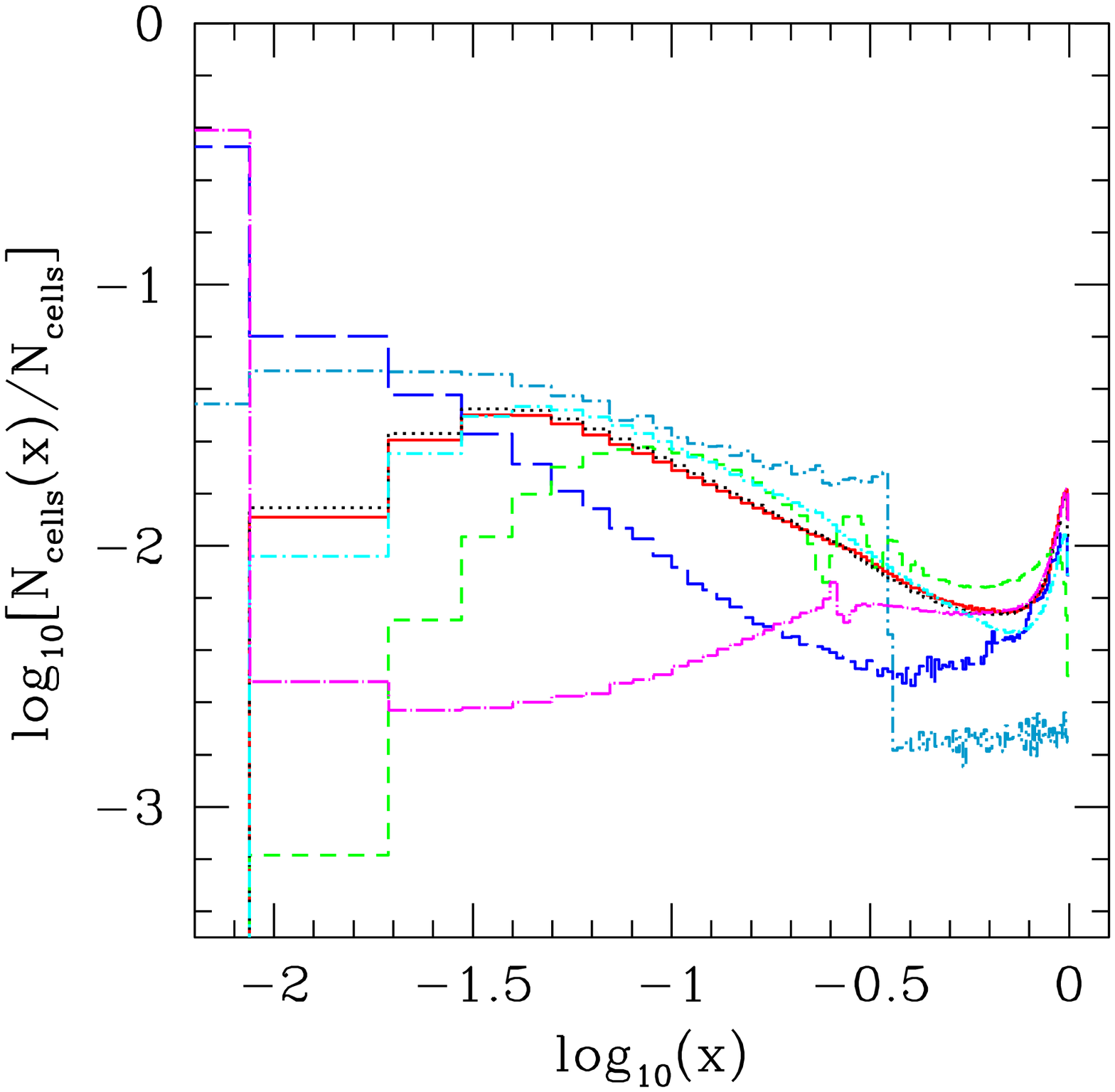}
\caption{Test 2 (H~II region expansion in an uniform gas with varying 
temperature): Fraction of cells with a given ionized fraction, $x$, at
   times (left) $t=10$ Myr, (middle) 100 Myr and (right) 500 Myr. 
\label{T2_hist_fig}}
\end{center}
\end{figure*}

In Figure~\ref{T2_hist_fig} we show the histograms of the fraction of cells 
with a given ionized fraction $x$ at the same times as the radial profiles
above. During the early, fast expansion phase all codes agree well except the
OTVET code, which finds a slightly thinner I-front, but an otherwise same histogram
distribution shape, and CRASH, whose I-front is a bit thicker. IFT finds a 
different ionized fraction distribution, again as a consequence of its 
equilibrium chemistry, which is not correct at the I-front transition. Later,
when the I-front slows down ($t=100$~Myr) the same trends hold, but in
addition the FTTE results start diverging significantly from the rest, finding
notably smaller ionized region and a quite different shape distribution in
the largely-neutral regions. This reflects its much sharper I-front with
little spectrum hardening, as noted above. Finally, the ionized fraction 
histogram corresponding to the Str\"omgren sphere ($T=500$~Myr) shows similar
differences. The $C^2$-Ray, ART and RSPH codes again agree very closely, and
CRASH also finds a similar distribution, but with a thicker I-front and
correspondingly fewer neutral cells. The OTVET distribution follows a roughly 
similar shape but with a thinner front transition and more neutral cells, while
FTTE agrees well with the other ray-tracing codes in the highly-ionized
region, but still diverges considerably at the I-front and ahead of it. 

\begin{figure*}
\begin{center}
  \includegraphics[width=2.2in]{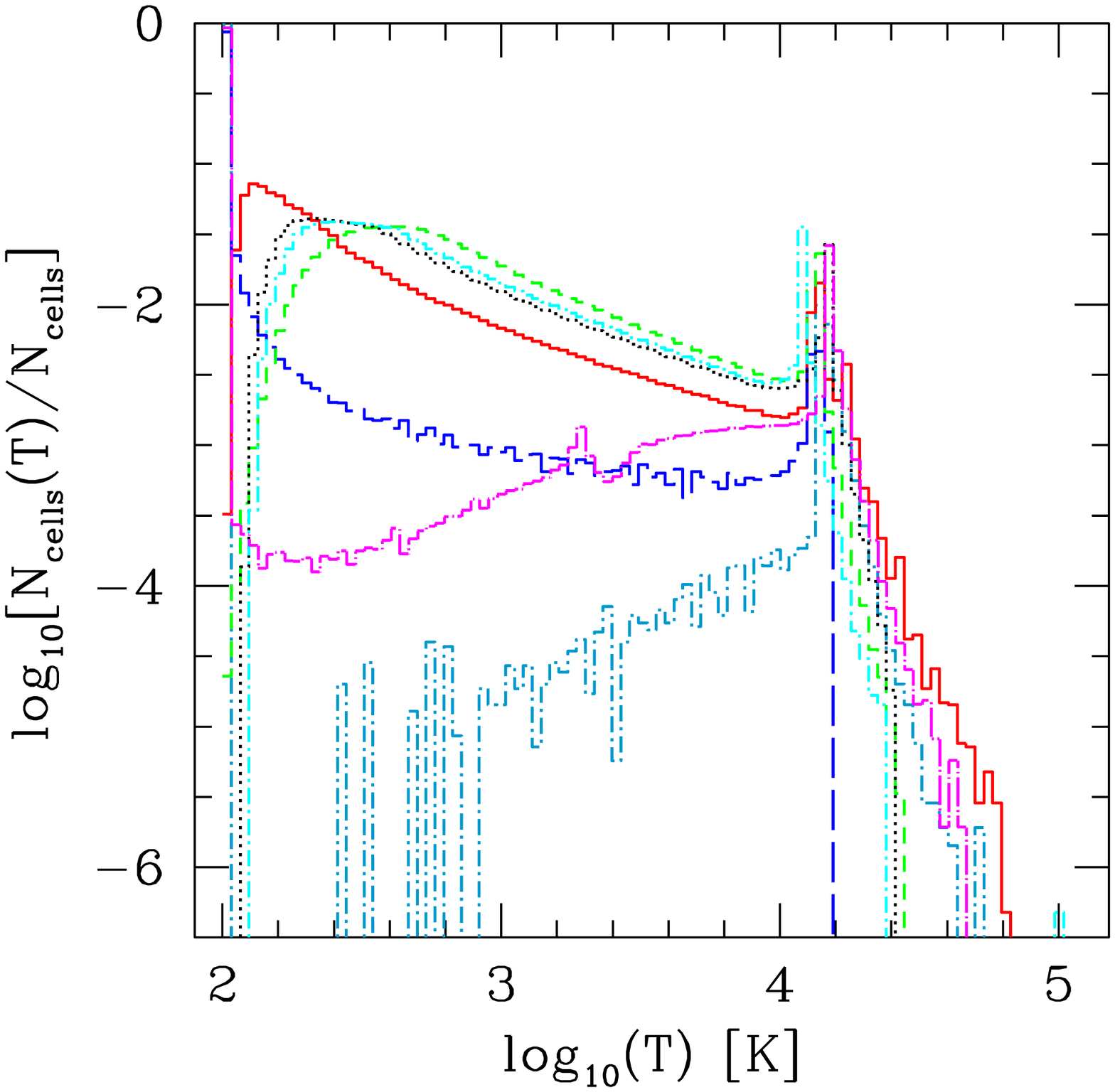}
  \includegraphics[width=2.2in]{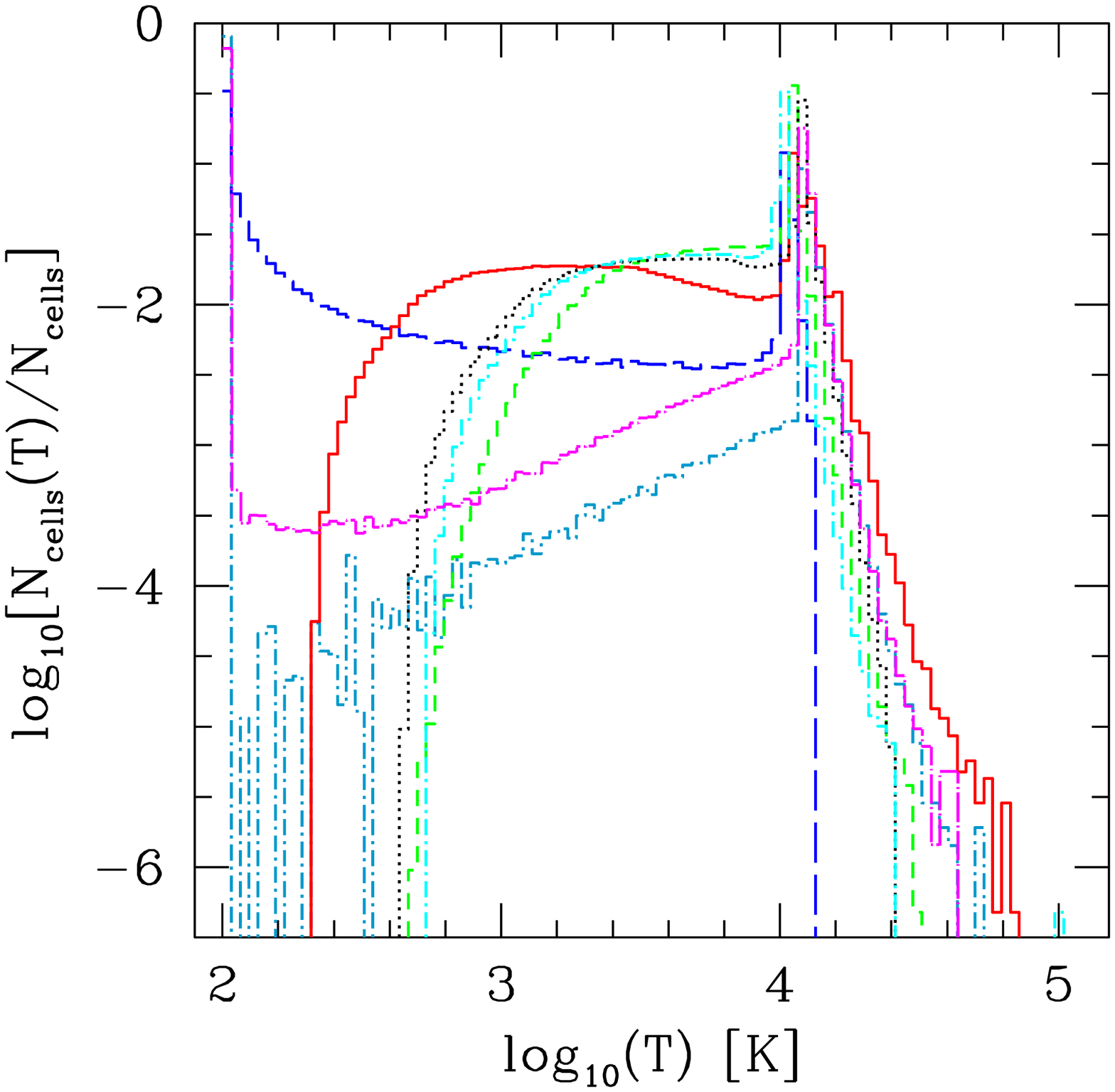}
  \includegraphics[width=2.2in]{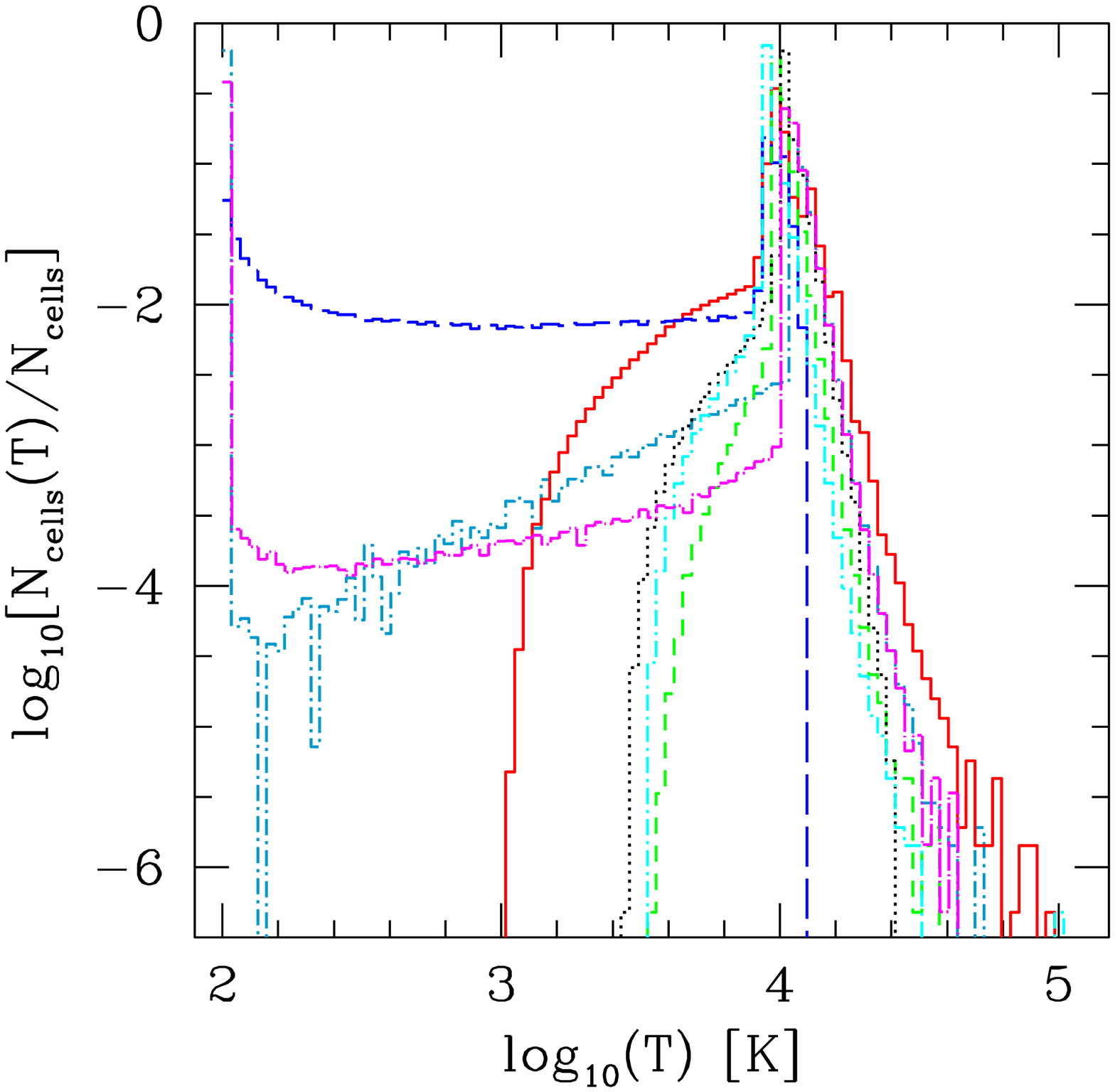}
\caption{Test 2 (H~II region expansion in an uniform gas with varying 
temperature): Fraction of cells with a given temperature $T$ at
   times (left) $t=10$ Myr, (middle) 100 Myr and (right) 500 Myr. 
\label{T2_histT_fig}}
\end{center}
\end{figure*}
\begin{figure}
  \includegraphics[width=3.5in]{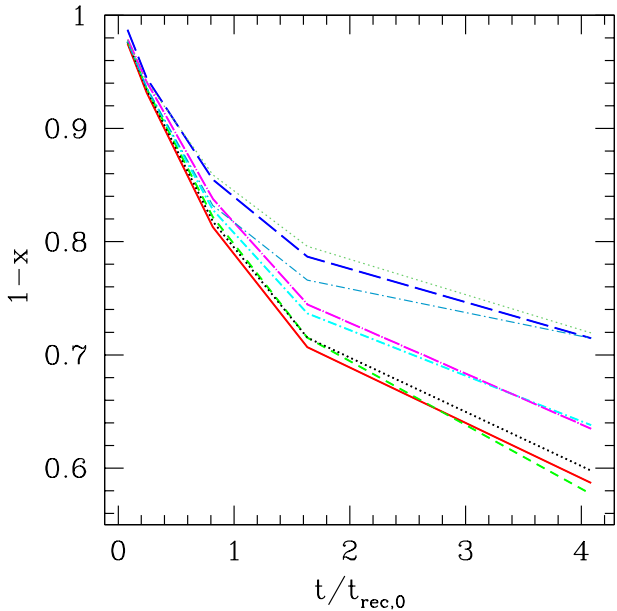}
\caption{Test 2 (H~II region expansion in an uniform gas with varying 
temperature): Evolution of the total neutral gas fraction. 
\label{T2_fracs_fig}}
\end{figure}
\begin{figure*}
  \includegraphics[width=3.5in]{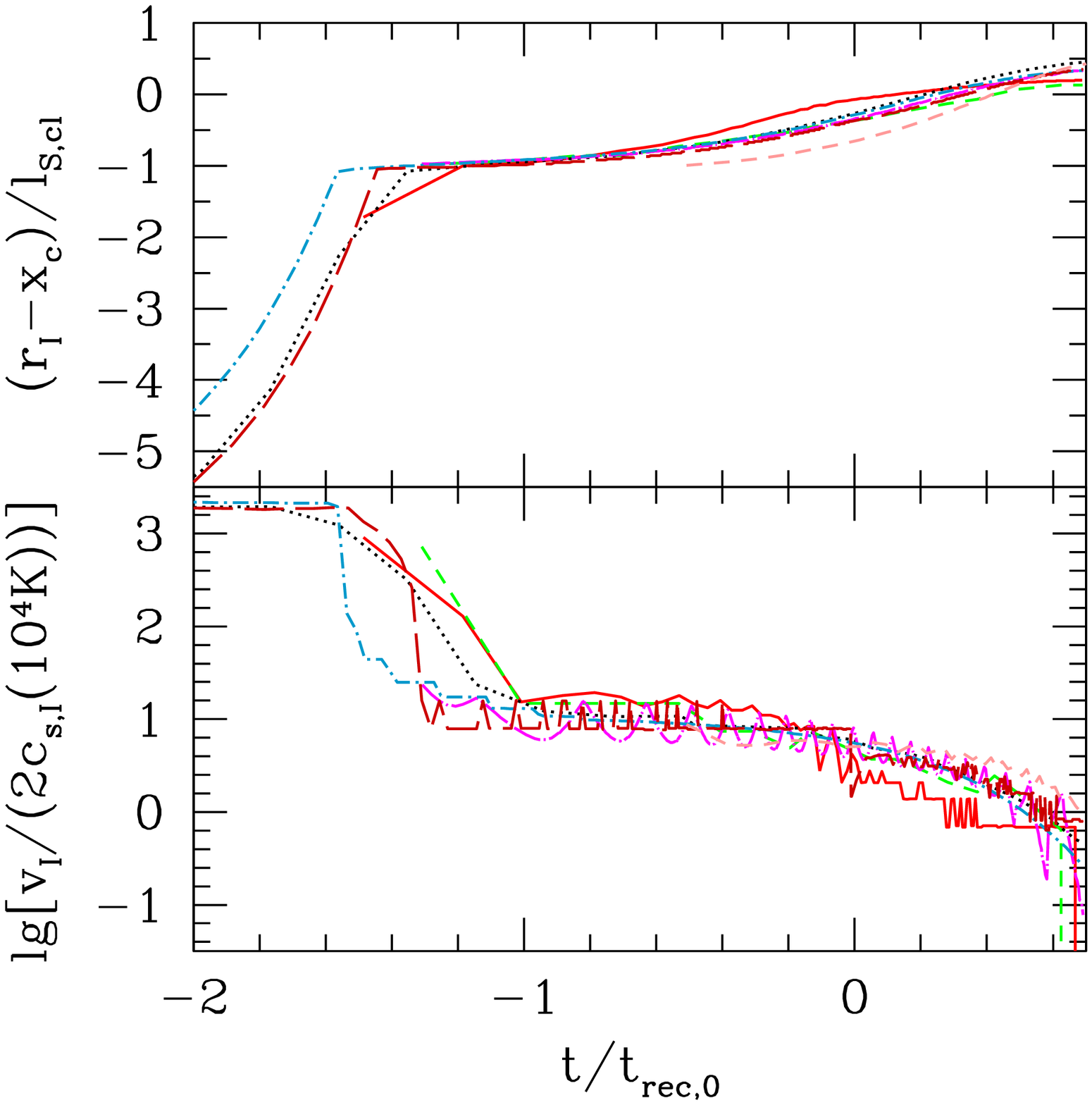}
\caption{Test 3 (I-front trapping in dense clump): The evolution of the 
position and velocity of the I-front. 
\label{T3_Ifront_evol_fig}}
\end{figure*}

The corresponding histograms of temperature are shown in
Figure~\ref{T2_histT_fig}. The ionized gas temperatures found by all codes
have a strong peak at slightly above $10^4$~K, with only small variations in
the peak position between the different codes. This peak was to be expected,
as a consequence of the combination of photoheating and hydrogen line cooling
(which peaks around $10^4$~K). At higher temperatures (corresponding to cells 
close to the ionizing source) some differences emerge. The ray-tracing codes
($C^2$-Ray, CRASH, ART, RSPH, IFFT, IFT) largely agree among themselves, with
only $C^2$-ray finding slightly larger fraction of hot cells. OTVET, on the
other hand does not predict any cells with temperature above $\sim16,000$~K,
due to missing the very hot proximity region of the source, as noted above.
Below $\sim8,000$~K, on the other hand, the differences between results are
more significant, reflecting the variations in the I-front thickness and
spectral hardening noted above. The results from CRASH, ART and RSPH agree
well at all times, while the $C^2$-Ray histograms have similar shape, but with
some offset, due to its current somewhat simplified handling of the energy
input which uses a single bin in frequency. The OTVET, FTTE and IFT codes
find much smaller pre-heating ahead of the I-front, and thus different
distributions. 

Finally, in Figure~\ref{T2_fracs_fig} we show the evolution of the total
neutral gas fraction. All codes agree well on the final neutral fraction, 
within $\sim25\%$ or better. The differences are readily understood in 
terms of the different recombination rates, mostly as a consequence of 
the somewhat different temperatures found inside the H~II region, in 
addition to the small differences in the recombination rate fits used.  

\subsection{Test 3: I-front trapping in a dense clump and the formation of a
  shadow.}

\begin{figure*}
   \includegraphics[width=2.2in]{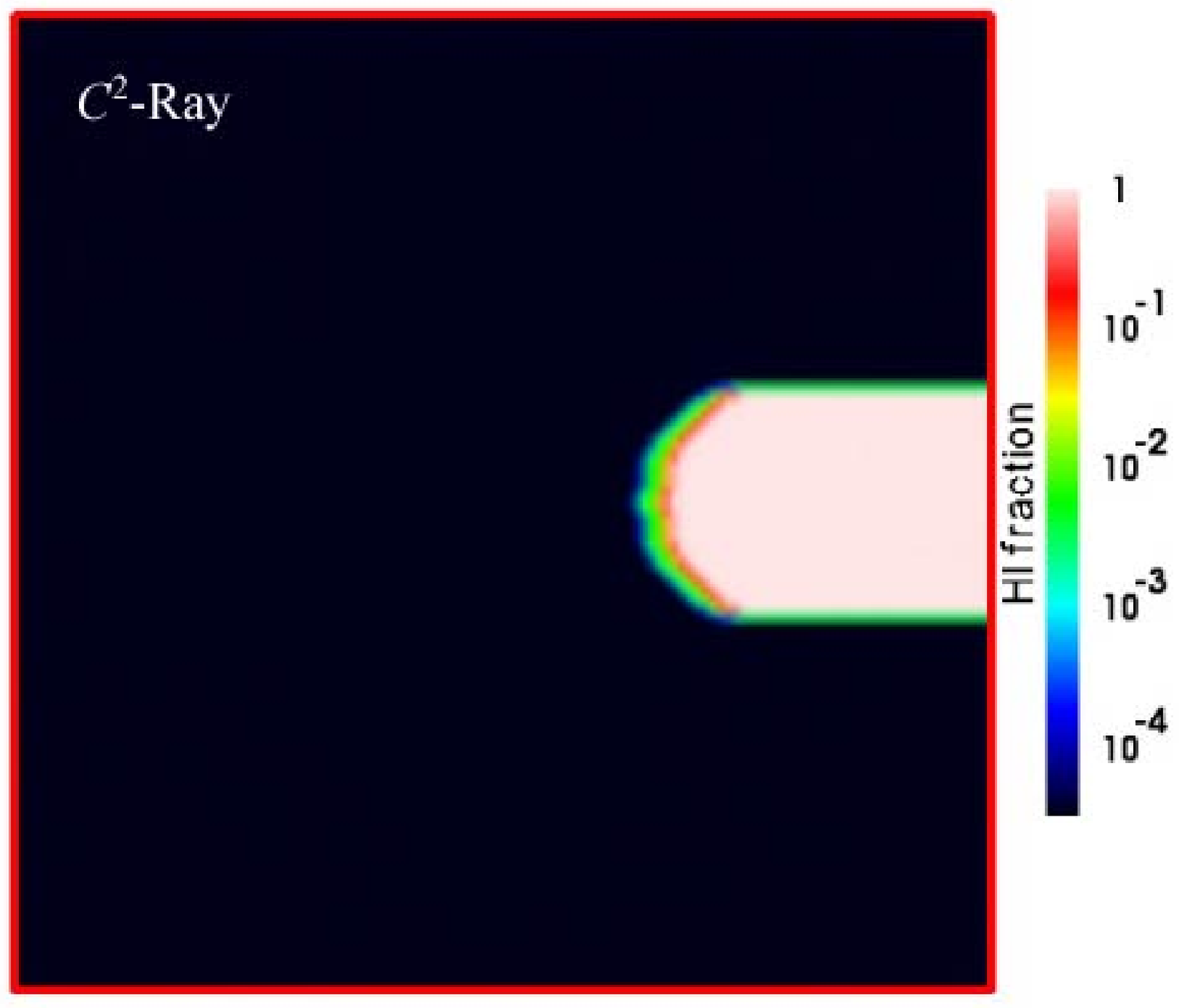}
   \includegraphics[width=2.2in]{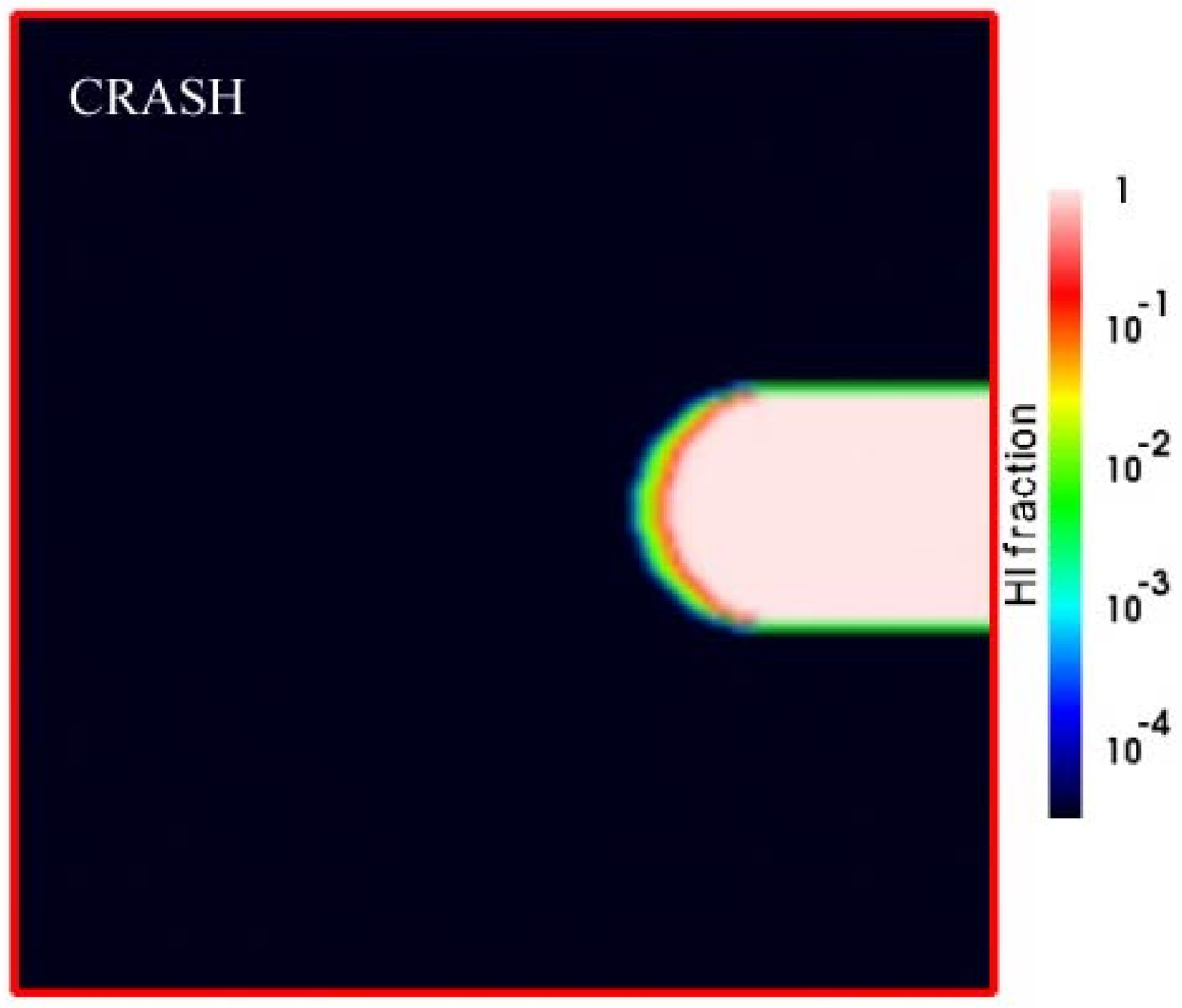}
   \includegraphics[width=2.2in]{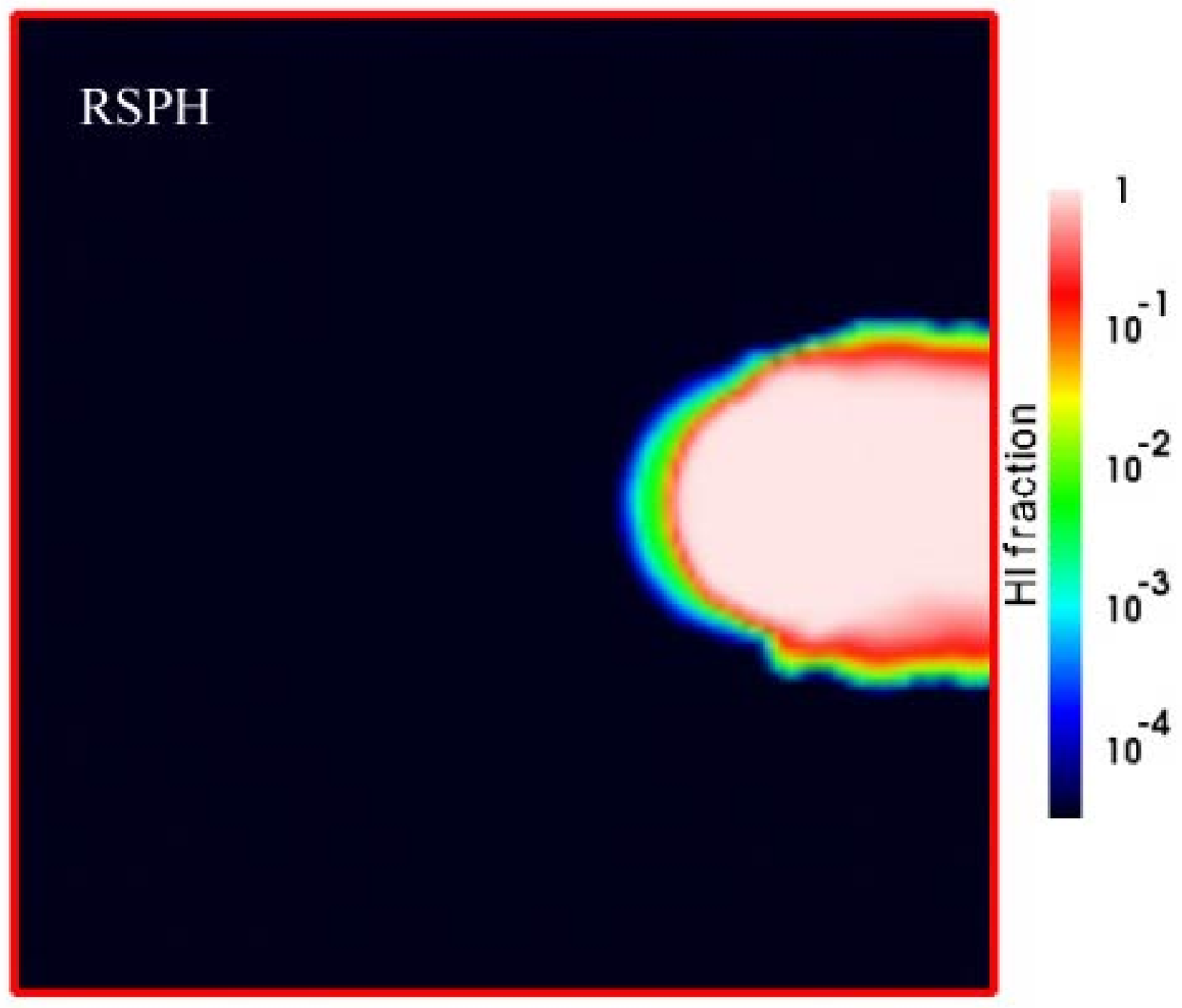}
   \includegraphics[width=2.2in]{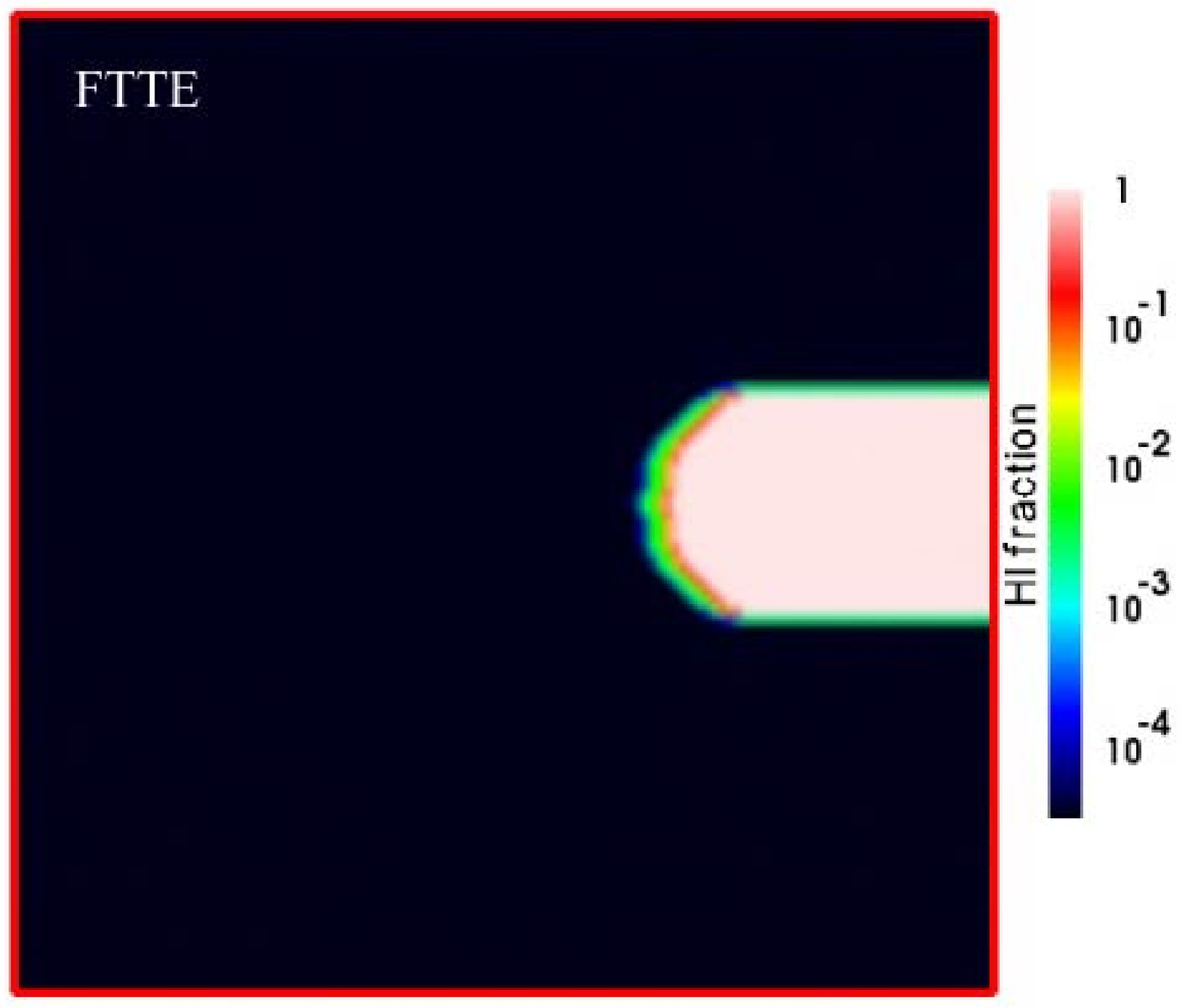}
   \includegraphics[width=2.2in]{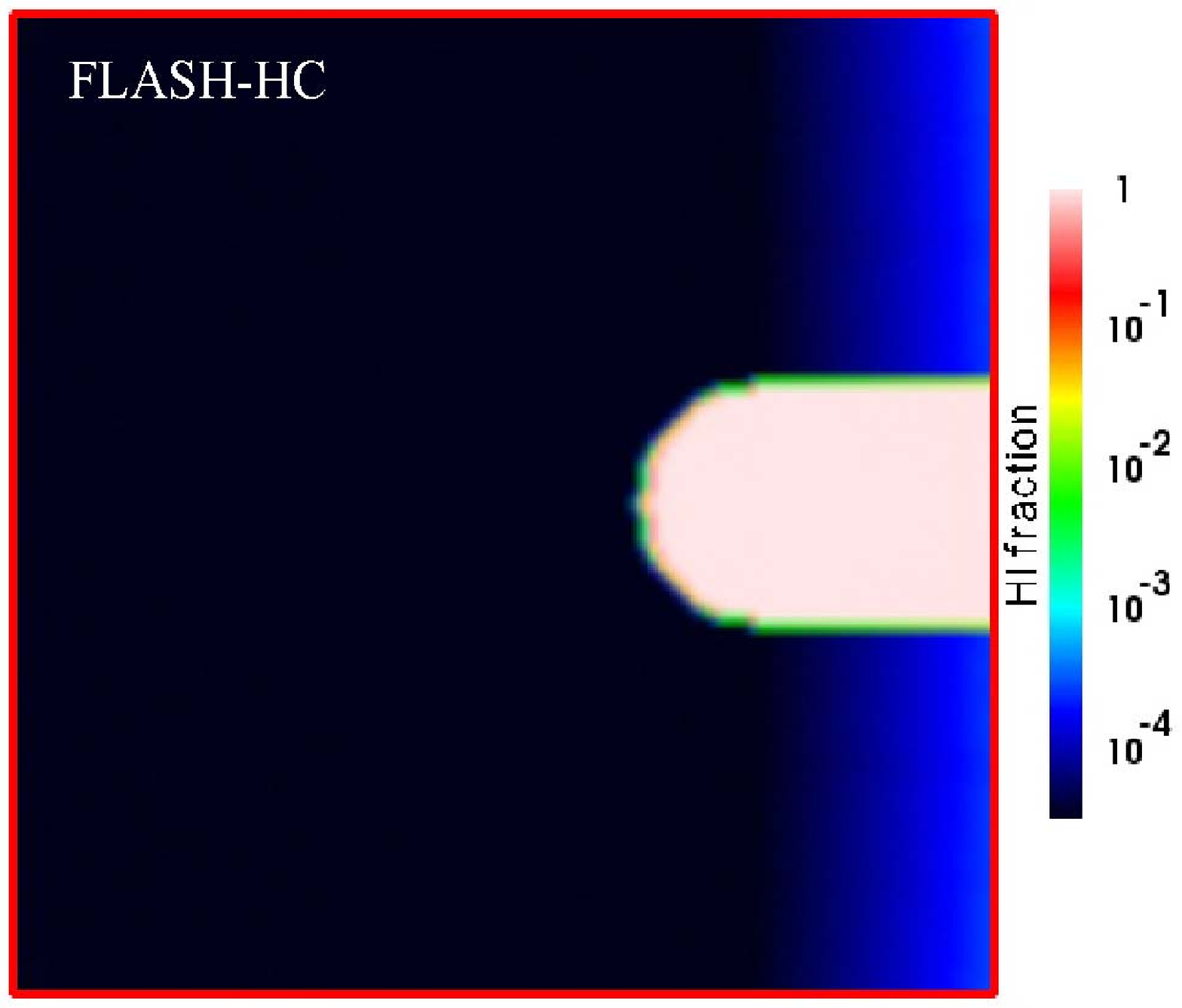}
   \includegraphics[width=2.2in]{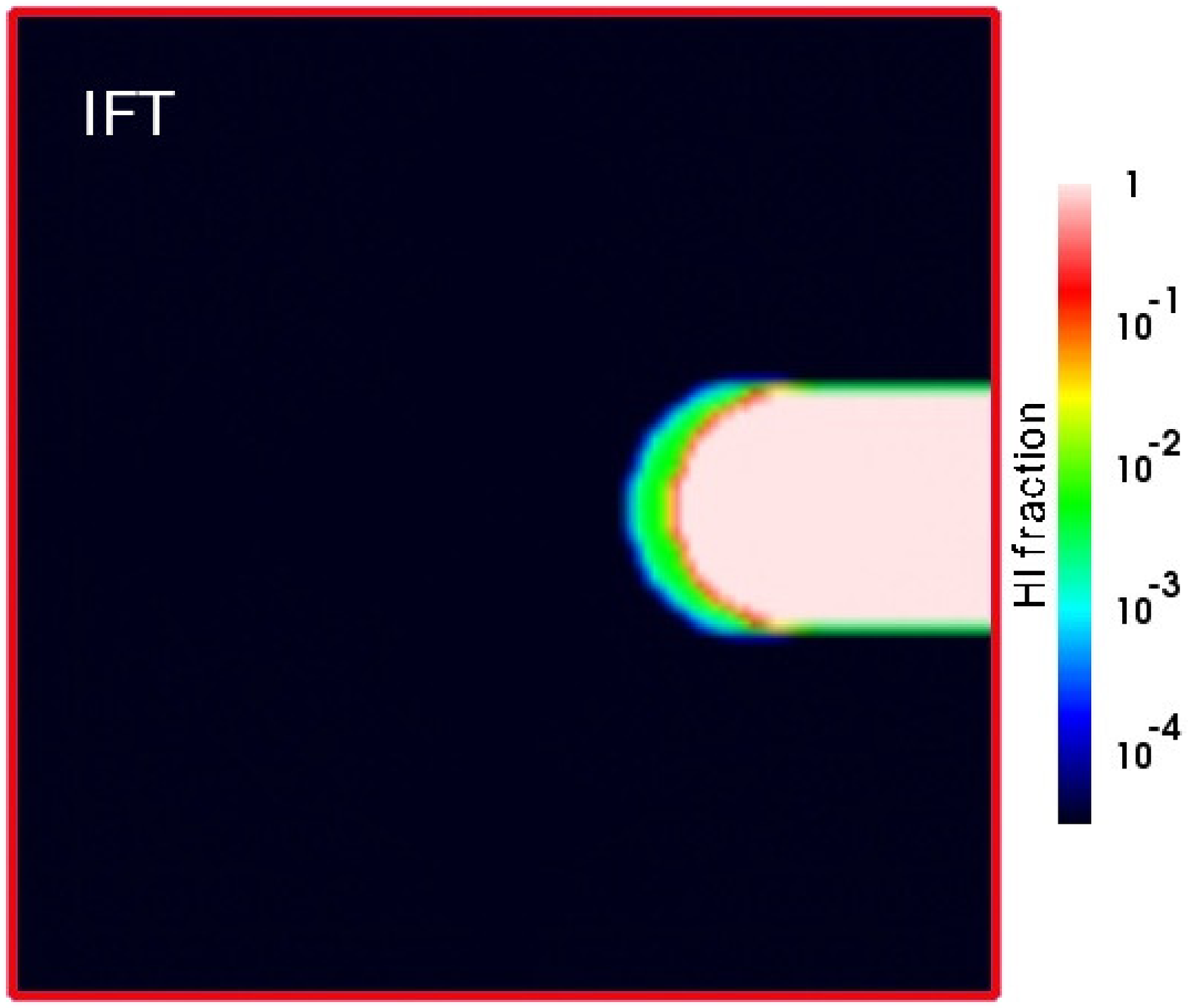}
   \includegraphics[width=2.2in]{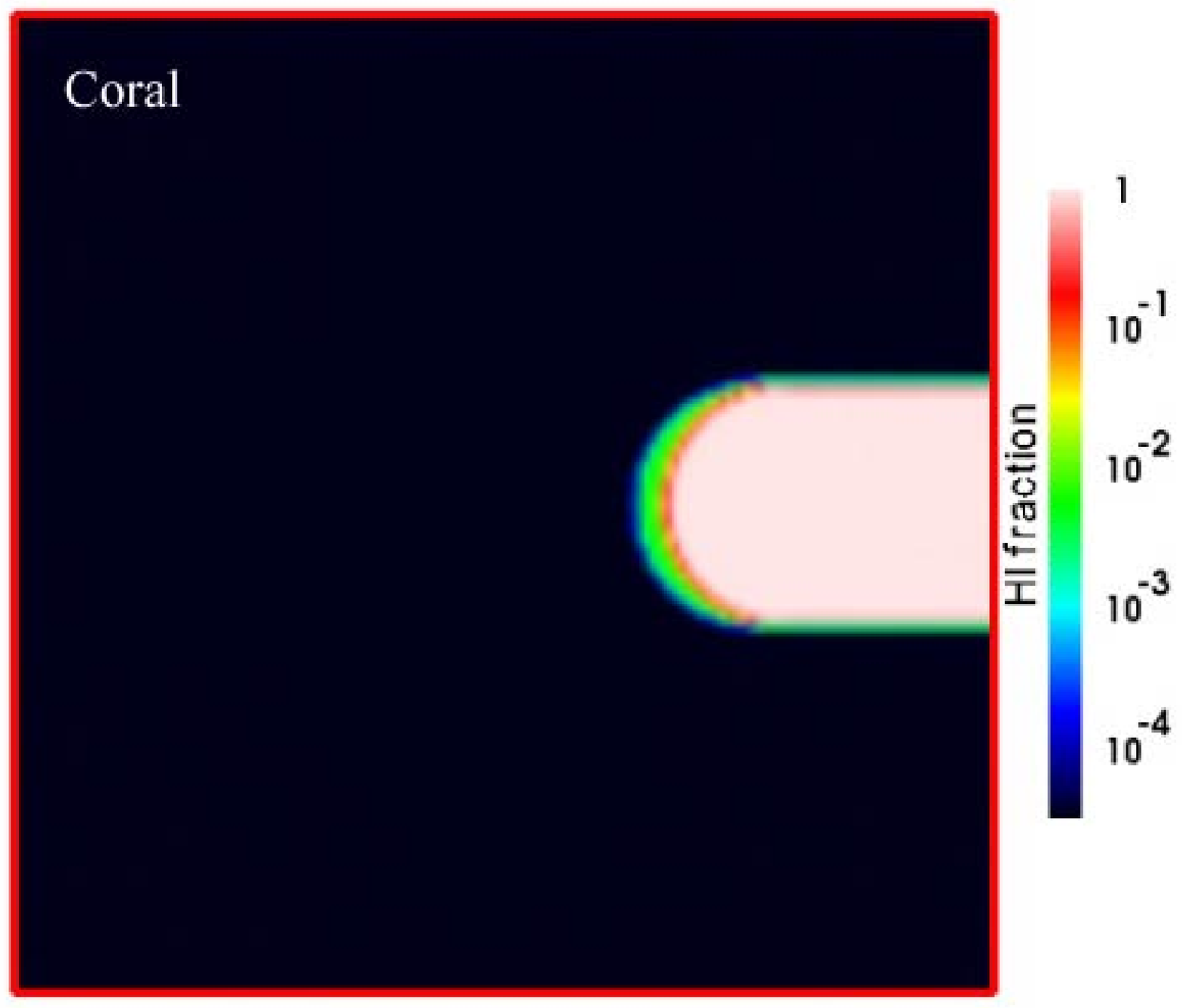}
\caption{Test 3 (I-front trapping in dense clump): Images of the H~I 
fraction, cut through the simulation volume at midplane at time $t=1$~Myr 
for $C^2$-Ray, CRASH, RSPH, FTTE, Flash-HC, IFT and Coral.
\label{T3_images1_HI_fig}}
\end{figure*}

\begin{figure*}
   \includegraphics[width=2.2in]{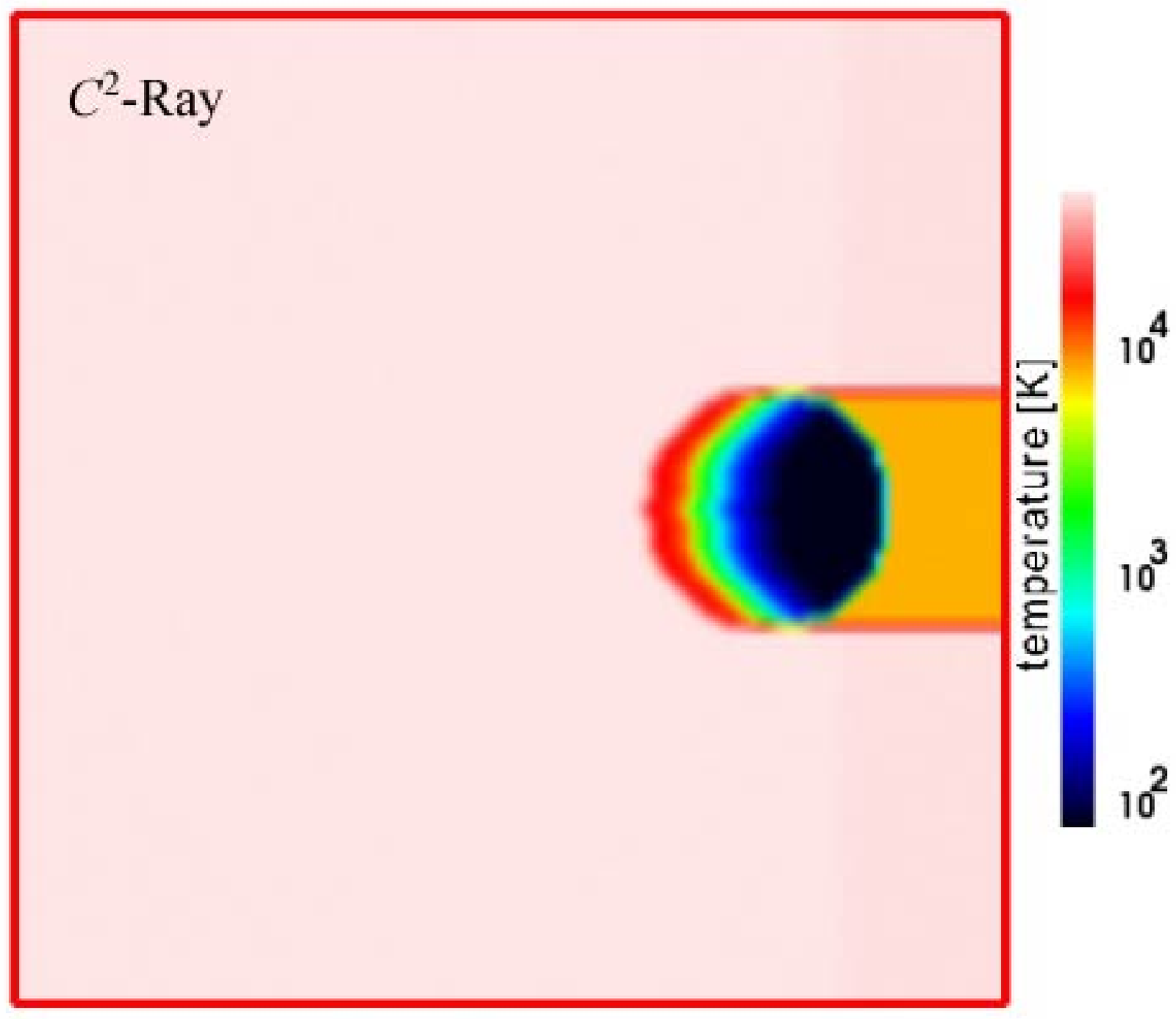}
   \includegraphics[width=2.2in]{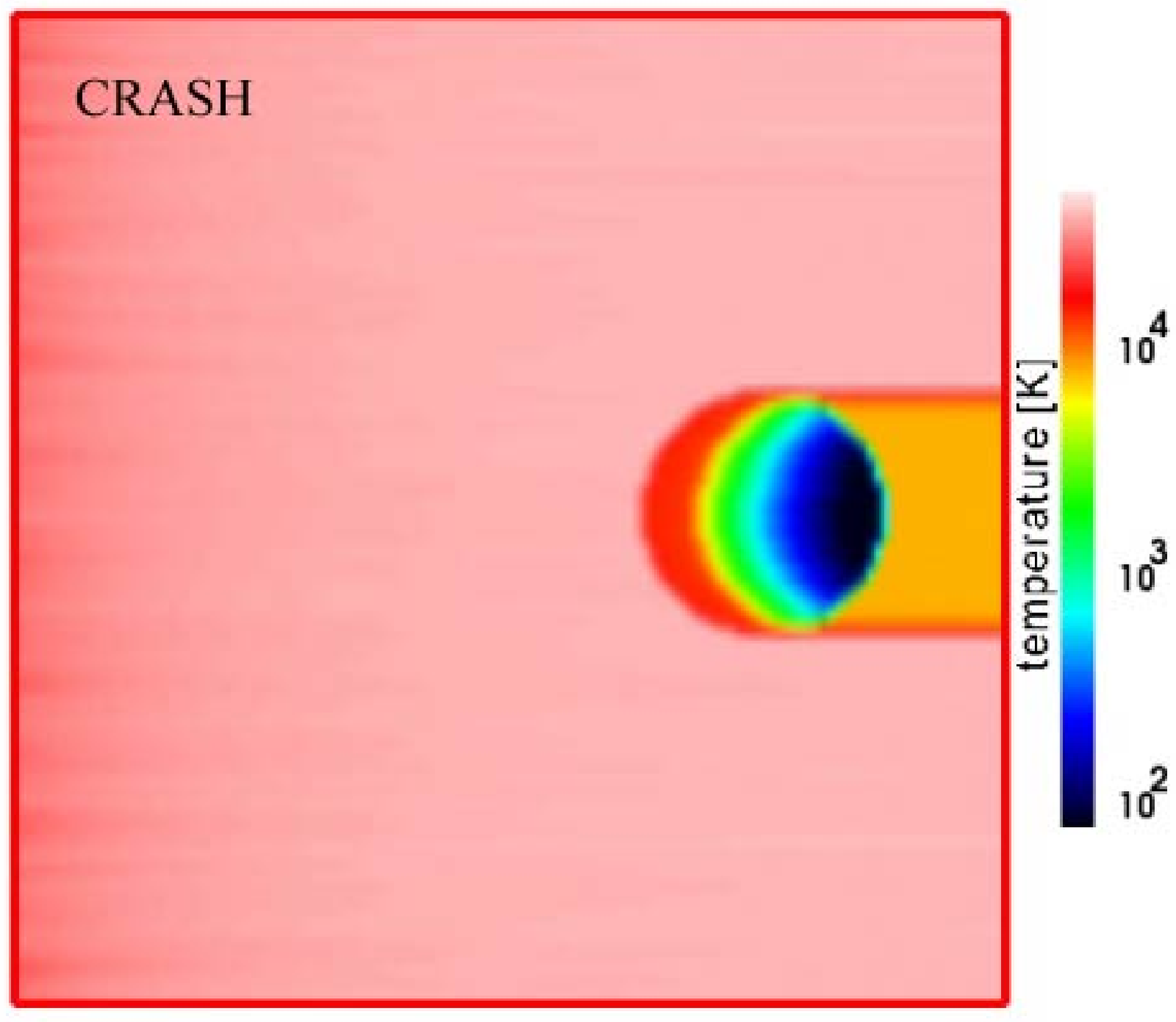}
   \includegraphics[width=2.2in]{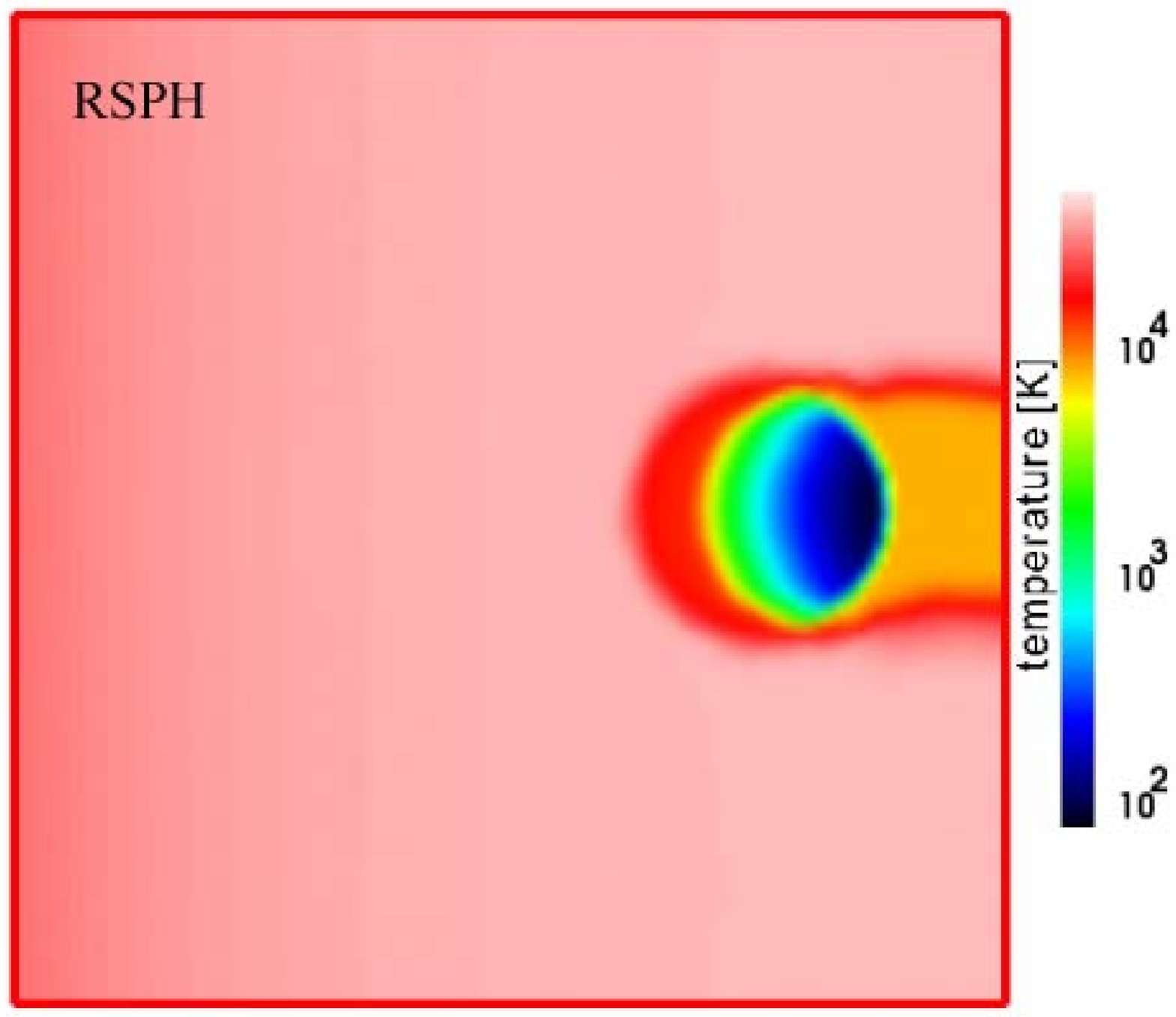}
   \includegraphics[width=2.2in]{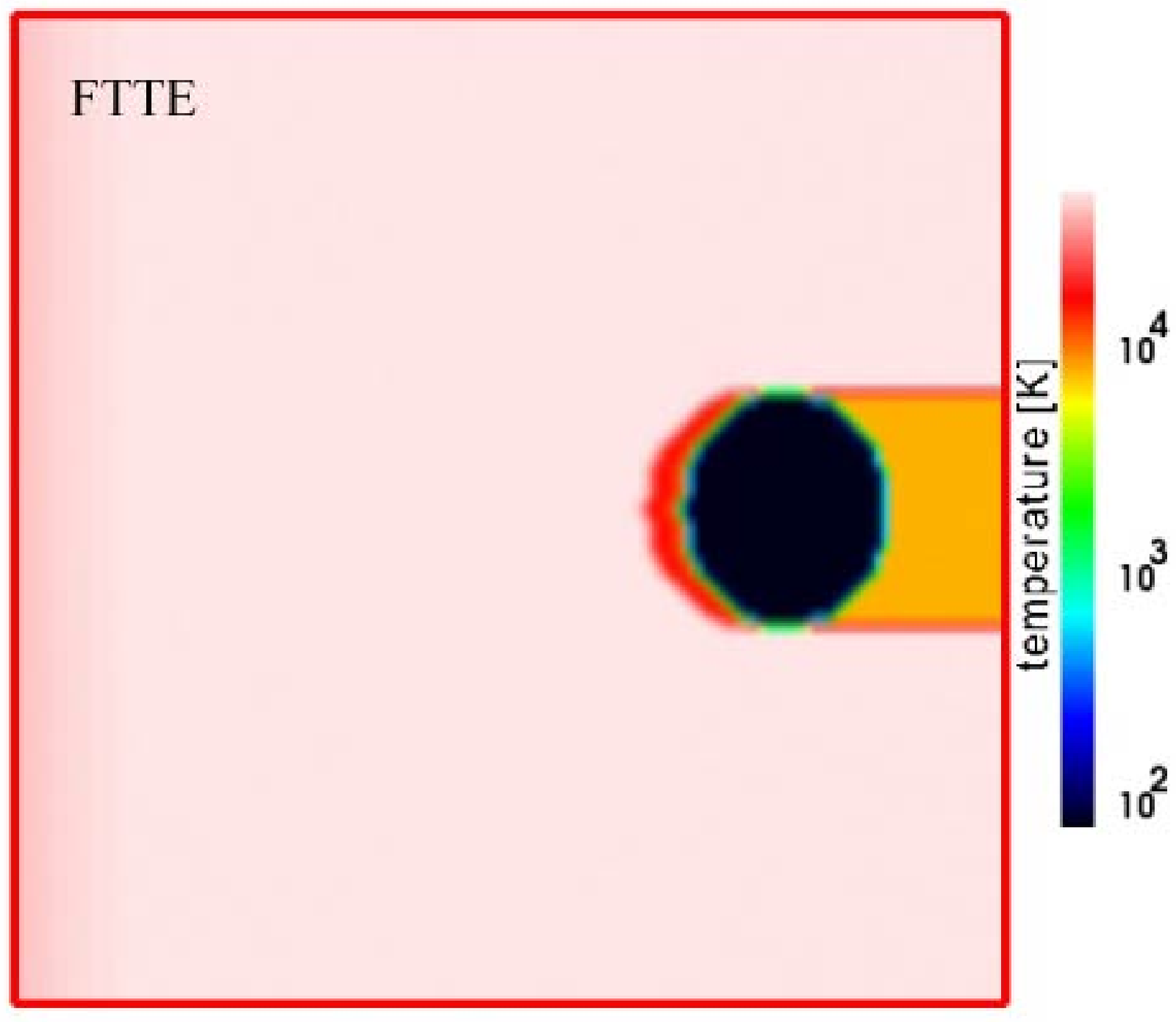}
   \includegraphics[width=2.2in]{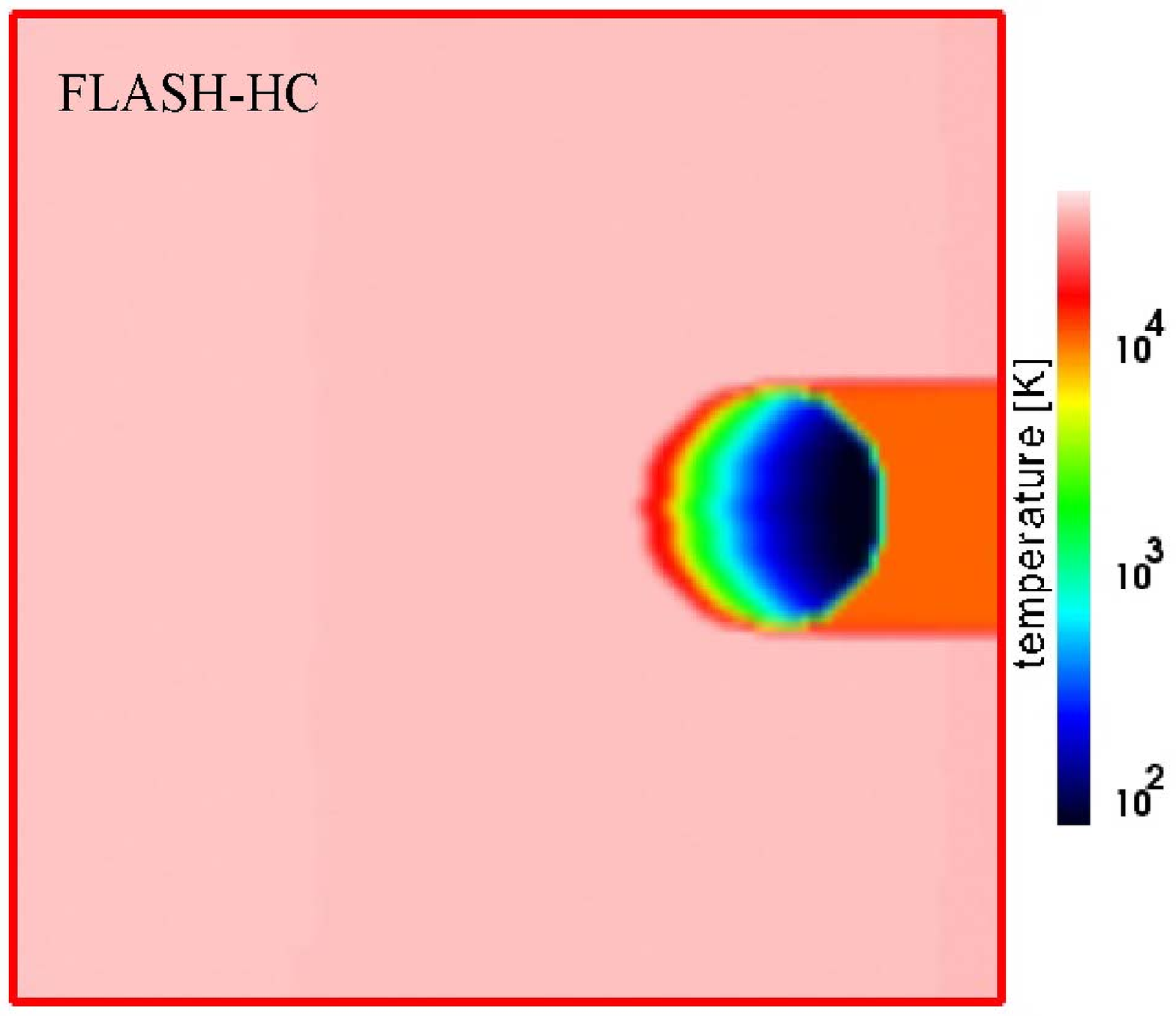}
   \includegraphics[width=2.2in]{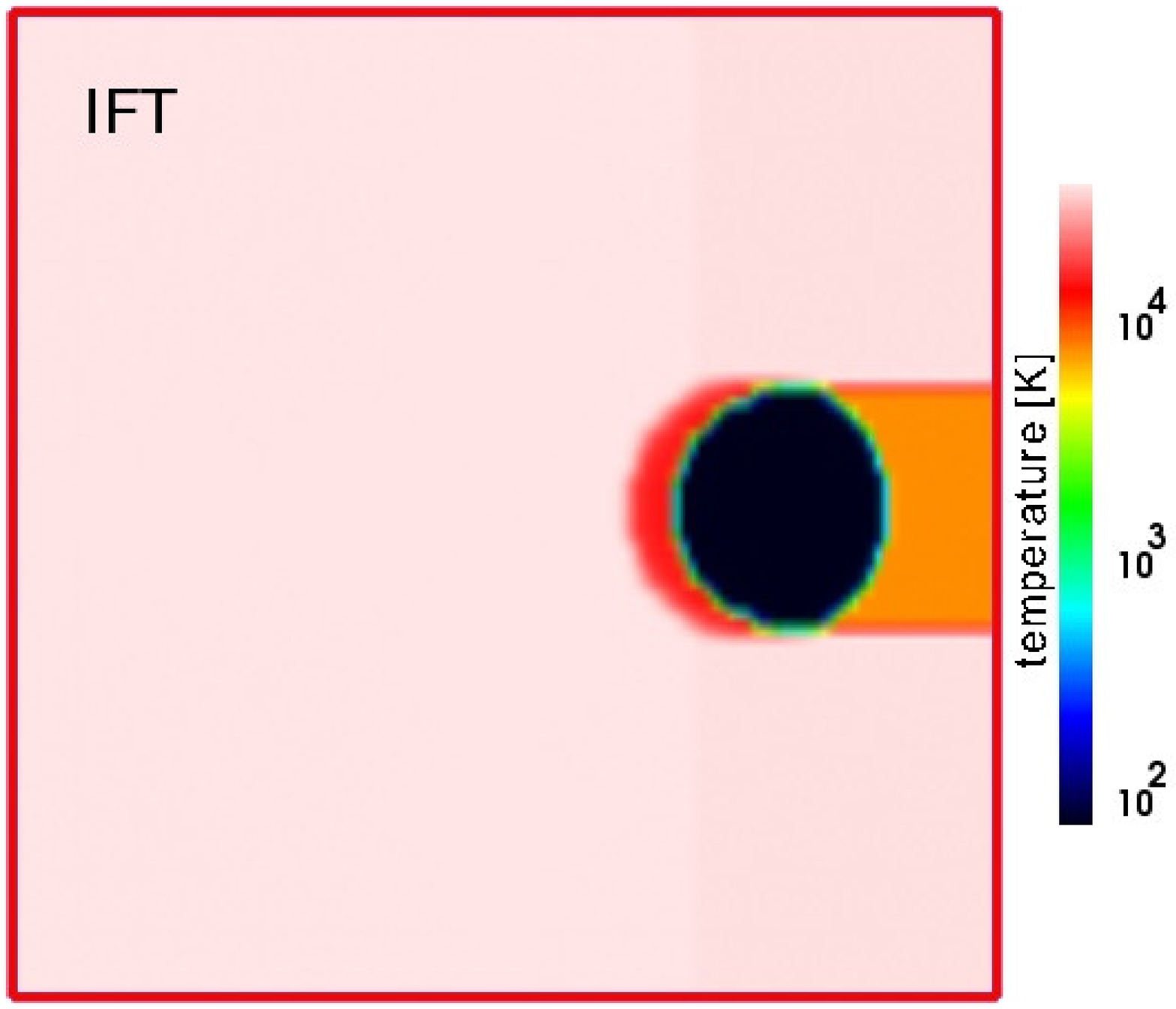}
   \includegraphics[width=2.2in]{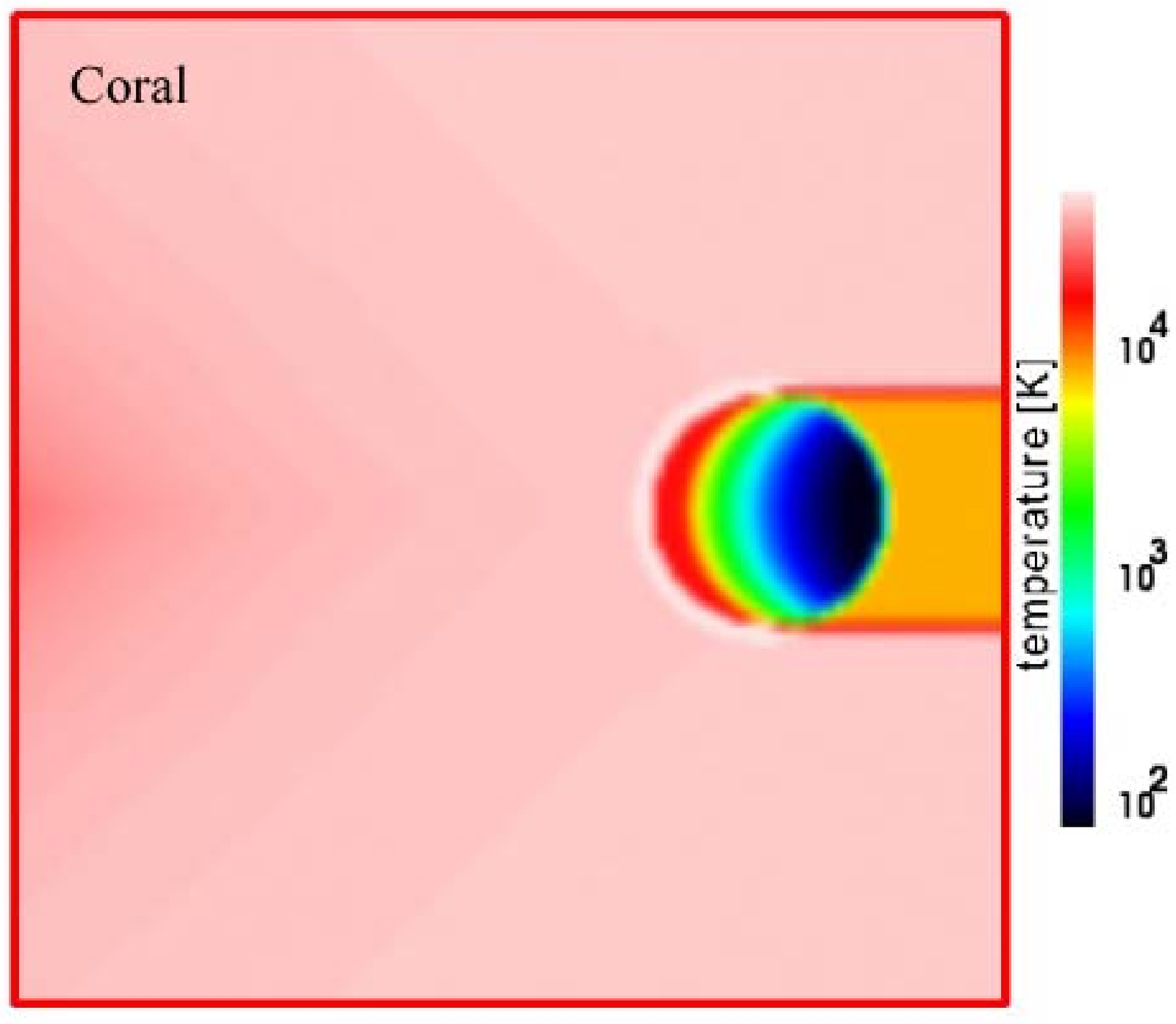}
\caption{Test 3 (I-front trapping in dense clump): Images of the 
temperature, cut through the simulation volume at midplane at time 
$t=1$~Myr for $C^2$-Ray, CRASH, RSPH, FTTE, Flash-HC, IFT 
and Coral.
\label{T3_images1_T_fig}}
\end{figure*}

\begin{figure*}
\begin{center}
   \includegraphics[width=2.2in]{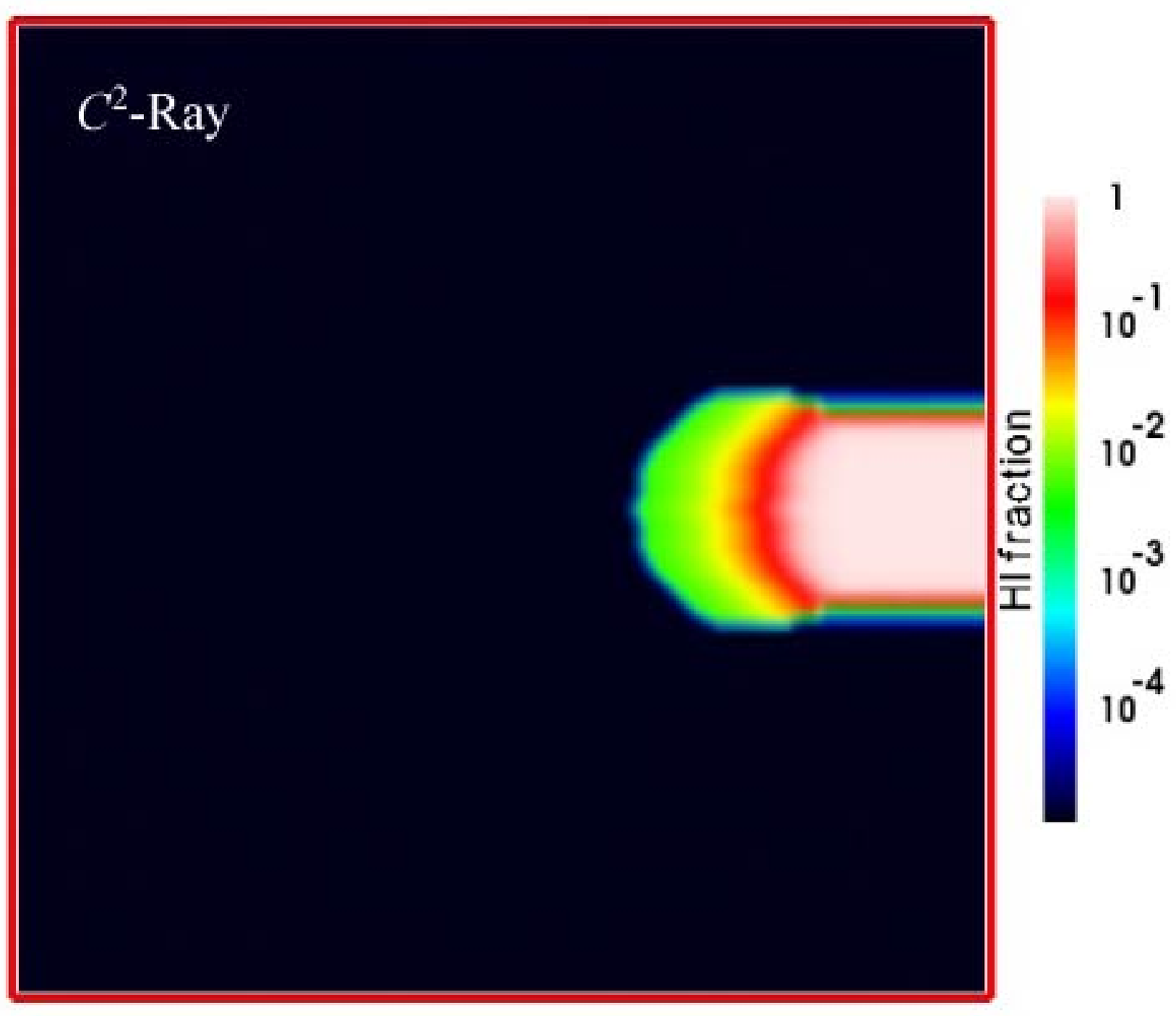}
   \includegraphics[width=2.2in]{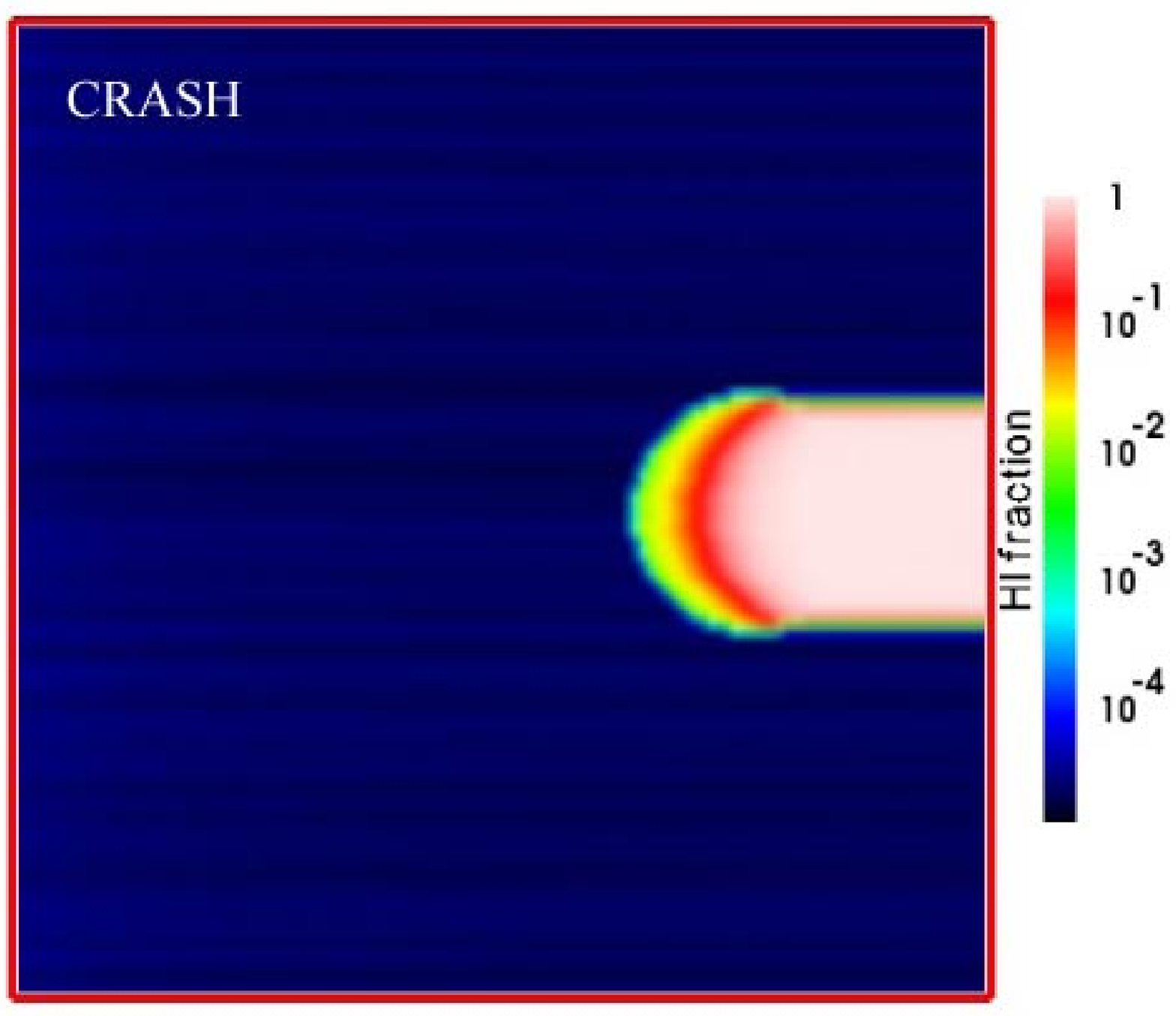}
   \includegraphics[width=2.2in]{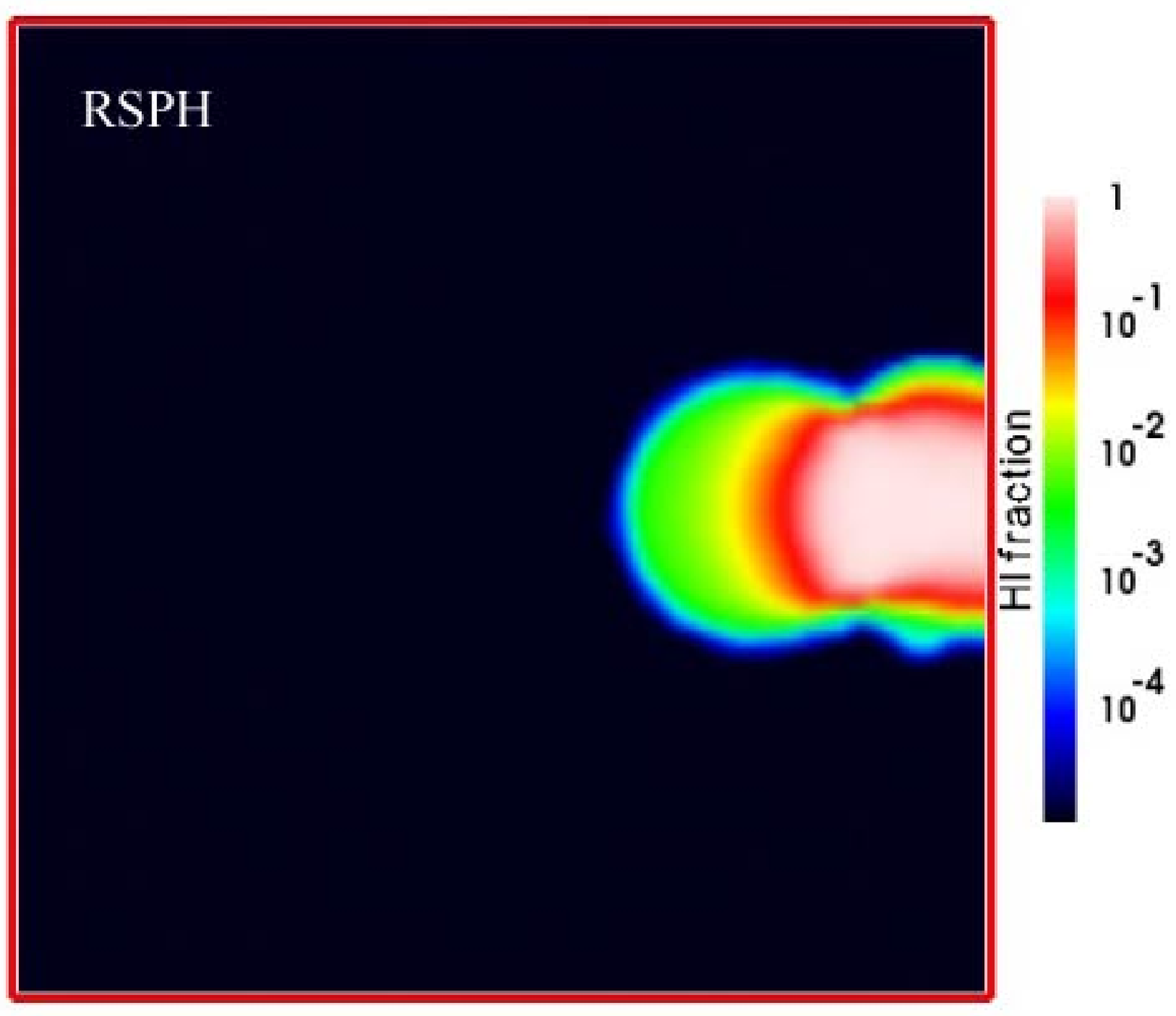}
   \includegraphics[width=2.2in]{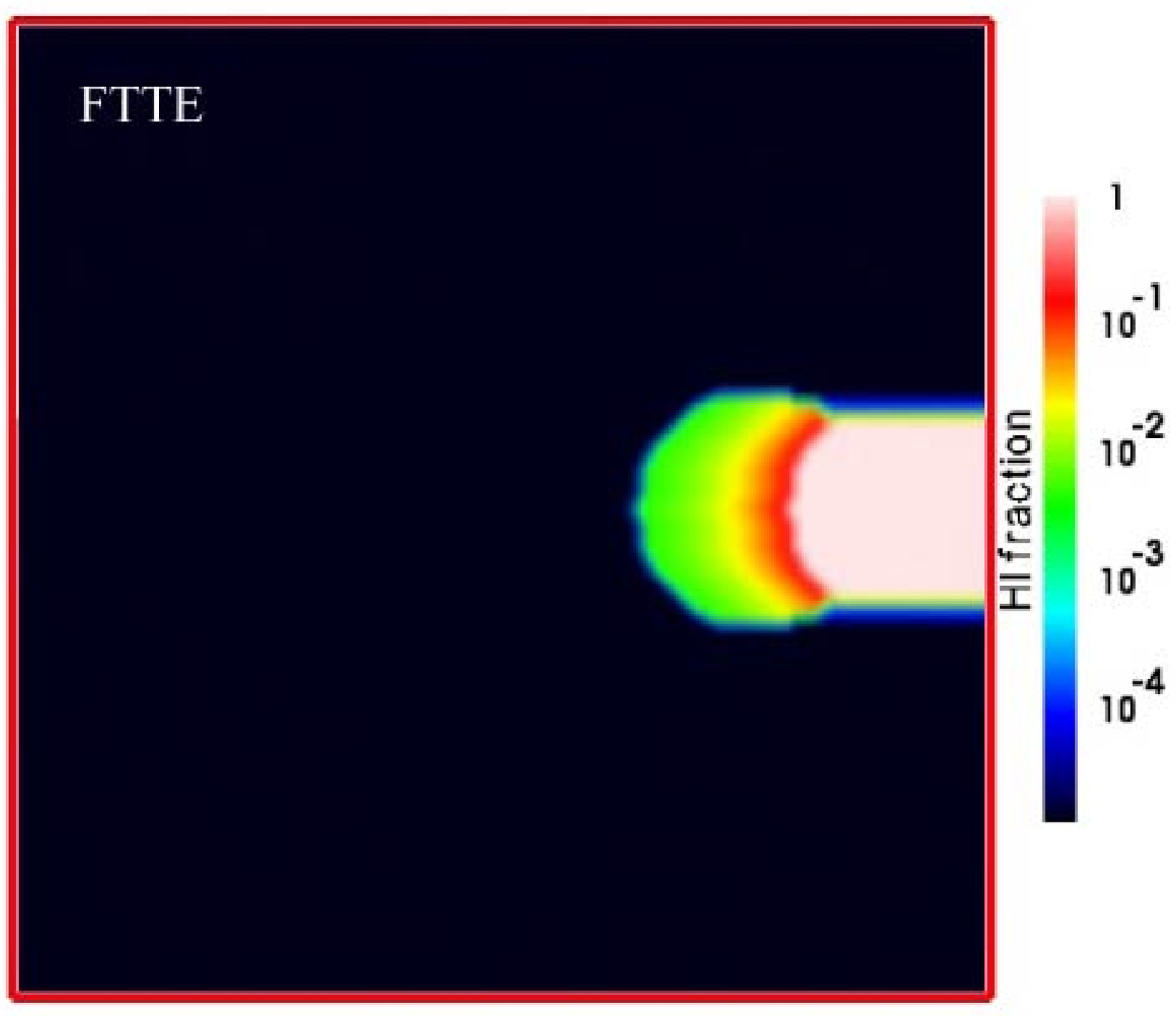}
   \includegraphics[width=2.2in]{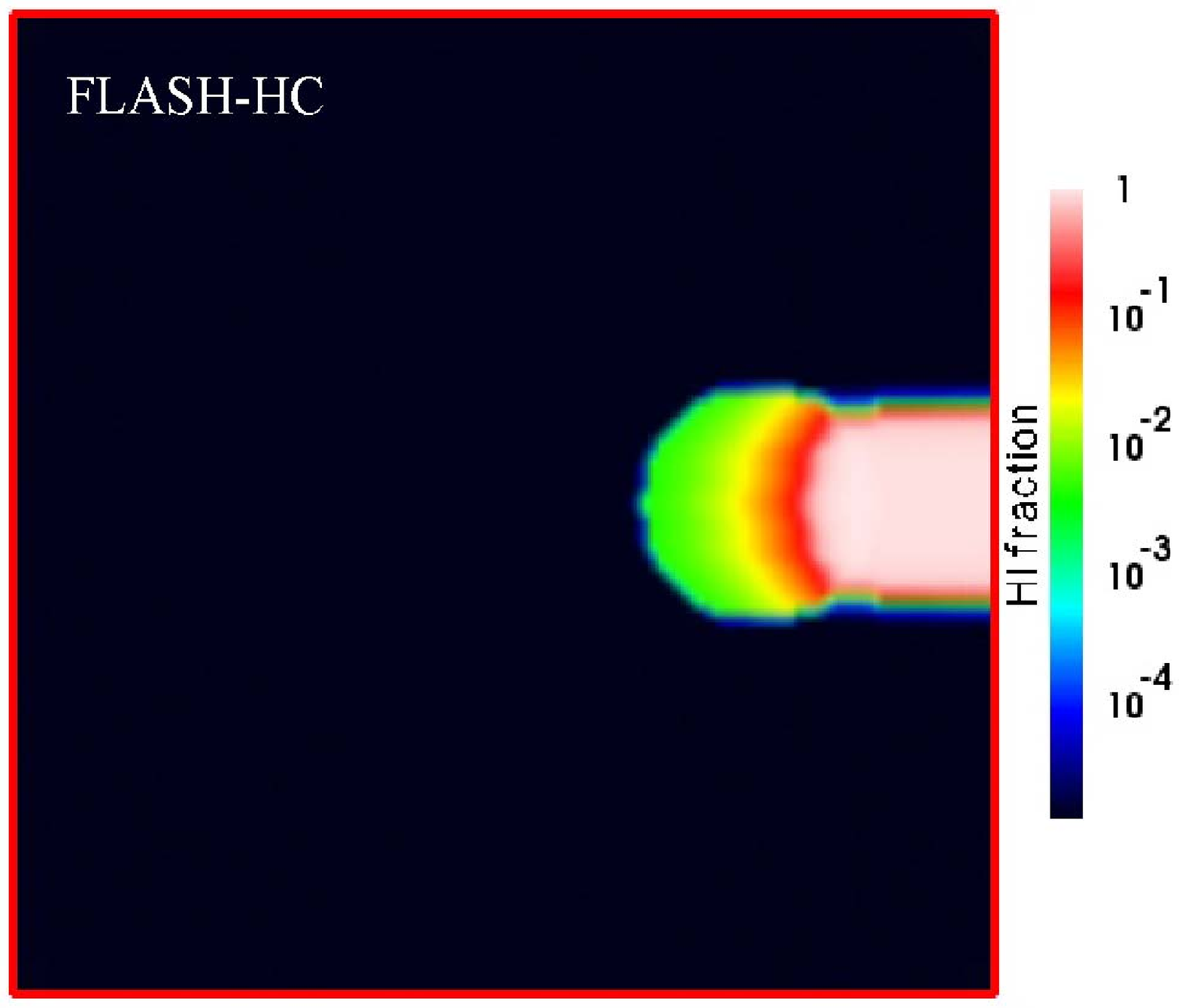}
   \includegraphics[width=2.2in]{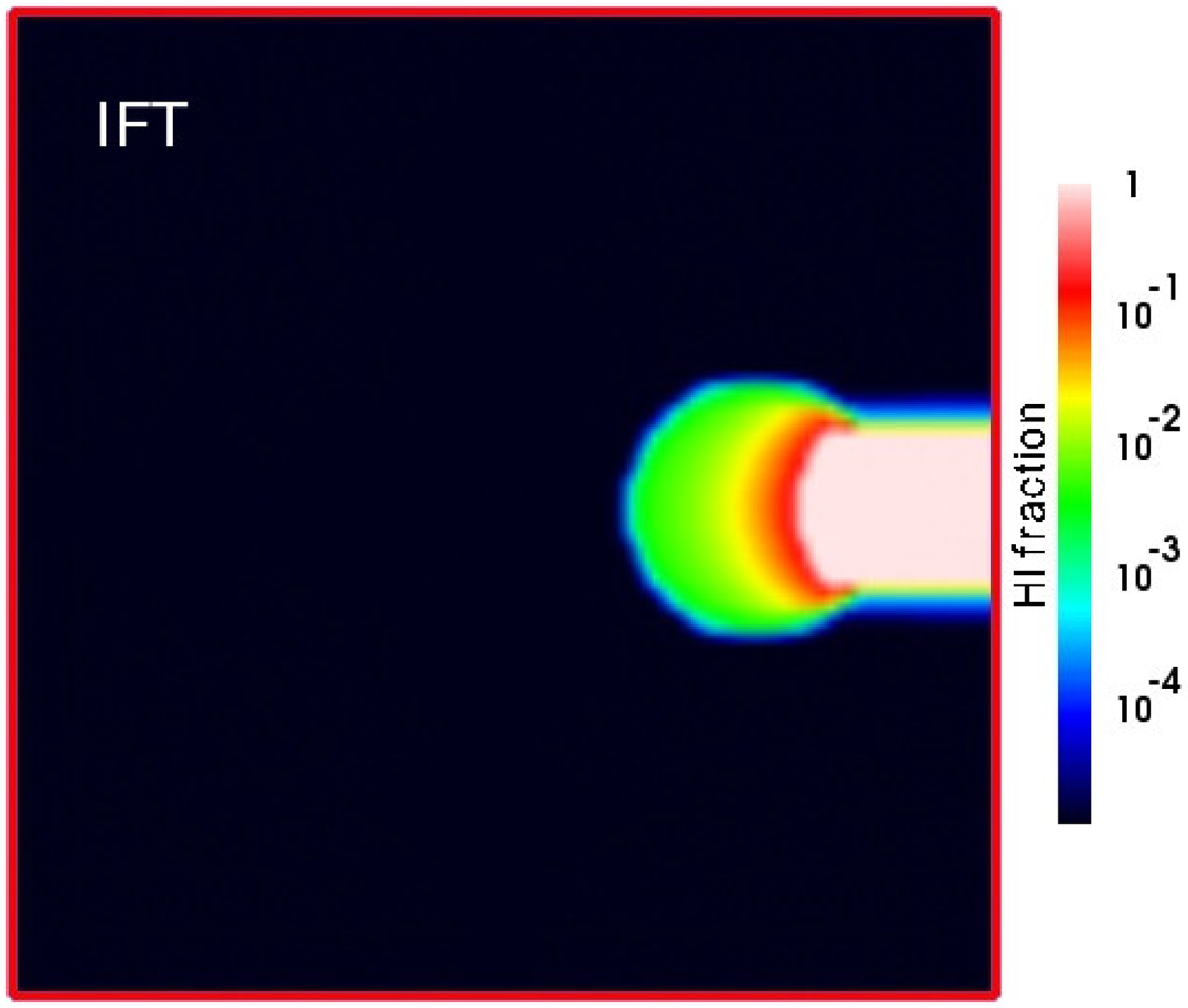}
   \includegraphics[width=2.2in]{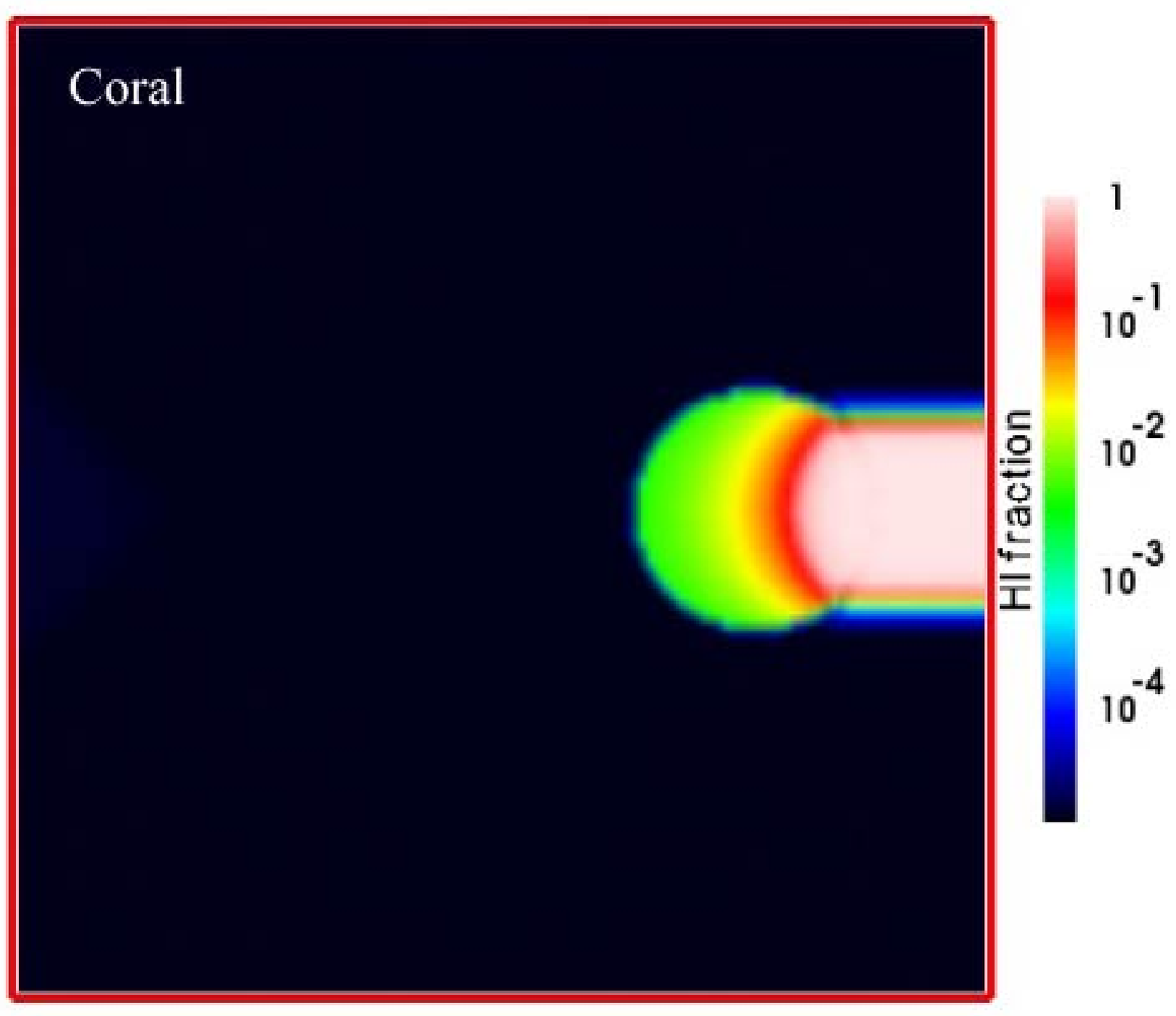}
\caption{Test 3 (I-front trapping in dense clump): Images of the H~I 
fraction, cut through the simulation volume at midplane at time $t=15$ Myr 
for $C^2$-Ray, CRASH, RSPH, FTTE, Flash-HC, IFT and Coral.
\label{T3_images5_HI_fig}}
\end{center}
\end{figure*}

\begin{figure*}
\begin{center}
   \includegraphics[width=2.2in]{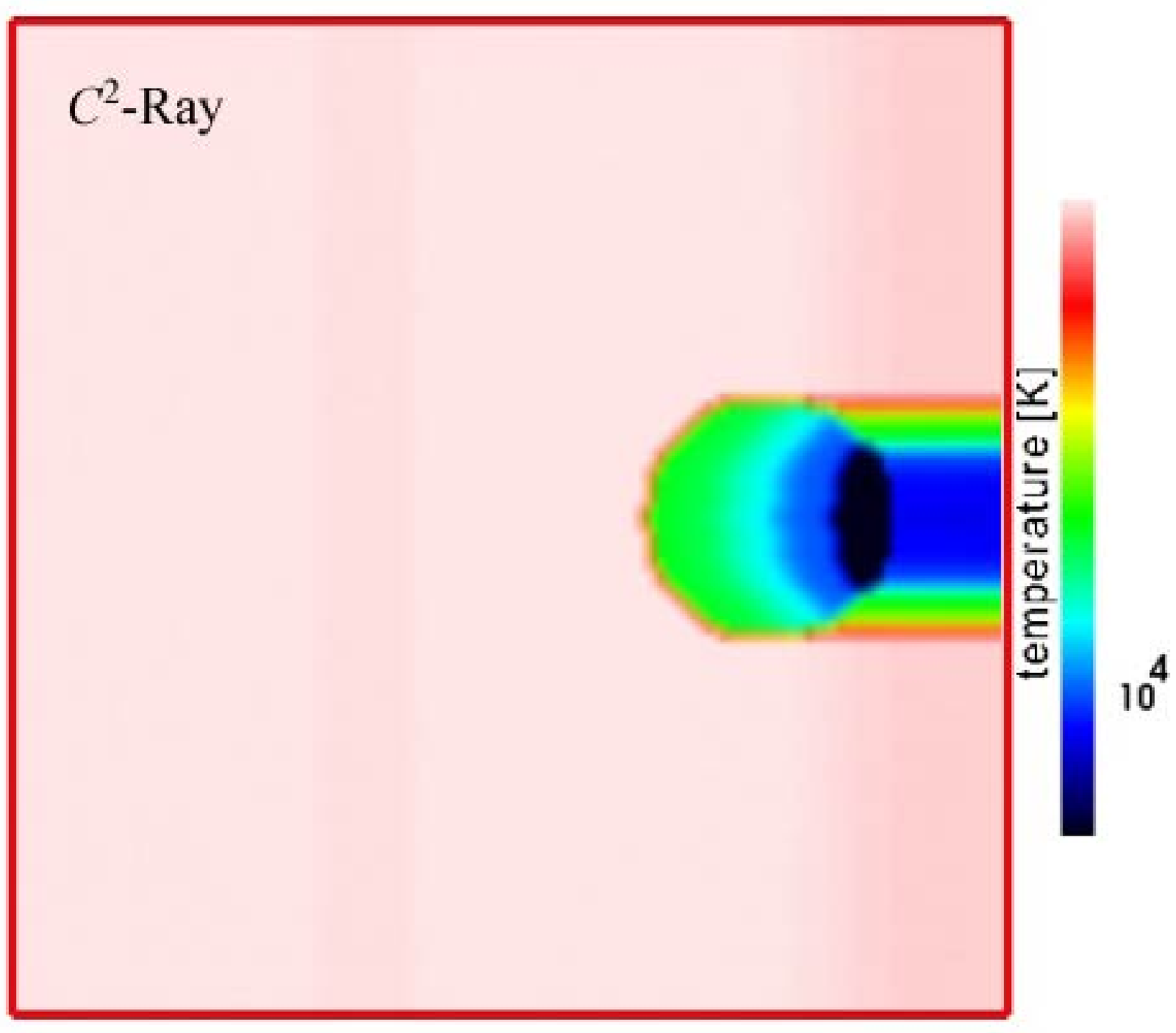}
   \includegraphics[width=2.2in]{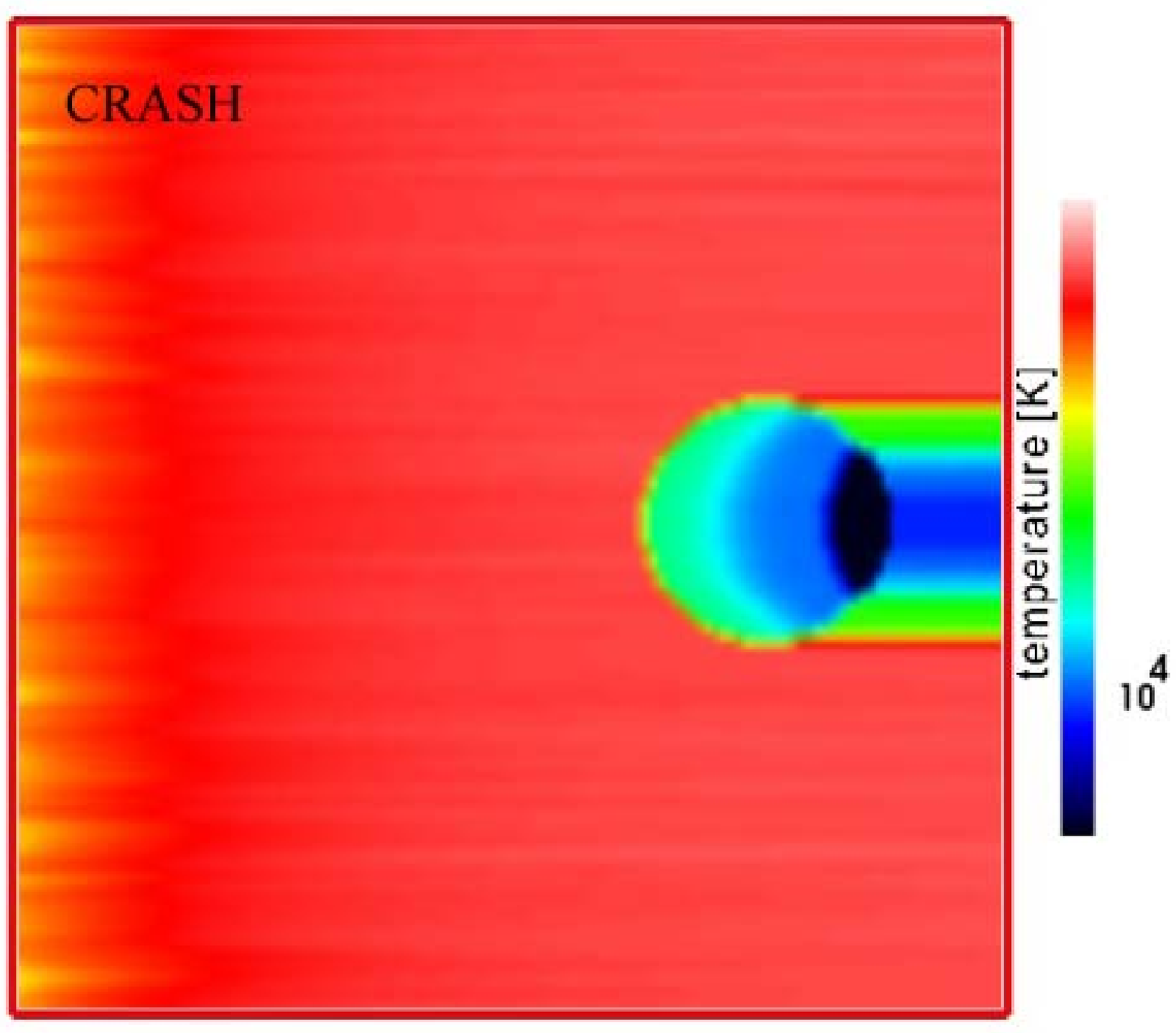}
   \includegraphics[width=2.2in]{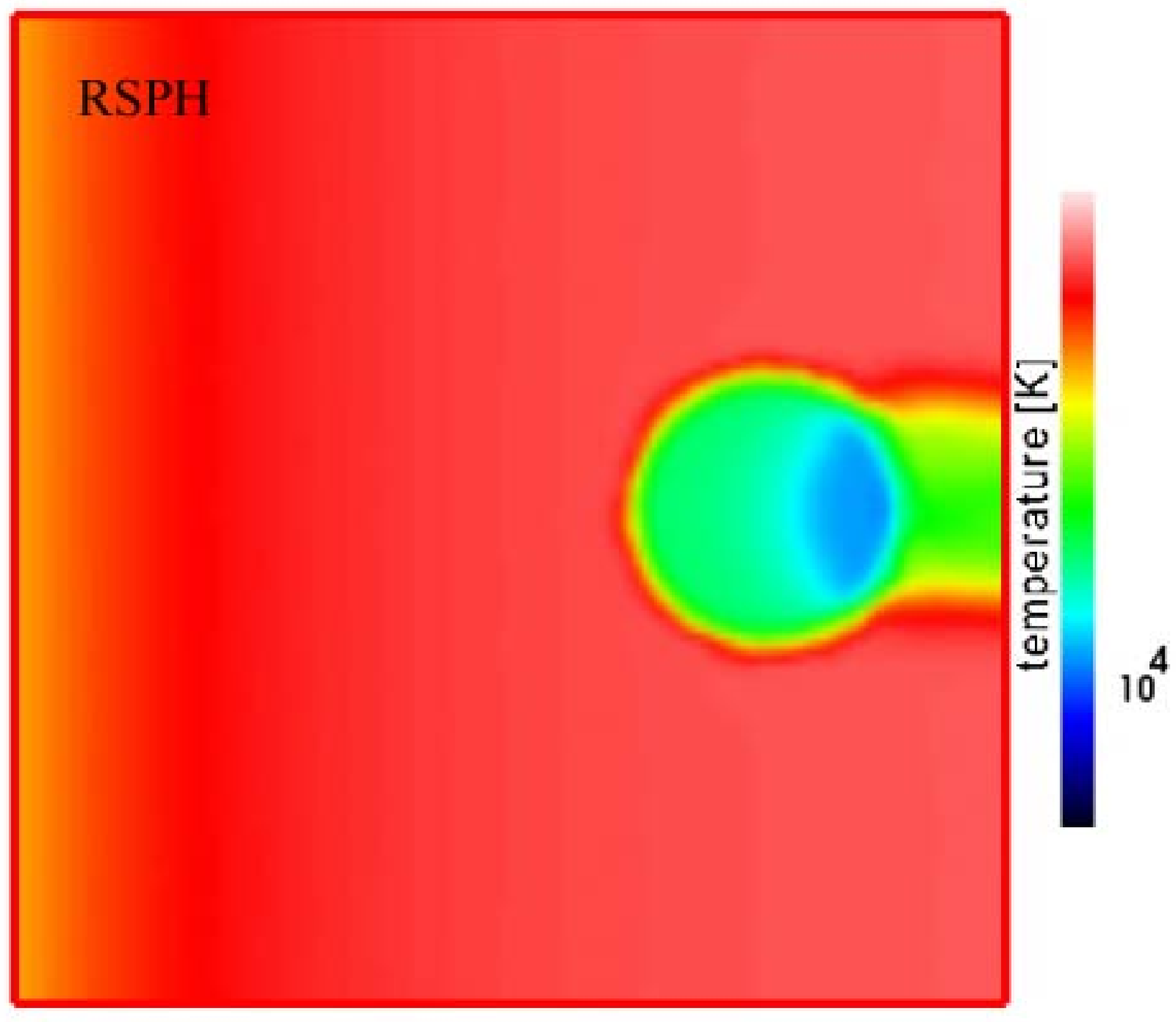}
   \includegraphics[width=2.2in]{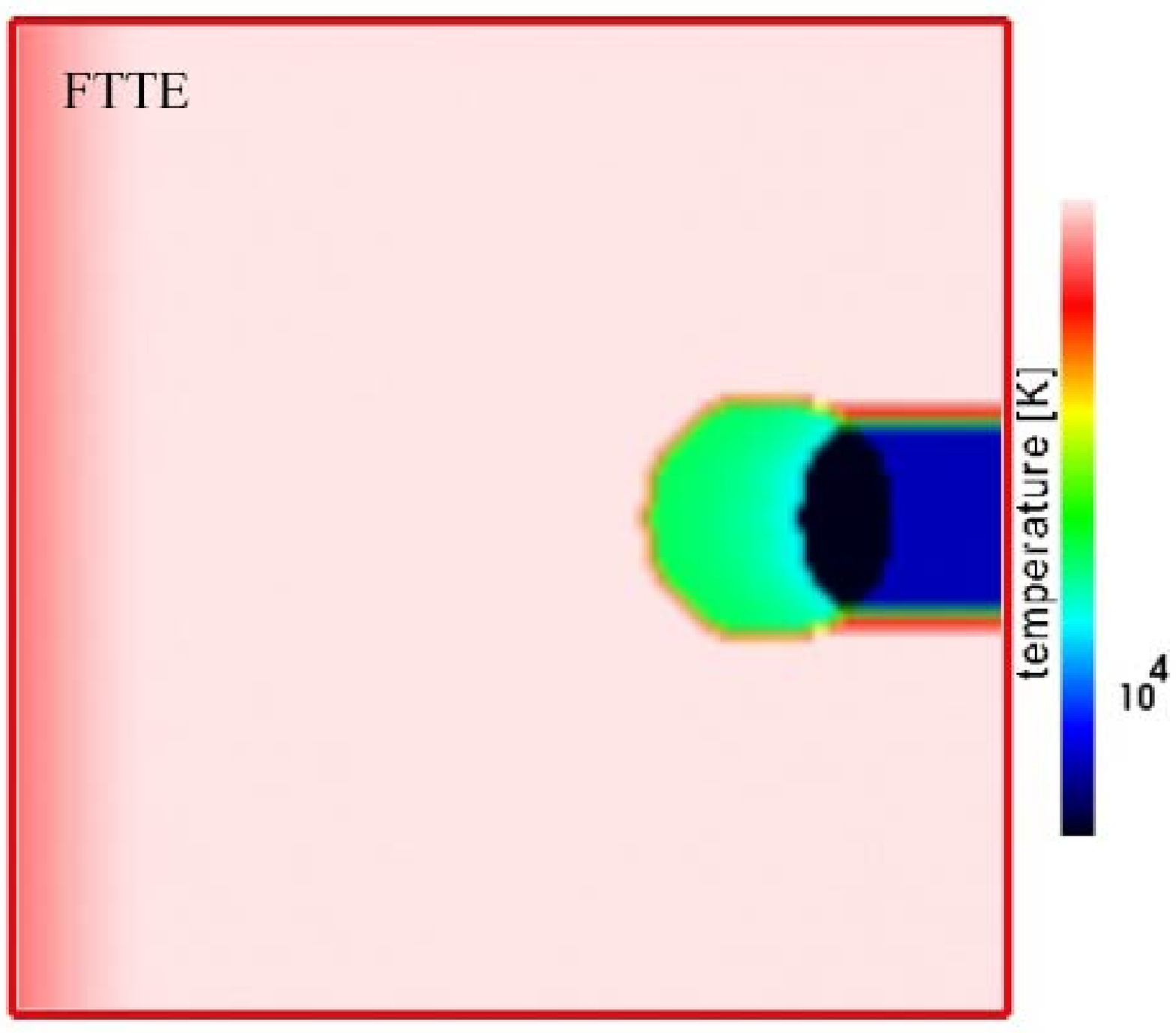}
   \includegraphics[width=2.2in]{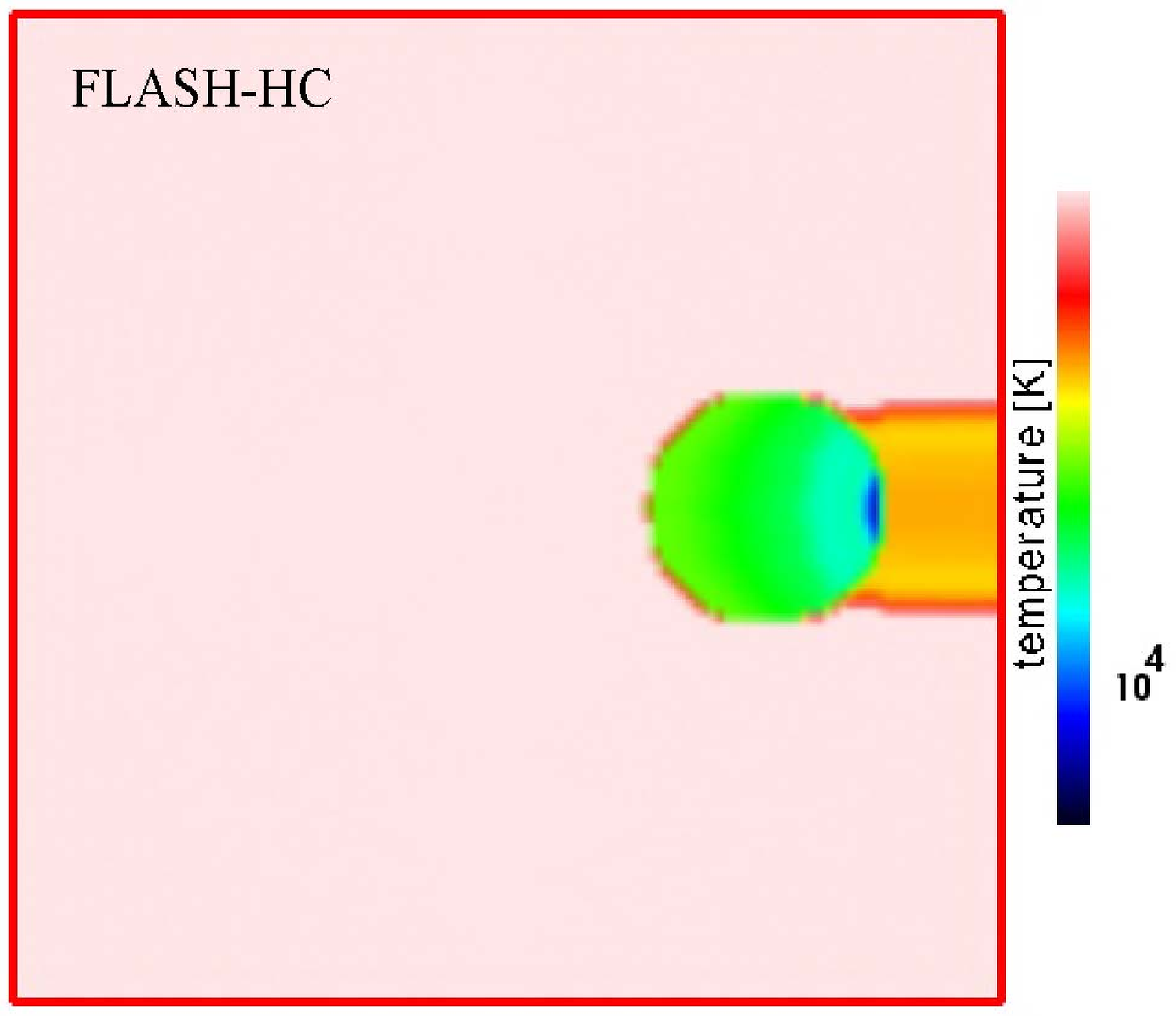}
   \includegraphics[width=2.2in]{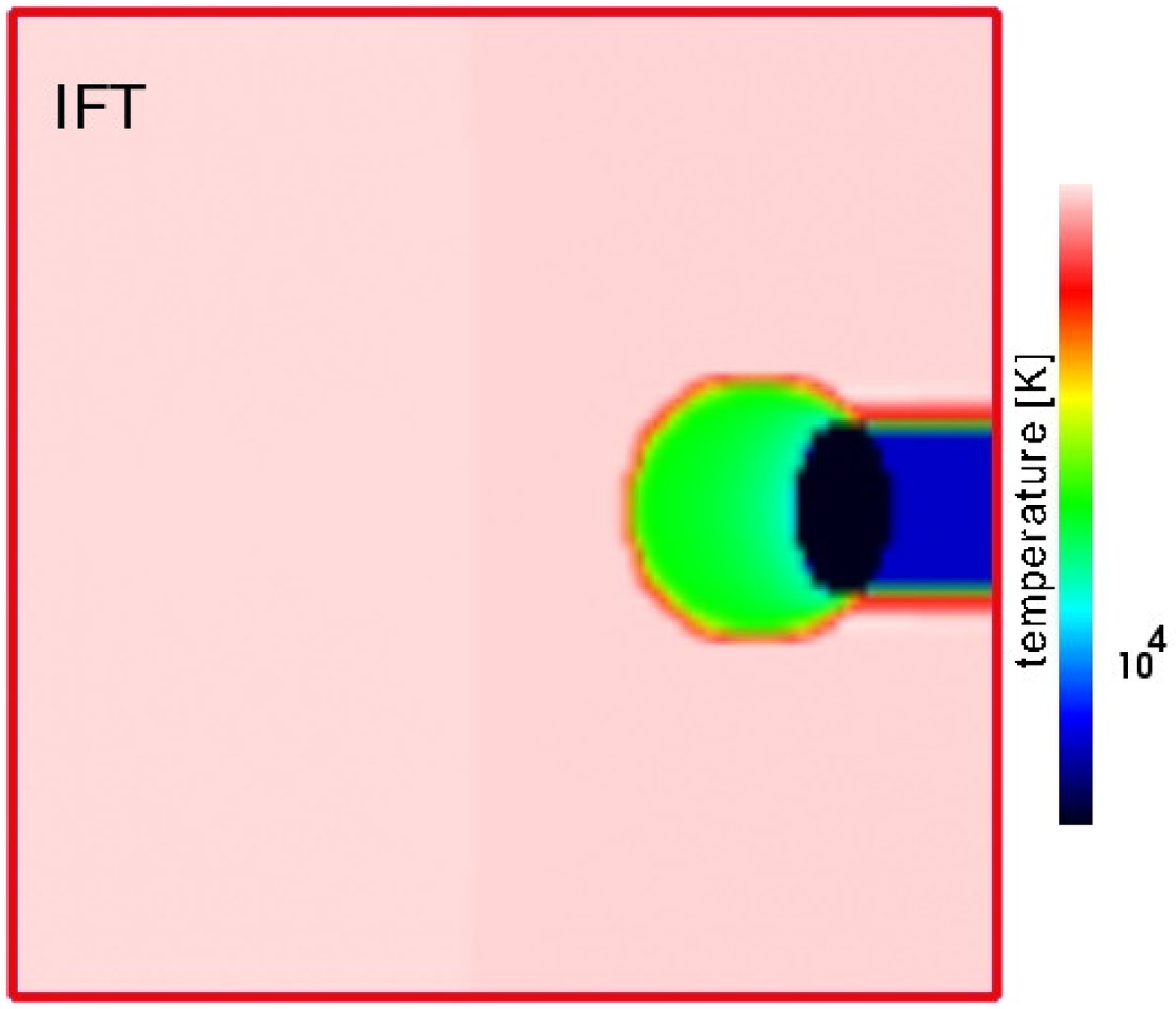}
   \includegraphics[width=2.2in]{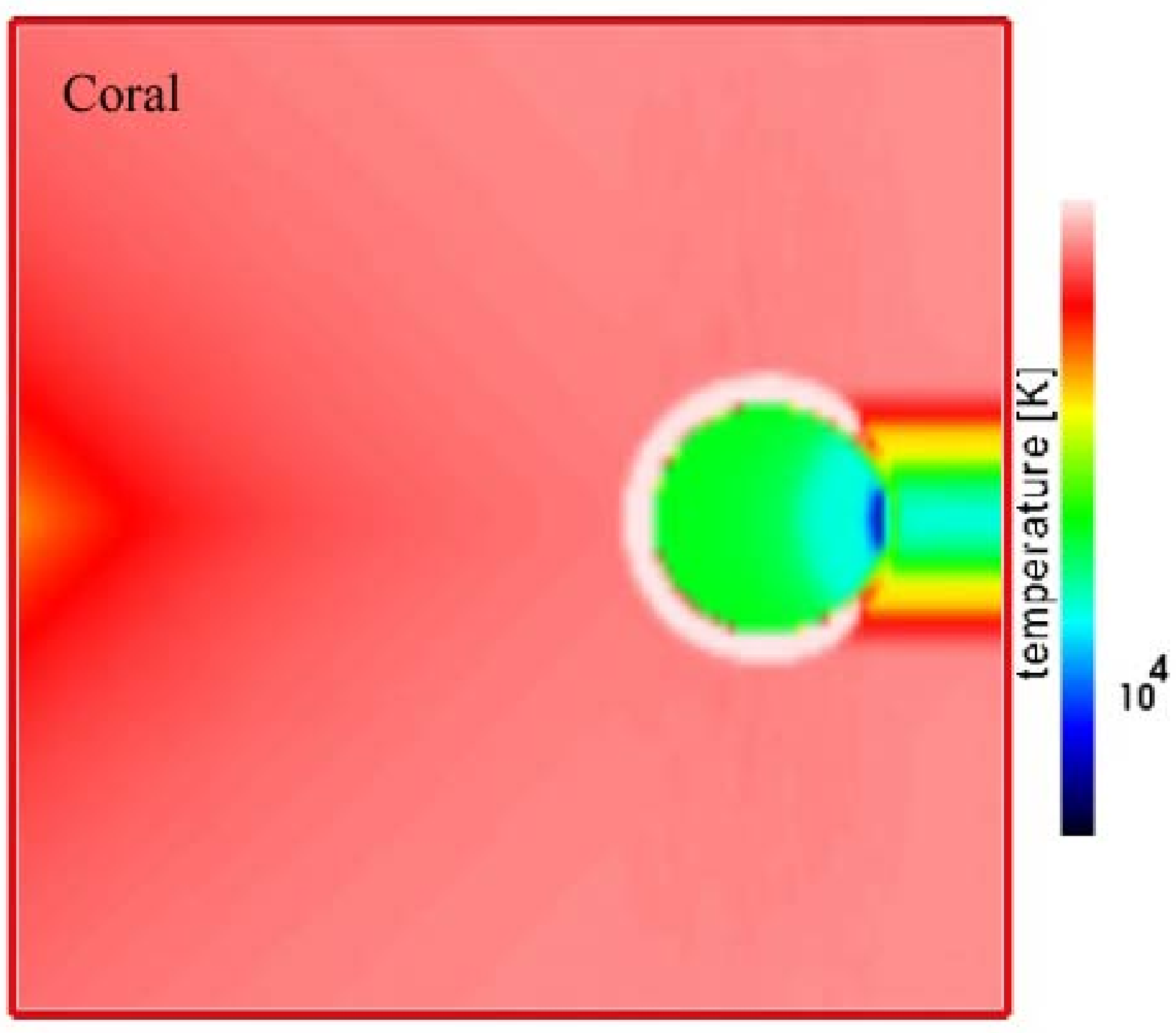}
\caption{Test 3 (I-front trapping in dense clump): Images of the 
temperature, cut through the simulation volume at midplane at time 
$t=15$ Myr for $C^2$-Ray, CRASH, RSPH, FTTE, Flash-HC, IFT and Coral.
\label{T3_images5_T_fig}}
\end{center}
\end{figure*}

Test 3 examines the propagation of a plane-parallel I-front and its trapping 
by a dense, uniform, spherical clump. The condition for an I-front to be 
trapped by a clump of gas with number density $n_H$ can be derived as 
follows \citep{2004MNRAS.348..753S}.  Let us define the Str\"omgren 
length $\ell_S(r)$ at impact parameter $r$ from the clump center using 
equation~(\ref{strom}), but in this case following lines-of-sight for 
each impact parameter. We then can define the ``Str\"omgren number'' for the 
clump as $L_S\equiv2r_{\rm clump}/\ell_S(0)$, where $r_{\rm clump}$ is the 
clump radius and $\ell_S(0)$ is the Str\"omgren length at zero impact 
parameter. Then, if $L_S>1$ the clump is able to trap the I-front, while if 
$L_S<1$, the clump would be unable to trap the I-front and instead would be 
flash-ionized by its passage.

For a uniform-density clump equation~(\ref{strom}) reduces to 
\be
\ell_S=\frac{F}{\alpha_H^{(2)} n_H^2},
\label{l_S_uniform}
\ee
and the Str\"omgren number is given by
\be
L_S=\frac{2r_{\rm clump}\alpha_B(T) n_H^2}{F}.
\ee

The numerical parameters for Test 3 are as follows: the spectrum
is a black-body with effective temperature $T_{eff}=10^5$~K and constant
ionizing photon flux, $F=10^6\rm\,s^{-1}cm^{-2}$, incident to the $y=0$ box side; the
hydrogen number density and initial temperature of the environment are 
$n_{\rm out}=2\times10^{-4}\rm\, cm^{-3}$ and $T_{\rm out,init}=8,000$ K, 
while inside the clump they are 
$n_{\rm clump}=200n_{\rm out}=0.04\rm\, cm^{-3}$ and 
$T_{\rm clump,init}=40$~K. The box size is $L=6.6$ kpc, the radius of
the clump is $r_{\rm clump}=0.8$ kpc, 
and its center is at $(x_c,y_c,z_c)=(5,3.3,3.3)$ kpc, or
$(x_c,y_c,z_c)=(97,64,64)$ cells, and the evolution time is 15~Myr.
For these parameters and assuming for simplicity that the Case B recombination
coefficient is given by $\alpha_B(T)=2.59\times10^{-13}(T/10^4\, K)^{-3/4}$,
we obtain $\ell_S\approx0.78(T/10^4\, K)^{3/4}$ kpc, and
$L_S\approx2.05(T/10^4\, K)^{-3/4}$; thus, along the axis of symmetry the 
I-front should be trapped approximately at the center of the clump for
$T=10^4$~K. In reality, the temperature could be expected to be somewhat 
different and spatially-varying, but to a rough first approximation this
estimate should hold.

In fact, the I-front does get trapped as expected, slightly beyond the clump
center. In Figure~\ref{T3_Ifront_evol_fig} we plot $(r_I-x_c)/\ell_S$, the
evolution of the position of the I-front with respect to the clump center in 
units of the Str\"omgren length (top panel) and the corresponding velocity
evolution (in units of $2c_{s,I}(T=10^4\rm K)=2(p/\rho)^{1/2}$, twice the 
isothermal sound speed in gas at temperature of $10^4$~K), both vs. 
$t/t_{\rm rec,0}$,
time in units of the recombination time inside the clump (which is $\sim3$~Myr
at $10^4$~K). The I-front is initially highly supersonic due to the low
density outside the clump. Once it enters the clump it shows down sharply, to
about 20 times the sound speed, by the same factor as the density jump at the
clump boundary. As it penetrates further into the clump it approaches its
(inverse, i.e. outside-in) Str\"omgren radius at time $t\sim t_{\rm rec,0}$,
at which point the propagation slows down even further until the I-front is
trapped after a few recombination times. The velocity drops below $2c_{s,I}$,
at which point if gas motions were allowed the I-front would become slow 
D-type (i.e. coupled to the gas motion, rather than much faster than
them). All codes capture these basic phases of the trapping process correctly
and agree well on both the front position and velocity. We note here that the 
FLASH-HC code currently does not have the ability to track fast I-fronts, so
its data starts only after the front has slowed down. The IFT method
assumes a sharp front, and thus does not allow pre-heating and partial 
ionization ahead of the front, which results in its being slowed down more 
abruptly than is the case for the other results. Due to some diffusion, the
RSPH code finds that the front slows down slightly before the I-front actually
enters the clump. There are also minor differences in the later stages of the
evolution, to be discussed in more detail below.

In Figure~\ref{T3_images1_HI_fig} we show the images of the neutral gas
fraction on the plane through the centre of the clump at time $t=1$~Myr, when
the I-front is already inside the clump, but still not trapped and moving 
supersonically. The ionizing source is far to the left of the box.
All results show a sharp shadow behind the clump, as
expected for such a dense, optically-thick clump. Only the RSPH code shows
diffusion at the shadow boundaries, due to the intrinsic difficulty of 
representing such a sharply-discontinuous density distribution with SPH particles
and the corresponding smoothing kernel. The FLASH-HC code derives a noticeably
sharper I-front in both the clump and the external medium 
(where it is the only result to still have some gas with neutral fraction 
above $\sim10^{-4}$). This is due to its current inability to correctly track 
fast I-fronts, as discussed above, which leads to somewhat incorrect early
evolution. The corresponding temperature image cuts
(Figure~\ref{T3_images1_T_fig}) show the same trends:  FLASH-HC and IFT
find a very sharp transition, while the rest of the codes agree
reasonably well, with only minor differences in the pre-heating region.  

In Figure~\ref{T3_images5_HI_fig} we show the images of the neutral fraction 
at the final
time of the simulation, $t=15$~Myr. All codes except CRASH find very similar
ionized structure inside the clump. The CRASH result has significantly higher 
neutral fraction inside the clump, and correspondingly larger shadow behind 
the clump, as well as slightly higher neutral gas fraction in the low-density 
gas. This could be due to the fact that, as mentioned in \S~\ref{T2_sect},
CRASH follows multiple bins in frequency over a wider frequency range
with respect to the other codes; this results in a higher ionizing power
at high frequencies, which also have smaller photo-ionization cross-sections. 
This in turn could be the origin of the lower ionization state of the clump 
and of the low-density gas.
The RSPH result again exhibits significant diffusion around the edges of
the shadow. The corresponding temperature structures, on the other hand,
(Figure~\ref{T3_images5_T_fig}) show some differences, which stem from the
different treatments of the energy equation and spectral hardening by the
codes. The FTTE and IFT codes find almost no pre-heating in the shielded
region and the shadow behind it. $C^2$-Ray and CRASH get smaller self-shielded
regions and some hard photons penetrating into the sides of the
shadow. Finally, RSPH, FLASH-HC and Coral find almost no gas that is
completely self-shielded, but still find sufficient column densities to create
temperature-stratified shadows similar to the ones found by $C^2$-Ray and
CRASH codes, albeit at higher temperature levels. The Coral result also has a
thin, highly-heated shell at the source side of the clump, resulting from
this code's problems in properly finding the temperature state in the first 
dense, optically-thick cells encountered by its rays, which leads to their
overheating. In production runs this problem was corrected by increasing the
resolution and decreasing the cell size so that cells are not as
optically-thick, and by the gas motions, which quickly cool the gas down as it
expands. The low-density gas outside the clump is somewhat cooler in the CRASH
and RSPH results compared to the other codes.

\begin{figure*}
\begin{center}
  \includegraphics[width=3.5in]{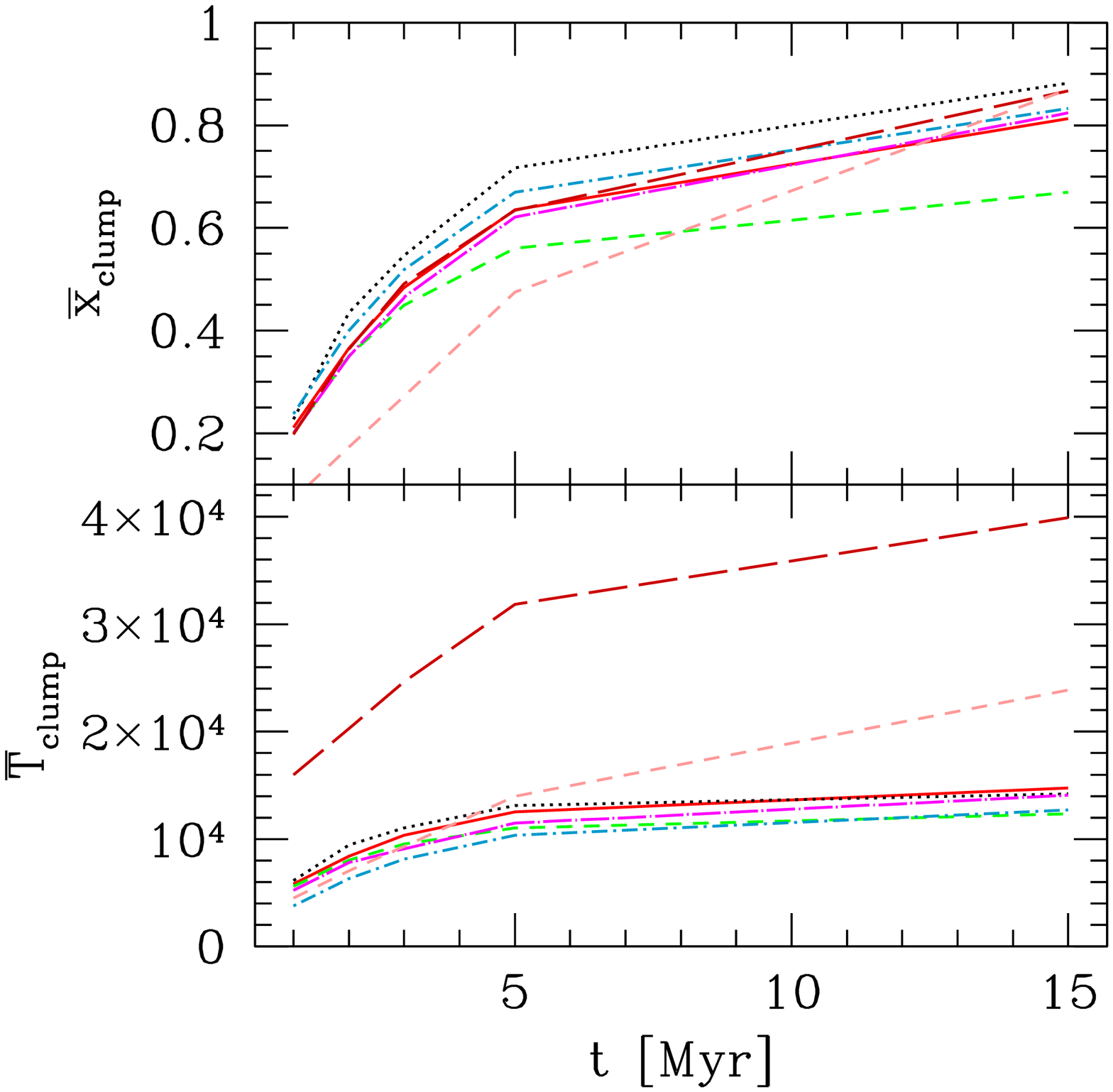}
\caption{Test 3 (I-front trapping in dense clump): The evolution of the 
ionized fraction (top) and the mean temperature (bottom) inside the dense 
clump.
\label{T3_x_T_evol_fig}}
\end{center}
\end{figure*}

These observations are confirmed by the evolution of the mean
ionized fraction and the mean temperature inside the clump shown in
Figure~\ref{T3_x_T_evol_fig}. All codes agree very well on the evolution of
the mean ionized fraction, except for CRASH, which finds about $\sim25\%$
lower final ionized fraction, and for FLASH-HC, which early-on finds a lower
ionized fraction, but catches up with the majority of the codes as the I-front
becomes trapped. In terms of mean temperature, Coral, and to a lesser extent
FLASH-HC find higher mean temperature due to the overheating of some cells
mentioned above.    

 \begin{figure*}
\begin{center}
  \includegraphics[width=2.2in]{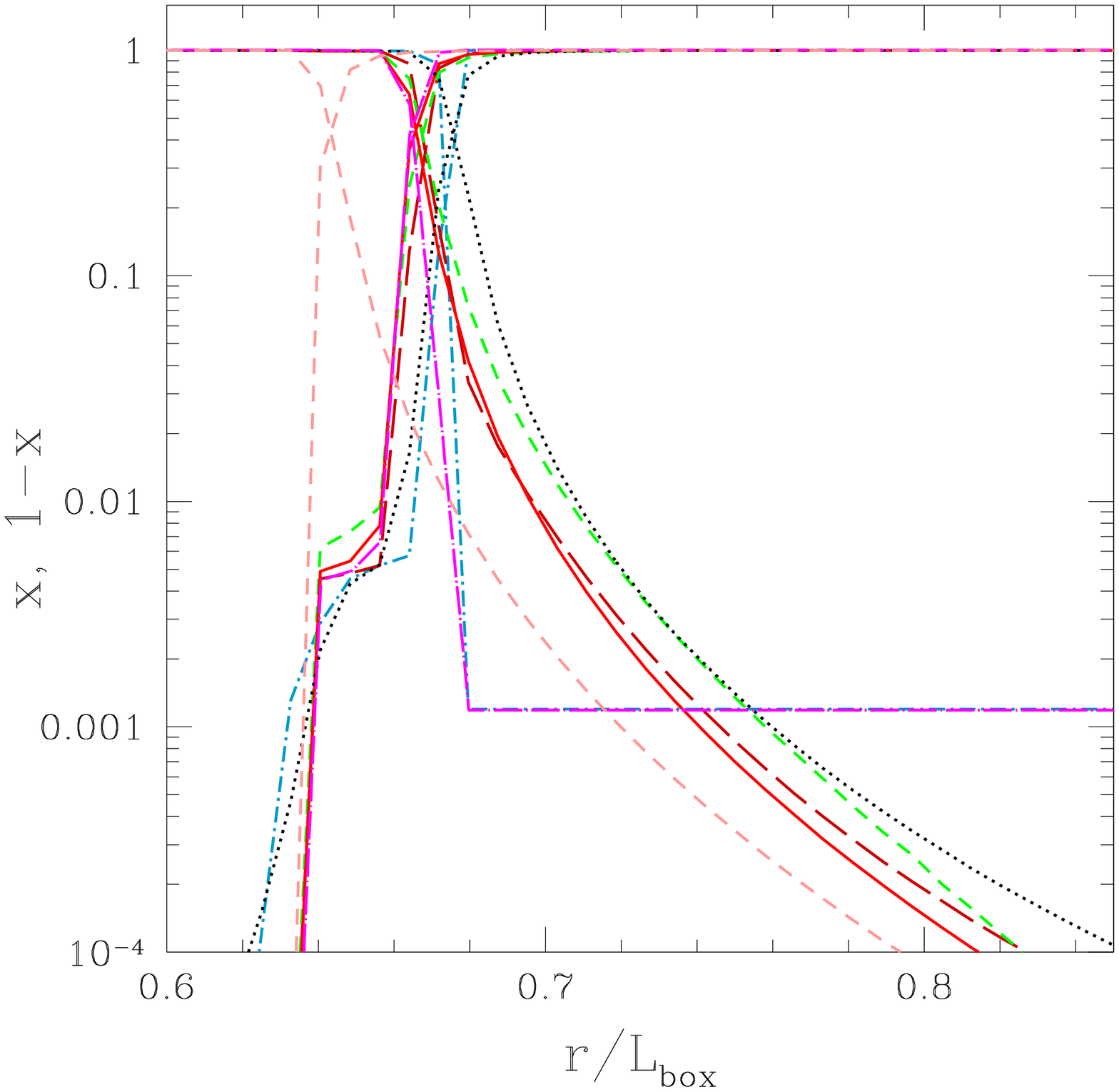}
  \includegraphics[width=2.2in]{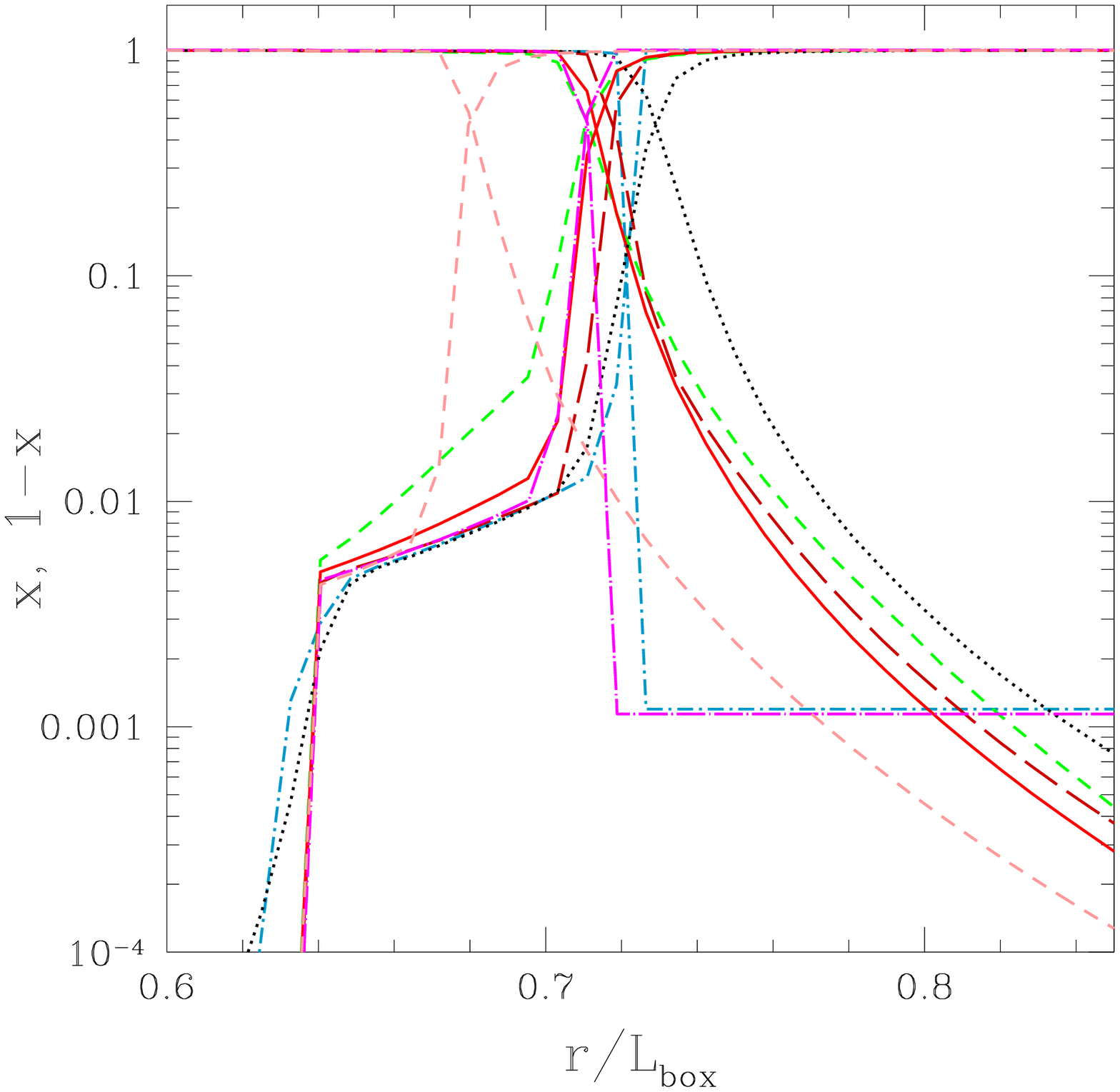}
  \includegraphics[width=2.2in]{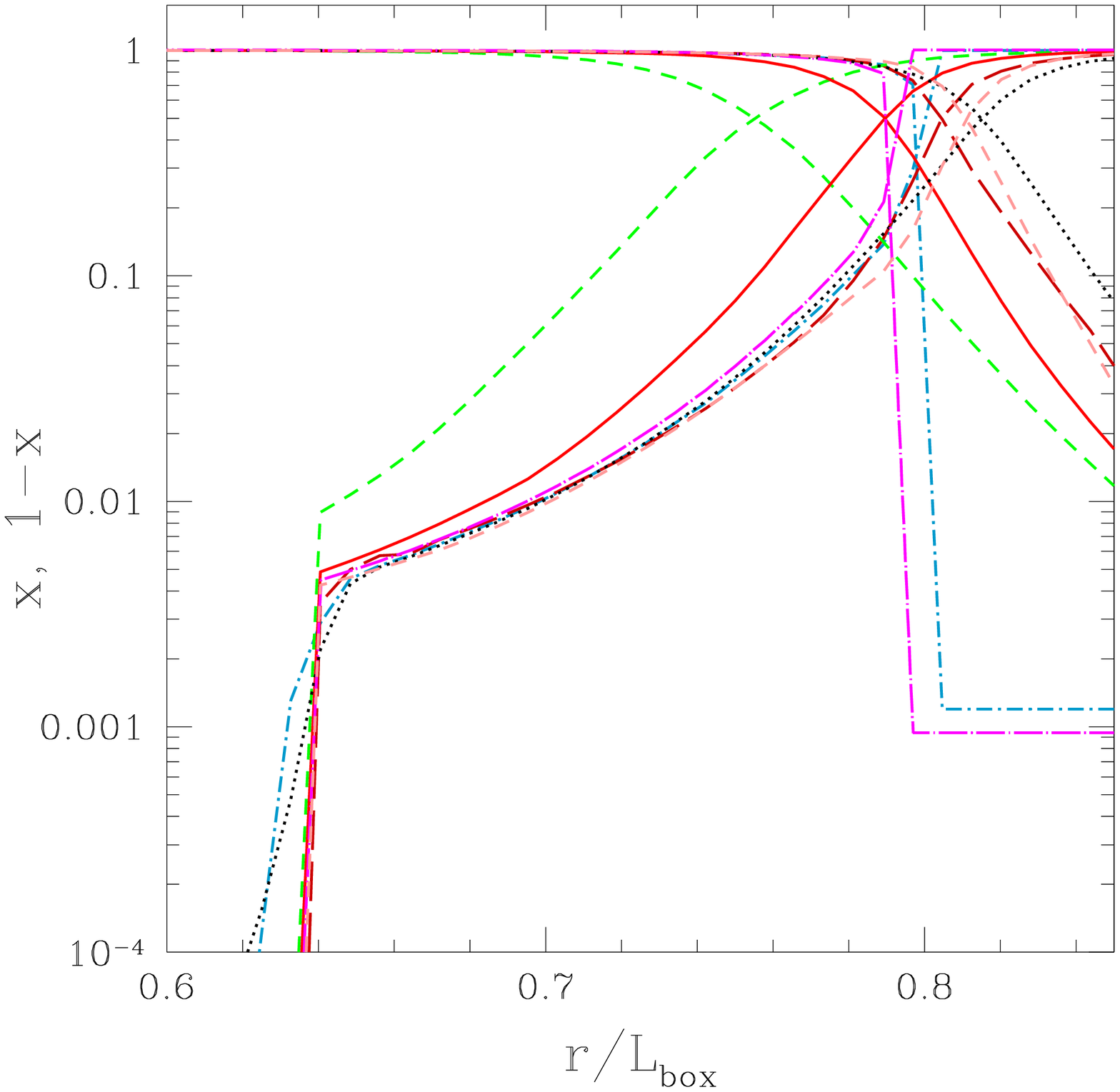}
\caption{Test 3 (I-front trapping in dense clump): Line cuts of the 
ionized and neutral fraction along the axis of symmetry through the 
center of the clump at times $t=1$ Myr, 3 Myr and 15 Myr. 
\label{T3_profs_fig}}
\end{center}
\begin{center}
  \includegraphics[width=2.2in]{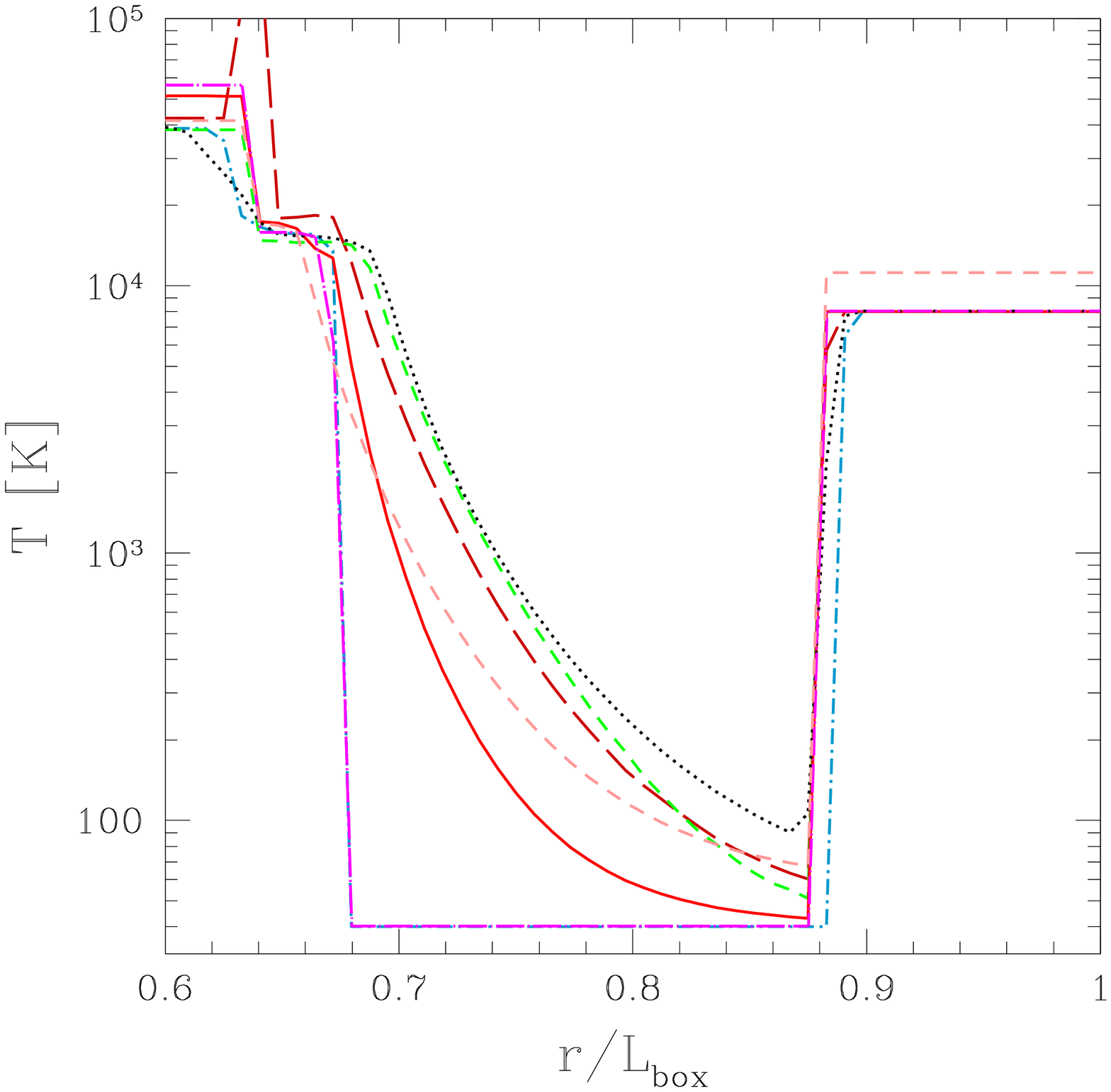}
  \includegraphics[width=2.2in]{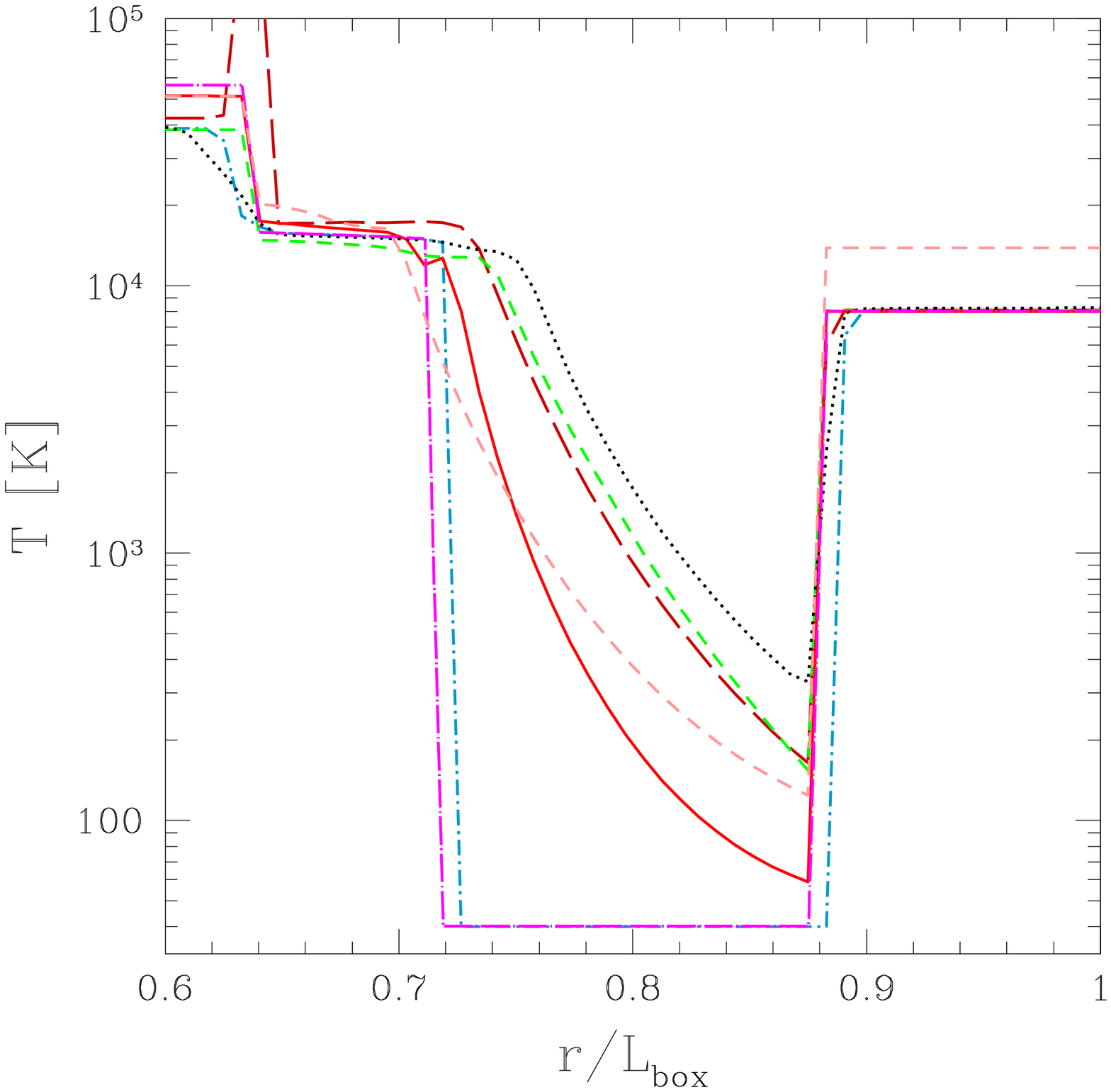}
  \includegraphics[width=2.2in]{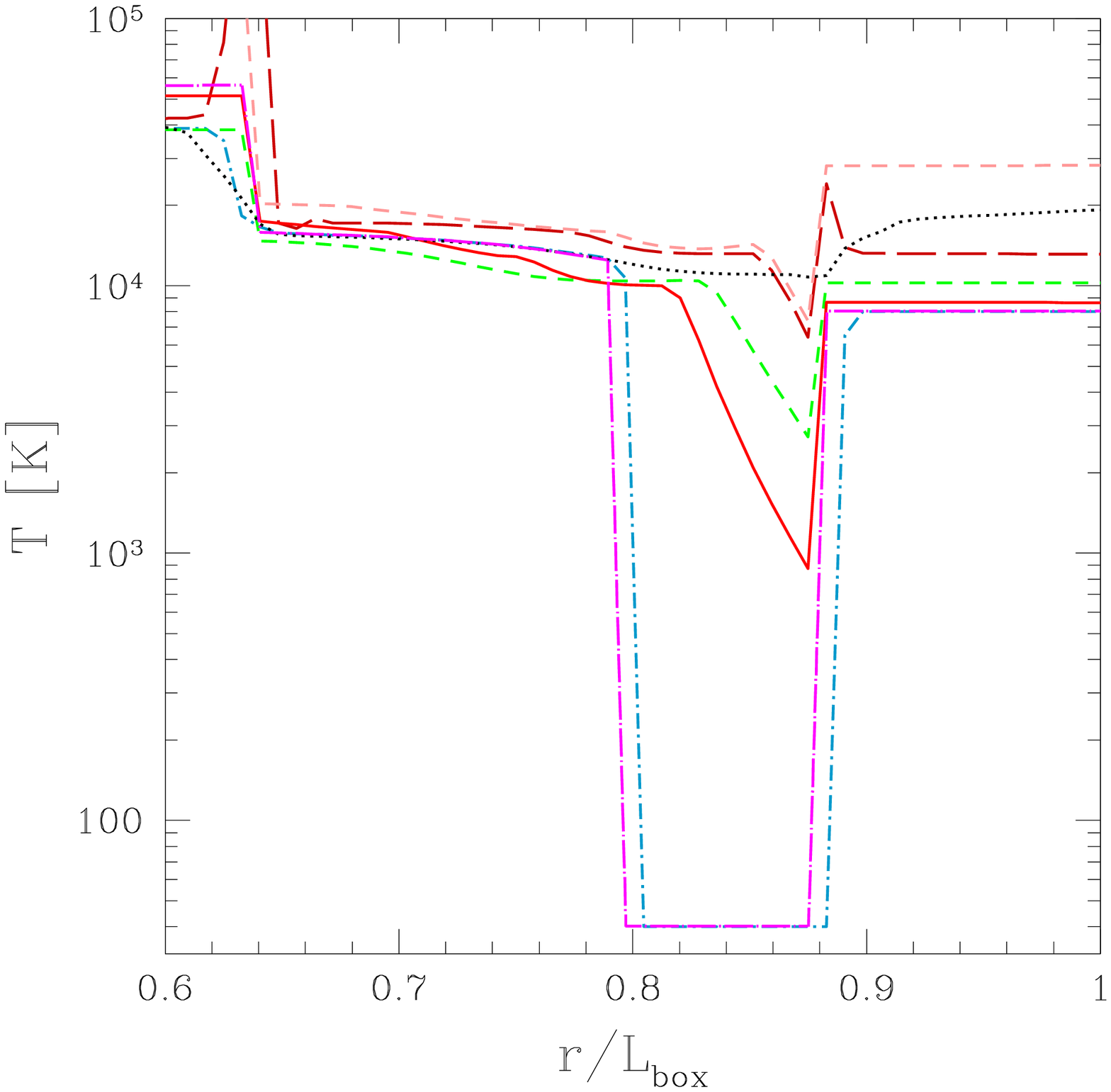}
\caption{Test 3 (I-front trapping in dense clump): Line cuts of the 
temperature along the axis of symmetry 
through the center of the clump at times $t=1$ Myr, 3 Myr and 15 Myr. 
\label{T3_profsT_fig}}
\end{center}
\end{figure*}

\begin{figure*}
\begin{center}
  \includegraphics[width=2.9in]{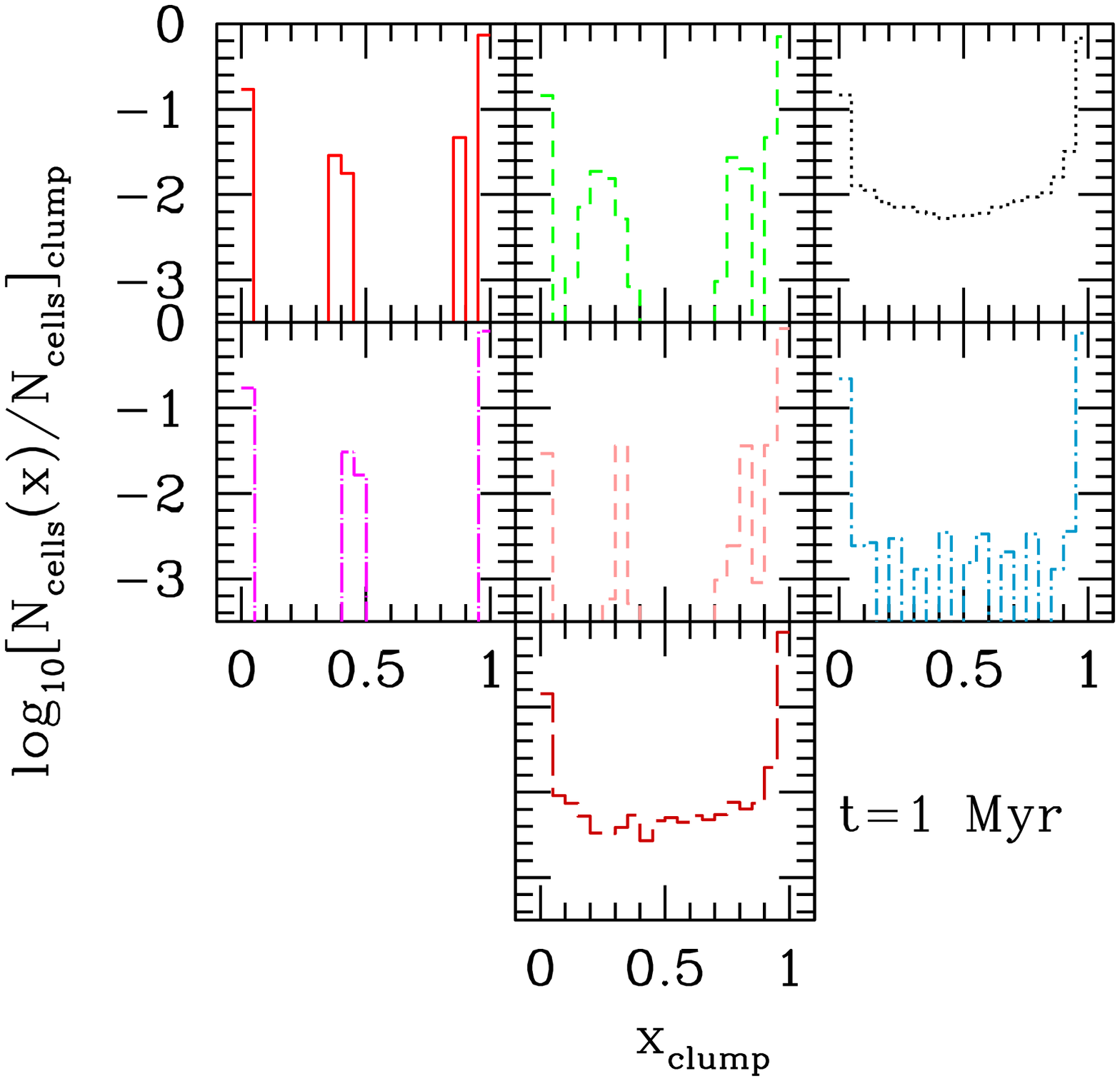}
  \includegraphics[width=2.9in]{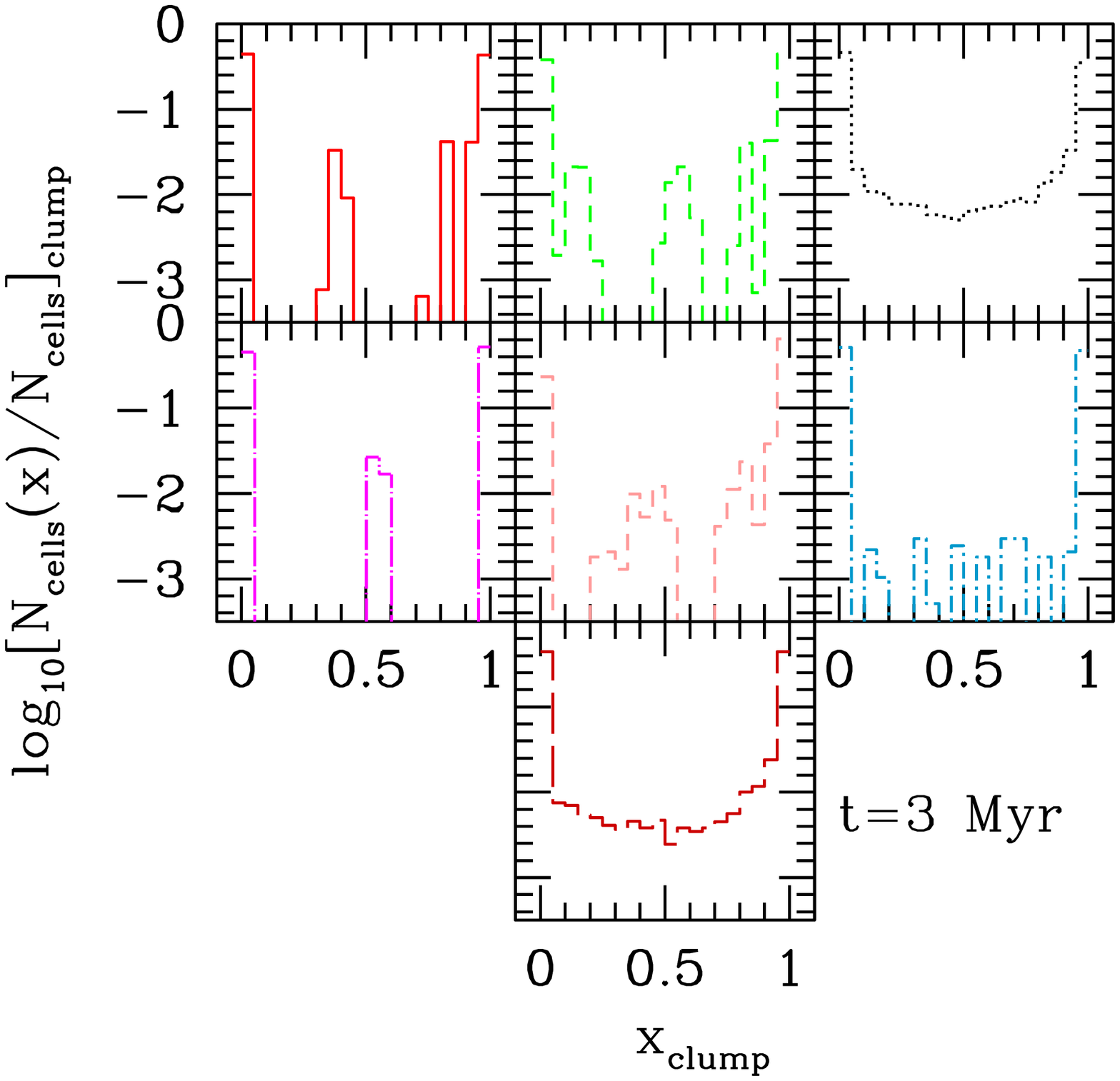}
  \includegraphics[width=2.9in]{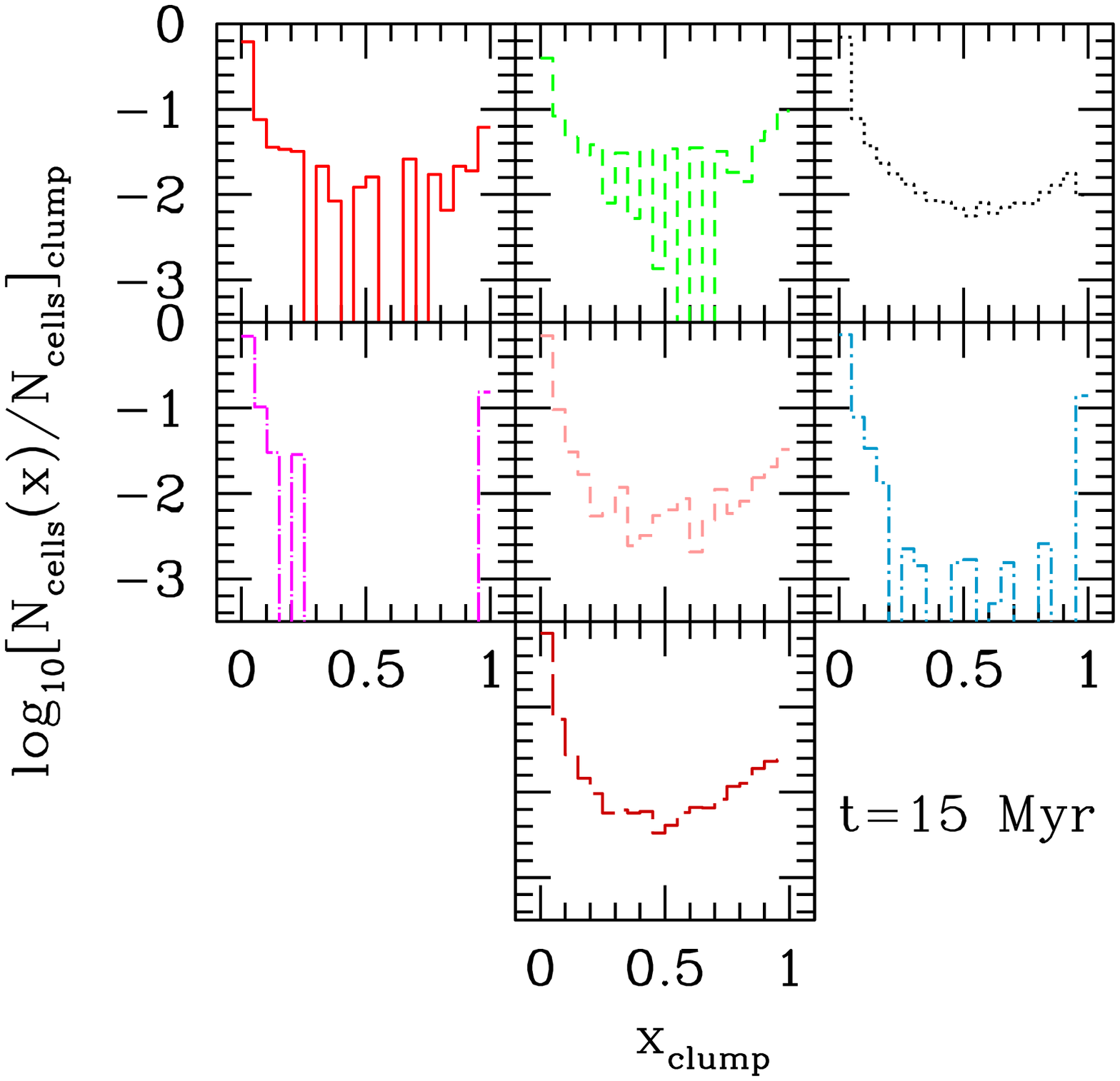}
\caption{Test 3 (I-front trapping in dense clump): Histograms of the 
ionized fraction inside the clump at times $t=1$, 3 and 15 Myrs.
\label{T3_hist_fig}}
\end{center}
\end{figure*}

In Figure~\ref{T3_profs_fig} we show the ionized and neutral fraction profiles
along the axis of symmetry at three stages of the evolution - early
($t=1$~Myr), during the slow-down due to recombinations ($t=3$~Myr, about one
recombination time in the clump) and late ($t=15$~Myr). Only the region inside
and around the clump is plotted in order to show details. In the
pre-ionization, spectrum-hardening zone ahead of the main I-front all 
profiles agree fairly well at all times, except that the IFT profiles have a
sharp I-front and no hardening by definition, and the FTTE current method
appears to produce no hardening, either. Otherwise, these two codes agree well 
with the others in the post-front region. Some differences emerge in the 
position of the I-front (defined as the point of 50\% ionized fraction) and
the neutral fraction profiles behind the I-front. During the initial, fast
propagation phase of the front all codes agree on its position. However, CRASH
and, to a lesser extent, FLASH-HC both find consistently higher neutral 
fractions in the ionized part of the clump than the rest of the codes. As a
consequence, they show that the I-front is trapped closer to the surface of the dense clump. The
reason for this can be seen in the corresponding temperature profiles
(Figure~\ref{T3_hist_fig}). Both CRASH and FLASH-HC obtain a slightly lower
temperature for the dense ionized gas, resulting in a higher recombination rate
there. The RSPH and IFT results show some diffusion at the source-side clump
boundary (at $r/L_{\rm box}\sim0.64$), resulting in a less sharp
transition there, regardless of the sharp discontinuity of the gas density.
The temperatures in the post-front fraction of the clump otherwise agree quite
well, with the exception of the first cells on the source side, where Coral
finds very high temperatures, as was already mentioned. In the pre-front
region the temperature profile results also agree fairly well. Only IFT and 
FTTE differ there, again producing a very sharp I-front. $C^2$-Ray finds a
bit less pre-heating in this region due to its single frequency bin method.
Until $t=3$~Myr all results except the FLASH-HC one agree that the shadow 
region right behind the clump is completely shielded and remains at the 
initial temperature (8,000 K). At the final time, however, more differences 
emerge. At this time only IFT and FTTE still find no pre-heating, and 
$C^2$-Ray and CRASH find only little heating of the shadow. The Coral, RSPH
and FLASH-HC results all find significant pre-heating, albeit at different
levels.  RSPH is the only code that at the final time finds no partially 
shielded gas at the back of the clump, the presence of which is indicated by 
the temperature dip at $r/L_{\rm box}\sim0.88$ for all the other results.     

\begin{figure*}
\begin{center}
  \includegraphics[width=2.9in]{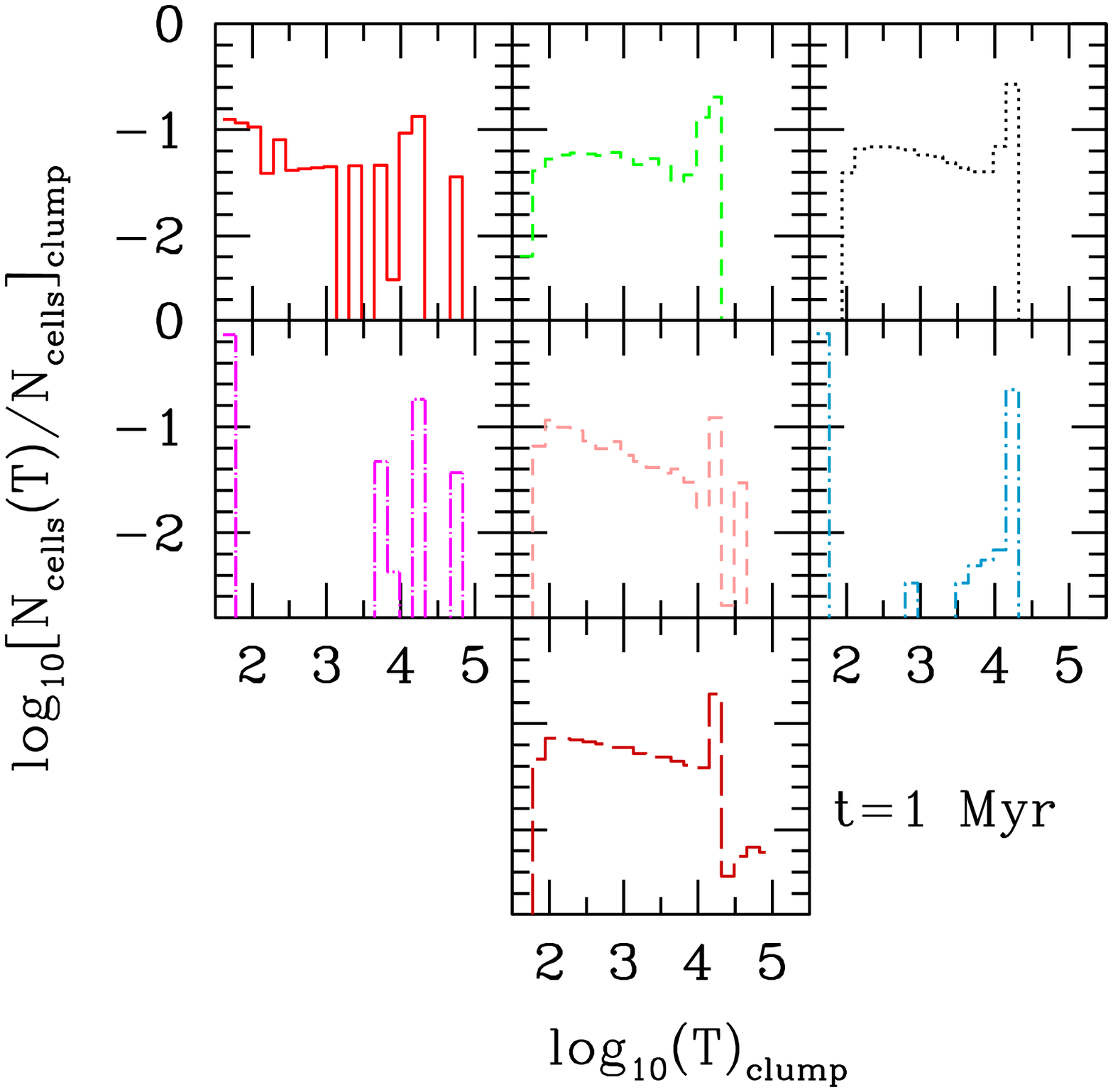}
  \includegraphics[width=2.9in]{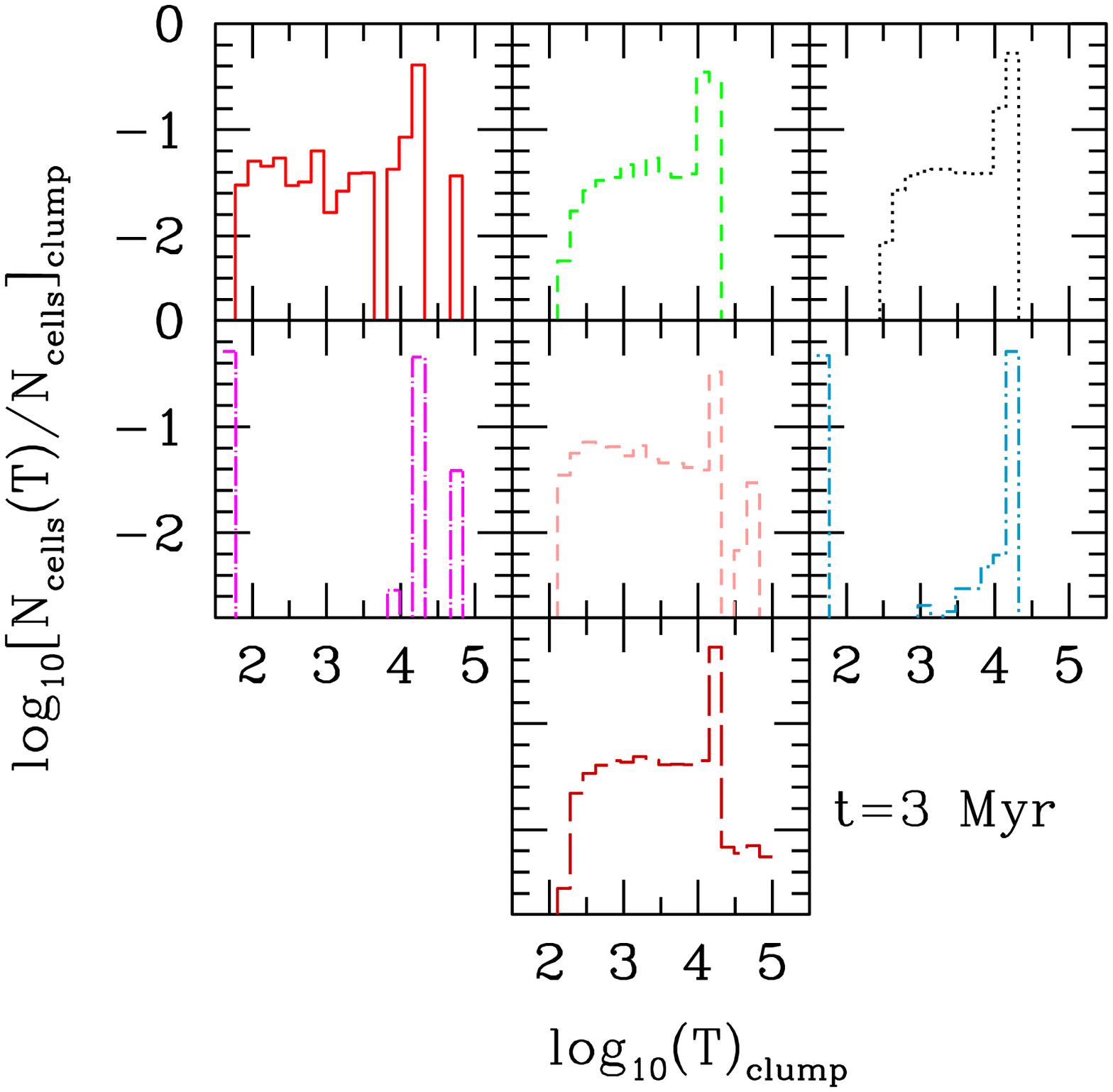}
 \includegraphics[width=2.9in]{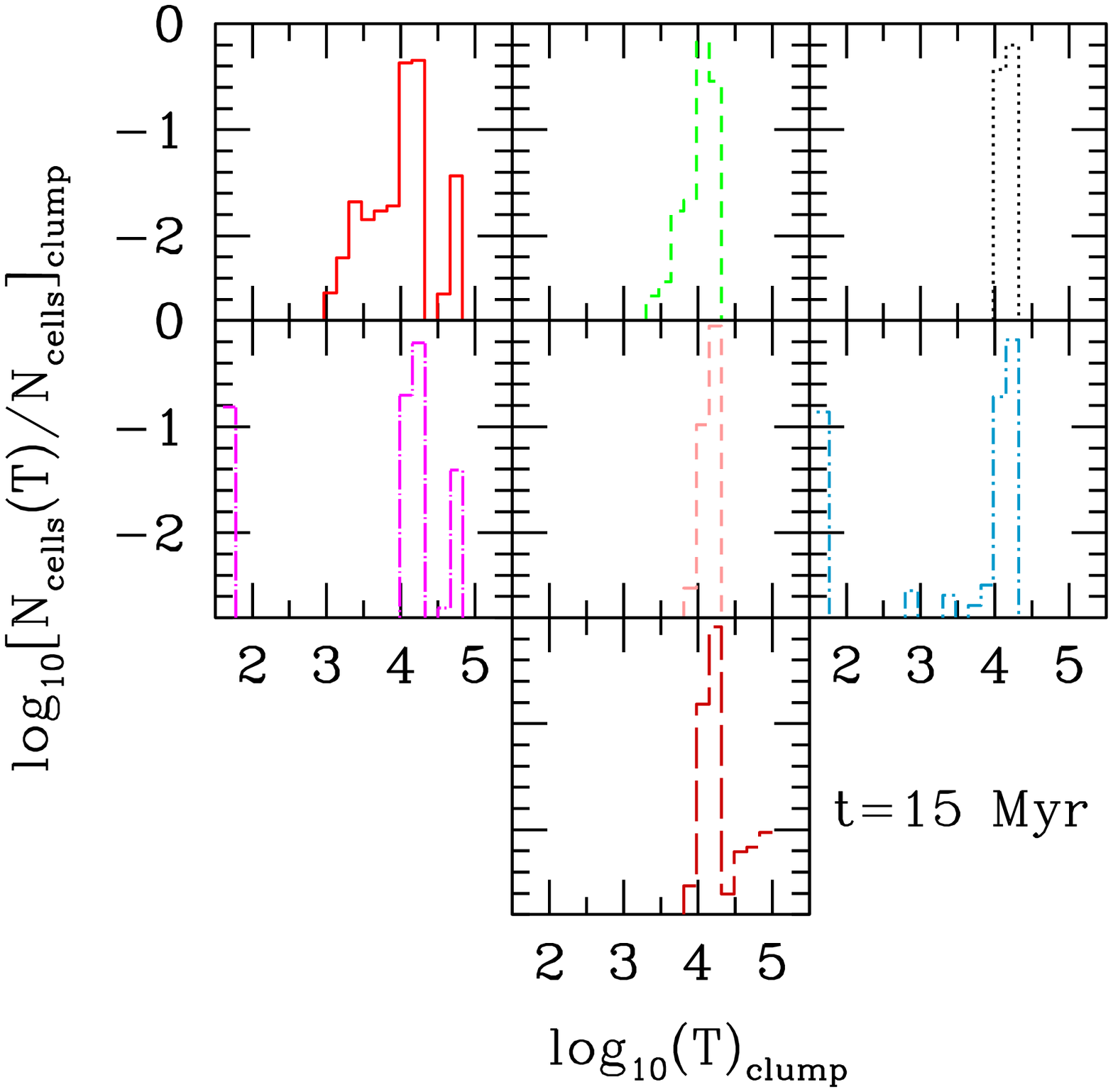}
\caption{\vspace{-1cm} 
Test 3 (I-front trapping in dense clump): Histograms of the 
temperature inside the clump at times $t=1$, 3 and 15 Myrs.
\label{T3_hist_T_fig}}
\end{center}
\end{figure*}

The final results we show for this test are histograms of the fraction of
cells inside the clump with a given ionized fraction
(Figure~\ref{T3_hist_fig}) and with a given temperature
(Figure~\ref{T3_hist_T_fig}) at the same three evolutionary stages discussed 
above. These histograms reflect the differences in the thickness and the 
internal structure of the I-fronts inside the clump. During the initial fast
propagation and the slow-down phases the RSPH and Coral codes consistently
find somewhat thicker I-front transitions than the other codes. The results of
$C^2$-Ray, CRASH and FLASH-HC start with the front somewhat thinner than the
ones for  RSPH and Coral, but the distribution changes as the I-front gets
trapped. At the final time, the ionized fraction distributions are very
similar, with only the one for CRASH being slightly thicker. Finally, FTTE and
IFT find significantly thinner front transitions and significant self-shielded
gas fractions, as was noted before. The corresponding temperature distributions
all show a strong peak at a similar temperature, a few tens of thousands of
degrees, typical for gas heated by photoionization. This temperature is well
above $10^4$~K because of the hot black-body spectrum of the source. At
temperatures lower than this peak value, which largely correspond to the
pre-heated zone ahead of the I-front, there again is broad agreement, apart
from the FTTE and IFT codes which have a sharp front and little
pre-heating. At the high-temperature end of the distribution $C^2$-Ray, FTTE,
FLASH-HC and Coral diverge from the rest, by finding a small fraction of 
very hot cells, while the majority of codes find essentially no cells hotter
than the distribution peak. It should be noted, however, that the fraction of
these cells is only 0.1-1\% of the total.    

\subsection{Test 4: Multiple sources in a cosmological density field}

Test 4 involves the propagation of I-fronts from multiple sources in a
static cosmological density field. The initial condition is 
provided by a time-slice (at redshift $z=9$) from a cosmological N-body 
and gas-dynamic simulation performed using the cosmological PM+TVD code 
by D. Ryu \citep{1993ApJ...414....1R}. The simulation box size is 
$0.5\,h^{-1}$ comoving Mpc, the resolution is $128^3$ cells, 
$2\times64^3$ particles. The halos in the simulation box were found 
using friends-of-friends halo finder with linking length of 0.25. 
For simplicity the initial temperature is fixed at $T=100$ K everywhere. 
The ionizing sources are chosen so as to correspond to the 16 most 
massive halos in the box. We assume that these have a black-body spectrum 
with effective temperature $T_{\rm eff}=10^5$~K. The ionizing photon 
production rate for each source is constant and is assigned assuming 
that each source lives $t_s=3$ Myr and emits $f_\gamma=250$ ionizing 
photons per atom during its lifetime, hence
\be
\dot{N}_\gamma=f_\gamma \frac{M\Omega_b}{\Omega_0 m_pt_s},
\ee
where $M$ is the total halo mass, $\Omega_0=0.27$, $\Omega_b=0.043$, 
$h=0.7$. 

For simplicity all sources are assumed to switch on at the same time. 
The boundary conditions are transmissive (i.e. photons leaving the 
computational box are lost, rather than coming back in as in periodic 
boundary conditions).  The evolution time is $t=0.4$ Myr.

\begin{figure*}
\begin{center}
  \includegraphics[width=2.2in]{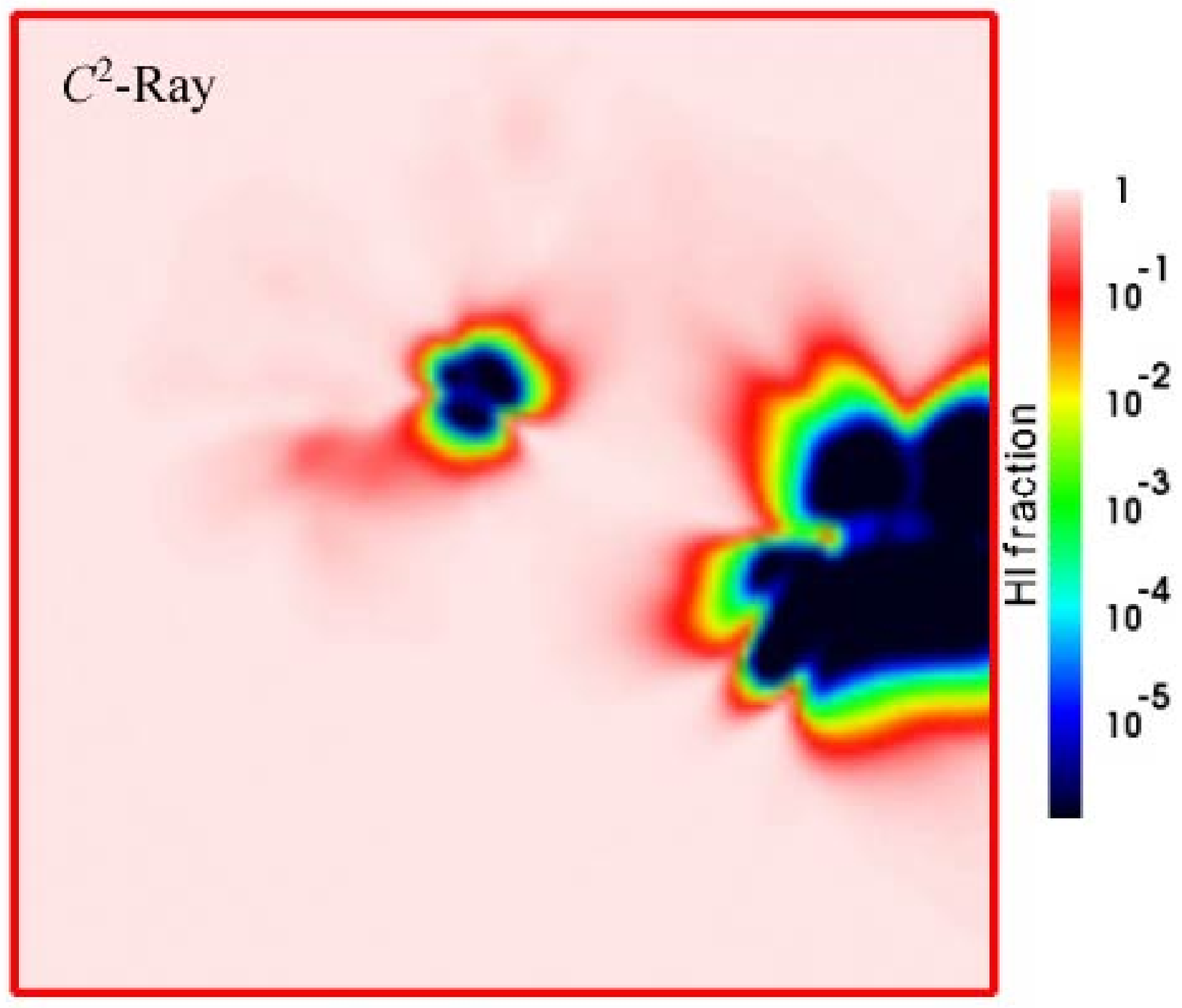}
  \includegraphics[width=2.2in]{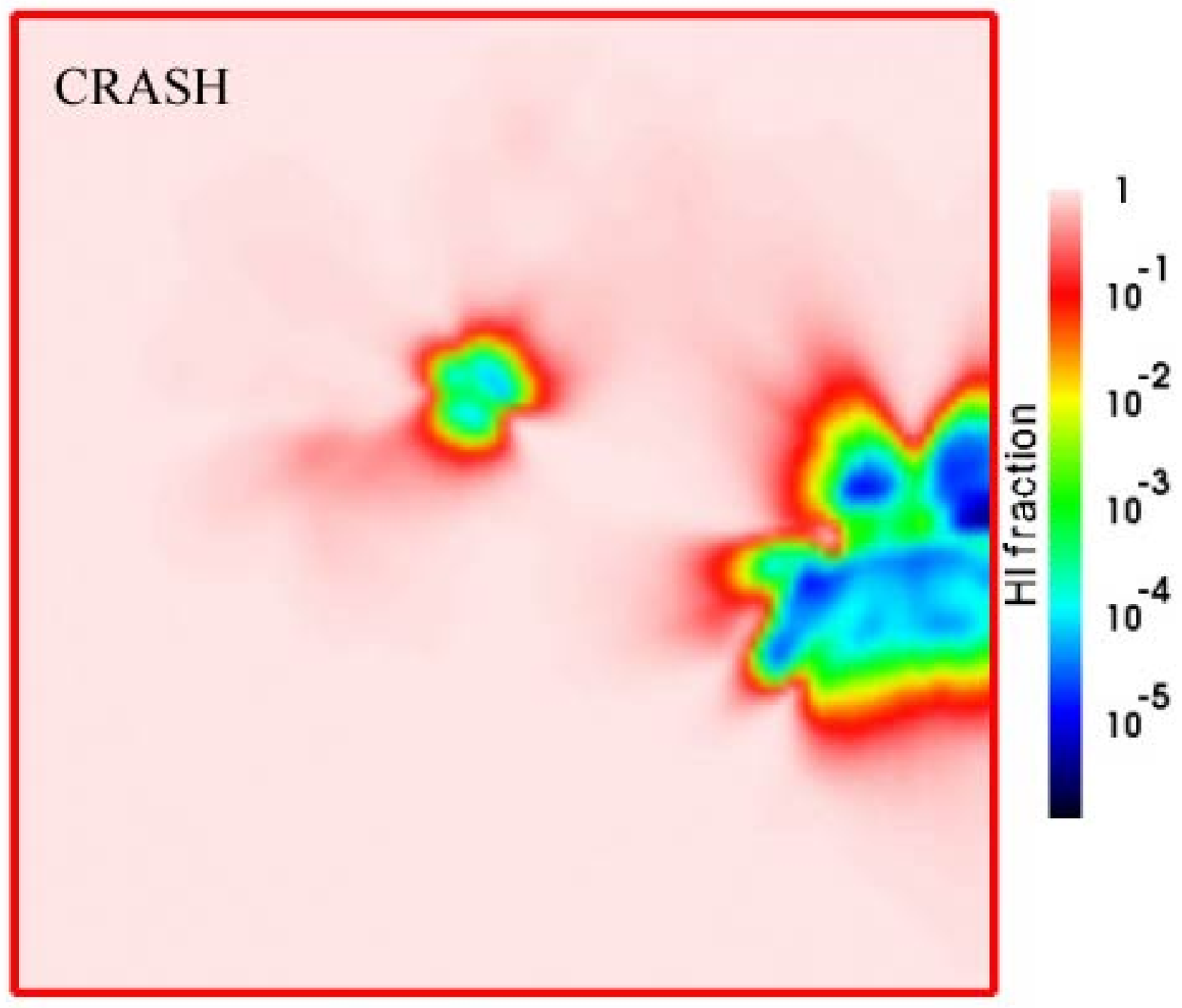}
  \includegraphics[width=2.2in]{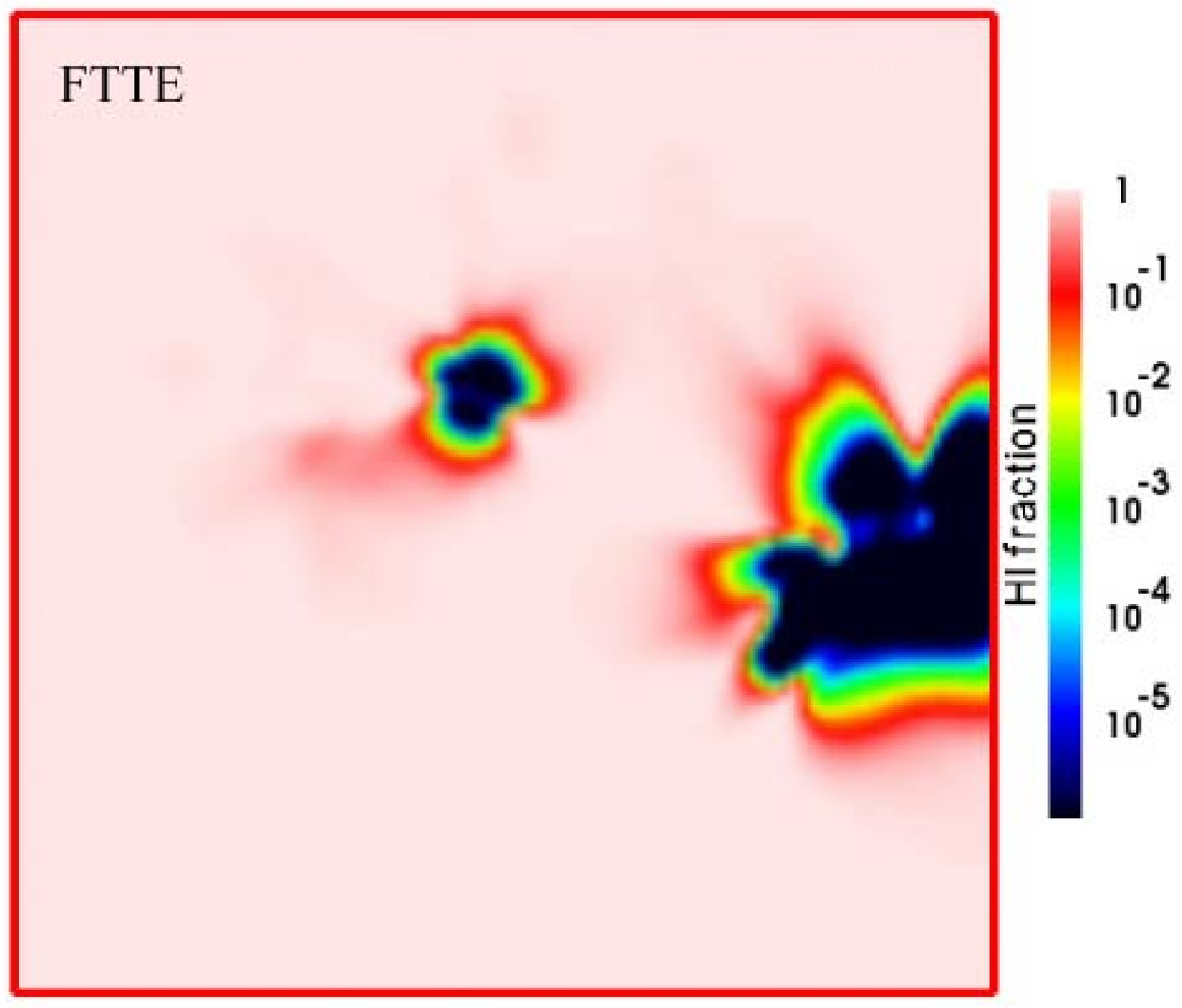}
  \includegraphics[width=2.2in]{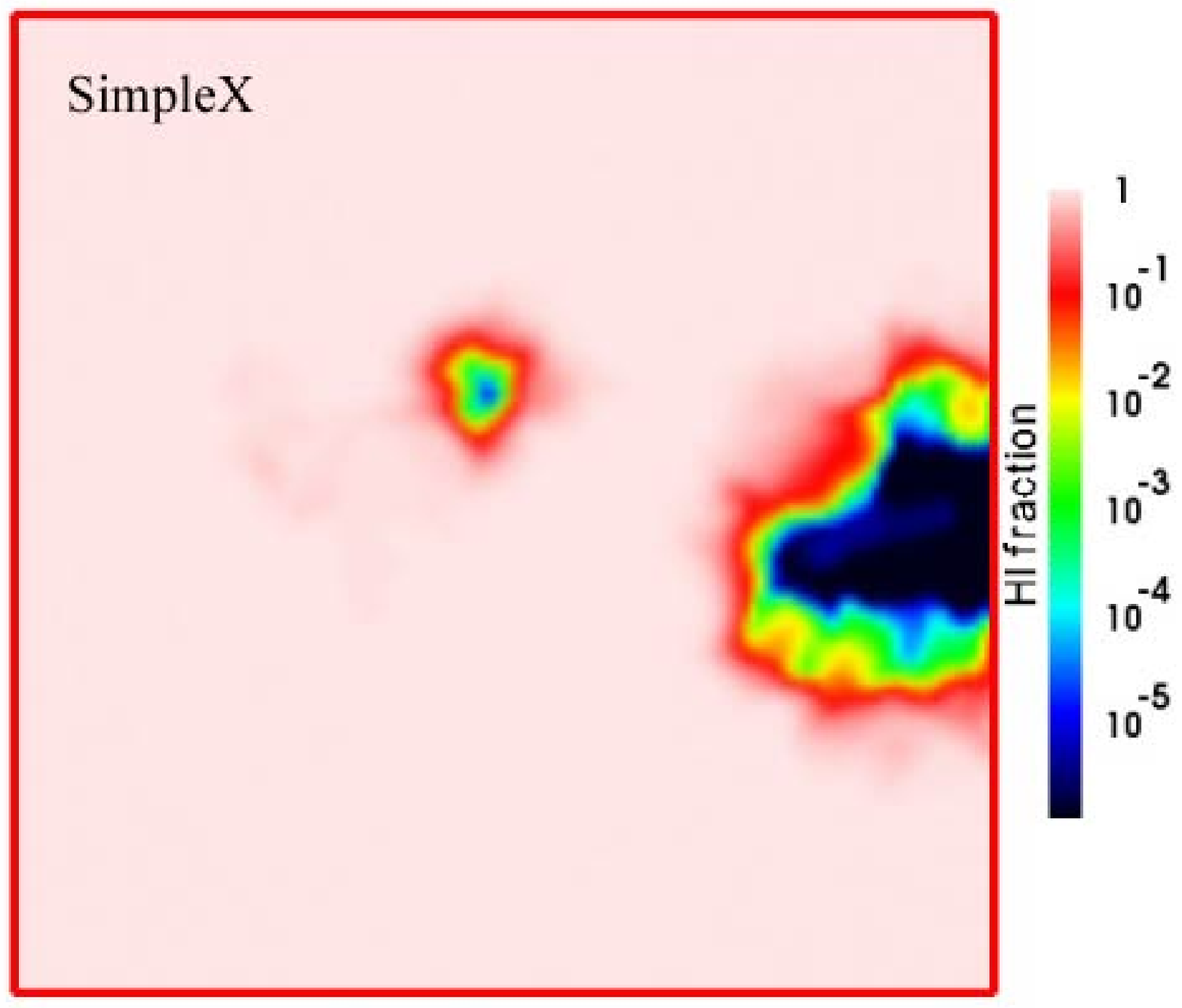}
\caption{Test 4 (reionization of a cosmological density field): Images of 
the H~I fraction, cut through the simulation 
volume at coordinate $z=z_{\rm box}/2$ and time $t=0.05$ Myr for $C^2$-Ray 
(left), CRASH (middle), FTTE (right), and SimpleX (bottom). The black-body 
spectrum has an effective temperature $T_{\rm eff}=10^5$~K.
\label{T4_images1_HI_fig}}
\includegraphics[width=2.2in]{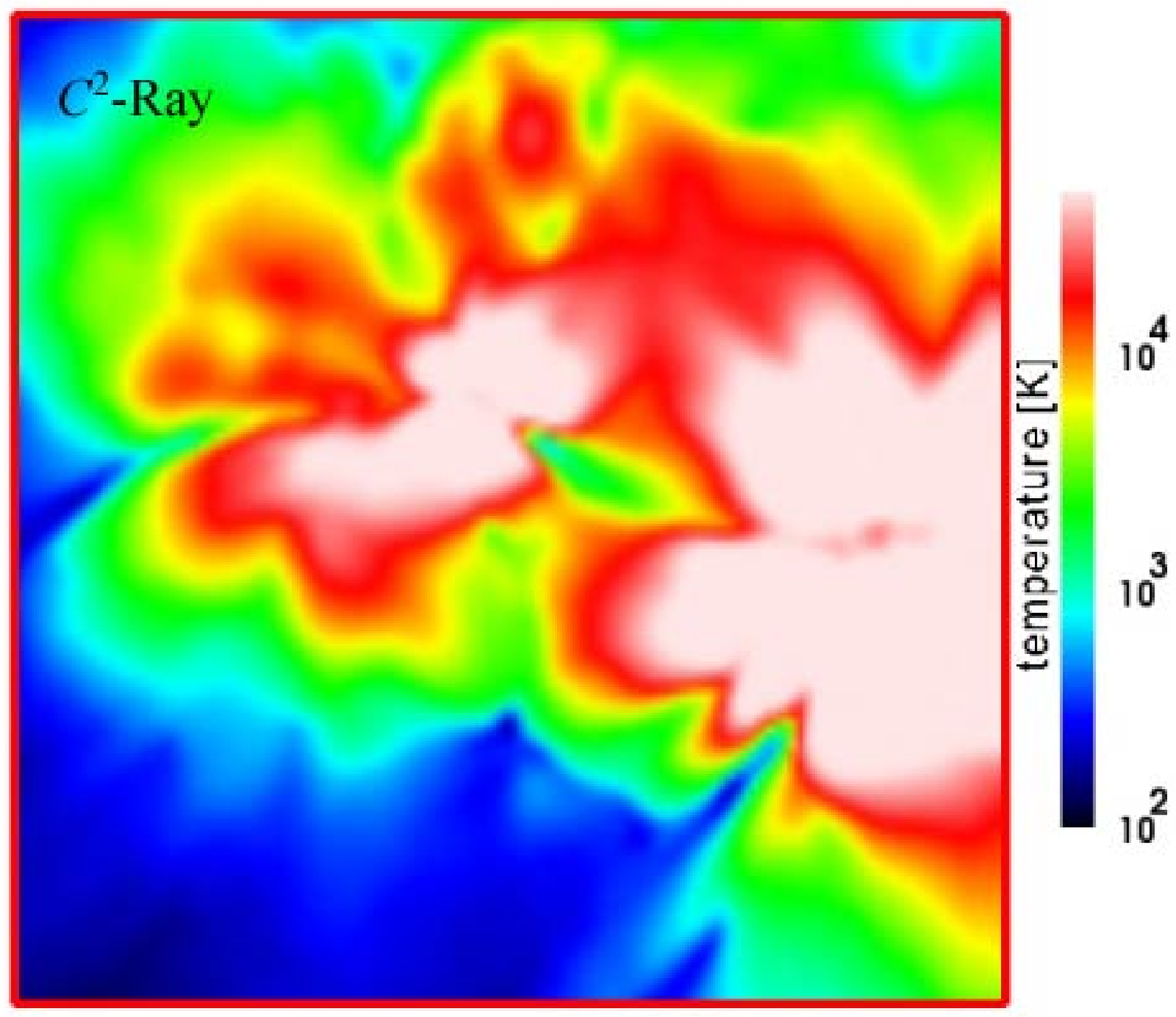}
\includegraphics[width=2.2in]{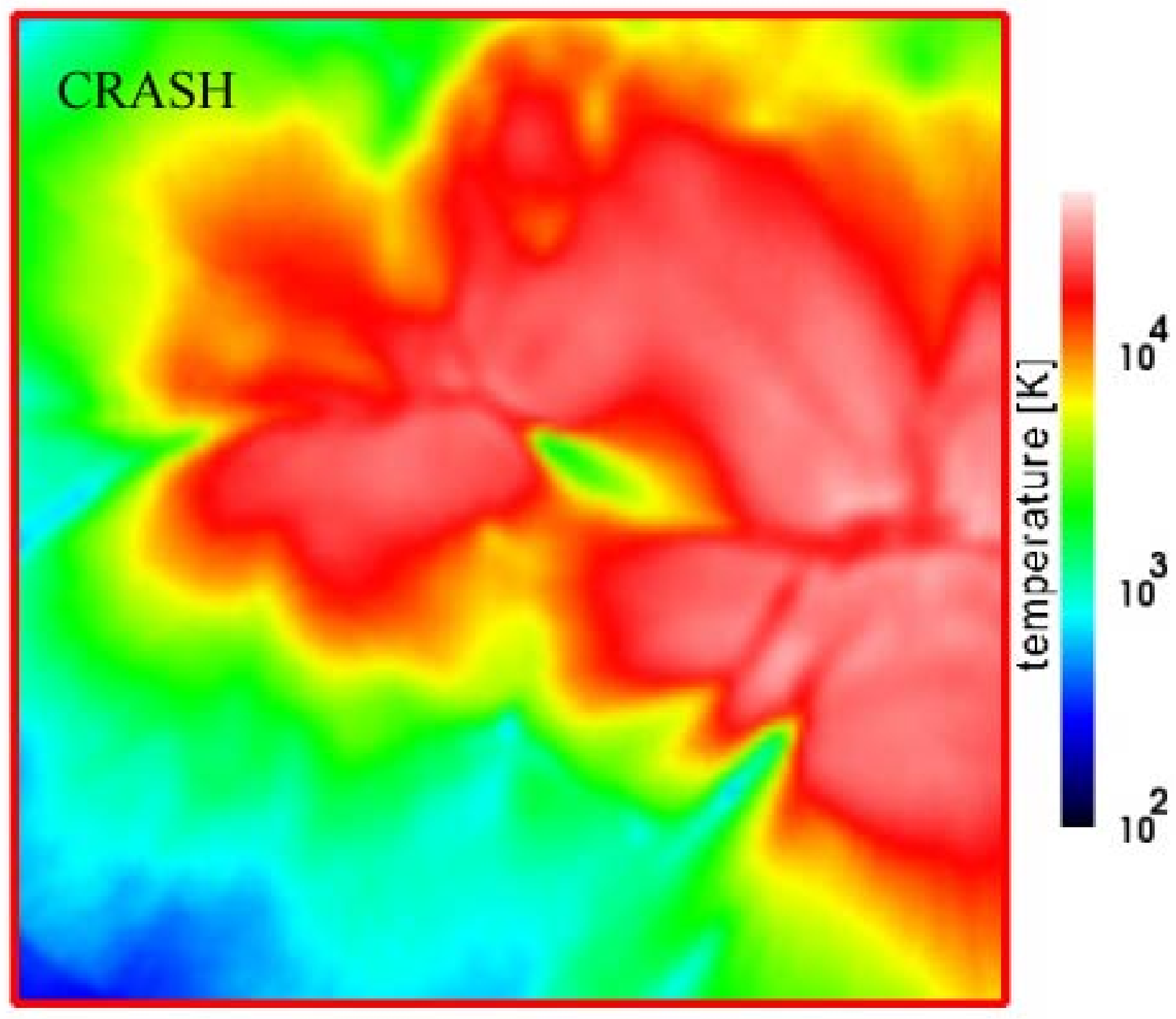}
\includegraphics[width=2.2in]{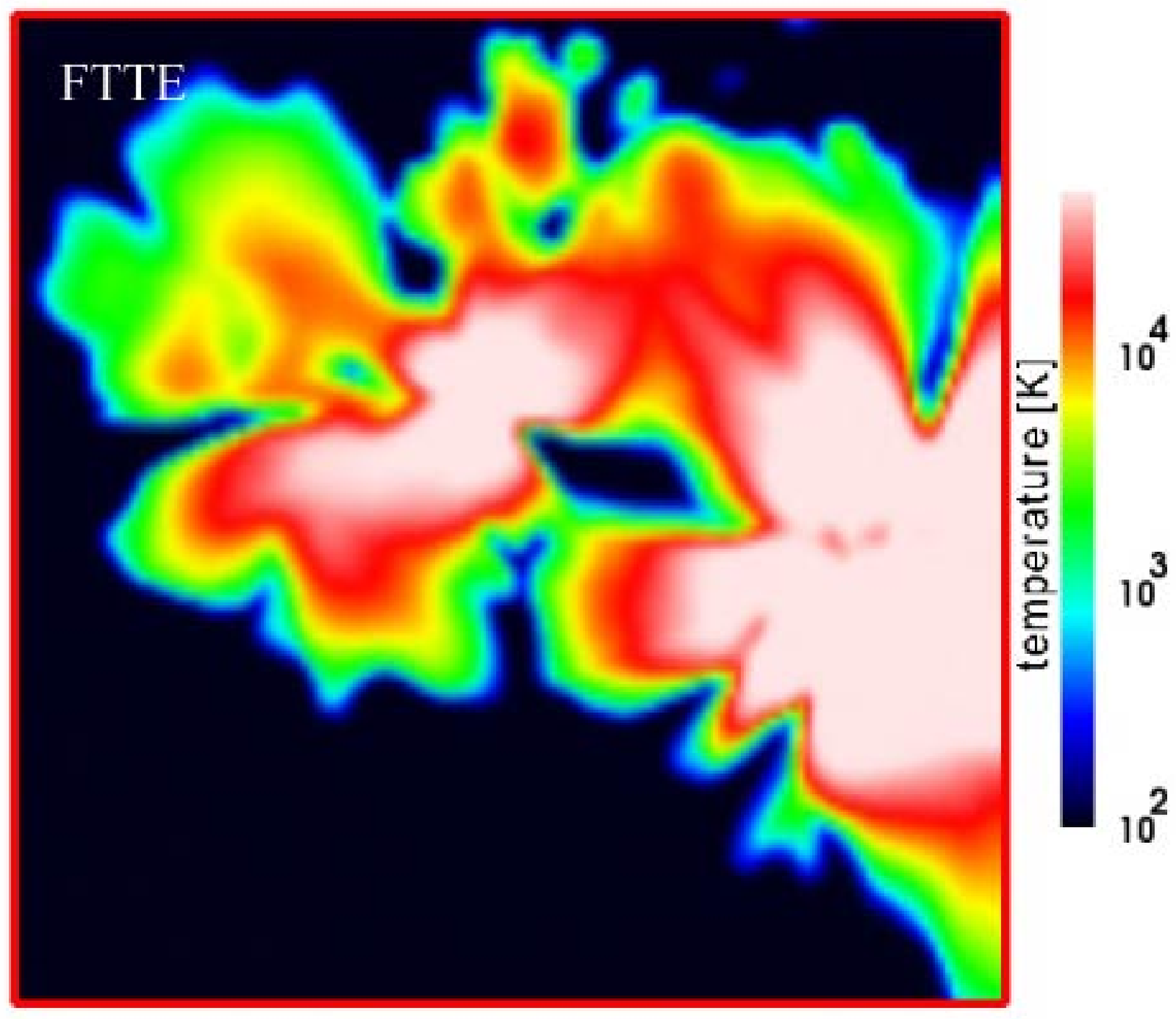}
\caption{Test 4 (reionization of a cosmological density field): Images of the
  temperature, cut through the simulation  
volume at coordinate $z=z_{\rm box}/2$ and time $t=0.05$ Myr for 
$C^2$-Ray (left), CRASH (middle), and FTTE (right). The black-body spectrum
has an effective temperature $T_{\rm eff}=10^5$~K.
\label{T4_images1_T_fig}}
\end{center}
\end{figure*}

\begin{figure*}
\begin{center}
  \includegraphics[width=2.2in]{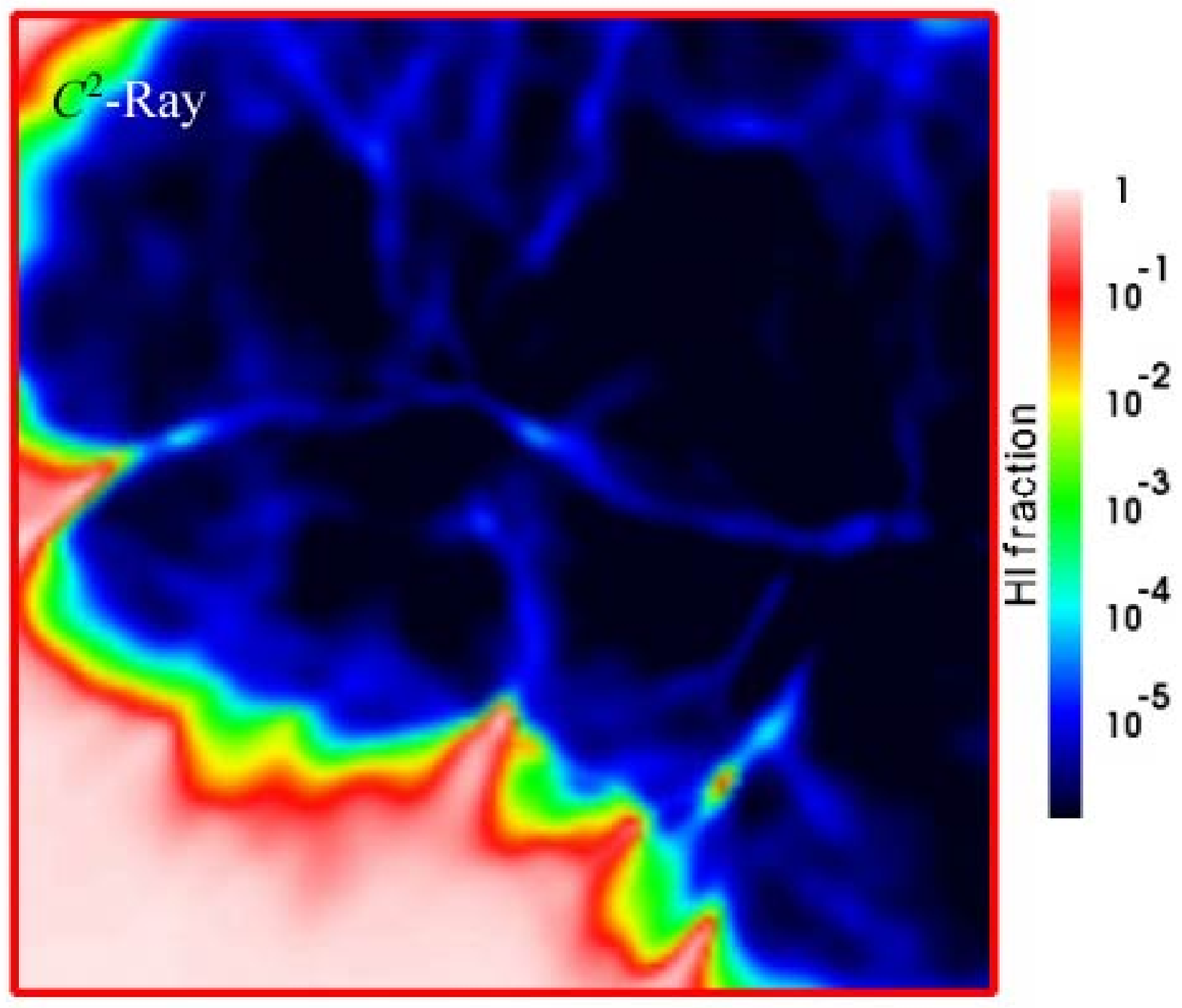}
  \includegraphics[width=2.2in]{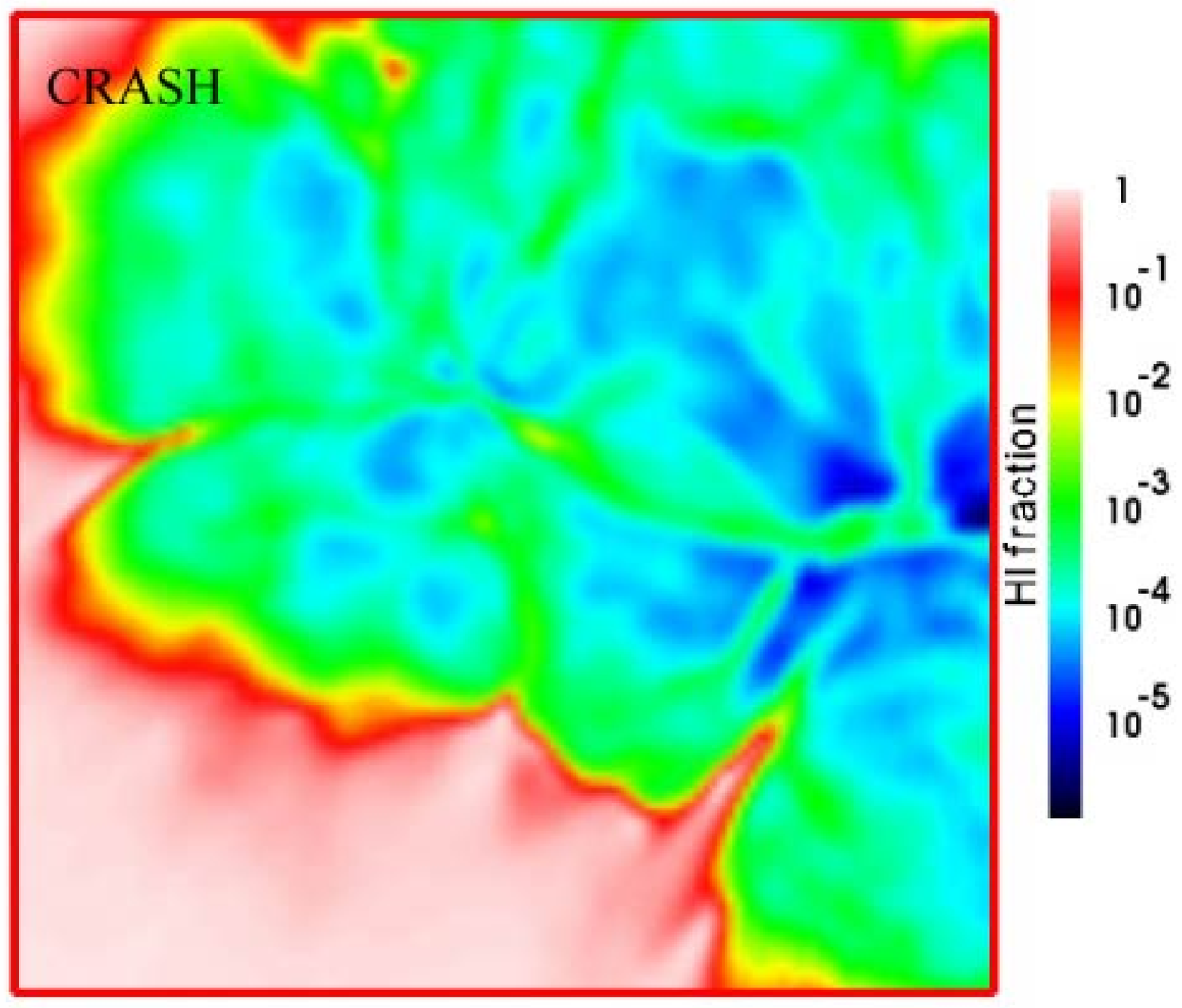}
  \includegraphics[width=2.2in]{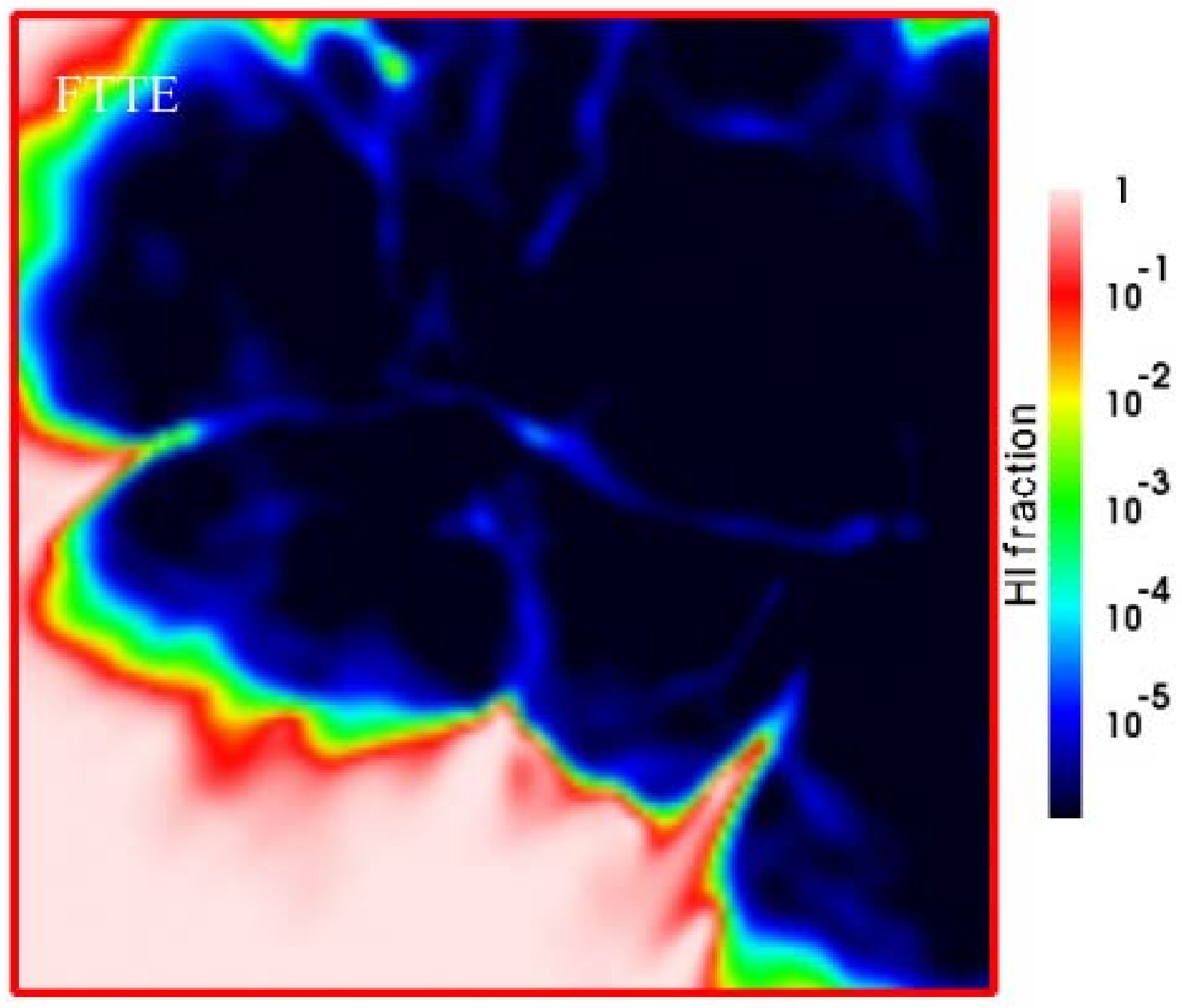}
  \includegraphics[width=2.2in]{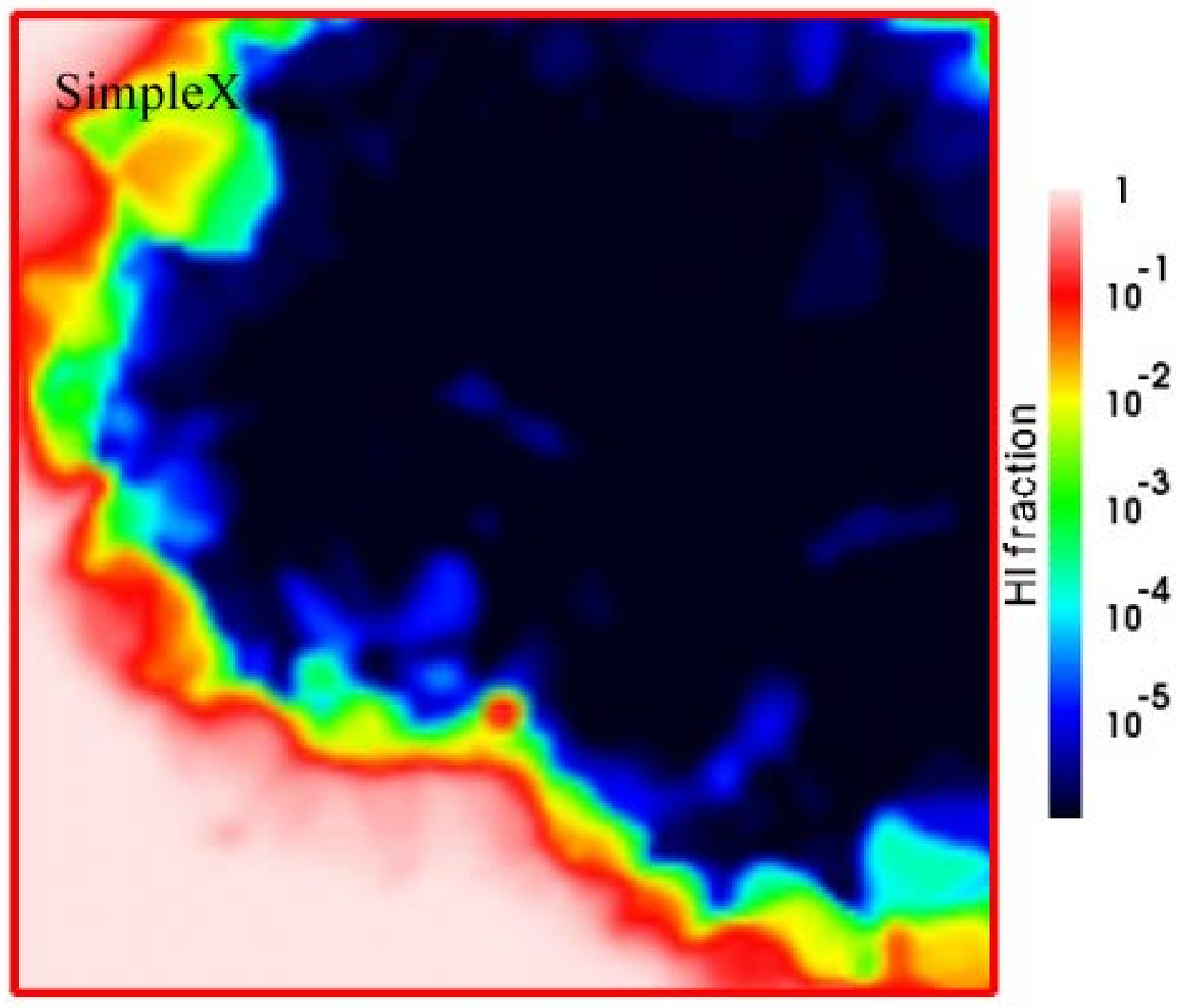}
\caption{Test 4  (reionization of a cosmological density field): Images of the
  H~I fraction, cut through the simulation volume at
  coordinate $z=z_{\rm box}/2$ and time $t=0.2$ Myr for $C^2$-Ray (left),
  CRASH (middle), FTTE (right), and SimpleX (bottom). The black-body spectrum
has an effective temperature $T_{\rm eff}=10^5$~K.
\label{T4_images3_HI_fig}}
\includegraphics[width=2.2in]{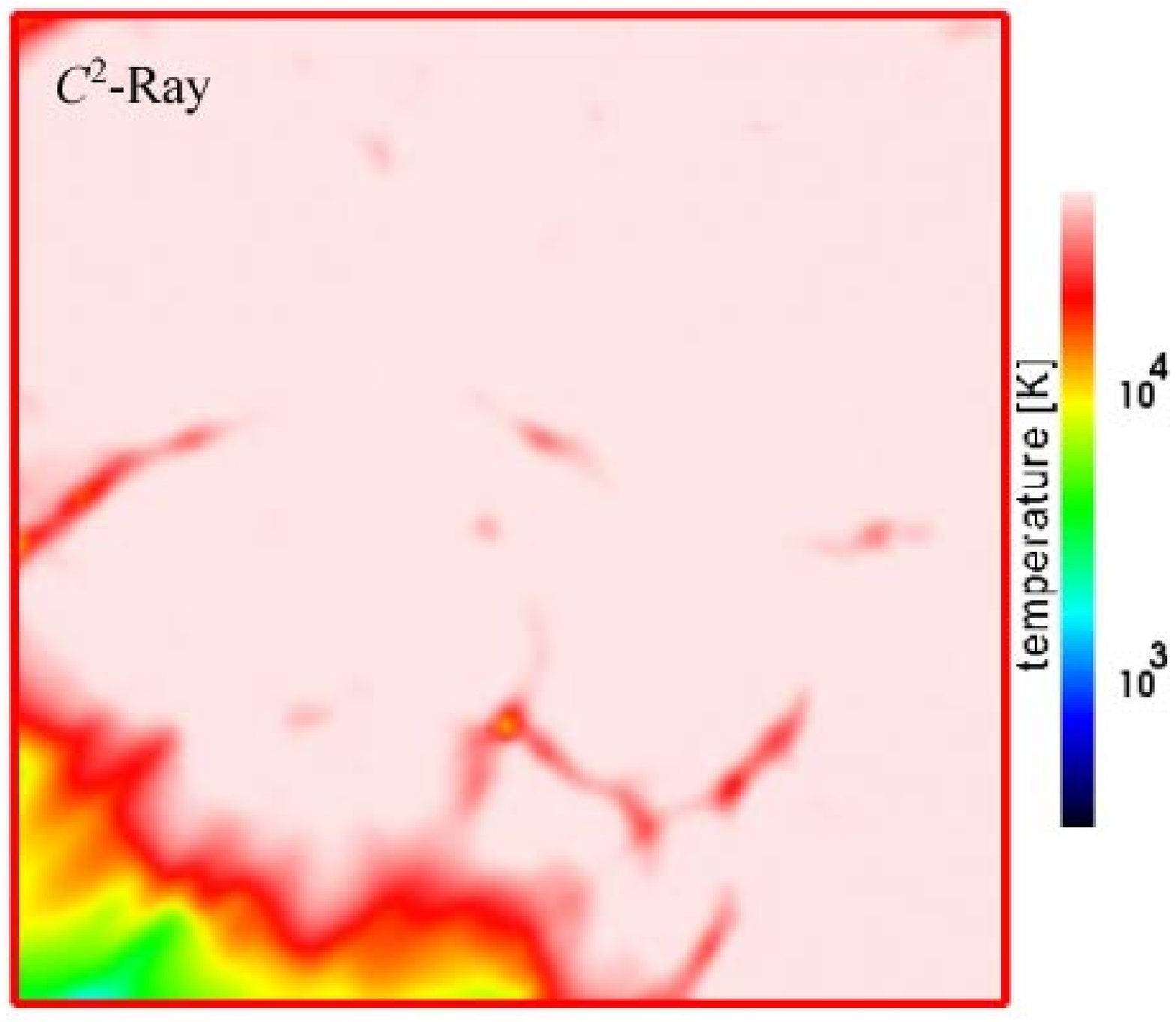}
\includegraphics[width=2.2in]{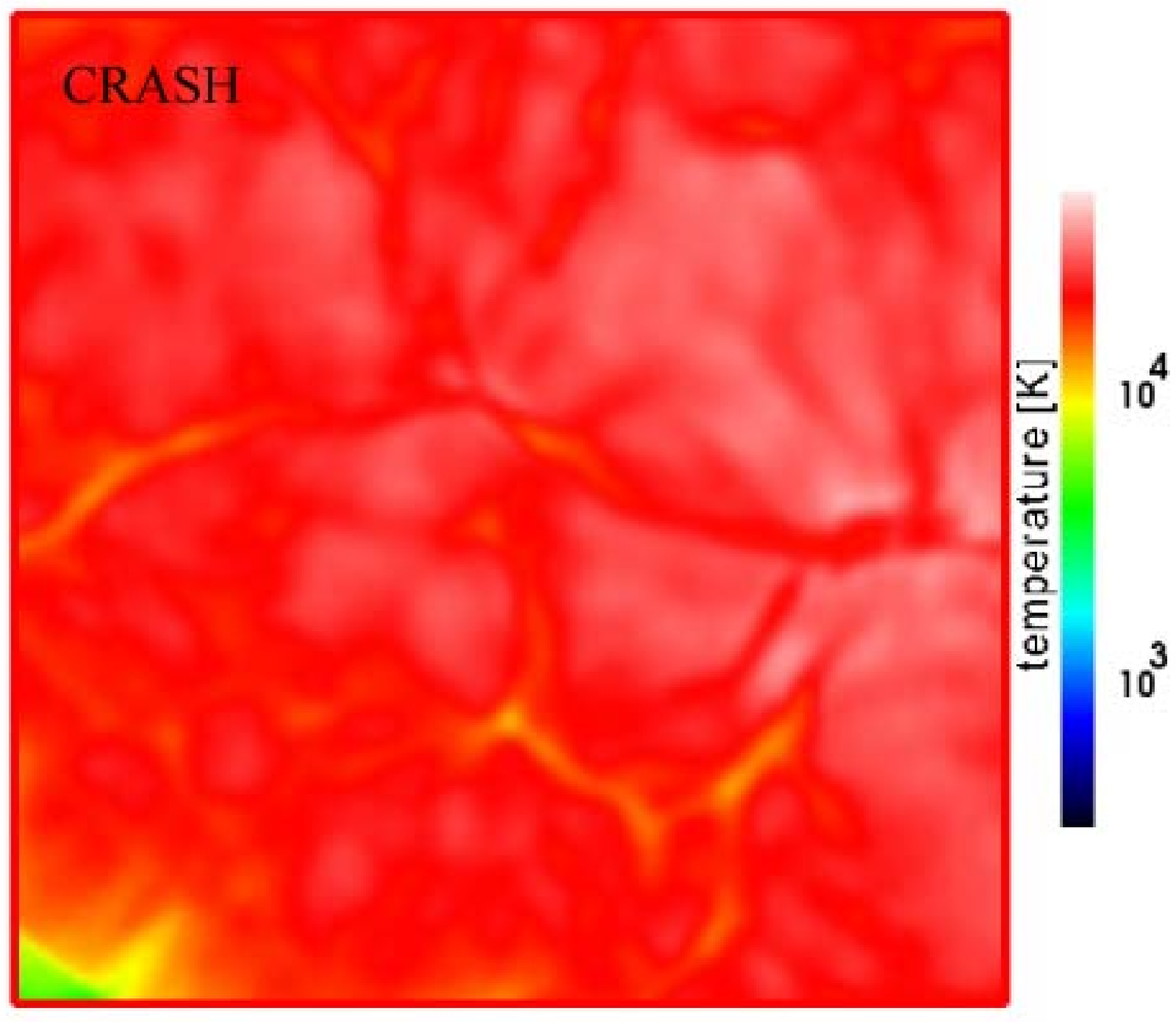}
\includegraphics[width=2.2in]{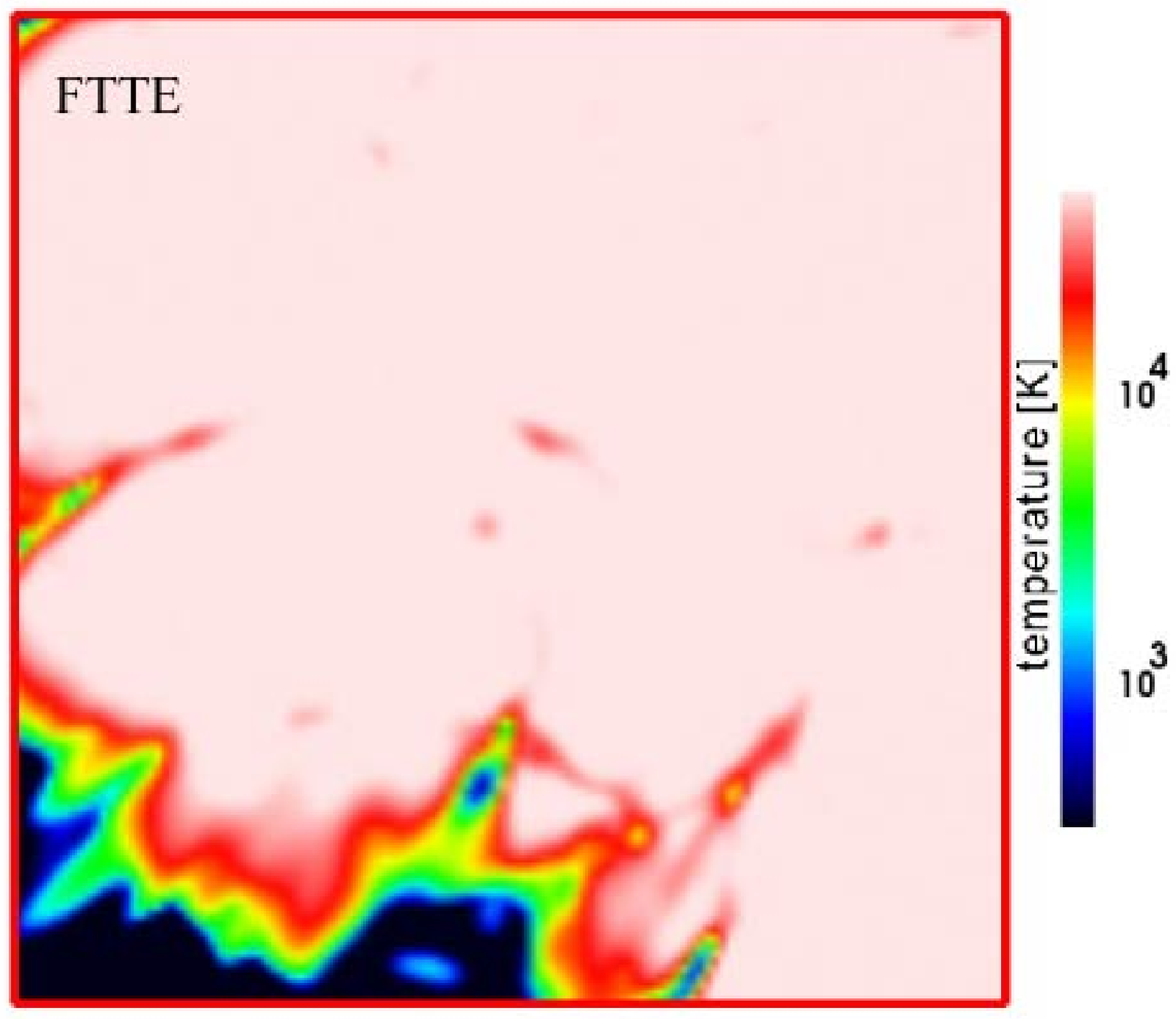}
\caption{Test 4  (reionization of a cosmological density field): Images of the
  temperature, cut through the simulation 
volume at coordinate $z=z_{\rm box}/2$ and time $t=0.2$ Myr for
$C^2$-Ray (left), CRASH (middle), and FTTE (right). The black-body spectrum
has an effective temperature $T_{\rm eff}=10^5$~K.
\label{T4_images3_T_fig}}
\end{center}
\end{figure*}

\begin{figure*}
\begin{center}
  \includegraphics[width=3.2in]{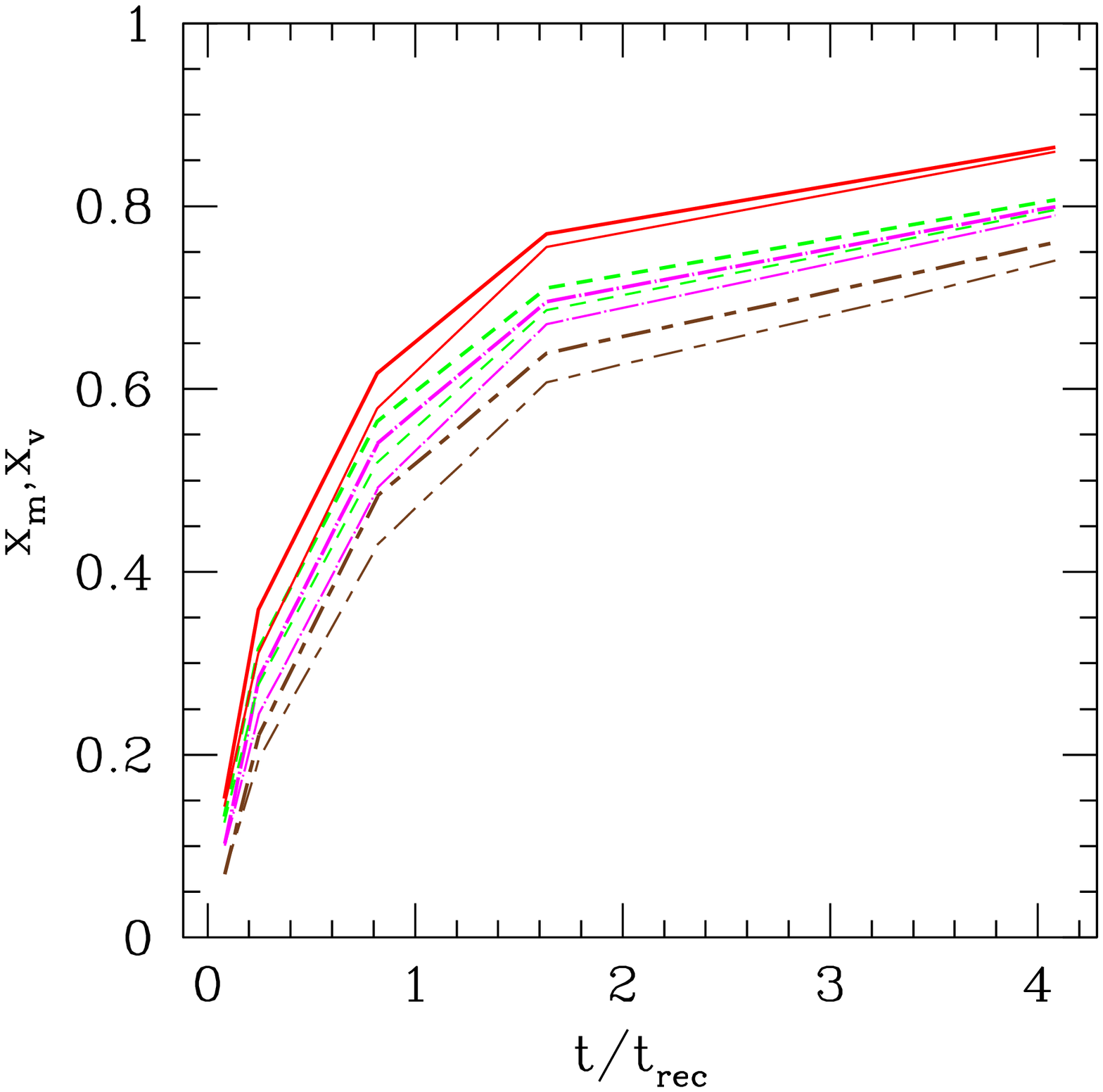}
  \includegraphics[width=3.2in]{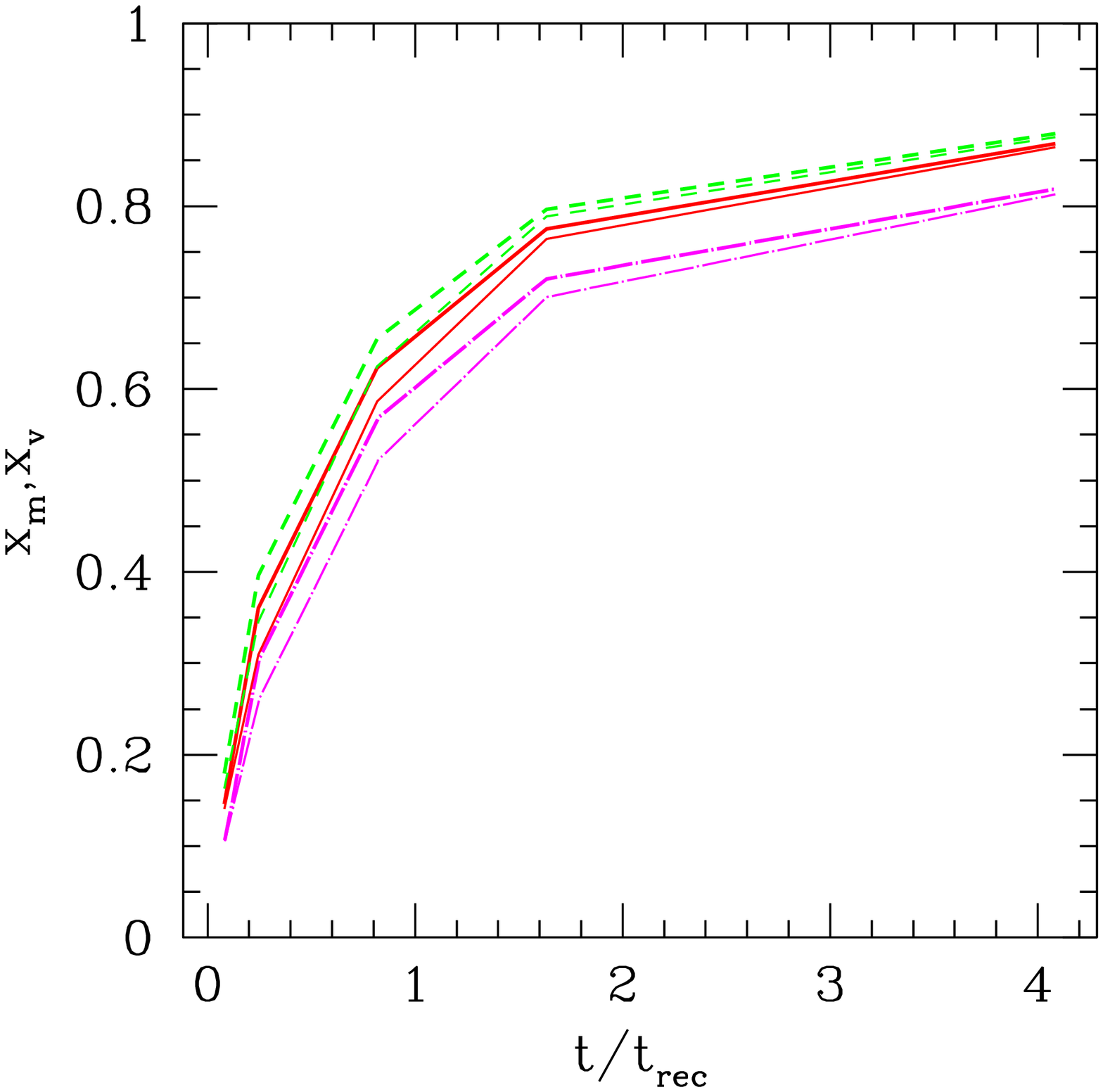}
\caption{Test 4  (reionization of a cosmological density field): The evolution
  of the volume- and mass-weighted ionized 
  fractions, $x_v$ (thin lines) and $x_m$ (thick lines), for a black-body 
source spectra with $T_{\rm eff}=10^5$ K (left) and 
$T_{\rm eff}=3\times10^4$ K (right).
\label{T4_x_evol_fig}}
\end{center}
\end{figure*}

\begin{figure*}
  \includegraphics[width=2.2in]{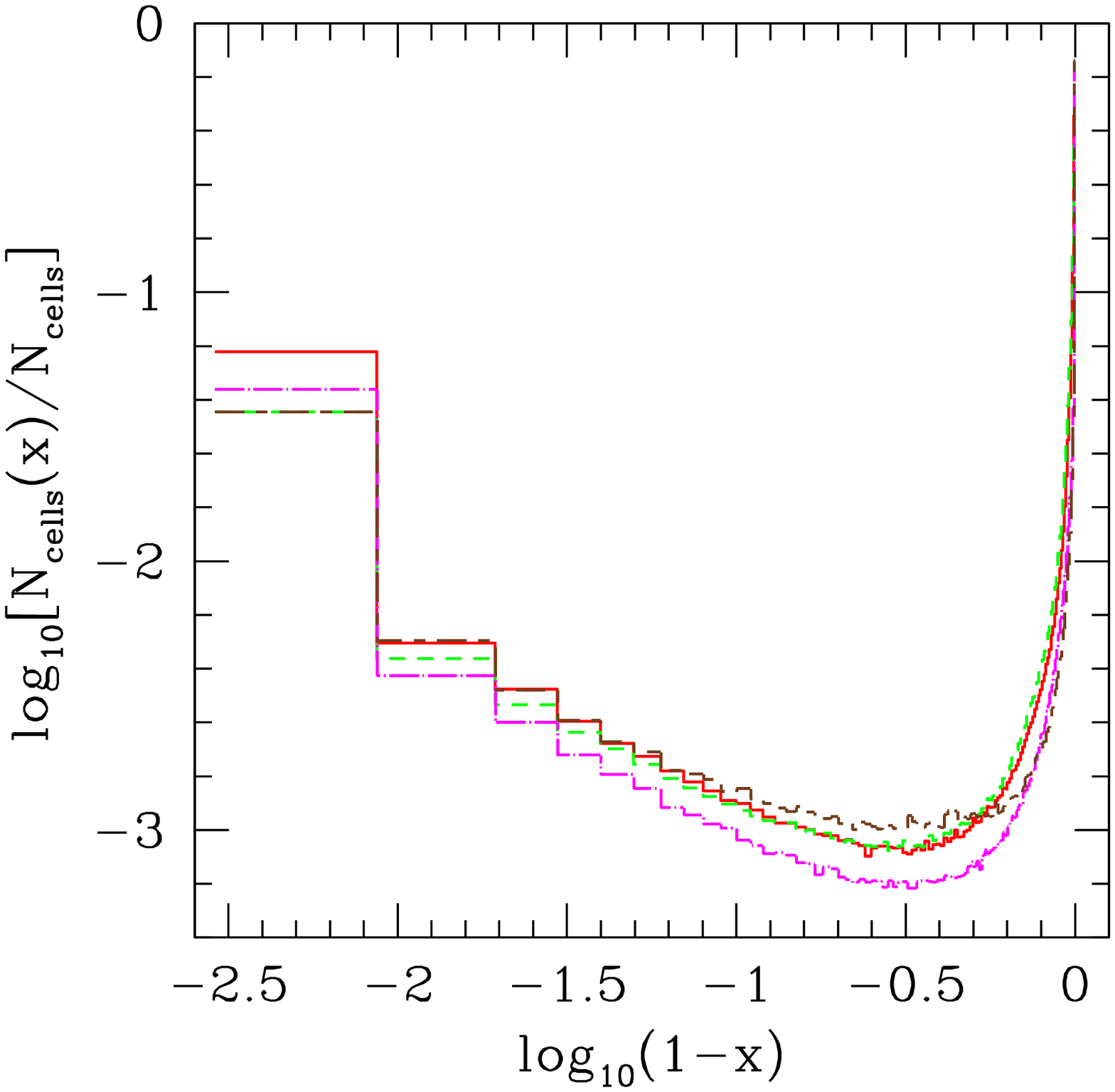}
  \includegraphics[width=2.2in]{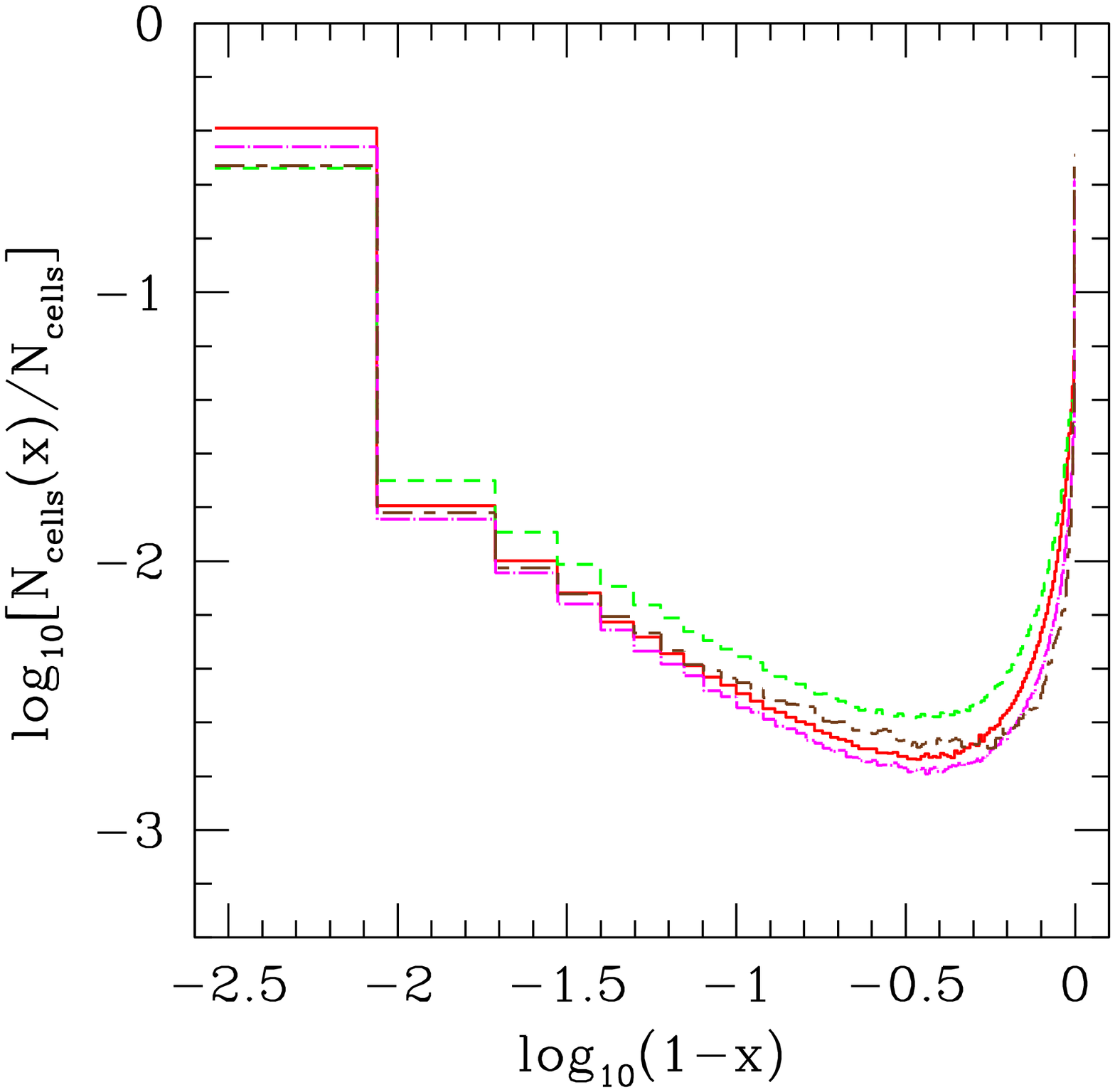}
  \includegraphics[width=2.2in]{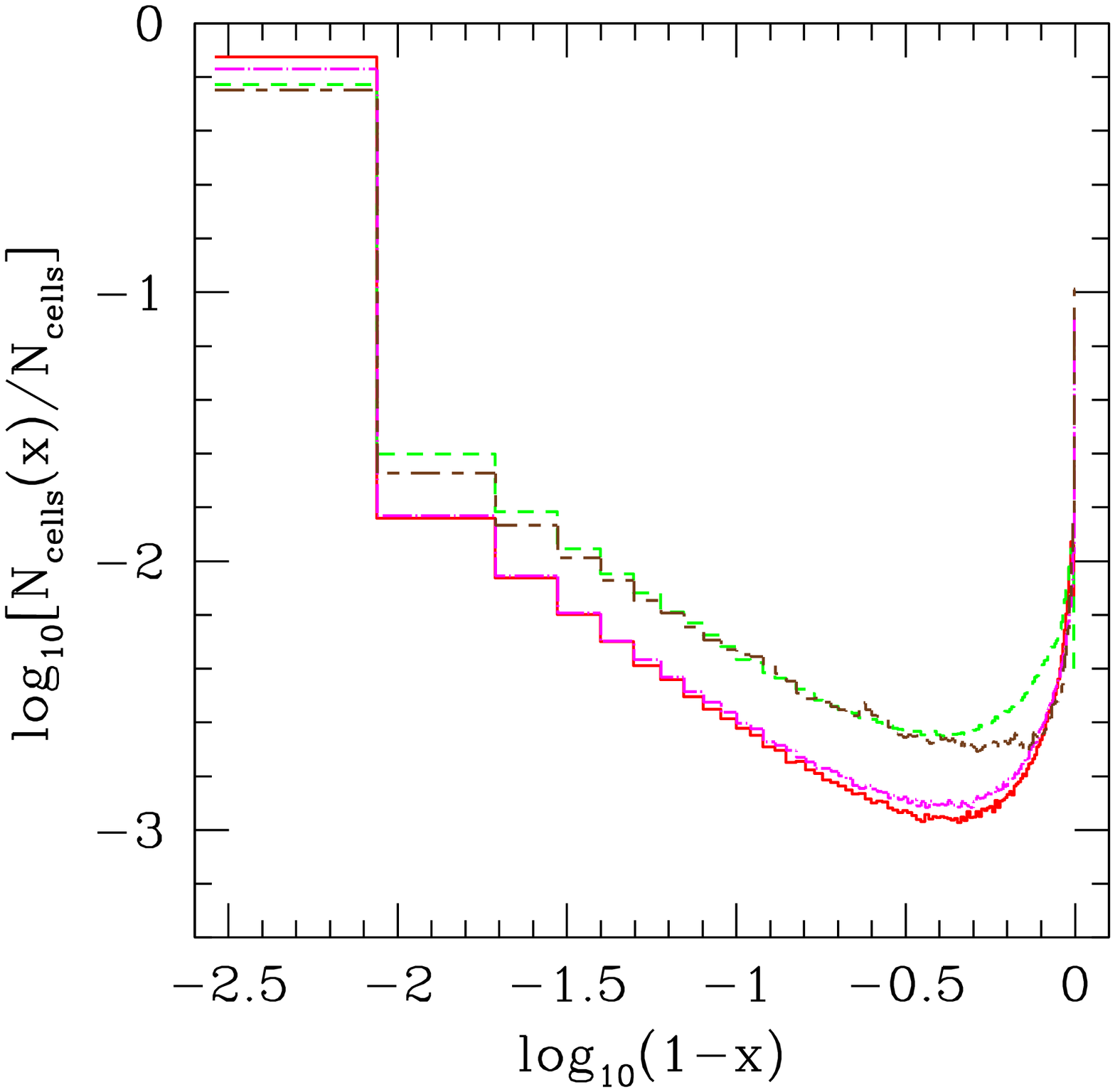}
\caption{Test 4  (reionization of a cosmological density field): Histograms of
  the neutral fraction at times $t=0.05$,  
  0.2 and 0.4 Myrs for a black-body source spectrum with $T_{\rm eff}=10^5$ K.
\label{T4_hist_H_fig}}
  \includegraphics[width=2.2in]{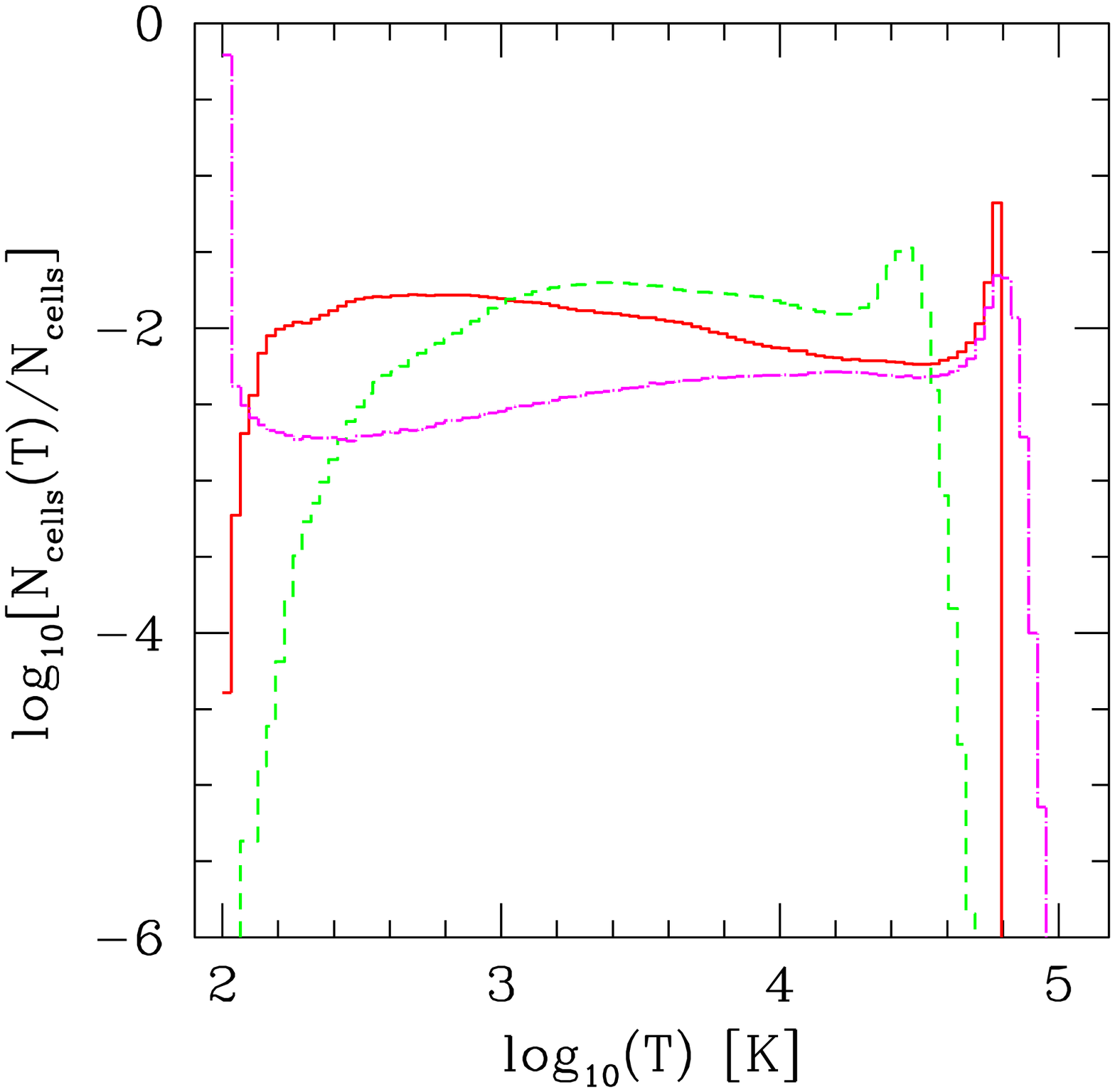}
  \includegraphics[width=2.2in]{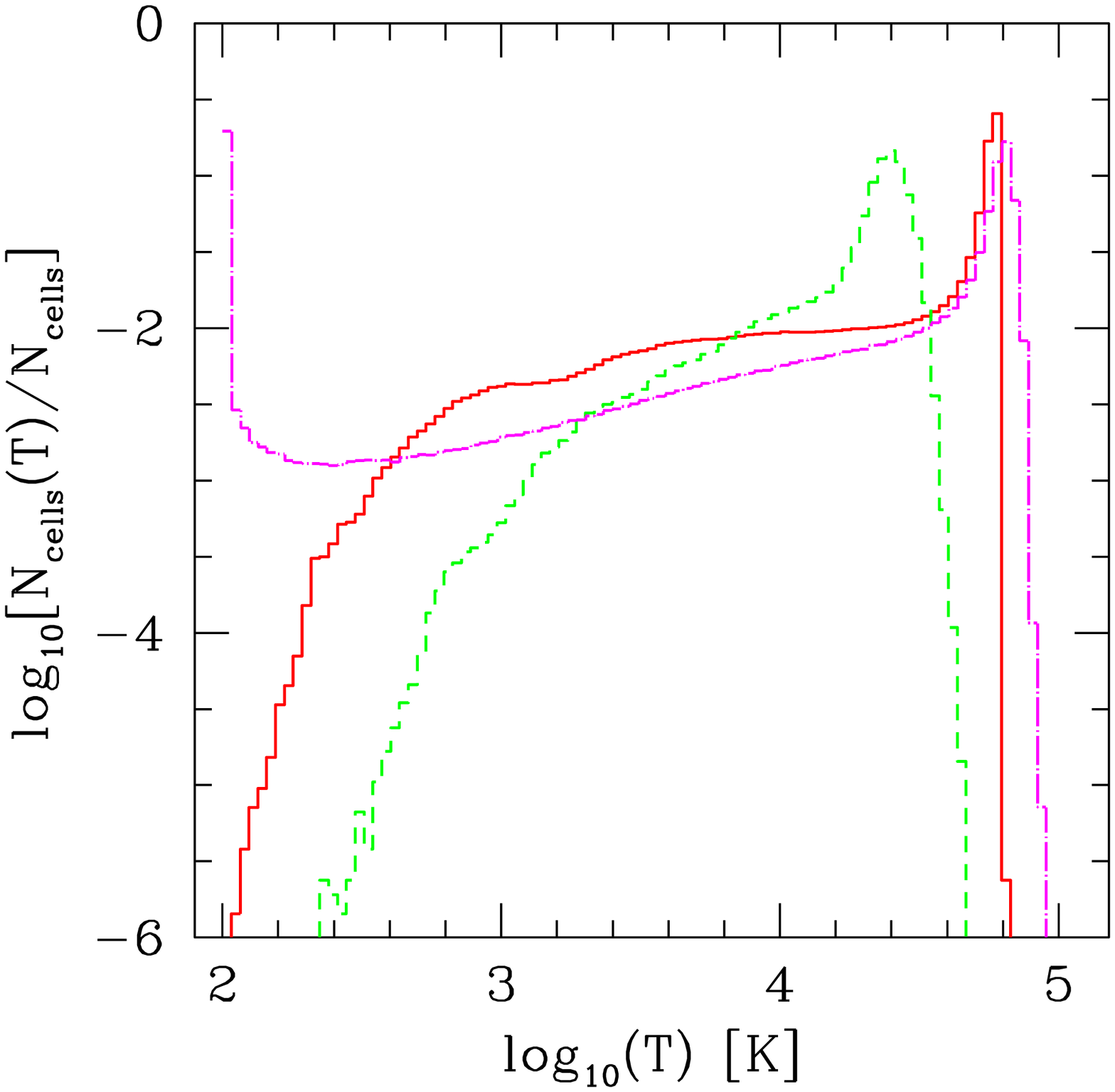}
  \includegraphics[width=2.2in]{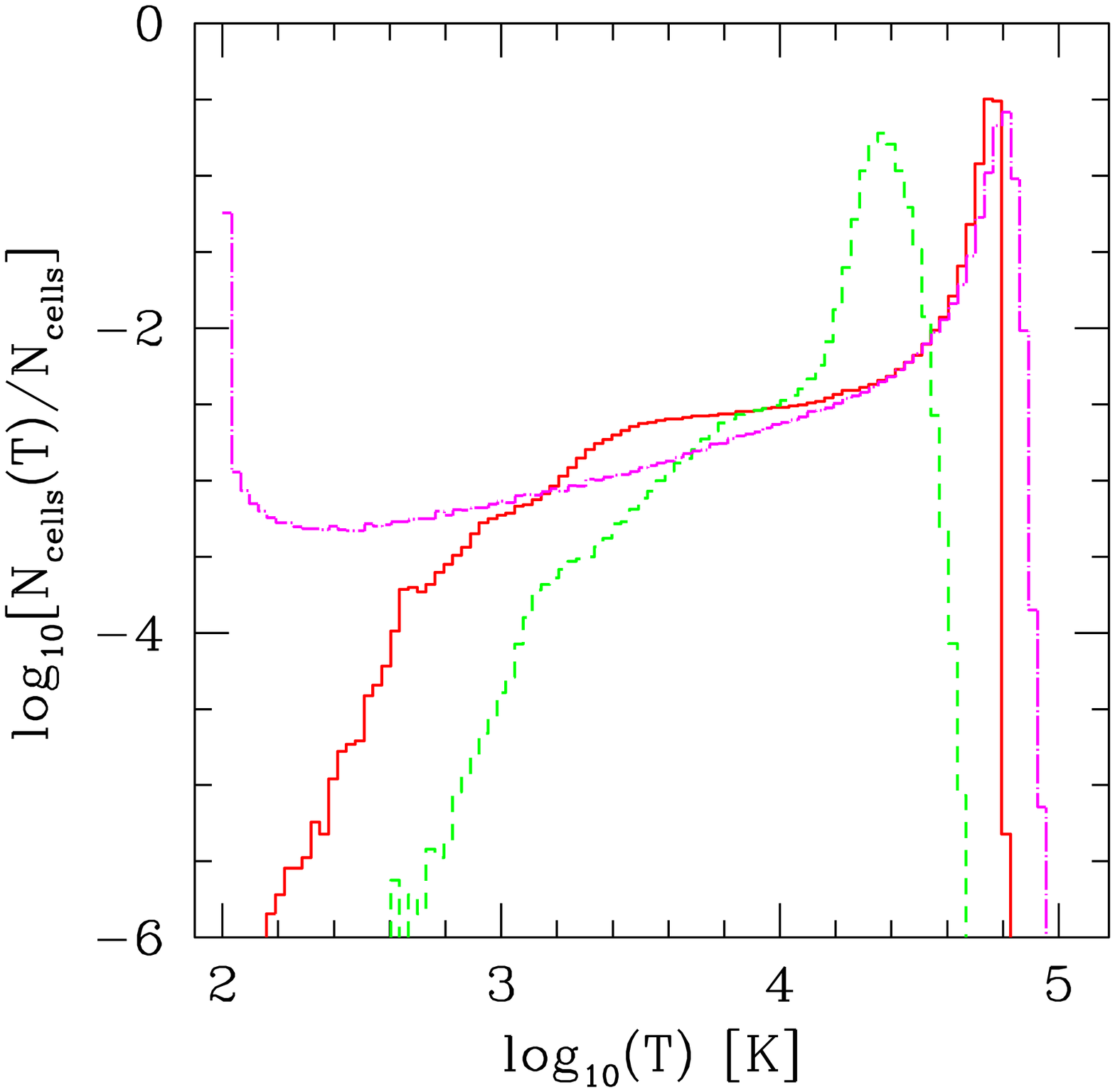}
\caption{Test 4  (reionization of a cosmological density field): Histograms of the temperature  at times $t=0.05$, 
  0.2 and 0.4 Myrs  for a black-body source spectrum with $T_{\rm eff}=10^5$ K.
\label{T4_hist_T_fig}}
\end{figure*}

\begin{figure*}
  \includegraphics[width=2.2in]{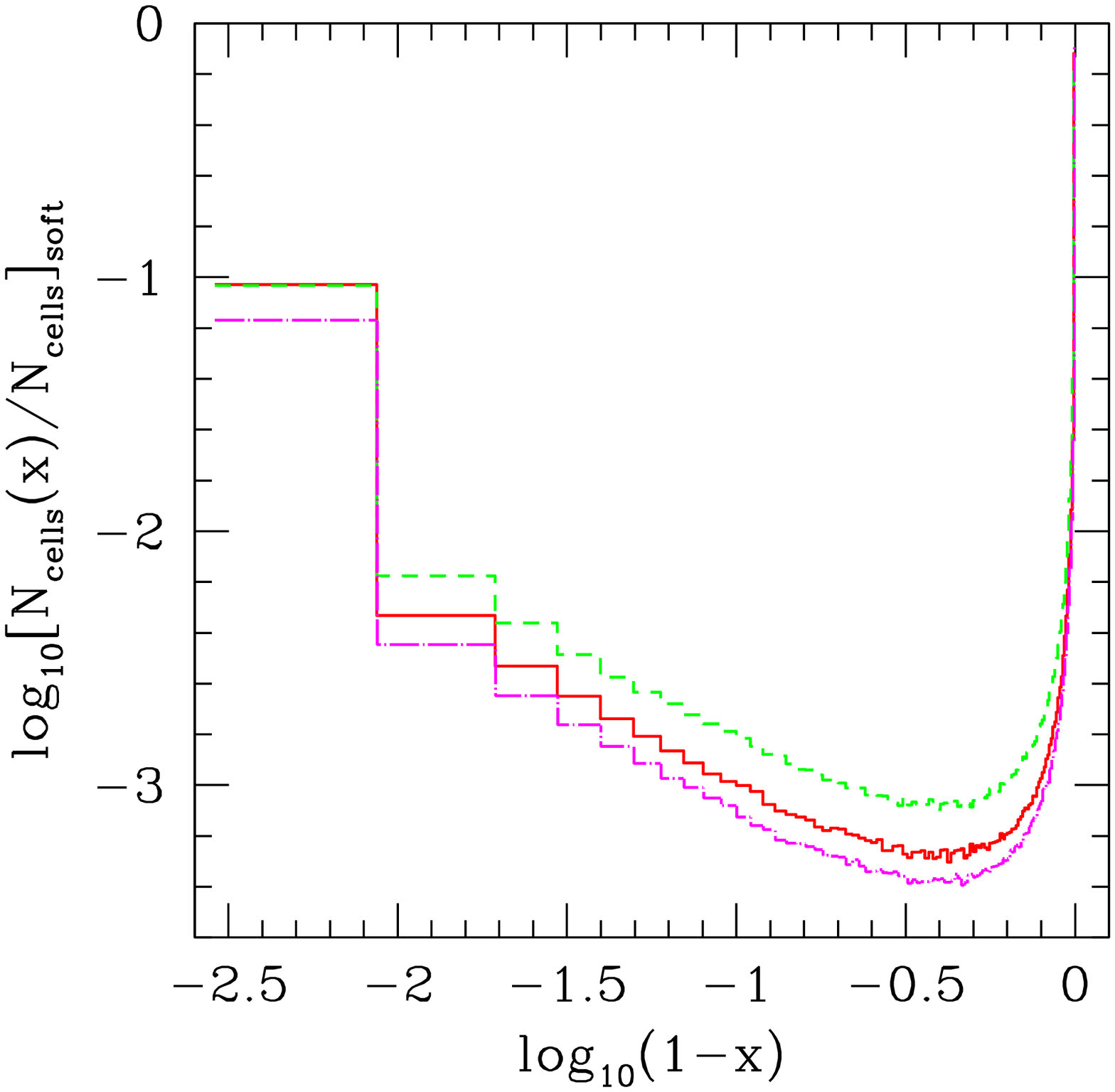}
  \includegraphics[width=2.2in]{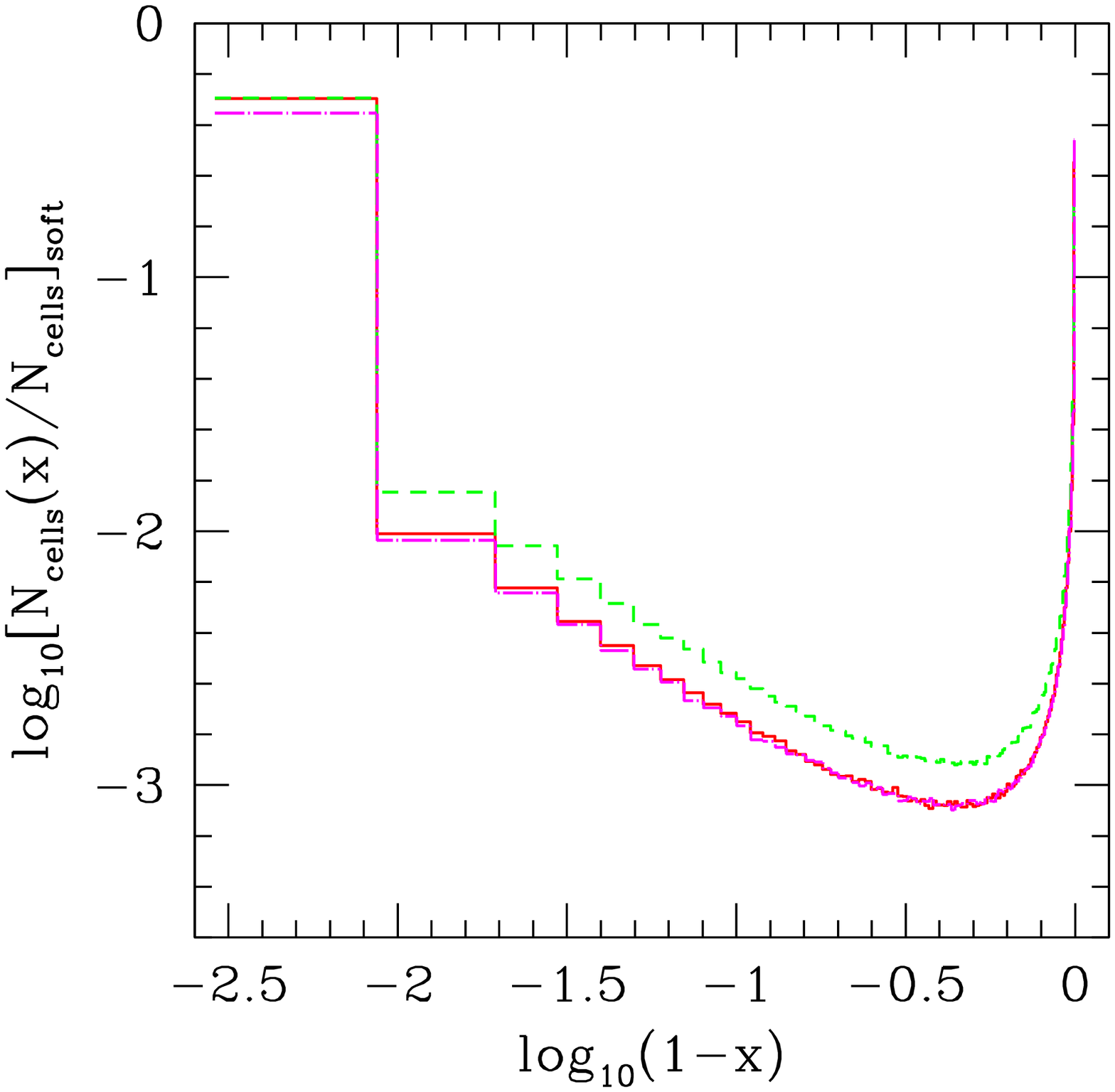}
  \includegraphics[width=2.2in]{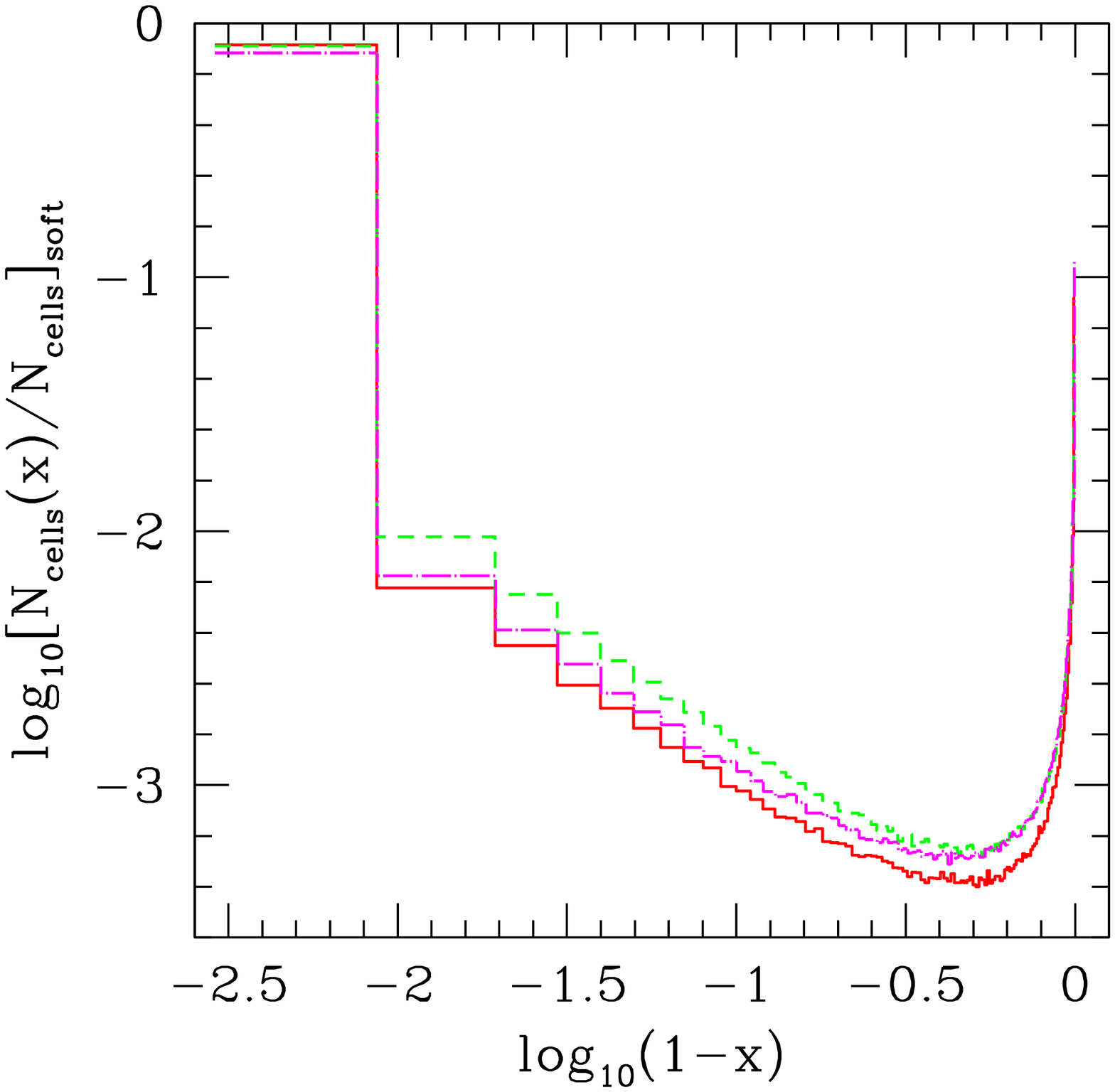}
\caption{Test 4  (reionization of a cosmological density field): Histograms of
  the ionized fraction at times $t=0.05$,  
  0.2 and 0.4 Myrs  for a black-body source spectrum with 
$T_{\rm eff}=3\times10^4$ K.
\label{T4_hist_H_soft_fig}}
  \includegraphics[width=2.2in]{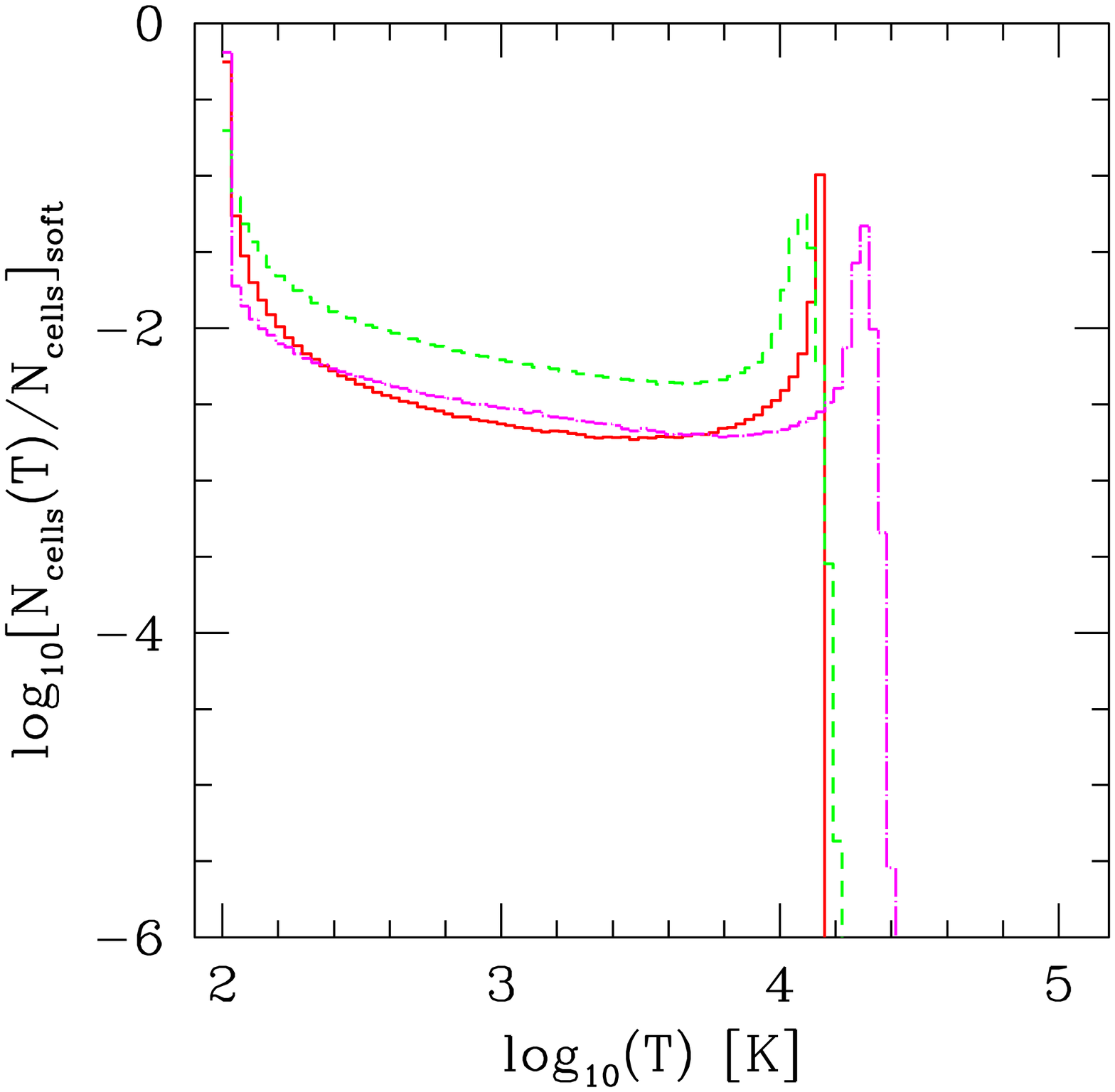}
  \includegraphics[width=2.2in]{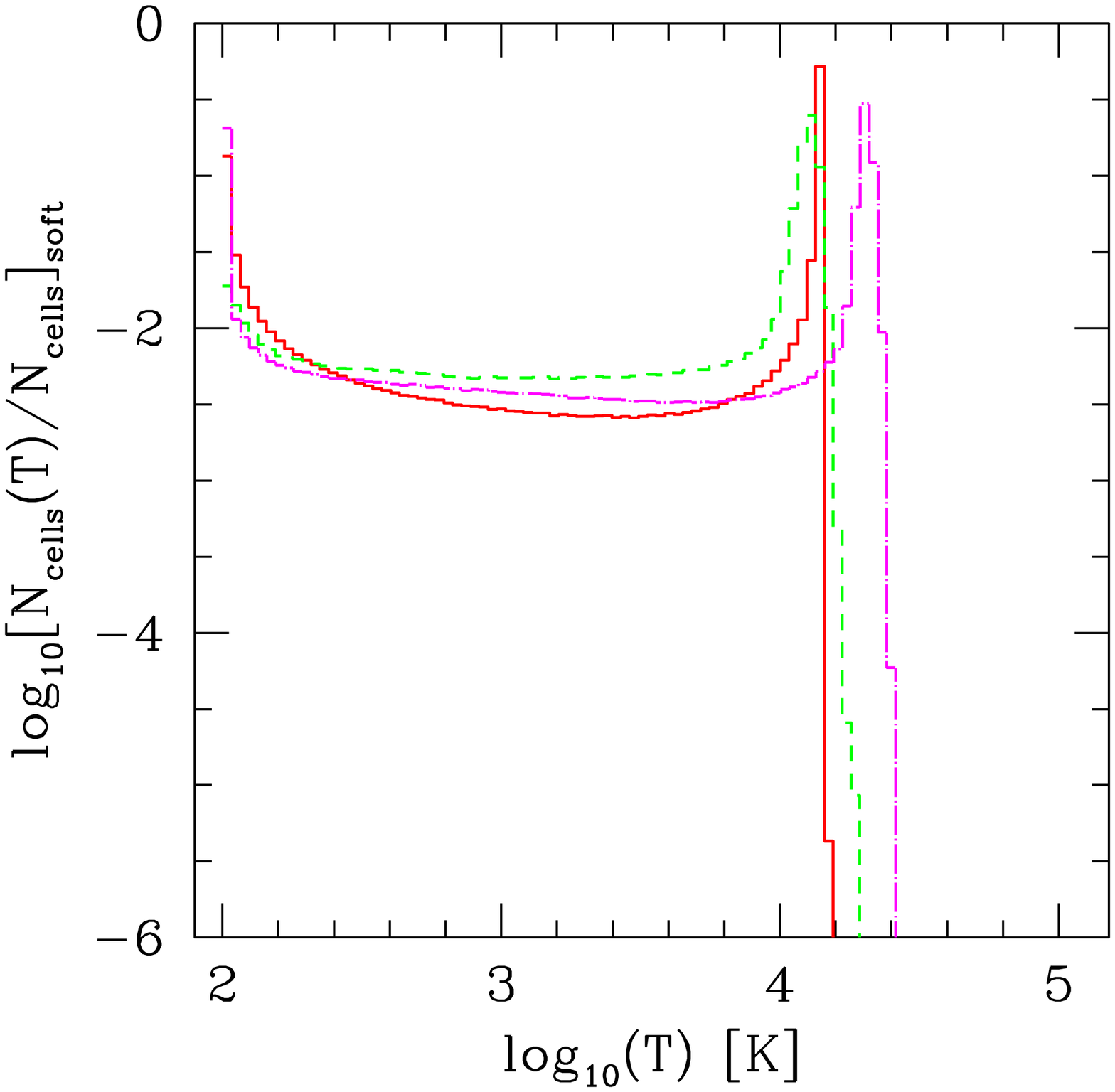}
  \includegraphics[width=2.2in]{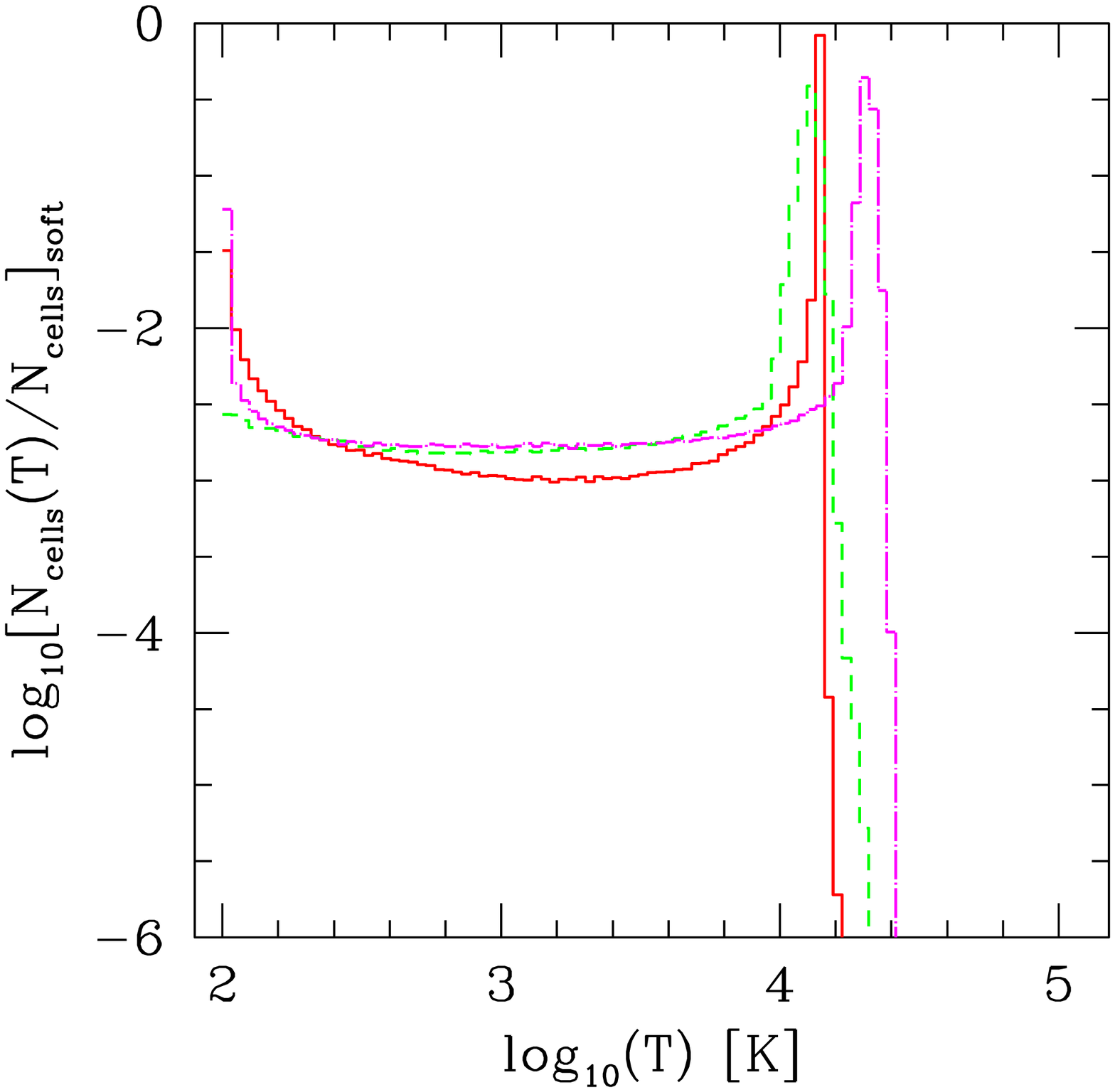}
\caption{Test 4  (reionization of a cosmological density field): Histograms of
  the temperature  at times $t=0.05$,  
  0.2 and 0.4 Myrs  for a black-body source spectrum with 
 $T_{\rm eff}=3\times10^4$ K.
\label{T4_hist_T_soft_fig}}
\end{figure*}

In Figure~\ref{T4_images1_HI_fig} we show slices of the H~I fraction cut 
through the simulation box at coordinate $z=z_{\rm box}/2$ and time 
$t=0.05$~Myr, and in Figure~\ref{T4_images1_T_fig} we show the corresponding 
temperature distributions the same time. SimpleX does not appear in the
temperature maps since currently the code does not follow the temperature
evolution self-consistently, but instead assumes a constant temperature. 
In Figures~\ref{T4_images3_HI_fig} and~\ref{T4_images3_T_fig} we show the same as
above but at time $t=0.2$~Myr. Some discrepancies are already evident from a 
visual inspection, which shows somewhat different morphologies, but still 
general agreement.  
In Figure~\ref{T4_x_evol_fig} (left panel) we present the temporal evolution of 
the volume- (thin lines) and mass- (thick lines) weighted ionized fractions. 
While CRASH and FTTE find comparable ionized fractions,
$C^2$-Ray and SimpleX produce slightly higher and lower values, respectively. 
The lower value in SimpleX is obtained as a consequence of the temperature
being fixed to $10^4$~K, which is a little lower than that obtained by the
other codes, resulting in a higher recombination rate.

The finer sampling in CRASH of the high energy tail of the spectrum allows 
a higher resolution of its hardening. As the total energy is distributed 
differently from the other codes (more energy is in the hard photons), 
this results in less ionization/heating closer to the sources and more 
ionization/heating further away and a lower mean ionized fraction. 

In Figures~\ref{T4_hist_H_fig} and~\ref{T4_hist_T_fig} we show histograms 
of the ionized fraction and the temperature at times $t=0.05$, 0.2 and 
0.4~Myr are shown. While $C^2$-Ray and FTTE agree quite well, especially 
at later times, CRASH, as explained above, produces a thicker I-front
transitions. The thickness of the I-front as found by SimpleX appears to 
oscillate with time, initially starting thick, then becoming thinner, and
thick again towards the end of the simulation.  This might have been caused 
by the interpolation of its unstructured grid to the regular grid required for
the results. In terms of temperature, CRASH finds a systematically lower peak
($\sim 1.6 \times 10^4$~K compared to $\sim 6.3 \times 10^4$~K for $C^2$-Ray
and FTTE, which agree well there) inside the ionized regions and a higher 
value in the lower density regions. This is once again due to the spectrum
hardening effects discussed above. The peak produced by FTTE at very low 
temperatures arises because of the lack of spectrum hardening and sharp
I-fronts consistently produced by this code, in which case no photons 
propagate ahead of the I-front and thus the gas away from the sources 
remains cold and neutral.

To understand whether the differences discussed above can be in fact 
attributed to photon hardening and the different treatment of the high 
energy tail of the spectrum, we have repeated the same simulations with 
a softer black-body spectrum with effective temperature 
$T_{\rm eff}=3 \times 10^4$~K, in which case the spectrum hardening and
I-front spreading should be minimized. We note that this part was not done by
the SimpleX code since currently it does not treat different spectra. Our 
results are shown in Figures~\ref{T4_x_evol_fig} (right panel), 
~\ref{T4_hist_H_soft_fig} and~\ref{T4_hist_T_soft_fig}. The averaged ionized 
fraction produced by $C^2$-Ray and CRASH now have a much better agreement,
as far fewer hard photons are present. However, FTTE still produces a lower 
value, although by only $\sim5$\%. CRASH still obtains a somewhat thicker 
ionizing front, due to the inherent adopted method, but the agreement is now 
better, especially at later times, as high energy photons are not present in
this case. This is even more evident from an analysis of the temperature 
behaviour, where the agreement between $C^2$-Ray and CRASH is now very good at
all times, while FTTE consistently finds higher temperatures inside the
ionized regions. These higher temperatures seem to self-contradict the lower
ionized fractions found by this code, while the recombination rate used by
this code is consistent with the others, indicating a possible (modest) problem 
with photon conservation. 

\section{Summary and Conclusions}
We have presented a detailed comparison of a large set of cosmological 
radiative transfer methods on several common tests. The participating codes 
represent the full variety of existing methods, multiple ray-tracing and one
moment method, which solve the radiative transfer on regular, adaptive or 
unstructured grids, even with no grid at all, but instead using particles 
to represent the density field. The comparison is a collaborative project 
involving most of the cosmological RT community. The results from this comparison will be 
publicly available for testing of future codes during their development. 

We began by comparing our basic physics, like chemistry, cooling rates 
and photoionization cross-sections, which came from a variety of sources, and 
evaluated the effects these have on the propagation of I-fronts. We concluded
that even quite approximate rates generally result in relatively modest 
divergences in the results. The discrepancies were limited to $\sim5-6\%$ in 
I-front position but were somewhat larger in velocity and the internal
structure of the H~II region (but never exceeding 20-40\%, and usually
much smaller).

Then, we turned to some simple, but instructive and cosmologically-interesting
problems which tested the different aspects encountered in realistic applications.
In summary, the results showed that at fixed temperature and for monochromatic
ionizing spectrum all of these methods track I-fronts well, to within a few 
per cent accuracy. There are some differences in the inherent thickness (due 
to finite mean free path) and the internal structure of the I-fronts, related 
to some intrinsic diffusivity of some of the methods.   

Somewhat greater differences emerge when the temperature is allowed
to vary, although even in this case the agreement is very good, typically
within 10-20\% in the ionized fraction. Other variations between the
results are due to the variety of multi-frequency radiative transfer methods
and consequent spectrum hardening and pre-heating ahead of the
I-fronts. Some of the algorithms consistently find quite sharp I-fronts, with 
little spectrum hardening and pre-heating, while the majority of the codes
handle these features more precisely and in reasonable agreement among
themselves. 

We conclude that the various approximations employed by the different methods
perform quite well and generally produce consistent and reliable results. 
There are, however, certain differences between the available methods and 
the most appropriate method for any particular problem would vary. 
Therefore, the code employed should be chosen with care depending on the 
specific questions whose answer is sought. For example, for problems in which the
I-front structure is not resolved (due to coarse simulation resolution 
compared to the characteristic front width), most methods considered here 
should perform well and the main criterion should be their numerical
efficiency. On the other hand, in problems where the I-front width and 
spectral hardening matters significantly, attention should be paid to the 
accurate multi-frequency treatment and the energy equation.

\section*{Acknowledgments} 
We are very grateful to the Canadian Center for Theoretical Astrophysics
(CITA) and the Lorentz Center at Leiden University for their hospitality
to the two workshops in 2005 which made this project possible, and to NSERC
for funding support. 
This work was partially supported by NASA Astrophysical Theory Program grants
NAG5-10825 and NNG04GI77G to PRS. GM acknowledges support from the Royal 
Netherlands Academy of Art and Sciences. MAA is grateful for the support of 
a DOE Computational Science Graduate Fellowship.
The software used in this work was in part developed by the DOE-supported 
ASC/Alliance Center for Astrophysical Thermonuclear Flashes at the University 
of Chicago. DW was funded in part by the U.S.\ Dept.\ of Energy through its 
contract W-7405-ENG-36 with Los Alamos National Laboratory.

\bibliographystyle{mn} \bibliography{../refs}

\end{document}